\pdfoutput=1

\documentclass{aa}   

\usepackage{graphicx,natbib,url,twoopt}
\usepackage[varg]{txfonts}           
\usepackage{hyperref}   
\usepackage{url}            
\usepackage{pdfcomment}              
\usepackage{acronym} 
\usepackage{epstopdf}                
\usepackage{bm}
\usepackage{longtable}
\usepackage{caption}

\hypersetup{
  colorlinks=true,   
  urlcolor=blue,     
  linkcolor=red,     
}

\bibpunct{(}{)}{;}{a}{}{,}    

\makeatletter
\newcommand{\bibnote}[2]{\global\@namedef{#1note}{#2}}
\newcommand{\biblink}[2]{\global\@namedef{#1link}{#2}}
\makeatother

\makeatletter
 \newcommandtwoopt{\citeads}[3][][]{%
   \nonstopmode
   \href{http://adsabs.harvard.edu/abs/#3}%
        {\def\hyper@linkstart##1##2{}%
         \let\hyper@linkend\@empty\citealp[#1][#2]{#3}}
   \biblink{#3}{\href{http://adsabs.harvard.edu/abs/#3}{ADS}}%
   \errorstopmode}            
 \newcommandtwoopt{\citepads}[3][][]{%
   \nonstopmode
   \href{http://adsabs.harvard.edu/abs/#3}%
        {\def\hyper@linkstart##1##2{}%
         \let\hyper@linkend\@empty\citep[#1][#2]{#3}}
   \biblink{#3}{\href{http://adsabs.harvard.edu/abs/#3}{ADS}}%
   \errorstopmode}            
 \newcommandtwoopt{\citetads}[3][][]{%
   \nonstopmode
   \href{http://adsabs.harvard.edu/abs/#3}%
        {\def\hyper@linkstart##1##2{}%
         \let\hyper@linkend\@empty\citet[#1][#2]{#3}}
   \biblink{#3}{\href{http://adsabs.harvard.edu/abs/#3}{ADS}}%
   \errorstopmode}            
 \newcommandtwoopt{\citeyearads}[3][][]{%
   \nonstopmode
   \href{http://adsabs.harvard.edu/abs/#3}%
        {\def\hyper@linkstart##1##2{}%
         \let\hyper@linkend\@empty\citeyear[#1][#2]{#3}}
   \biblink{#3}{\href{http://adsabs.harvard.edu/abs/#3}{ADS}}%
   \errorstopmode}            
\makeatother

\newacro{ADS}{Astrophysics Data System}
\newacro{NLTE}{non-local thermodynamic equilibrium}
\newacro{NASA}{National Aeronautics and Space Administration}

\begin{document}


              

\title{Probing cosmic isotropy with a new X-ray galaxy cluster sample through the $L_{\text{X}}-T$ scaling relation}

\author{K. Migkas$^1$, G. Schellenberger$^2$, T. H. Reiprich$^1$, F. Pacaud$^1$, M. E. Ramos-Ceja$^1$ and L. Lovisari$^2$}
\institute{$^1$ Argelander-Institut f{\"u}r Astronomie, Universit{\"a}t Bonn, Auf dem H{\"u}gel 71, 53121 Bonn, Germany \\ $^2$ Center for Astrophysics | Harvard \& Smithsonian, 60 Garden Street, Cambridge, MA 02138, USA \\ \email{kmigkas@astro.uni-bonn.de}
}

\date{Received date} 

\abstract{The isotropy of the late Universe and consequently of the X-ray galaxy cluster scaling relations is an assumption greatly used in astronomy. However, within the last decade, many studies have reported deviations from isotropy when using various cosmological probes; a definitive conclusion has yet to be made. New, effective and independent methods to robustly test the cosmic isotropy are of crucial importance. In this work, we use such a method. Specifically, we investigate the directional behavior of the X-ray luminosity-temperature ($L_{\text{X}}-T$) relation of galaxy clusters.  A tight correlation is known to exist between the luminosity and temperature of the X-ray-emitting intracluster medium of galaxy clusters. While the measured luminosity depends on the underlying cosmology through the luminosity distance $D_{\text{L}}$, the temperature can be determined without any cosmological assumptions. By exploiting this property and the homogeneous sky coverage of X-ray galaxy cluster samples, one can effectively test the isotropy of cosmological parameters over the full extragalactic sky, which is perfectly mirrored in the behavior of the normalization $A$ of the $L_{\text{X}}-T$ relation. To do so, we used 313 homogeneously selected X-ray galaxy clusters from the Meta-Catalogue of X-ray detected Clusters of galaxies (MCXC). We thoroughly performed additional cleaning in the measured parameters and obtain core-excised temperature measurements for all of the 313 clusters. The behavior of the $L_{\text{X}}-T$ relation heavily depends on the direction of the sky, which is consistent with previous studies. Strong anisotropies are detected at a $\gtrsim 4\sigma$ confidence level toward the Galactic coordinates $(l,b)\sim (280^{\circ}, -20^{\circ})$, which is roughly consistent with the results of other probes, such as Supernovae Ia. Several effects that could potentially explain these strong anisotropies were examined. Such effects are, for example, the X-ray absorption treatment, the effect of galaxy groups and low redshift clusters, core metallicities, and apparent correlations with other cluster properties, but none is able to explain the obtained results. Analyzing $10^5$ bootstrap realizations confirms the large statistical significance of the anisotropic behavior of this sky region.
Interestingly, the two cluster samples previously used in the literature for this test appear to have a similar behavior throughout the sky, while being fully independent of each other and of our sample. Combining all three samples results in 842 different galaxy clusters with luminosity and temperature measurements. Performing a joint analysis, the final anisotropy is further intensified ($\sim 5\sigma$), toward $(l,b)\sim (303^{\circ}, -27^{\circ})$, which is in very good agreement with other cosmological probes. The maximum variation of $D_L$ seems to be $\sim 16\pm 3\%$ for different regions in the sky. This result demonstrates that X-ray studies that assume perfect isotropy in the properties of galaxy clusters and their scaling relations can produce strongly biased results whether the underlying reason is cosmological or related to X-rays. The identification of the exact nature of these anisotropies is therefore crucial for any statistical cluster physics or cosmology study.}

\keywords{cosmology: observations -- X-rays:galaxies:clusters -- (cosmology:) large-scale structure of Universe -- galaxies: clusters: general -- methods: statistical -- catalogs}

\titlerunning{Strong $L_{\text{X}}-T$ anisotropies from a new X-ray galaxy cluster sample}
\authorrunning{K. Migkas et al. }

\maketitle

\section{Introduction} \label{intro}


The isotropy of the Universe on sufficiently large scales is a fundamental pillar of the standard model of cosmology. The most important consequence of isotropy is that the expansion rate of the Universe as well as the physical properties of all astronomical objects must be the same regardless of the direction in the sky. Due to the high significance of this hypothesis, it is necessary that it is robustly scrutinized and tested against different cosmological probes using the latest data samples. 

What was initially introduced as a repercussion of general relativity and the Friedmann-Lema\^{i}tre-Robertson-Walker (FLRW) metric was later supported by observations. The most crucial of them is arguably the cosmic microwave background (CMB) as observed by Cosmic Background Explorer \citep[\textit{COBE}, ][]{COBE}, the Wilkinson Microwave Anisotropy Probe \citep[\textit{WMAP}, ][]{bennett}, and the \textit{Planck} \citep{planck} telescopes.
CMB shows a remarkable isotropy at small angular scales (high multipoles), whilst some anisotropies are still present in lower multipoles. The most prominent one is the so-called CMB dipole which, if one assumes its purely kinematic origin, is caused by the Doppler shift due to the motion of our Solar System with respect to the CMB rest frame. This indicates that the Solar System moves toward the Galactic coordinates $(l,b)\sim (264^{\circ}, +48^{\circ})$ with a peculiar velocity of $\sim 370$ km/s \citep{kogut,fixsen} with respect to the CMB rest frame. Another anisotropic feature present in the CMB is the dipole power asymmetry detected in both \textit{WMAP} and \textit{Planck} with a significance of $\sim 2-3.5\sigma$ toward $(l,b)\sim (230^{\circ}, -20^{\circ})$ \citep{eriksen,hanson,bennett11,akrami,planck13,planck}. Its nature still remains relatively unclear. The interpretation of the significance of these results differ between papers. Other potential challenges for the isotropy of the Universe found in the CMB is the parity asymmetry, $\sim 3\sigma$ toward $(l,b)=(264^{\circ}, -17^{\circ})$ and the unexpected quadrupole-octopole alignment and the existence of the Cold Spot at $(l,b)=(210^{\circ}, -57^{\circ})$ \citep{tegmark,vielva,kim,aluri,cai14,planck13,planck,schwarz16}.

The CMB is a great tool to study the behavior of the early Universe. However, it can be quite challenging to extract information about the directional behavior of the late Universe based on that probe. This problem intensifies when one considers that according to $\Lambda$CDM the late Universe is dominated by dark energy whose effects are not directly present in the CMB spectrum. Moreover, since the nature of dark energy is still completely unknown, one can only make assumptions about the isotropic (or not) behavior of dark energy. Consequently, it becomes clear that other cosmological probes, at much lower redshifts than the CMB, are needed in order to search for possible anisotropies in the late Universe. There are indeed many probes that have been used for such tests.

For instance, Type Ia Supernovae (SNIa) have been extensively used to test the isotropy of the Hubble expansion with many results reporting no significant deviation from the null hypothesis \citep{lin,andrade,sun,wang}. Other studies however, claim to detect mild-significance anisotropies in the SNIa samples that more often than not approximately match the direction of the CMB dipole \citep{schwarz,antoniou,colin2011,mariano,kalus,appleby-john,bengaly15,javanm15,migkas,colin}. Generally, the reported results from SNIa strongly depend on the used catalog. Moreover, the robustness of SNIa as probes for testing the isotropy of the Universe based on the current status of the relative surveys has been recently challenged \citep{colin,jimenez,rameez2}. This is mainly because of the highly inhomogeneous spatial distribution of the data (most SNIa in the latest catalogs lie close to the CMB dipole direction), their sensitivity to the applied kinematic flow models which readjust their measured heliocentric redshifts, as well as the assumptions that go into the calibration of their light curves. 

Other probes that have been used to pinpoint possible anisotropies or inconsistencies with the $\Lambda$CDM model are the X-ray background \citep{shafer, plionis99}, the distribution of optical \citep{javanm17,sarkar} and infrared galaxies \citep{yoon,rameez}, the distribution of distant radio sources \citep{condon,blake,singal,rubart,tiwari,bengaly18,colin2,bengaly19}, GRBs \citep{ripa,andrade19}, peculiar velocities of galaxy clusters \citep{kashl08,kashl10,watkins,kash11,atrio-bar} and of SNIa \citep{appleby-john} etc.  While some of them find no statistically significant challenges for the null hypothesis of isotropy, others provide results which are unlikely to occur within the standard cosmological model framework. The use of standard sirens for such tests in the future has also been proposed \citep[e.g.,][]{cai}.

Since the outcome of the search for a preferred cosmological direction remains ambiguous, new and independent methods for such tests should be introduced and applied to the latest data samples. In \citet{Migkas18} (hereafter M18), the use of the directional behavior of the galaxy cluster X-ray luminosity-temperature relation is described as a cosmological probe. It is well-known that galaxy clusters are the most massive gravitationally bound systems in the universe, strongly emitting X-ray photons due to the large amounts of hot gas they contain ($\sim 10$\% of their total mass) in their intra-cluster medium (ICM). Their physical quantities follow tight scaling relations, for which \citet{kaiser} provided mathematical expressions. Specifically, the correlation between the X-ray luminosity ($L_{\text{X}}$) and the ICM gas temperature ($T$) of galaxy clusters is of particular interest since it can be used to trace the isotropy of the Universe, which is a new concept for such cosmological studies. The general properties of the $L_{\text{X}}-T$ scaling relation have been extensively scrutinized in the past by several authors \citep[e.g.,][]{vikhl02,pacaud07,pratt,mittal,reichert,hilton,maughan,bharad,lovisari,giles,zou,Migkas18,xcs2}.

In a nutshell, the gas temperature, the flux and the redshift of a galaxy cluster do not require any cosmological assumptions in order to be measured \footnote{Only indirectly when the selection of the cluster relative radius within which the used spectra are extracted is based on a cosmological distance, such as the X-ray luminosity-mass scaling relation ($L_{\text{X}}-M$) and the conversion of the radius from Mpc to arcmin, where the luminosity and angular diameter distances enter. Nevertheless, even in these cases, the dependence happens to be very weak.}. Using the flux and the redshift together with the luminosity distance, through which the cosmological parameters come into play, one can obtain the luminosity of a cluster. The luminosity however can also be predicted (within an uncertainty range) based on the cluster gas temperature. Hence, adjusting the cosmological parameters, one can make the two luminosity estimations match. This can be repeatedly applied to different sky patches in order to test the consistency of the obtained values as a function of the direction. The full detailed physical motivation behind this is discussed in M18. There, it is shown that the directional behavior of the normalization of the $L_{\text{X}}-T$ relation strictly follows the directional behavior of the cosmological parameter values. This newly introduced method to test the Cosmological Principle (CP) could potentially prove very effective due to the very homogeneous sky coverage of many galaxy cluster samples (in contrast to SNIa samples), the plethora of available data as well as large upcoming surveys such as \textit{eROSITA} \citep{erosita16} which will allows us to measure thousands of cluster temperatures homogeneously \citep{borm}. For studying the isotropy of the Universe with the future surveys, it is of crucial importance that any existing systematic biases that could potentially affect the $L_{\text{X}}$ and $T$ measurements of galaxy clusters would have been identified and taken into account by then.

In this paper, we construct and use a new galaxy cluster sample in order to identify regions that share a significantly different $L_{\text{X}}-T$ relation compared to others. This could lead to pinpointing an anisotropy in the Hubble expansion or discover previously unknown factors which could potentially affect X-ray measurements of any kind. Except for the high quality observations and measurements, another advantage of our sample is the small overlap with the XCS-DR1 \citep{mehrt} and ACC \citep{horner} samples used in M18. There are only three common clusters between our sample and XCS-DR1 ($<1\%$) and only a 30\% overlap with ACC. Since we reanalyze the XCS-DR1 and ACC sample in this paper as well, using the same methods we use for our sample in order to perform a consistent comparison, all the common clusters between the different catalogs are excluded from XCS-DR1 and ACC. Our cluster sample does not suffer from any strong archival biases in contrast to ACC. Throughout this paper we use a $\Lambda$CDM cosmology with $H_0=70\  \text{km}\  \text{s}^{-1} \text{Mpc}^{-1}$, $\Omega_{\text{m}}=0.3$ and $\Omega_{\Lambda}=0.7$ unless stated otherwise. 

The paper is organized as follows: In Sect. \ref{eehifl} we describe the construction of the sample and how we derive the properties of the clusters. In Sect. \ref{spec_fit} we explain the steps we follow for the data reduction and the spectral analysis of the observations. In Sect. \ref{analysis_method} we present the modeling of the $L_{\text{X}}-T$ relation together with the parameter fitting procedure. We also explain how we identify possible spatial anisotropies and assign their statistical significance. In Sect. \ref{mainresults} we present the first results, including the overall $L_{\text{X}}-T$ results and the 1-dimensional (1D) and 2-dimensional (2D) anisotropies of our sample. In Sect. \ref{possible_causes} we investigate several X-ray and cluster-related causes that could possibly produce the observed anisotropic signal in our sample. In Sect. \ref{cosmology} we examine the case where the anisotropies in our sample have a cosmological origin instead, assuming there are no systematics associated with X-ray photons or cluster properties. In Sect. \ref{joint_an} we combine our results with those obtained from ACC and XCS-DR1. We express these joint-analysis anisotropies in cosmological terms. In Sects. \ref{discuss} and \ref{conclusions} we discuss our findings and their implications, compare with other studies and summarize.



\section{Sample selection} \label{eehifl}

The cluster sample used in this work is a homogeneously selected one based on the Meta-Catalogue of X-ray detected Clusters of galaxies \citep[MCXC,][]{mcxc}. Initially it consisted of the 387 galaxy clusters above an unabsorbed flux cut of $f_{0.1-2.4\ \text{keV}}\geq 5 \times 10^{-12}$ ergs/s/cm$^2$ excluding the Galactic plane ($|b|\leq 20^{\circ}$), the Magellanic clouds and the Virgo cluster area. The flux for every MCXC cluster is found based on the given X-ray luminosity and redshift, combined with a reversed K-correction (the necessary temperature input was found by the $L_{\text{X}}-T$ relation of \citet{reichert}. The MCXC luminosities (and therefore the calculated fluxes on which our sample is based) were corrected for absorption based on HI measurements (see Sects. \ref{nh_sec} and \ref{lum_estim}).

The parent catalogs of the clusters we use are The ROSAT extended Brightest Cluster Sample \citep[eBCS,][]{ebcs}, The Northern ROSAT All-Sky (NORAS) Galaxy Cluster Survey \citep{noras} and the ROSAT-ESO Flux-Limited X-Ray (REFLEX) Galaxy Cluster Survey Catalog \citep{reflex}. They are all based on the ROSAT All-Sky Survey \citep[RASS,][]{rass}. 

Another selection criterion was for the clusters to have good quality \textit{Chandra} \citep{chandra} or \textit{XMM-Newton} \citep{xmm} public observations (as of July 2019). This criterion is satisfied for 331 clusters. The rest 56 clusters for which such observations were not available have a sparse sky distribution and similar average properties with the 331 clusters (as discussed later in the paper) and therefore their inclusion is not expected to alter the results.

We reduced these 331 available observations, analyzed them and extracted cluster properties as described below. This sample includes a large fraction ($\sim 85\%$) of the eeHIFLUGCS  \citep[extremely expanded HIghest X-ray FLUx Galaxy Cluster Sample][Pacaud et al. in prep.]{eeHIF} sample clusters. eeHIFLUGCS is a complete, purely X-ray flux-limited sample with similar selection criteria.

\begin{figure*}[hbtp]
\centering
               \includegraphics[width=0.95\textwidth, height=7.5cm]{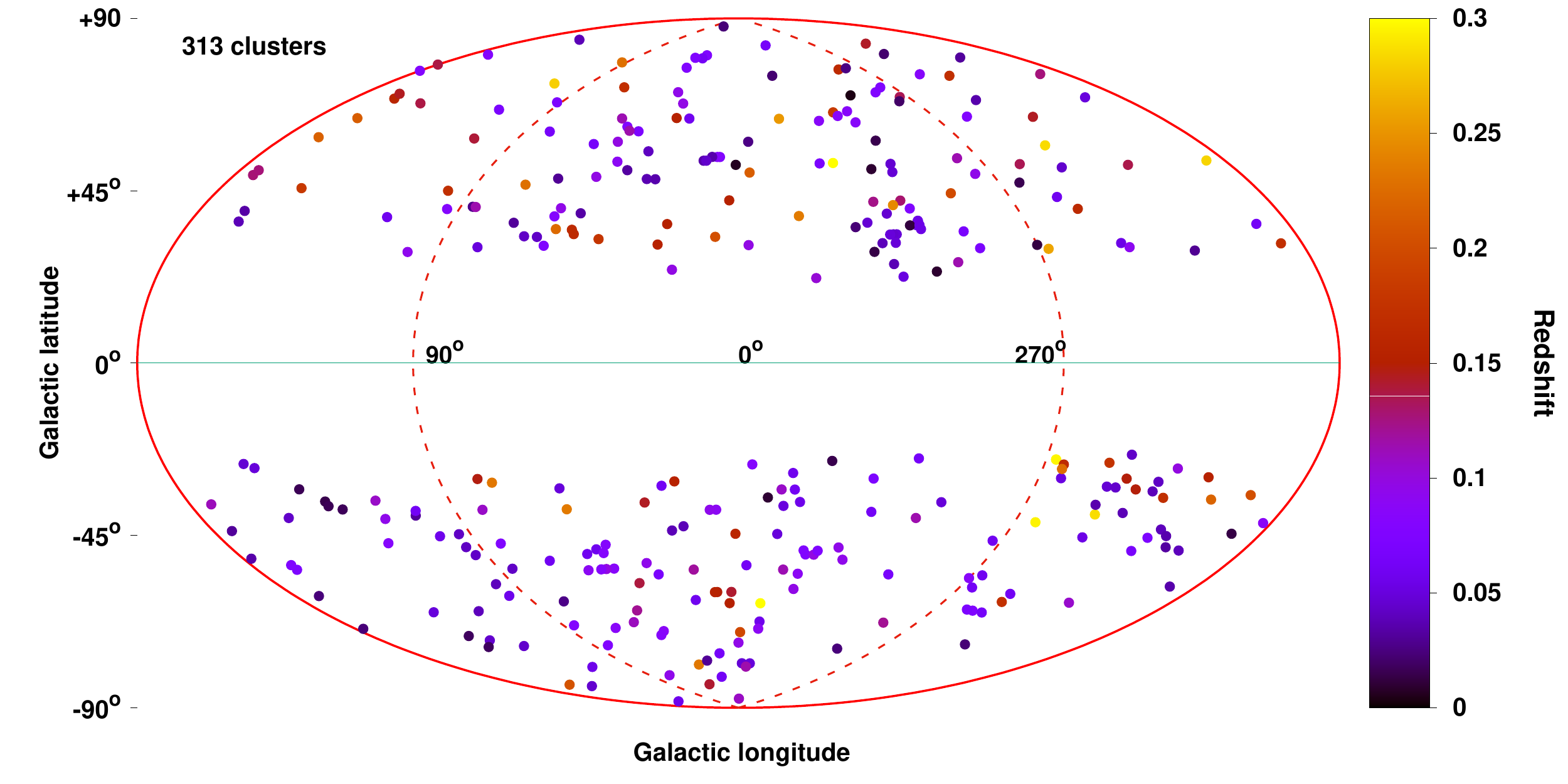}
               \caption{Sky distribution of the 313 clusters in Galactic coordinates, with the colorbar indicating their redshift. There are two clusters at $z>0.3$ but the scale is set in such way so the color contrast is optimal.}
        \label{fig1}
\end{figure*}

The final sample we use for this paper consists of 313 galaxy clusters. The other 18 clusters are not used because of the following reasons. Firstly, we excluded 11 clusters from our analysis that we identified as apparent multiple systems (out of a total of 15). This is due to the fact that clusters in rich environments tend to be systematically fainter than single clusters (M18 and references therein), thus biasing the final results. Moreover, some of these seemingly multiple systems are located at different redshifts but projected in our line of sight as real double and triple systems \citep[see][hereafter R19]{miriam}. When these different components are accounted as one system in MCXC the flux of the "single cluster" is overestimated. As a result, some of these systems falsely overcome the selected flux limit while none of their true individual components has the necessary flux to be included in our catalog. On the other hand, there are cases where one of the individual extended components has enough flux to be kept in our sample and at the same time it is located at a different redshift than the other components of the system (it does not belong to the same rich environment). In this case, these extended sources were kept in the catalog and their $L_{\text{X}}$ values were adjusted correspondingly (see Sect. \ref{lum_estim}) while the other component(s) were excluded. Since we use \textit{Chandra} and \textit{XMM-Newton} observations while the MCXC $L_{\text{X}}-T$ values come from \textit{ROSAT} observations, some minor inconsistencies between these values and our own measurements are expected. In order to account for this, we consider it as an extra source of uncertainty and adjust the confidence levels of the final cluster fluxes and luminosities accordingly as described later in the paper.

Furthermore, we identify nine clusters to be strongly contaminated by point sources, likely Active Galactic Nuclei (AGN). We confirm that by the absence of significant extended emission around the suspected point sources and by fitting a power-law (constraining the power index and the normalization) and an \textit{apec} model to the spectra of the bright part of the point source using XSPEC \citep{xspec}. If the \textit{pow} model returns a better fit than the \textit{apec} model we mark the source as an AGN. Moreover, we search the literature for known stars and AGNs at these positions. For three of these nine clusters (\object{A2055}, \object{A3574E}, \object{RXCJ1840.6-7709}) the point sources are located close to the (bright) cores of the clusters and cannot be deblended. Thus, we chose to exclude these clusters since their MCXC $L_{\text{X}}$ values would be overestimated and would add extra bias to our analysis. For another cluster (\object{A1735}) there is a strong AGN source and a galaxy cluster with extended emission with an angular separation of $\sim 7'$. This system has been identified as one cluster in the MCXC catalog centered at the AGN position, which has the higher contribution to the X-ray flux. Therefore, this system was also excluded since the single extended emission component does not surpass the necessary flux limit. For the rest five clusters, using the \textit{Chandra} and \textit{XMM-Newton} images the point sources are easily distinguishable from the cluster emission while the MCXC objects are centered close to the extended emission centers. For these five systems we calculate the flux of the point source using its spectra from one of the two aforementioned telescopes and a \textit{pow} model, and subtract it from the MCXC flux in order to see if the extended emission alone overcomes the flux limit. This procedure results in the exclusion of three clusters (\object{A0750}, \object{A0901}, \object{A2351}), while the other two (\object{A3392}, \object{S0112}) stay above the desired flux limit and are considered in our analysis after appropriately decreasing their MCXC luminosity values (Sect. \ref{lum_estim}). 

Since $\sim 50$\% of the clusters included in our sample have been observed by both \textit{Chandra} and \textit{XMM-Newton}, we decide to analyze these common clusters with the former. This is due to the fact that \textit{Chandra} data are generally less flared than \textit{XMM-Newton} data. As a result, 237 clusters are analyzed using \textit{Chandra} observations while 76 clusters are processed using \textit{XMM-Newton} observations. For both telescopes, we extract and fit the spectra within the energy range of  $0.7$ to $7$ keV. The cross calibration of the two satellites is discussed in Sect. \ref{temperat}.
Using \textit{Chandra} data for $\sim 75$\% of our sample offers another advantage as well. As mentioned before, in M18 two samples are used: ACC which consists only of ASCA observations, and XCS-DR1 which consists only of \textit{XMM-Newton} observations. Subsequently, mostly using a third independent telescope to built our sample and study the anisotropy of the $L_{\text{X}}-T$ eliminates any systematics that might occur in the results because of telescope-specific reasons.

Consequently, the sample with which this analysis is performed consists of 313 single galaxy clusters. For these clusters we have self-consistently measured their gas temperatures $T$ and their uncertainties, as well as their metallicities $Z$ and their X-ray redshifts $z$. Furthermore, we know their optical spectroscopic $z$, their fluxes $f_X$ and their luminosities $L_{\text{X}}$ within the 0.1-2.4 keV energy range together with their uncertainties, their Galactic $(l,b)$ and Equatorial ($RA,\ Dec$) coordinates and the atomic and molecular hydrogen column density $N_{\text{Htot}}$ in their direction. The exact information for every parameter and where it comes from is described in the following subsections. Their spatial distribution together with the redshift value used for each cluster can be seen in Fig. \ref{fig1}. The vast majority of these 313 galaxy clusters are included in the eeHIFLUGCS sample. 

\subsection{Redshift} \label{redshifts}

For 264 out of the 313 clusters the given MCXC redshifts are used. We have checked that all these clusters have at least seven galaxies with optical spectroscopic redshifts in the NASA/IPAC Extragalactic Database (NED) and that agree with the assigned MCXC redshift. The median number of galaxies per cluster is 52 for these cases. For seven other clusters we reassigned a redshift based on the already-existing optical spectroscopic data when the offset between the apparently correct redshift value and the MCXC redshift is $\Delta z \geq 0.007$, corresponding to $\sim 2000$ km/s. 

The remaining 42 clusters either do not have enough optical spectroscopic data in order to trust the given redshift or the distribution of the galaxy redshifts of the cluster is inconclusive. In that case, the redshift of the cluster was determined from the available X-ray data. For that, we extracted and fit the spectra within the $0-0.2\ R_{500}$\footnote{$R_{500}$=the radius within which the mean density of the cluster is 500 times greater than the critical density of the Universe} circle and within the $0.2-0.5\ R_{500}$. Using two \textit{apec} models (one for each cluster region) with the temperature and metallicity parameters free to vary for both, the redshift is also fit simultaneously but linked for the two regions (same $z$ for both regions). In Sect. \ref{spec_fit} the technical details of the spectral fitting process are discussed.

\begin{figure}[hbtp]
               \includegraphics[width=0.49\textwidth, height=6cm]{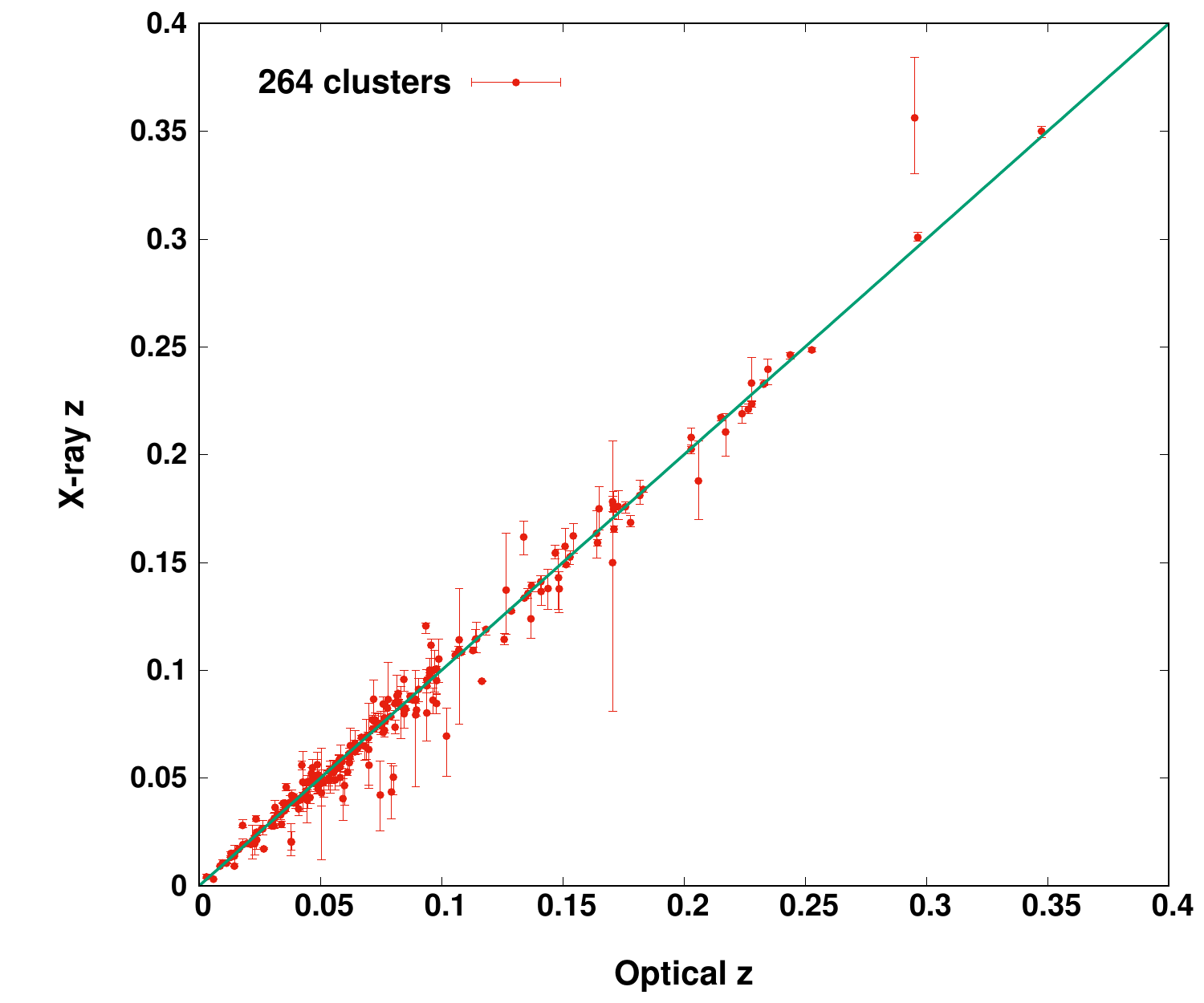}
               \includegraphics[width=0.49\textwidth, height=6cm]{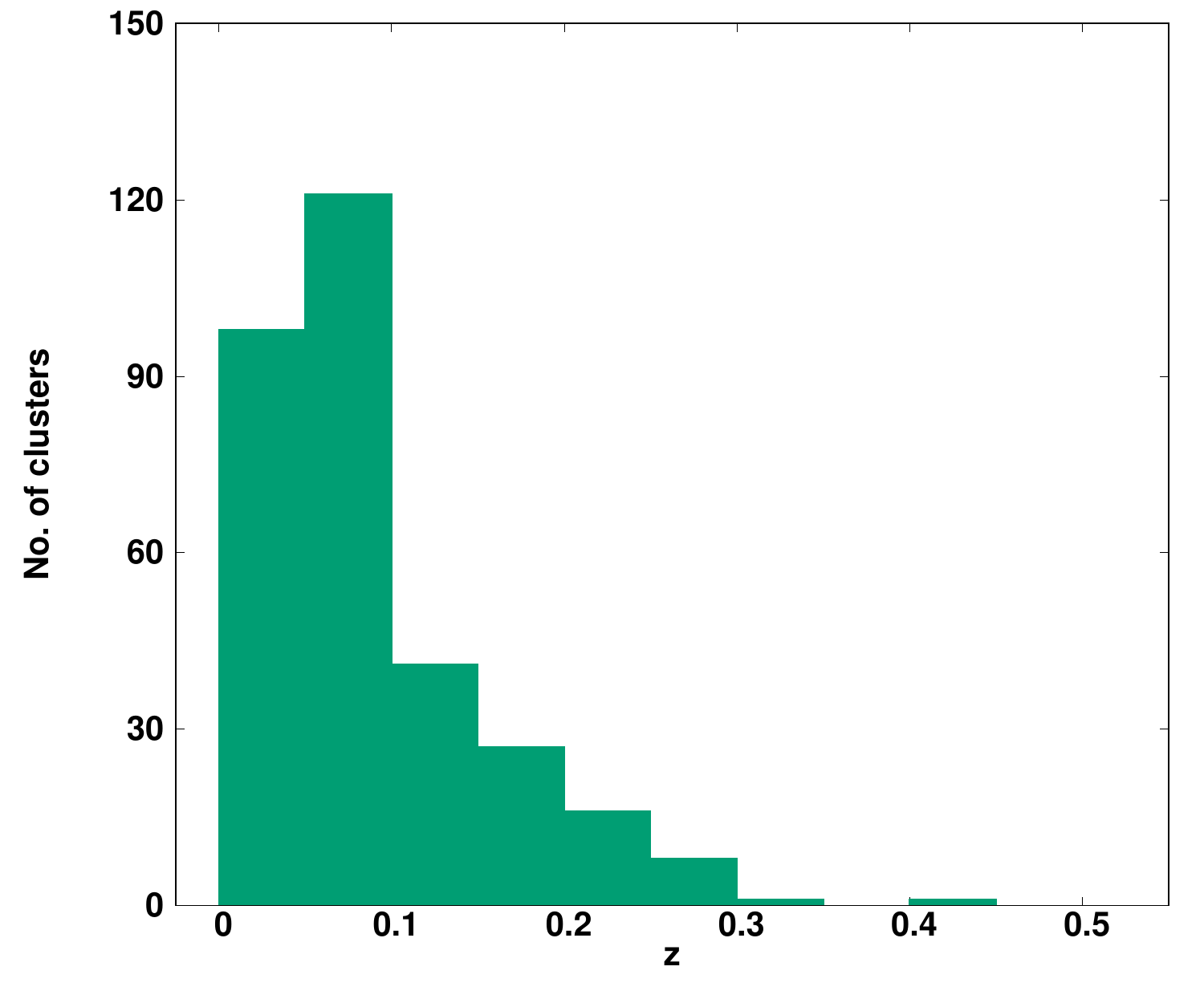}
               \caption{\textit{Top}: Comparison between the safe optical redshifts and the X-ray redshifts obtained from this analysis. \textit{Bottom}: Redshift distribution of the final sample.}
        \label{fig2}
\end{figure}

In order to make sure that the obtained X-ray redshifts are trustworthy, we also determined them for the 264 clusters with safe optical redshifts. The comparison of the optical and X-ray redshifts is displayed in the top panel of Fig. \ref{fig2}. By comparing the well-known optical redshifts with the X-ray ones, we see that there is only a very small intrinsic scatter of $\Delta z=0.0014$ ($\sim 420$ km/s) with the agreement being remarkable. It is noteworthy that only 10\% of the clusters have a deviation of more than $\Delta z/(1+z_{\text{opt}})\geq 0.01$ between the X-ray and the good quality optical redshift, whilst only 3\% deviate by more than $\Delta z/(1+z_{\text{opt}})\geq 0.02$. Therefore, using the X-ray determined redshifts for the 42 clusters without optically spectroscopic redshifts seems to introduce no bias. The final redshift distribution is shown in the bottom panel of Fig. \ref{fig2}. Moreover, in the Appendix is further shown that using or not the clusters with X-ray redshifts one derives very similar results. The redshift distribution of our sample covers the $z=(0.003-0.45)$ range while the median redshift of the sample is $z=0.075$ ($z=0.072$ for the excluded clusters). All the aforementioned redshifts are heliocentric. The clusters for which we changed their $z$ values compared to the ones from MCXC are displayed in Table \ref{tab2} with a star (*) next to their names.

\subsection{Hydrogen column density} \label{nh_sec}

The value of $N_{\text{H}}$ enters in both the determination of the $L_{\text{X}}$ (as done in the parent catalogs) and in the $T$ determination that is performed in this analysis. Hence, an inaccurate treatment of the input $N_{\text{H}}$ values could potentially bias both parameters mostly in the opposite direction and eventually affect our results. 

\begin{figure}[hbtp]
               \includegraphics[width=0.45\textwidth, height=6cm]{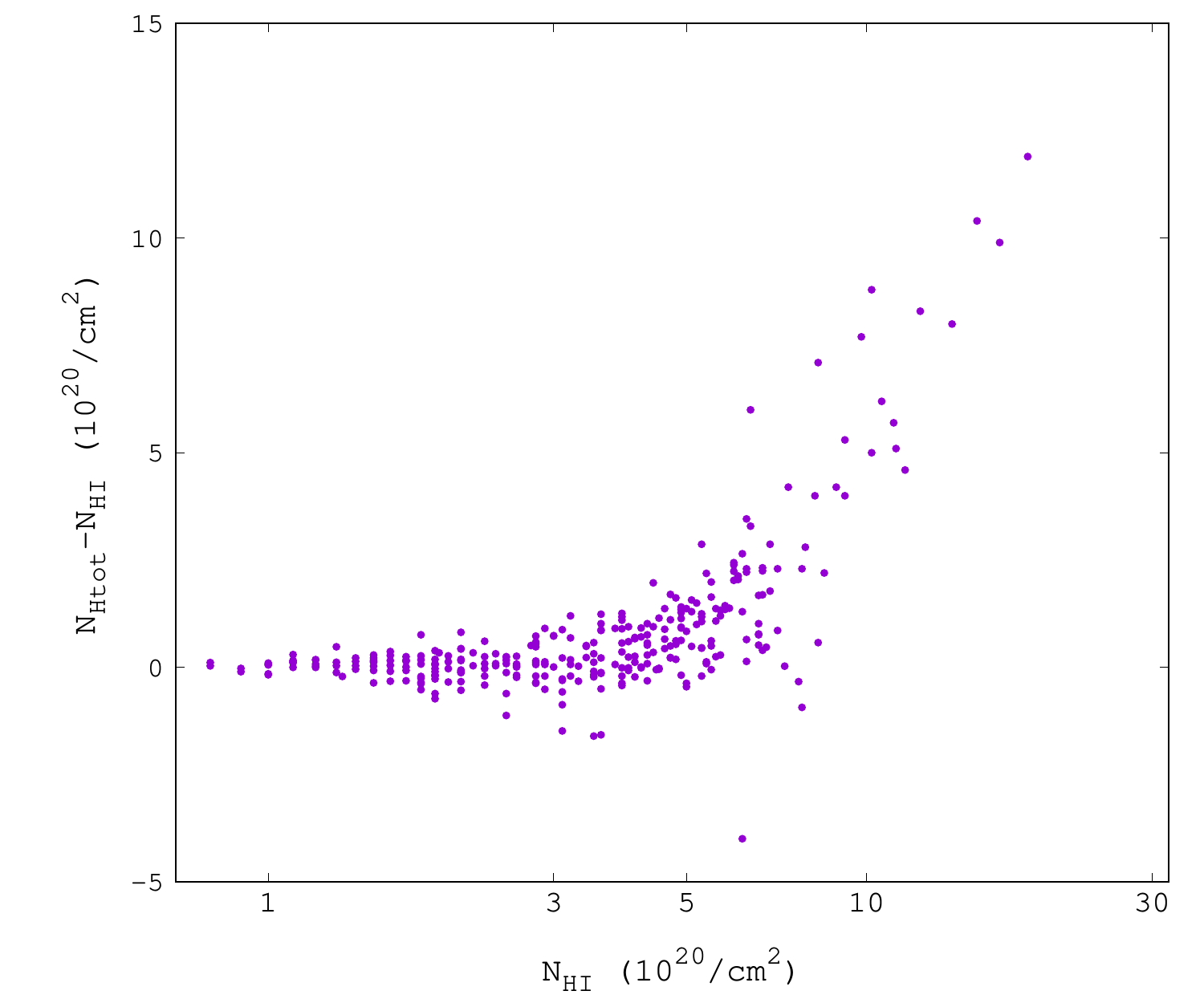}
               \caption{Comparison between the atomic hydrogen column density $N_{\text{HI}}$ as given in DL90 and the total one $N_{\text{Htot}}$ as given in W13 for the 313 clusters.}
        \label{fig3}
\end{figure}

In the calculation of the $L_{\text{X}}$ of every cluster, the REFLEX and NORAS catalogs used the neutral hydrogen $N_{\text{HI}}$ values coming from \citet{dickey} (thereafter DL90), while the (e)BCS catalog used again $N_{\text{HI}}$ values but as given in \citet{stark}. As shown in \citet{baumgartner} and \citet{schellenberger} (S15 hereafter) the total hydrogen column density $N_{\text{Htot}}$ (neutral+molecular hydrogen) starts to get significantly larger than $N_{\text{HI}}$ for $N_{\text{HI}}\geq 6\times 10^{20}/$cm$^2$. If this is not taken into account it would result in a misinterpretation of the total X-ray absorption due to Galactic material and hence to underestimating $L_{\text{X}}$ while generally overestimating $T$. In order to account for this effect, we used the $N_{\text{Htot}}$ values as given by \citet{willingale} (hereafter W13) in all the spectra fittings we performed and for correcting the MCXC $L_{\text{X}}$ values as described in the following subsections. The comparison between the $N_{\text{HI}}$ used in the parent catalogs and the $N_{\text{Htot}}$ values we use is displayed in Fig. \ref{fig3}.

One can see that for certain clusters $N_{\text{Htot}}<N_{\text{HI}}$, something that seems counter-intuitive. This happens because in the calculation of $N_{\text{Htot}}$, W13 use the $N_{\text{HI}}$ from the LAB survey \citep{LAB} and not from DL90. The LAB survey has a better resolution and tends to give slightly lower $N_{\text{HI}}$ values than DL90 for the same sky positions, something that can create these small inconsistencies for clusters where the molecular hydrogen is not yet high enough. As stated above, we corrected these inconsistencies for all the 313 clusters. The LAB survey covers the velocity range of ($-450$ km/s, $+400$ km/s), within which all neutral hydrogen is supposed to be detected. This velocity range naturally propagates to the $N_{\text{Htot}}$ values we use and any amount of hydrogen outside of this velocity range is not accounted for. The median $N_{\text{Htot}}$ for the 313 clusters is $3.81\times 10^{20}/$cm$^2$ ($4.35\times 10^{20}/$cm$^2$ for the excluded clusters). 

\subsection{Luminosity} \label{lum_estim}

We chose to use \textit{ROSAT} luminosity measurements for the simple reason that the entire $R_{500}$ area of the clusters is observed in the RASS. On the contrary, the field of view (FOV) of \textit{XMM-Newton} and (especially) \textit{Chandra} does not cover the full $R_{500}$ for most of our clusters\footnote{For this particular study, another advantage of using \textit{ROSAT} measurements is that we excluded the possibility of a \textit{XMM-Newton}-related anisotropic bias, since such anisotropies have already been detected for XCS-DR1, a sample constructed purely by \textit{XMM-Newton} observations (see M18).}. It has been shown though that the \textit{ROSAT} $L_{\text{X}}$ values are fully consistent with the ones from \text{XMM-Newton} within the \textit{ROSAT} $L_{\text{X}}$ uncertainties \citep{bohrin_rosat_xmm, yuying} (see Sect. \ref{rosat_vs_xmm} in Appendix for further tests)

Our estimates for the cluster X-ray luminosities used the reported X-ray luminosities in the MCXC catalog as a baseline. These luminosities were homogenized for systematics between the different parent catalogs, and were aperture-corrected to reflect the flux within $R_{500}$ (for more details see \citet{mcxc}, Chapter 3.4.1). For the relative luminosity uncertainty $\sigma_{L_{\text{X}}}$ we assumed $\sigma_{L_{\text{X}}}\sim \dfrac{1}{\sqrt{C_N}}$, where $C_N$ is the RASS counts from the parent catalogs. We applied further corrections to the MCXC cluster luminosities.

Firstly, we calculated K-correction factors to account for the redshifted source spectrum when observed in the observer reference frame. We derived these factors in two iterative steps in XSPEC using the scaling relations by \citet{reichert} for the input temperatures, and the 49 updated redshifts. The changes that occurred are much smaller than $\sigma_{L_{\text{X}}}$.
Secondly, the $L_{\text{X}}$ values were adjusted accordingly for the 49 clusters for which we used new redshifts. The uncertainties of the fitted X-ray $z$ (both statistical and intrinsic) were propagated to the $L_{\text{X}}$ uncertainties and added in quadrature to the already existing ones. Next, we corrected the MCXC $L_{\text{X}}$ for changes in the soft-band X-ray absorption by using the combined molecular and neutral hydrogen column density values as described in Sect. \ref{nh_sec}. We first derived an absorbed $L_{\text{X}}$ by reversing the absorption correction from \citet{mcxc}, and then derived updated unabsorbed $L_{\text{X}}$ values in XSPEC employing the cluster temperatures and metallicities derived in this work.
Finally, the redshift-derived distances of nearby clusters might be biased by peculiar velocities. For the five most nearby clusters ($50h^{-1}$ Mpc, $z\leq 0.0116$) we used redshift independent distance measurements from NED (published within the last 20 years) to derive the $L_{\text{X}}$ from the unabsorbed, k-corrected flux. The standard deviation of these distance measurements was propagated to the uncertainty of the luminosity. The average change in the distance compared to the redshift distances is $\sim 7\%$. For one cluster (S0851), no redshift independent distance was available and thus we adopted the redshift-derived distance but added an uncertainty of 250 km/s\footnote{Average difference between recession velocity as obtained by redshift-independent distances and measured heliocentric velocities for the other four clusters.} due to possible peculiar motions, which propagated in the $\sigma_{L_{\text{X}}}$ as well.

\begin{figure}[hbtp]
               \includegraphics[width=0.49\textwidth, height=6cm]{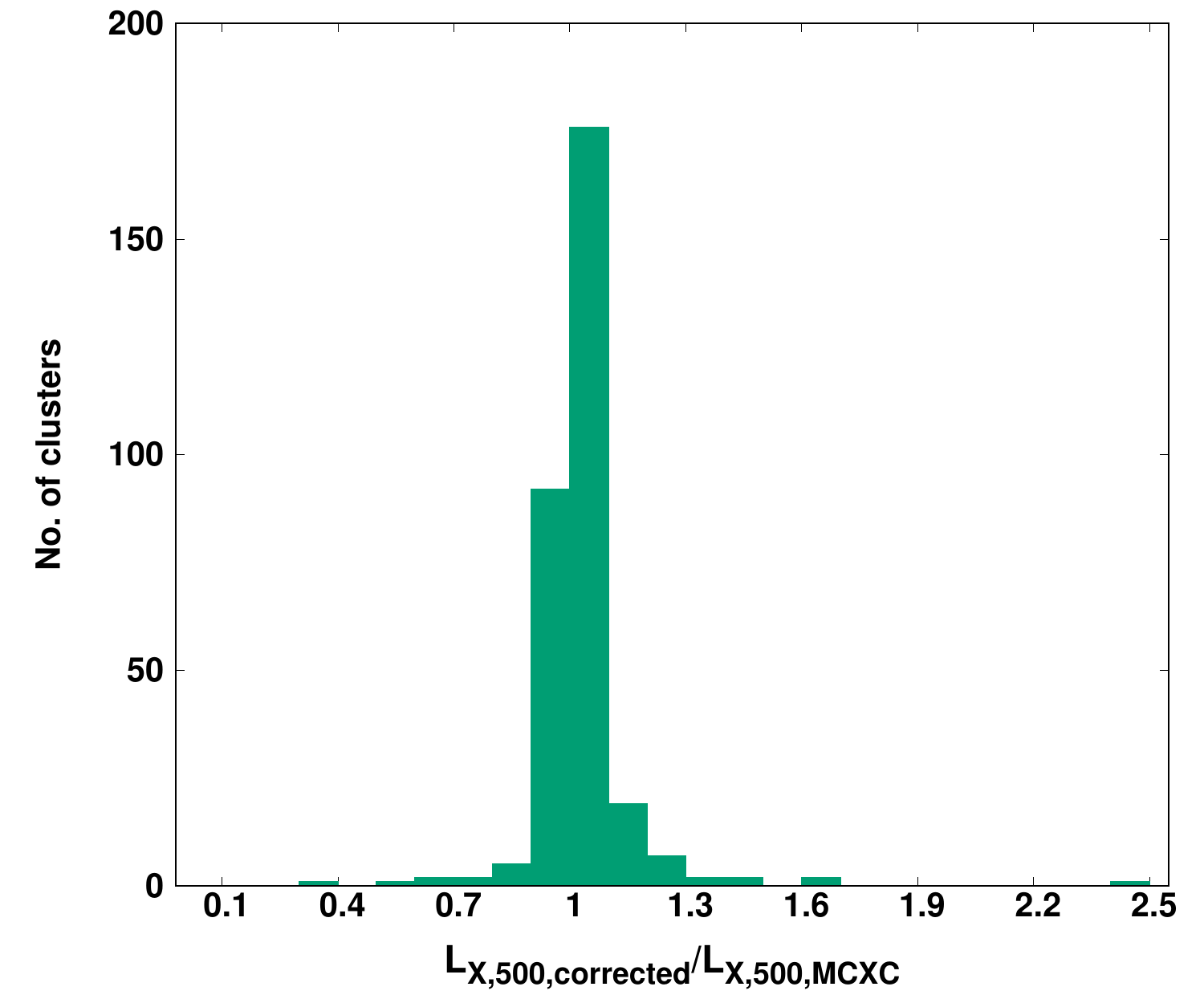}
               \caption{Comparison between the $L_{\text{X}}$ values after all the corrections we applied and the MCXC values for the 313 clusters.}
        \label{lum_comp}
\end{figure}

The comparison between the MCXC $L_{\text{X}}$ values and the values used in this analysis is shown in Fig. \ref{lum_comp}. As seen there, for 301 clusters (96\% of the sample) the change in $L_{\text{X}}$ is $\leq 25\%$. Since the intention of this paper is to look for spatial anisotropies of the $L_{\text{X}}-T$ relation in the sky we need to ensure that we do not introduce any directional bias through all of our $L_{\text{X}}$ corrections. To this end, we compare the fraction $L_{\text{X,ours}}/L_{\text{X,MCXC}}$ (the latter is the value given by MCXC) throughout the sky and we find it to be consistent within $\pm 4\%$. Similar results are obtained if one considers the fraction $L_{\text{X,MCXC}}/L_{\text{X,parent}}$, where $L_{\text{X,parent}}$ is the value given in the parent catalogs (more details in Sect. \ref{Lx_fraction} of the Appendix).

The final $L_{\text{X}}$ range of the clusters we use is $L_{\text{X}}=(1.15\times 10^{42}-3.51\times 10^{45}$) erg/s while the median value is $1.45\times 10^{44}$ erg/s ($1.10\times 10^{44}$ erg/s for the excluded objects). The median $\sigma_{L_{\text{X}}}$ is 10.6\%.

\subsection{Cluster radius}

The radii of the clusters were used for selecting the region within which the temperature and metallicity were measured.
The $R_{500}$ values of the clusters as given in MCXC were determined using the MCXC $L_{\text{X}}$ value and the X-ray luminosity-mass scaling relation $L_{\text{X}}-M$ as given in \citet{arnaud2}, which results in $L_{\text{X}}\sim R_{500}^{4.92}$. Since we applied certain changes to the MCXC $L_{\text{X}}$ values it is expected that also the respective $R_{500}$ should (slightly) change. Therefore, we used the same scaling relation to calculate the new $R_{500}$ in Mpc units. After that, the appropriate conversion to arcmin units was required, using the angular diameter distance ($D_A$). Since there is only a weak $R_{500}$  dependency on $L_{\text{X}}$ when the latter changes only because of alternations in the $N_{\text{H}}$ absorption, the new $R_{500}$ value does not significantly differ from the MCXC one (all changes $\leq 5\%$, since $D_A$ remains fixed). However, when $L_{\text{X}}$ changes because of a modification in the used $z$ value then also $D_A$ changes, as well as the normalized Hubble parameter $E(z)=[\Omega_\text{m}(1+z)^3+\Omega_{\Lambda}]^{1/2}$ and the critical density of the Universe $\rho_c(z)$ ,which are also included in the $L_{\text{X}}-R_{500}$ relation. This has a stronger impact on the final $R_{500}^{\text{arcmin}}$ ($R_{500}$ in arcmin units) than the absorption case alone. 
Nonetheless, the new $R_{500}^{\text{arcmin}}$ eventually changes by more than 10\% only for 5 out of 313 clusters. At the same time, 294 clusters ($>94\%$ of the sample) show a $\leq 4\%$ relative change in $R_{500}^{\text{arcmin}}$. Therefore, the direct use of the MCXC $R_{500}^{\text{arcmin}}$ values is practically equivalent to our new values.

\subsection{Temperature} \label{temperat}

We determined the temperature of each cluster within the $0.2-0.5$ R$_{500}$ annulus of every cluster in order to have self-consistent temperature measurements that reflect the other cluster properties (e.g., $L_{\text{X}}$) in a similar way. The cores of the clusters are excluded due to the presence of cool-cores, which significantly bias the temperature measurement and potentially increase the scatter of the $L_{\text{X}}-T$ relation \citep[e.g.,][]{hudson}. As previously stated, here we used the $N_{\text{Htot}}$ values in order to fit the spectra and obtain $T$. Since the new $R_{500}$ values do not considerably vary compared to the MCXC ones, we used the latter for the spectra extraction with one exception: when the difference between the two values was $>10\%$ (only five clusters as stated above) then we used the redetermined $R_{500}$ value. Generally, the temperature shows only a weak dependance to such small changes in $R_{500}$ since the vast majority of the spectral extraction regions remains unchanged. The relative difference in the obtained temperature for these five clusters when we use both ours and the MCXC $R_{500}$ is by average $\sim 8\%$.

It has been shown that the \text{Chandra} and \textit{XMM-Newton} telescopes have systematic differences in the constrained temperature values for the same clusters (S15). Thus, one has to take into account these biases when using temperature measurements from both telescopes. To this end, for the 76 clusters in our sample for which we use \textit{XMM-Newton} data since we do not have \textit{Chandra} data, we converted their measured temperatures to \textit{Chandra} temperatures adopting the conversion relation found in S15. To further check the consistency of this conversion, we applied this test ourselves choosing 15 clusters in an \textit{XMM-Newton} temperature range of $1.2-8.5$ keV (same range as for the 76 \textit{XMM-Newton} clusters) which have been observed by both telescopes. We constrained their temperatures with both instruments and we find that the best-fit relation provided by S15 still returns satisfactory results (Fig. \ref{temp_corr} in Appendix).
The final temperature range of these 313 clusters is $T=(0.83-19.24)$ keV with the median value being $T=4.5$ keV, while the median uncertainty is $\sigma _T=4.9$\%. Clearly the temperature range is considerably wide which significantly helps the purposes of our study. The suspiciously high temperature of 19.23 keV occurs for the galaxy cluster Abell 2163 (\object{A2163}) when the \citet{aspl} abundance table is used. \object{A2163} also lies in a high absorption region with $N_{\text{Htot}}=2\times 10^{21}$/cm$^2$. For other abundance tables, A2163 returns a temperature of $\sim 13-16$ keV. This large difference is mainly driven by the \textit{phabs} absorption model and the change of the Helium and Oxygen abundances. Generally, the average difference between the obtained temperature values between different abundance tables do not vary by more than $\sim 3-5$\% and thus, this cluster is a special case. For consistency reasons we use the 19.23 keV value. Excluding this cluster from the sample does not affect our results significantly since it is not an outlier in the $L_{\text{X}}-T$ plane.

\subsection{Metallicity}

The metallicity of each cluster was determined simultaneously with the temperature. The two different telescopes are not shown to give systematically different metallicity values for the same clusters (although the intrinsic scatter of the comparison is relatively large), and thus no conversion between the \textit{XMM-Newton} and \textit{Chandra} values was needed. The metallicity range of the used sample is $Z=(0.04-0.87)\ Z_{\odot}$ except for two clusters with  $Z=1.53^{+1.50}_{-1.07}\ Z_{\odot}$ and  $Z=3.86^{+0.61}_{-0.71}\ Z_{\odot}$, where the metallicity determination of the latter is clearly inaccurate while the former is consistent with typical metallicity values within the $1\sigma$ uncertainties. Excluding these two clusters from our analysis has no significant effects on the derived results. Finally, the median value is $Z=0.37\ Z_{\odot}$ with the median uncertainty being $\sigma _Z=23.1$\%

\section{Data reduction and spectral fitting} \label{spec_fit}

\subsection{Data reduction}

The exact data reduction process slightly differs for the two instruments. For the \textit{Chandra} analysis, we followed the standard data reduction tasks using the CIAO software package (version 4.8, CALDB 4.7.6). A more detailed description is given in \citet{gerrit2} (S17). For the \textit{XMM-Newton} analysis, we followed the exact same procedure as described in detail by R19. In a nutshell, every observation was treated for solar flares, anomalous state of CCDs \citep{kuntz}, instrumental background and exposure correction. For both instruments, the X-ray emission peak was determined and used as the centroid for the spectral analysis\footnote{Good agreement with MCXC for the vast majority of clusters.}, while bright point sources (AGNs and stars), extended structures unrelated to the cluster of interest (e.g., background clusters) and extended substructure sources were masked automatically and later by hand in a visual inspection. For this analysis, the HEASOFT 6.20, XMMSAS v16.0.0 and XSPEC v12.9.1 software packages were used.

\subsection{Background modeling}

For the \textit{Chandra} clusters, complementary to the S17 process, the ROSAT All-Sky survey maps in seven bands \citep{snowden97} were used to better constrain the X-ray background components. The background value in each of the seven bands was determined within $1^{\circ}-2^{\circ}$ around the cluster. 

For the \textit{XMM-Newton} clusters, the only difference with the process described in R19 is the background spectra extraction region. The X-ray sky background was obtained when possible, from all the available sky region in the FOV outside of 1.6 $R_{500}$ from the cluster's center. In this case, no cluster emission residuals were added in the background modeling. This was done mostly for the clusters located at $z\gtrsim 0.1$ which have a small apparent angular size in the sky. For most clusters, a partial overlap of the background extraction area with the 1.6 $R_{500}$ circle is inevitable and thus, an extra \textit{apec} component to account for the cluster emission residuals was added during the spectral fitting, with its temperature and metallicity free to vary. The normalizations of the background model components were also left free to vary during the cluster spectra fitting as described in detail in R19.

\subsection{Spectral fitting}


For the spectral fitting, the same methods were used as in R19 (\textit{apec}$\times$\textit{phabs}+emission and fluorescence lines) with only some small differences which are described here. Firstly, the $0.7-7$ keV energy range was used for all spectral fittings for both instruments. This way we managed to exclude the emission lines close to $0.6$ keV which originate from the Solar Wind Charge Exchange and cosmic X-ray background (S15, R19 and references therein). Moreover, we avoid the events produced by the fluorescent lines at $7.5, 8$ and $8.6$ keV which appear in the spectra of the pn detector of \textit{XMM-Newton}. Furthermore, \textit{Chandra} has a small effective area for energies higher than 7 keV. For all the spectral fits the \citet{aspl} abundance table was used. Finally, for the 237 \textit{Chandra} clusters the best-fit parameters of the spectral model were determined from an MCMC chain within XSPEC, while for the 76 \textit{XMM-Newton} clusters the $\chi^2$-statistic was used.
 
\section{The $L_{\text{X}}-T$ scaling relation} \label{analysis_method}

For obtaining the best-fit values of the $L_{\text{X}}-T$ relation parameters and comparing them for clusters located in different directions in the sky, we use a similar approach to M18. Here the strong dependance of the $L_{\text{X}}$ on the cosmological parameters should be stressed again, combined with the fact that $T$ can be measured without any cosmological assumptions (see Appendix for the exact $R_{500}$ and $T$ dependance on the chosen cosmology).  

\subsection{Form of the $L_{\text{X}}-T$ scaling relation}

We adopt a standard power-law form of the $L_{\text{X}}-T$ scaling relation as shown below:

\begin{equation}
\frac{L_{\text{X}}}{10^{44}\ \text{erg/s}}\ E(z)^{-1}=A \times \left(\frac{T}{4\ \text{keV}}\right)^B,
\label{eq1}
\end{equation}
where the term $E(z)=[\Omega_\text{m}(1+z)^3+\Omega_{\Lambda}]^{1/2}$ scales $L_{\text{X}}$ accordingly to account for the redshift evolution of the $L_{\text{X}}-T$ scaling relation. The scaling of the temperature term was chosen to be close to the median $T=4.5$ keV. The exact constant scaling of the $L_{\text{X}}$ values ($10^{44}$ erg/s) is not important since it is only a multiplication factor of the normalization. The exact scaling correction for the redshift evolution of the $L_{\text{X}}-T$ relation [$E(z)$] is also not particularly significant for consistent redshift distributions and low-$z$ samples like our own, as discussed later in the paper.
In order to constrain the best-fit parameters the $\chi^2$-minimization method is used and applied to the logarithmic form of the $L_{\text{X}}-T$ relation,

\begin{equation}
\begin{aligned}
\log{L'_X}&=\log{A}+B\log{T'}.
\label{eq2}
\end{aligned}
\end{equation}
Here, $L'_{\text{X}}$ and $T'$ are defined as

\begin{equation}
L'_{\text{X}}=\dfrac{L_{\text{X}}}{10^{44}\ \text{erg/s}}\ E(z)^{-1}\quad \text{and}\quad T'=\dfrac{T}{4\ \text{keV}}.
\end{equation}

\subsection{Linear regression}\label{fitting}

The exact form of the $\chi^2$-statistic used to find the best-fit $A$, $B$, $\sigma_{\text{int}}$ and $H_0$ values is given by

\begin{equation}
\chi^2_L=\sum\limits_{i=1}^N\frac{\left(\log{[L'_{{\text{X}},\text{obs}}]}-\log{[L'_{{\text{X}},\text{th}}(T',\mathbf{p})]}\right)^2}{{\sigma _{\log{L},i}}^2+{B^2\times \sigma _{\log{T},i}}^2+{\sigma_{\text{int}}}^2},
\label{eq3}
\end{equation}
where $N$ is the number of clusters used for the fit, $L'_{\text{X,obs}}$ and $T'$ are the measured luminosity and temperature values respectively (scaled as explained above), $L'_{X,\text{th}}$ is the theoretically expected value for the luminosity based on the measured temperature in addition to the fitted parameters $\mathbf{p}$ ($A$ and $B$, or $H_0$). Furthermore, $\sigma _{\log{L,T,i}}$ are the Gaussian logarithmic uncertainties which are derived in the same way as in M18 \footnote{$\sigma_{\log{x}}=\log{(e)}\times \frac{x^+-x^-}{2x}$, where $x^+$ and $x^-$ are the upper and lower limits of the main value $x$ of a quantity, considering its 68.3\% uncertainty.}, while $\sigma_{\text{int}}$ (which was not included in M18) accounts for the intrinsic scatter of the relation. 

The latter is fitted iteratively, starting from $0$ and increasing step-by-step until there is a combination of $\mathbf{p}$ that gives $\chi^2_{\text{red}}\sim 1$, as in \citet{Maughan07, maughan, zou} etc. Under certain conditions, this procedure might return slightly underestimated $\sigma_{\text{int}}$ values. However, this should not be a concern since the exact values of $\sigma_{\text{int}}$ are not of particular importance for this analysis and they are only used to derive trustworthy parameter uncertainties from our $\chi^2$ model.

Additionally, the $1\sigma$ uncertainties of the fitted parameters are based on the standard $\Delta\chi^2=\chi^2-\chi^2_{\text{min}}$ limits ($\Delta\chi^2\leq 1$ or 2.3 for one or two fitted parameters respectively). In the case of the slope being free to vary, the projection of the $x$-axis uncertainties to the $y$-axis also varies. This fitting method is comparable to the BCES Y|X fitting method described by \citet{akritas}. 

Finally, we should stress that $H_0$ and $A$ cannot be simultaneously constrained since they are degenerate. One can put absolute constraints only on the product $A\times H_0^2$. Therefore, one needs to fix one of the parameters to investigate the behavior of the other. In Sect. \ref{mainresults} we use a fixed $H_0=70\  \text{km}\  \text{s}^{-1} \text{Mpc}^{-1}$ to investigate the behavior of $A$. In Sect. \ref{cosmology} we fix $A$ to its best-fit value and study the directional behavior of $H_0$ through the $\chi^2-$minimization procedure described above\footnote{This is equivalent to directly converting $A$ values to $H_0$ values, since the product $A\times H_0^2$ remains unchanged.}.

\subsection{Pinpointing anisotropies via sky scanning} \label{scan}

With the purpose of studying the consistency of the fitted parameters throughout the sky and identifying specific sky patches that seem to show a significantly different behavior than the rest, we follow the method described below. 
We consider a cone of a given radius $\theta$ (we use $\theta =45^{\circ}$, $60^{\circ}$, $75^{\circ}$ and $90^{\circ}$) and we only consider the clusters that lie within this cone. For instance, if we choose a $\theta=60^{\circ}$ cone centered at $(l,b)=(150^{\circ},30^{\circ})$ then the subsample of clusters consists of all the clusters with an angular separation of $\leq 60^{\circ}$ from these specific coordinates. By fitting the $L_{\text{X}}-T$ scaling relation to these clusters, we obtain the normalization (or $H_0$), slope and intrinsic and total scatter for these clusters. The extracted best-fit value for the fitted parameter is assigned at these coordinates.

Shifting this cone throughout the full sky in steps of $\Delta l=1^{\circ}$ and  $\Delta b=1^{\circ}$ in Galactic coordinates\footnote{$360\times 181=65160$ different cones}, we can obtain the desired parameter values for every region of the sky. We additionally apply a statistical weighting on the clusters based on their angular separation from the center of the cone. This is given by simply dividing their uncertainties by $\cos{\left(\dfrac{\theta_1}{\theta}\times 90^{\circ}\right)}$, where $\theta_1$ is the above-mentioned angular separation. Hence, the weighting $\cos$ term is calibrated in such way that it shifts from 1 to 0 as we move from the center of the cone to its boundaries, independently of the angular size of the cone. This enlargement of the uncertainties results in an artificial decrease of the $\sigma_{\text{int}}$ which is not of relevance here since, as explained before, $\sigma_{\text{int}}$ mostly acts as a nuisance parameter. Nevertheless, we perform tests to ensure this does not bias our results, as explained in \ref{boot_section}. 

All the $A$ maps are plotted based on the $A/A_{all}$ value, where $A_{all}$ is the best-fit $A$ when all the clusters are used independently of the direction. Finally, the maps have the same color scale for easier comparison, except for the $\theta=45^{\circ}$ cone maps for which the color scale is enlarged for better visualization. 

\subsection{Statistical significance and sigma maps} \label{sigma_maps}

With the desired best-fit values and their uncertainties for every sky region at hand, it is easy to identify the direction that shows the most extreme behavior and assess the statistical significance of their deviation. For quantifying the latter in terms of number of sigma for two different subsamples we use:

\begin{equation}
\centering
\text{No. of }\sigma=\dfrac{\mathbf{p_1}-\mathbf{p_2}}{\sqrt{\sigma^2_{\mathbf{p_1}}+\sigma^2_{\mathbf{p_2}}}},
\label{sigma_sig}
\end{equation}
where $\mathbf{p_{1,2}}$ are the best-fit values for the two different subsamples and $\sigma^2_{\mathbf{p_{1,2}}}$ are their uncertainties\footnote{This formulation assumes that $\mathbf{p_{1}}$ and $\mathbf{p_{2}}$ are independent, which is true when the two subsamples do not share any common clusters. This is mostly the case for our results with few exceptions of $\lesssim 10\%$ common clusters between some compared subsamples. However, this does not significantly affect the significance especially when one considers that the weighting of these clusters is different for each subsamples based on their distance from the center of the cone.}. 

Each time we constrain the anisotropic amplitude of the most extreme dipole in the sky, while we also compare the two most extreme regions in terms of the fitted parameter, regardless of their angular separation. This is done by calculating the statistical deviation (in terms of $\sigma$) between all the different cone subsamples. The two sky regions for which the largest deviation (highest no. of $\sigma$) is found between them, are the ones reported in the following sections as "the most extreme regions".
In addition, a percentage value (\%) is displayed next to each $\sigma$ deviation. This value comes from the difference of these two extreme regions over the best-fit value for the full sample.

In order to create the significance maps, we use Eq. \ref{sigma_sig} to compare the best-fit result $\mathbf{p_1}$ of every cone with the best-fit result $\mathbf{p_2}$ of the rest of the sky. The obtained sigma value is assigned to the direction at the center of the cone. All the significance maps have the same color scale for easier comparison. For the majority of cases, the two most extreme regions as defined above match the highest $\sigma$ regions in the significance maps.

Finally, as an extra test we also create $10^5$ realizations using the bootstrap resampling method in order to check the probability of the extreme results to randomly occur independently of the sky direction. The followed procedure is described in detail in Sect. \ref{boot_section}. Using all these estimates, one can determine the consistency with what one would expect in an isotropic universe.

\section{Results}\label{mainresults} 

\subsection{The $L_{\text{X}}-T$ scaling relation for the full sky}

Before we search for apparent anisotropies in the sky we constrain the behavior of the $L_{\text{X}}-T$ scaling relation for the full sample. We do not account for any selection biases in our analysis since we believe that their effects are not important for this work. This is because we wish to study the \emph{relative} $L_{\text{X}}-T$  differences between different sky regions (or from the overall best-fit line). If we indeed corrected for selection effects we would constrain the "true" underlying $L_{\text{X}}-T$ relation which would not represent our data (but the true distribution). This might cause wrong estimates for the relative $L_{\text{X}}-T$  differences. Therefore, we need to constrain the relation that describes our 313 clusters best. Nevertheless, in Sect. \ref{systemat_corr} we discuss the possible effects of selection systematics and find that there is no indication that they compromise our results.

We use the aforementioned 313 clusters and fit Eq. (\ref{eq1}) obtaining the best-fit normalization and slope of the $L_{\text{X}}-T$ relation as well as its intrinsic scatter. The results are:

\begin{equation}
A_{\text{all}}=1.114^{+0.044}_{-0.040}, \ B_{\text{all}}=2.102\pm 0.064, \ \text{and}\ \sigma_{\text{int}}=0.242\ \text{dex}.
\end{equation}

\begin{figure}[hbtp]
               \includegraphics[width=0.45\textwidth, height=6cm]{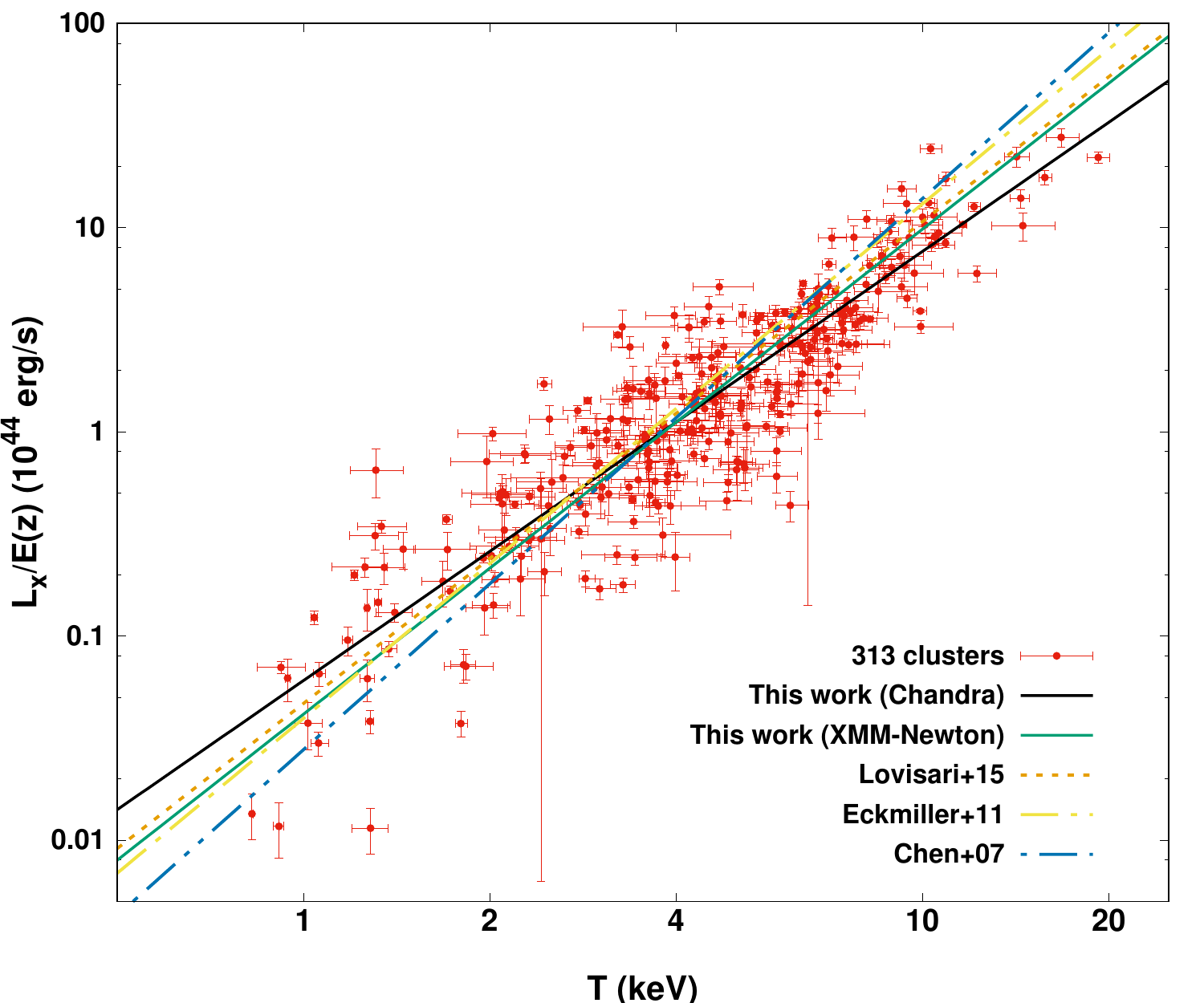}
               \includegraphics[width=0.45\textwidth, height=6cm]{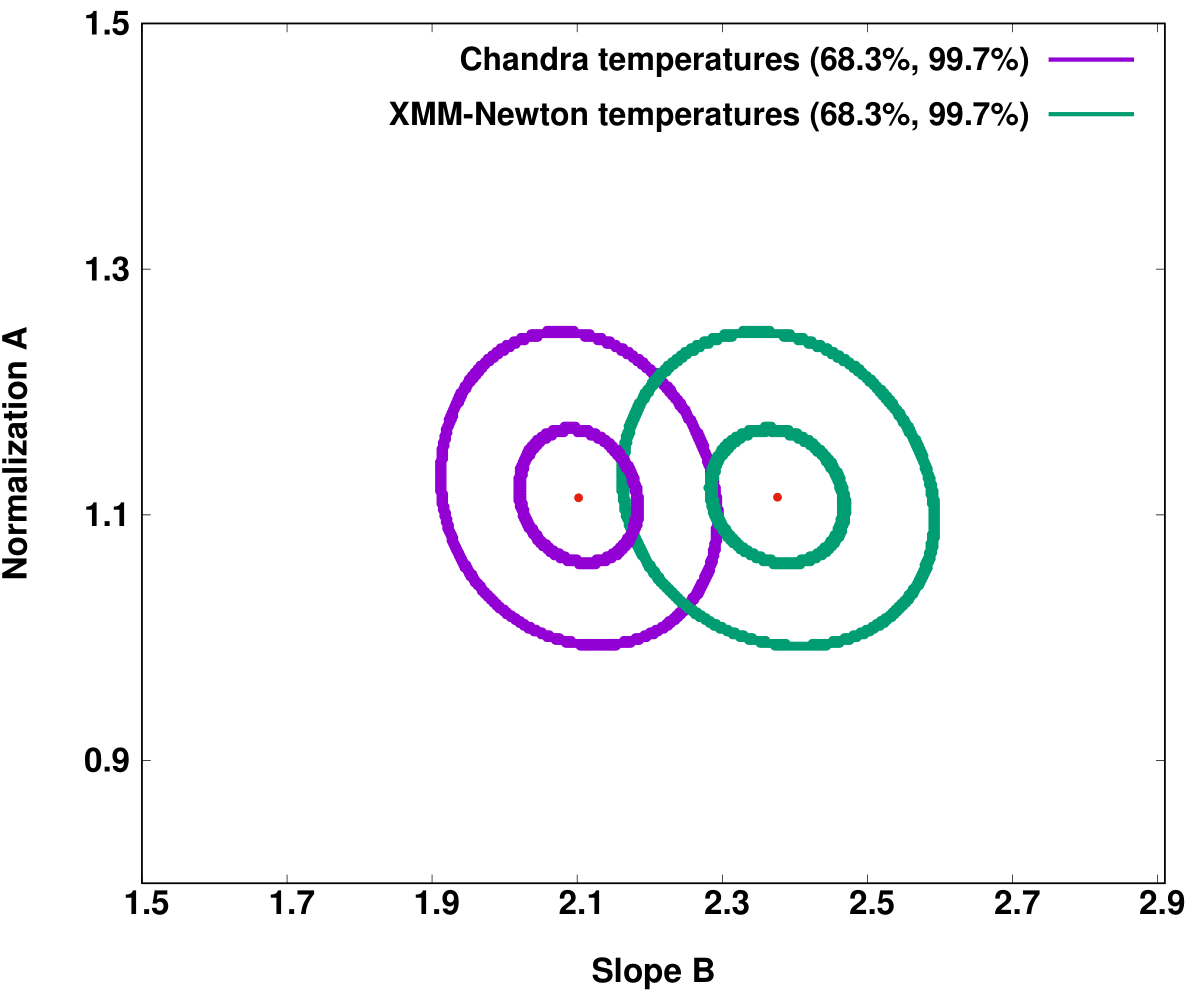}
               \caption{\textit{Top}: $L_{\text{X}}-T$ relation for the 313 clusters (red) with their best-fit model (black). $L_{\text{X}}$ is measured within the $0.1-2.4$ keV energy range. The best-fit models of other studies are displayed as well (dashed lines). The best-fit solution when \textit{XMM-Newton} temperatures are used is shown as well (solid green). \textit{Bottom}: $1\sigma$ (68.3\%) and $3\sigma$ (99.7\%) confidence levels of the normalization and slope of the $L_{\text{X}}-T$ relation as derived using \textit{Chandra}- or \textit{XMM-Newton}-converted temperatures for all 313 clusters (purple and green respectively). As shown the best-fit values for the same data can shift by $\sim 3\sigma$ depending on the instrument used.}
        \label{fig4}
\end{figure}

The statistical uncertainties for $A$ and $B$ are limited to $\sim 3-4\%$ which highlights the precision of our results based on the number and the quality of the data, combined with the large covered temperature range of the clusters. Moreover, the total scatter (statistical$+$intrinsic) of the $L_{\text{X}}-T$ is $\sigma_{\text{total}}=0.262$ dex, which means that the statistical uncertainties of the clusters contribute to only 7$\%$ of the total scatter. The $L_{\text{X}}-T$ fit of our sample is displayed in Fig. \ref{fig4} (top panel).

The best-fit slope is slightly steeper than the expected value from the self-similar model. Naively, the slope best-fit value ($\sim 2.1$) might seem surprising since most studies find a $L_{\text{X}}-T$ slope of $B\sim 2.3-3.6$ (see references in Sect. \ref{intro}). However, the exact value depends on multiple aspects such as the energy range for which $L_{\text{X}}$ was measured, the instrument used, the sample selection (when no bias correction is applied), the temperature distribution of the used sample, the cluster radius within which parameters were measured etc. Generally, it is expected that bolometric $L_{\text{X}}$ values return a steeper slope than soft band $L_{\text{X}}$ values, such as the $0.1-2.4$ keV band we use. This happens due to the fact that the bolometric emissivity $\epsilon$ of the ICM for thermal bremsstrahlung (which is the dominant emission process for $T\gtrsim 3$ keV) is $\epsilon \propto n_e^2 T^{0.5}$ (where $n_e$ is the electron density), while $\epsilon$ in the soft band (0.1-2.4 keV as used here) is rather independent of $T$ for $T\gtrsim 3$ keV ($\epsilon \propto n_e^2$). Therefore, one very roughly expects the slope of the $L_{\text{X}}-T$ relation to be smaller by $\sim 0.5$ in the 0.1-2.4 keV band; that is $L_{\text{X}}\sim T^{1.5}$ in the self-similar case. In general, the best-fit $L_{\text{X}}-T$ relation tends to change slightly when one corrects for selection biases (see references in Sect. \ref{intro} about the $L_{\text{X}}-T$ relation).


If we now convert all the measured temperatures to \textit{XMM-Newton} temperatures\footnote{The measurements of the 76 \textit{XMM-Newton} clusters are kept as they are while the measurements of the 237 \textit{Chandra} clusters are converted to \textit{XMM-Newton} temperatures based on S15.} using the relation given in S15, we obtain a slope of $B\sim 2.38$, shifting by $2.8\sigma$ compared to our main result, while the normalization remains the same. This result is consistent with the previously reported values that used the 0.1-2.4 keV luminosities \citep[e.g.,][]{chen,eckmiller,lovisari}\footnote{For Lovisari+15 we display the bias-uncorrected result when all the clusters and the Y|X fitting procedure are used. For Eckmiller+11 the shown result is for all the available clusters as well. The result from Chen+07 uses all the clusters and the hot temperature component as the $T$ value.}. In the bottom panel of Fig. \ref{fig4} the 68.3\% and $99.7\%$ confidence levels (1 and $3\sigma$ respectively) of the fitted parameters are shown for \textit{Chandra}-converted temperatures and \textit{XMM-Newton}-converted temperatures. It should be clear that this is done just for the sake of comparison and that the \textit{Chandra} temperatures are used for the rest of the paper.

A comparison between our results and the derived $L_{\text{X}}-T$ scaling relation from other works is also shown in the top panel of Fig. \ref{fig4}. We note that $L_{\text{X}}$ corresponds to the $0.1-2.4$ keV energy band for all the compared studies, while the results from the BCES (Y|X) fitting method were used when available. Additionally, the $L_{\text{X}}-T$ results for the full samples were used without any bias corrections. From this comparison, it is clear that all the derived results agree in the $L_{\text{X}}-T$ normalization value. In terms of the slope, our \textit{Chandra} fit is more consistent with the high-$T$ part of the distribution, while our \textit{XMM-Newton} fit is quite similar to the results of \citet{eckmiller} and \citet{lovisari}.

\subsection{1-dimensional anisotropies}

As a first test for the potentially anisotropic behavior of our galaxy cluster sample, we recreate the normalization against the Galactic longitude plot as presented in M18 (Fig. 3 in that paper). For this test, we consider regions centered at $l$ with a width of $\Delta l=90^o$. At the same time, the whole Galactic latitude range $b\in (-90^{\circ}, +90^{\circ})$ is covered by every region. 

\begin{figure}[hbtp]
\centering
               \includegraphics[width=0.49\textwidth, height=9cm]{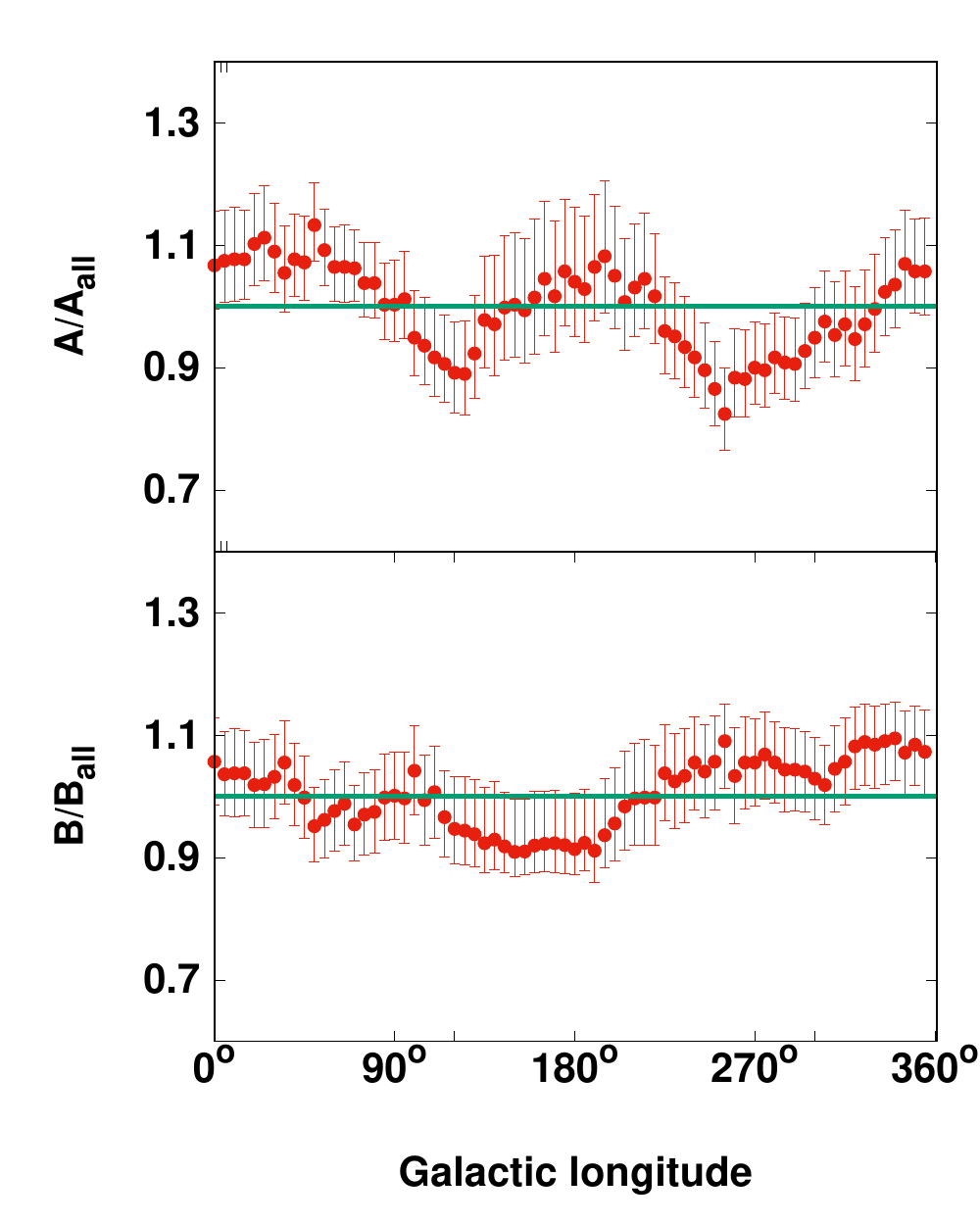}
               \caption{Best-fit normalization $A$ (top) and slope $B$ (bottom) of the $L_{\text{X}}-T$ relation for every sky region over the best-fit results for the full sample ($A_{\text{all}}$ and $B_{\text{all}}$) as functions of the Galactic longitude. The $1\sigma \ (68.3\%)$ uncertainties are also shown. Every region covers a sky area of $\Delta l=90^{\circ}$ and $\Delta b=180^{\circ}$. The $x$-axis values represent the central $l$ value for every bin.}
        \label{slope-glon}
\end{figure}

Firstly, we allow both $A$ and $B$ to vary simultaneously. The behavior of these two parameters as functions of the Galactic longitude are displayed in Fig. \ref{slope-glon}.
One sees that the slope remains relatively constant throughout the sky, varying only by 18\% from its lowest to highest value, and with a relatively low dispersion. Also, the largest deviation between any two independent sky regions is limited to $1.8\sigma$. No obvious systematic trend in the slope as a function of the galactic longitude can be seen since all the regions return slope values consistent with the full sample at $\leq 1.2\sigma$.  At the same time, this variation for the normalization reaches 31\% with a higher dispersion and a clear trend with galactic longitude, while the strongest tension between two independent sky regions appears to be $3.2\sigma$. 

\begin{figure*}[hbtp]
\centering
               \includegraphics[width=0.8\textwidth, height=8.5cm]{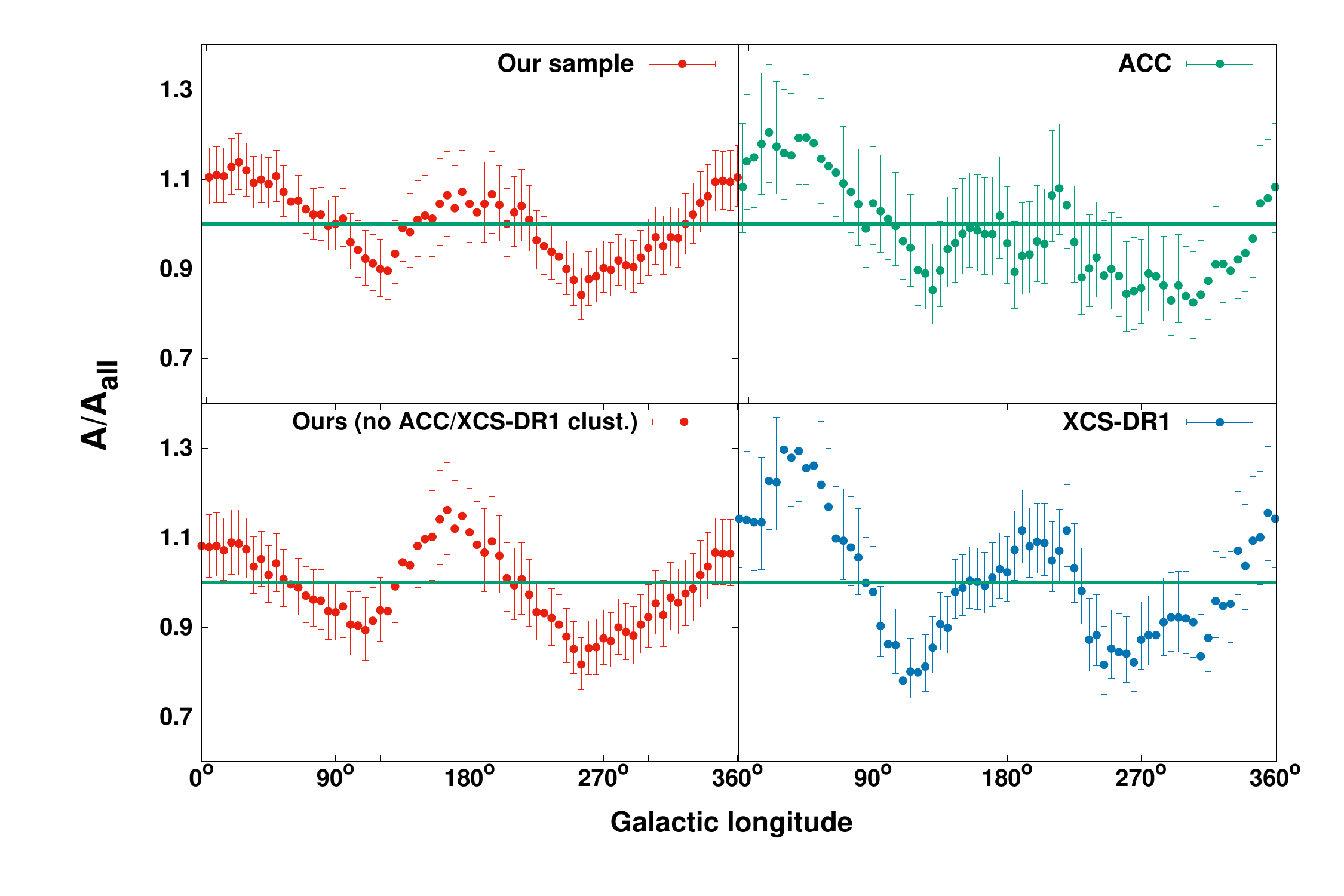}
               \caption{Best-fit normalization $A$ of the $L_{\text{X}}-T$ relation for every sky region over $A_{\text{all}}$  as a function of the Galactic longitude. The $1\sigma$ ($68.3\% $) uncertainties are also shown. The results correspond to this work's sample (top left), ACC (top right), this work's sample excluding all the common clusters with ACC and XCS-DR1 (bottom left) and XCS-DR1 (bottom right).}
        \label{3_samples}
\end{figure*}

Based on these results, the slope is kept fixed at the best-fit value for the whole sample and only the normalization of the $L_{\text{X}}-T$ relation is free to vary. In the top left panel of Fig. \ref{3_samples} the best-fit normalization value $A$ for every region is displayed with respect to the best-fit $A_{\text{all}}$ for the full sky (all 313 clusters). The same is also done for ACC and XCS-DR1 with the results displayed in the top and bottom left panels of Fig. \ref{3_samples} respectively. The only difference with the M18 results for these two samples is that here the intrinsic scatter term is taken into account as well during the fitting as shown in Eq. \ref{eq3}. 

Surprisingly enough, the pattern in the behavior of the $L_{\text{X}}-T$ normalization for our sample strongly resembles the results of both ACC and XCS-DR1, despite being almost independent with XCS-DR1, sharing only $\sim 30\%$ of the clusters with ACC and following different analysis strategies. Specifically, the region with the most anisotropic behavior compared to the rest of the sky ($2.9\sigma$ significance) is the one with the lowest $A$ lying within $l\in [210^{\circ}, 300^{\circ}]$. This region exactly matches the findings of M18 for XCS-DR1, while the lowest $A$ region for ACC is separated by 40$^{\circ}$. Here we should remind the reader that the 313 clusters we use share only three common clusters with XCS-DR1 and 104 with ACC as these samples were used in M18. The opposite most extreme behavior (highest $A$) is detected in $l\in [-20^{\circ}, 70^{\circ}]$ (same brightest region in ACC as well, 25$^{\circ}$ away from XCS-DR1's brightest region) with a deviation of $2.5\sigma$ compared to the rest of the sky and $3.4\sigma$ compared to the lowest-$A$ region, which is similar to the two other samples.

In order to verify that the observed behavior is not caused by the few common clusters between our sample and ACC or XCS-DR1, we exclude all 104 of them from our sample and repeat the analysis. The result is shown in the bottom left panel of Fig. \ref{3_samples}. One can see that this systematic trend persists and does not significantly depend on the common clusters between the two samples. The region with the largest deviation from the rest of the sky remains the same as for the full sample with an even higher significance of $3.1\sigma$. This striking similarity between the three different samples in the 1D search for anisotropies should be investigated in more depth in order for its exact reason to be identified.

\subsection{2-dimensional investigation} \label{2d_analysis}

In order to identify the exact regions with the highest degree of anisotropy, we should consider every possible direction in the sky. Different size regions should be considered as well, thus systematic behaviors can be detected. To this end, we use scanning cones (solid angles), as described in Sects. \ref{scan} and \ref{sigma_maps}. The slope is fixed to the best-fit value since the variations of the normalization are much stronger. This choice does not bias our results, as shown in Sect. \ref{free_slope}.

\subsubsection{$\theta=90^{\circ}$ cone}

\begin{figure*}[hbtp]
               \includegraphics[width=0.51\textwidth, height=5cm]{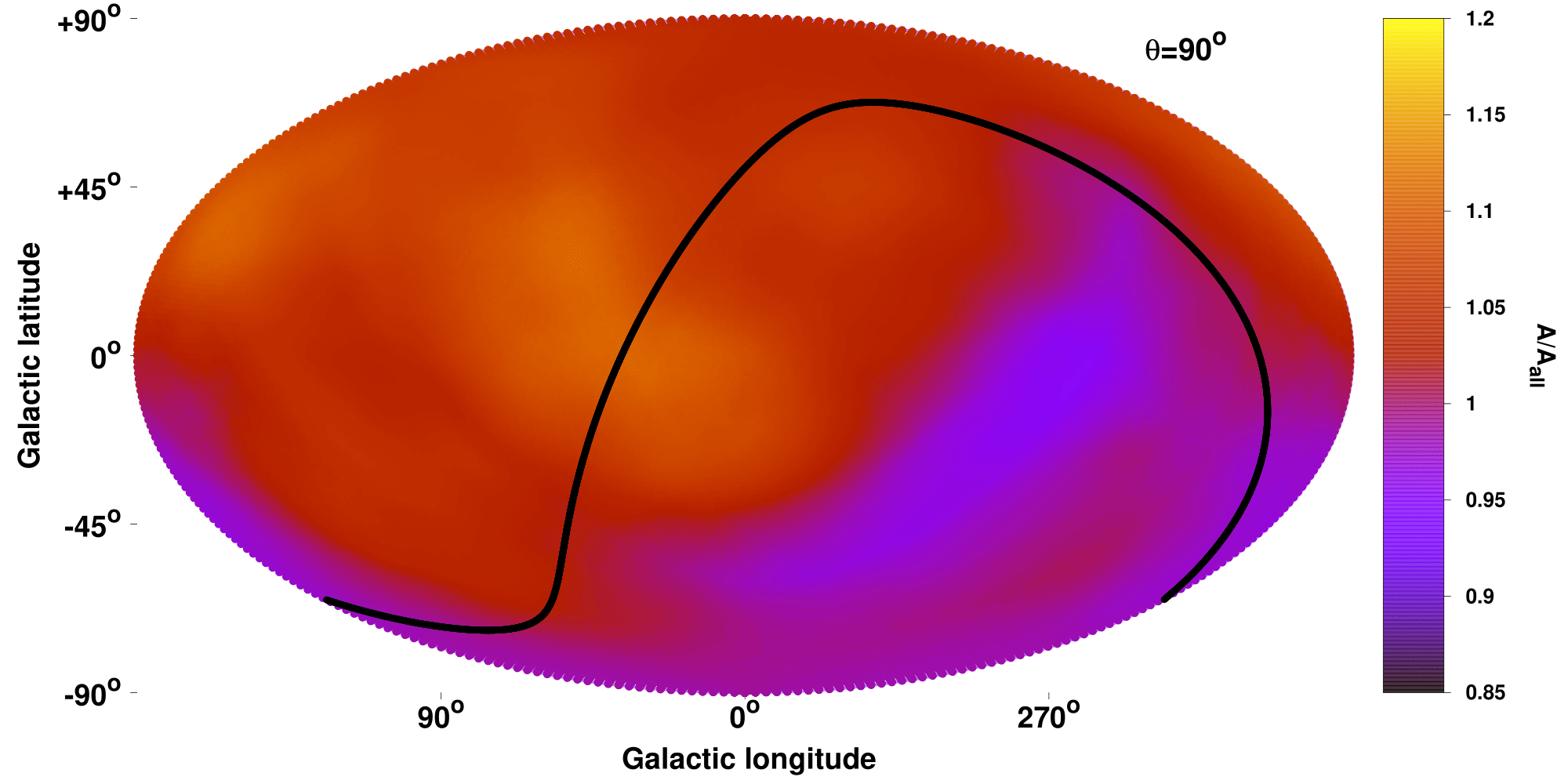}
               \includegraphics[width=0.51\textwidth, height=5cm]{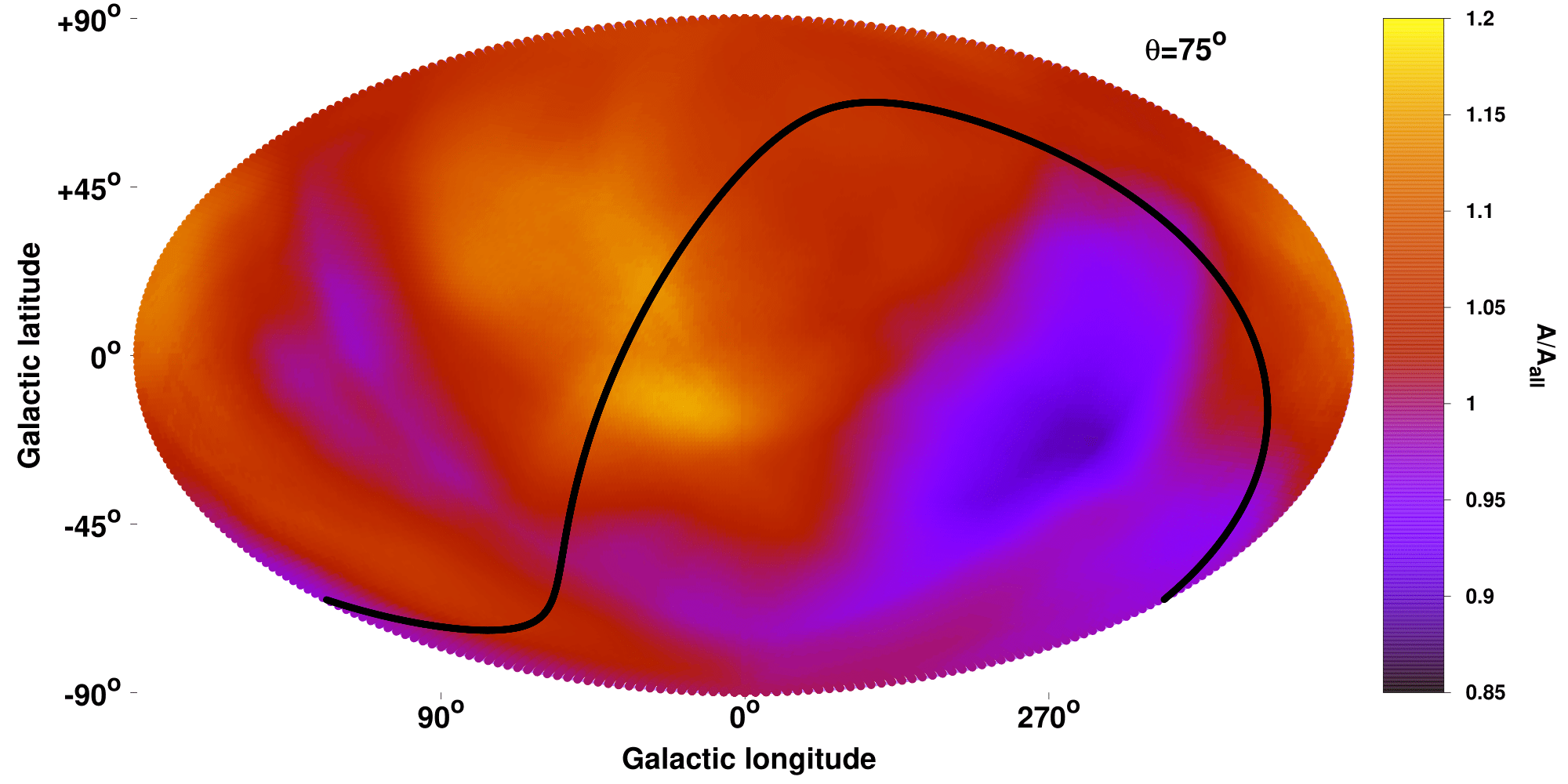}
               \includegraphics[width=0.51\textwidth, height=5cm]{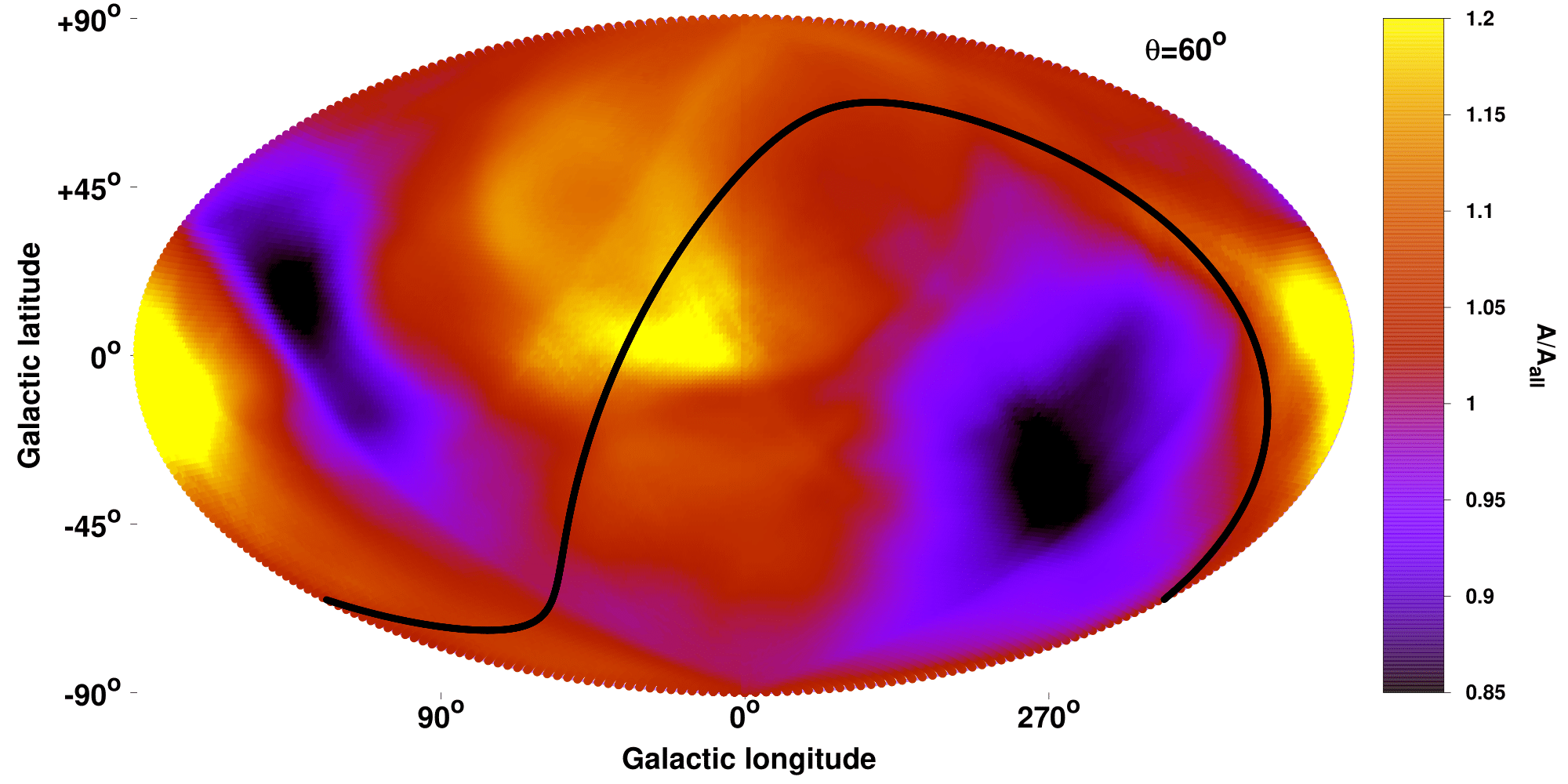}
               \includegraphics[width=0.51\textwidth, height=5cm]{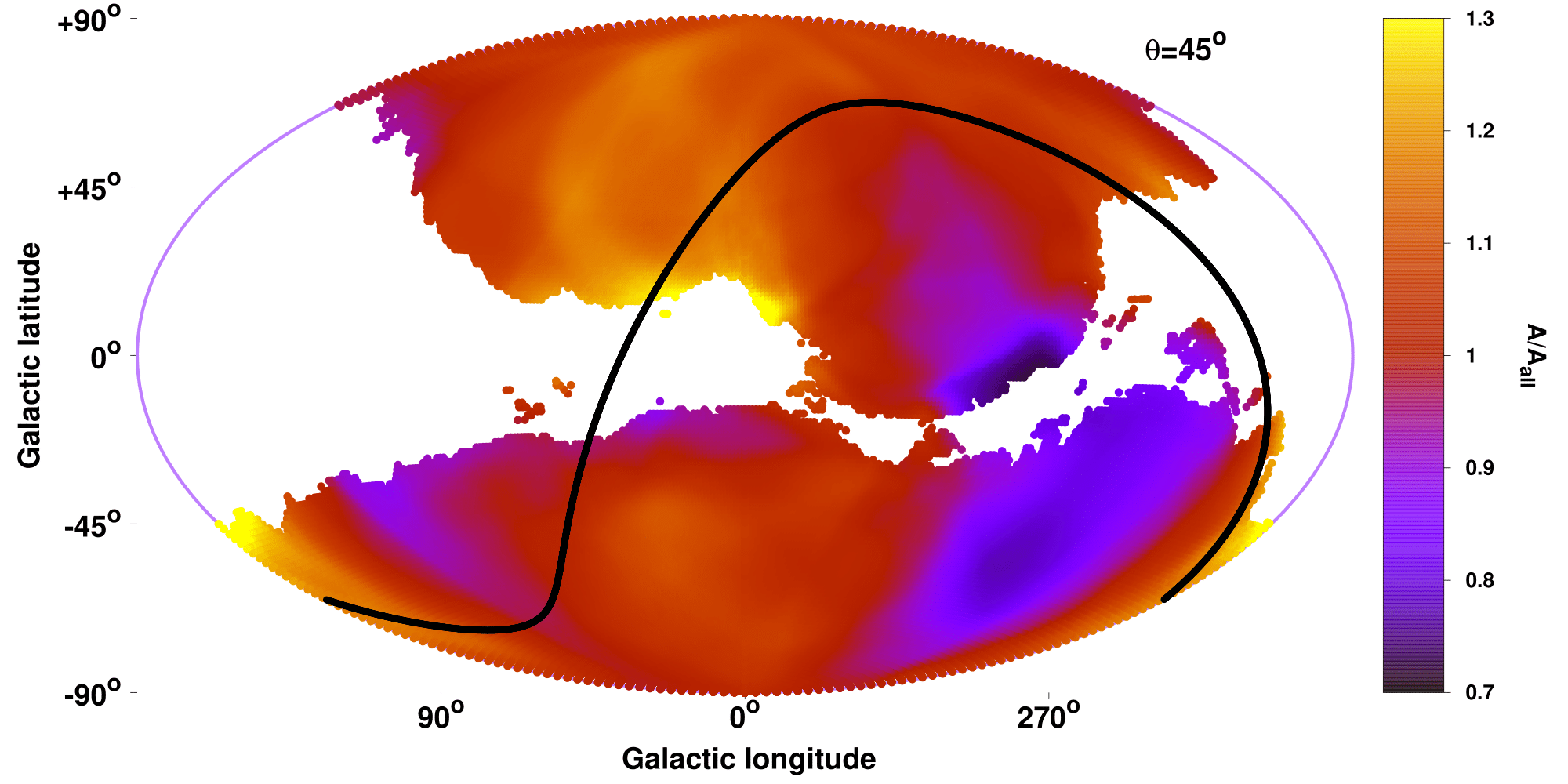}
               \caption{Best-fit normalization $A$ of the $L_{\text{X}}-T$ relation for every sky region over $A_{\text{all}}$  as a function of the position in the extragalactic sky. The maps are created with cones of $\theta =90^{\circ}$ (top left), $\theta=75^{\circ}$ (top right), $\theta =60^{\circ}$ (bottom left) and $\theta =45^{\circ}$ (bottom right, only region with $\geq 35$ clusters are shown). The first three maps have the same color scale ($85\%-120\%$), while the $\theta =45^{\circ}$ map has a wider color scale ($70\%-130\%$).}
        \label{fig7}
\end{figure*}

\begin{figure*}[hbtp]
               \includegraphics[width=0.51\textwidth, height=5cm]{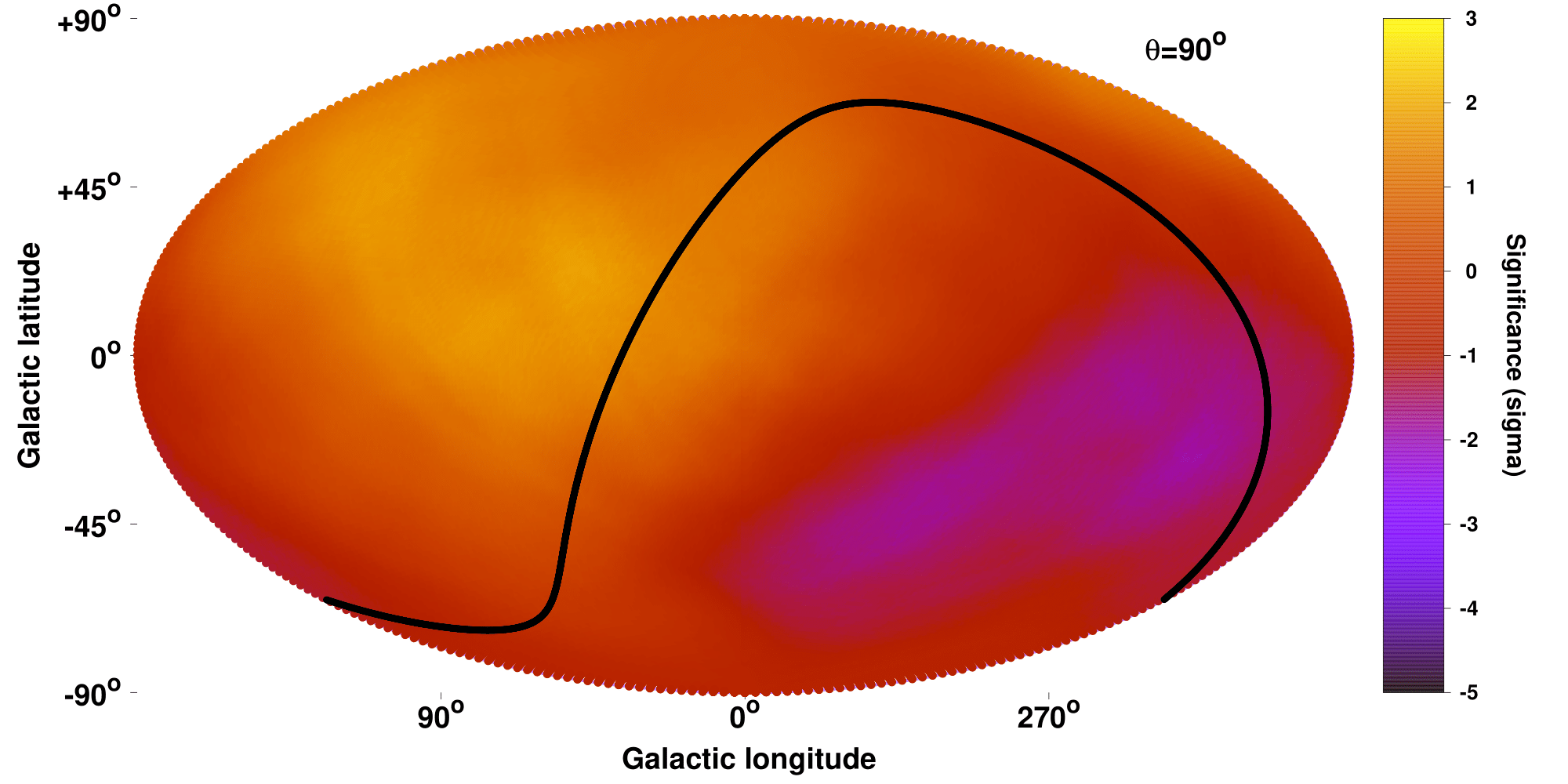}
               \includegraphics[width=0.51\textwidth, height=5cm]{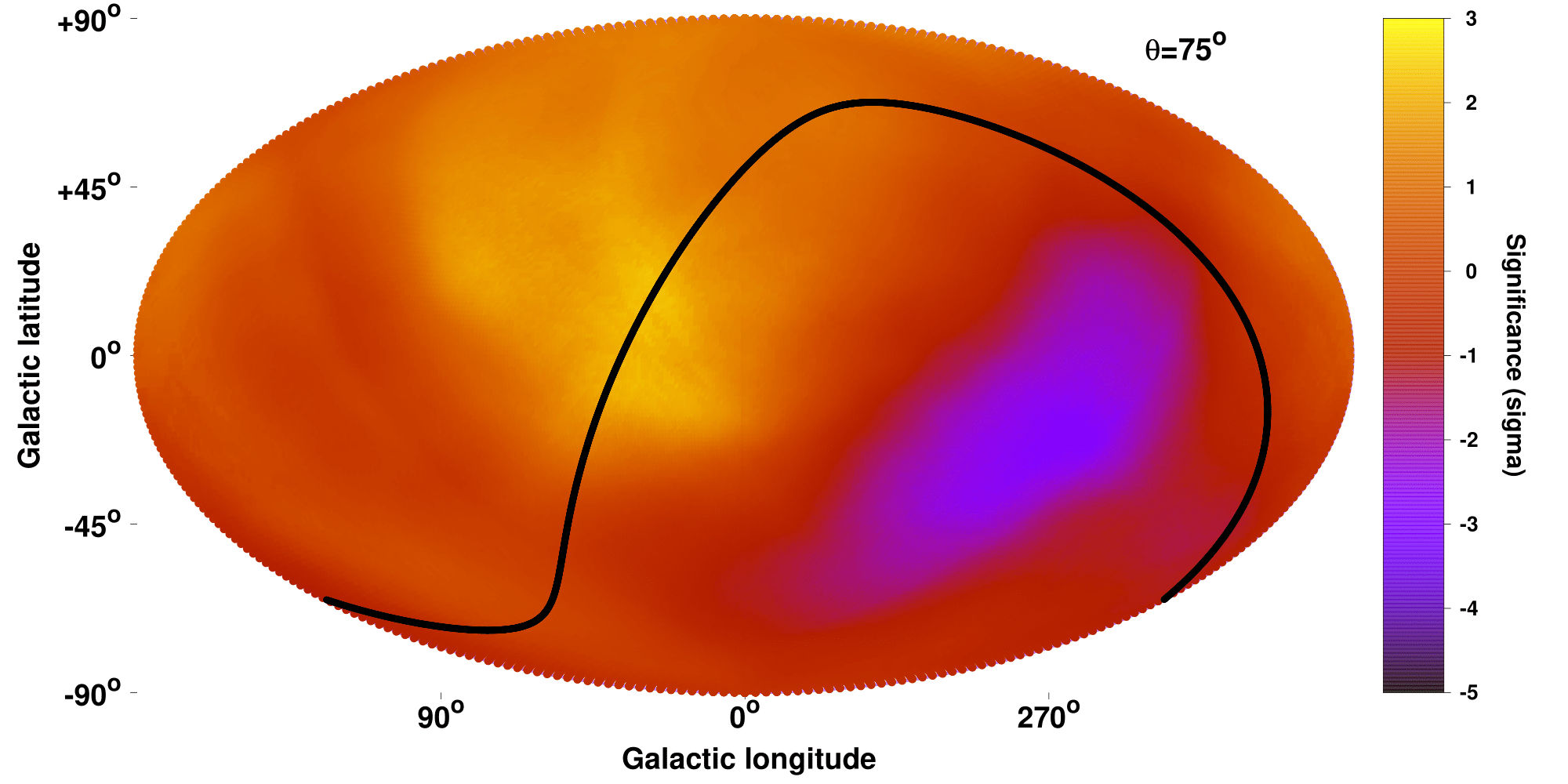}
               \includegraphics[width=0.51\textwidth, height=5cm]{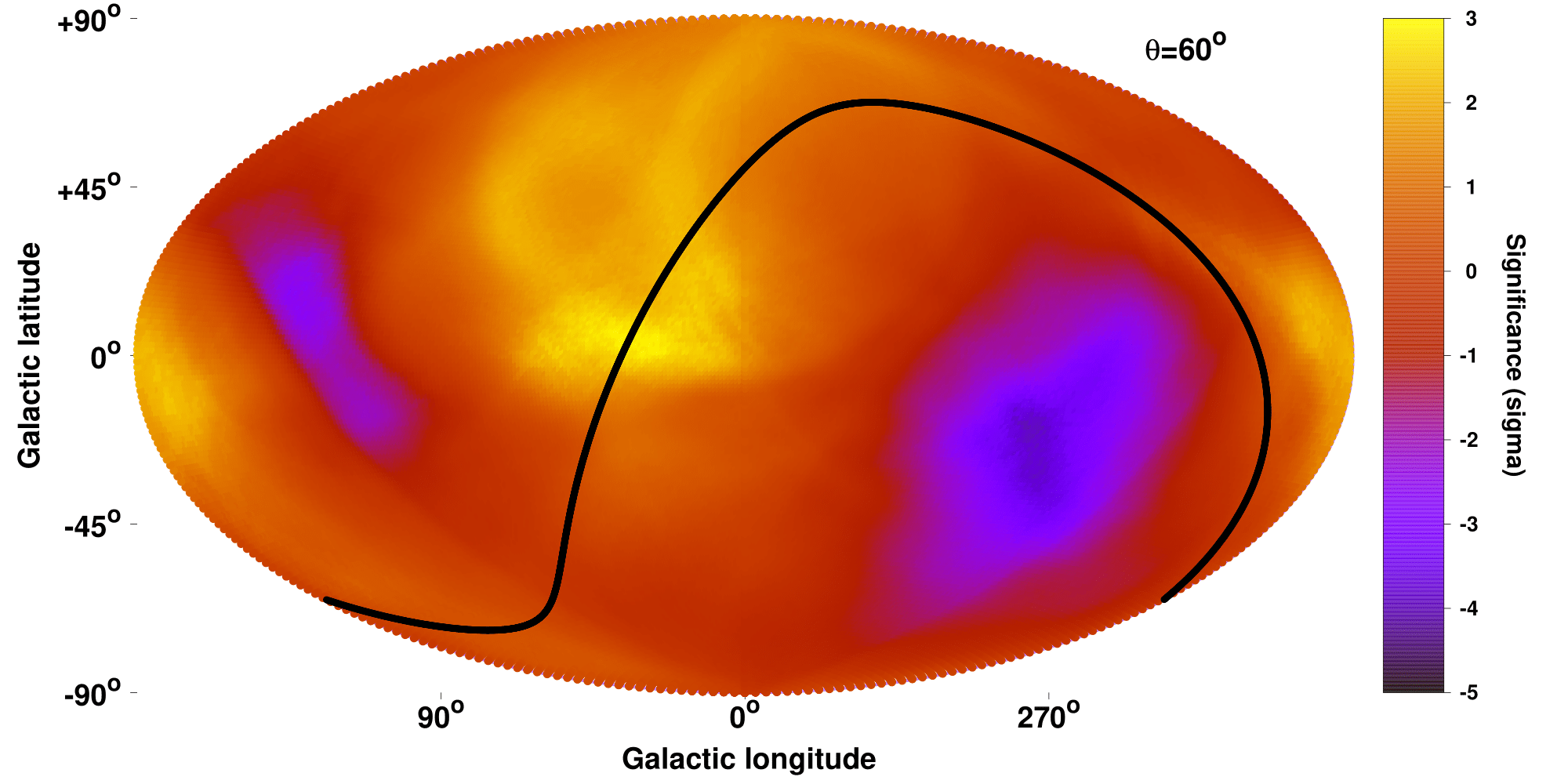}
               \includegraphics[width=0.51\textwidth, height=5cm]{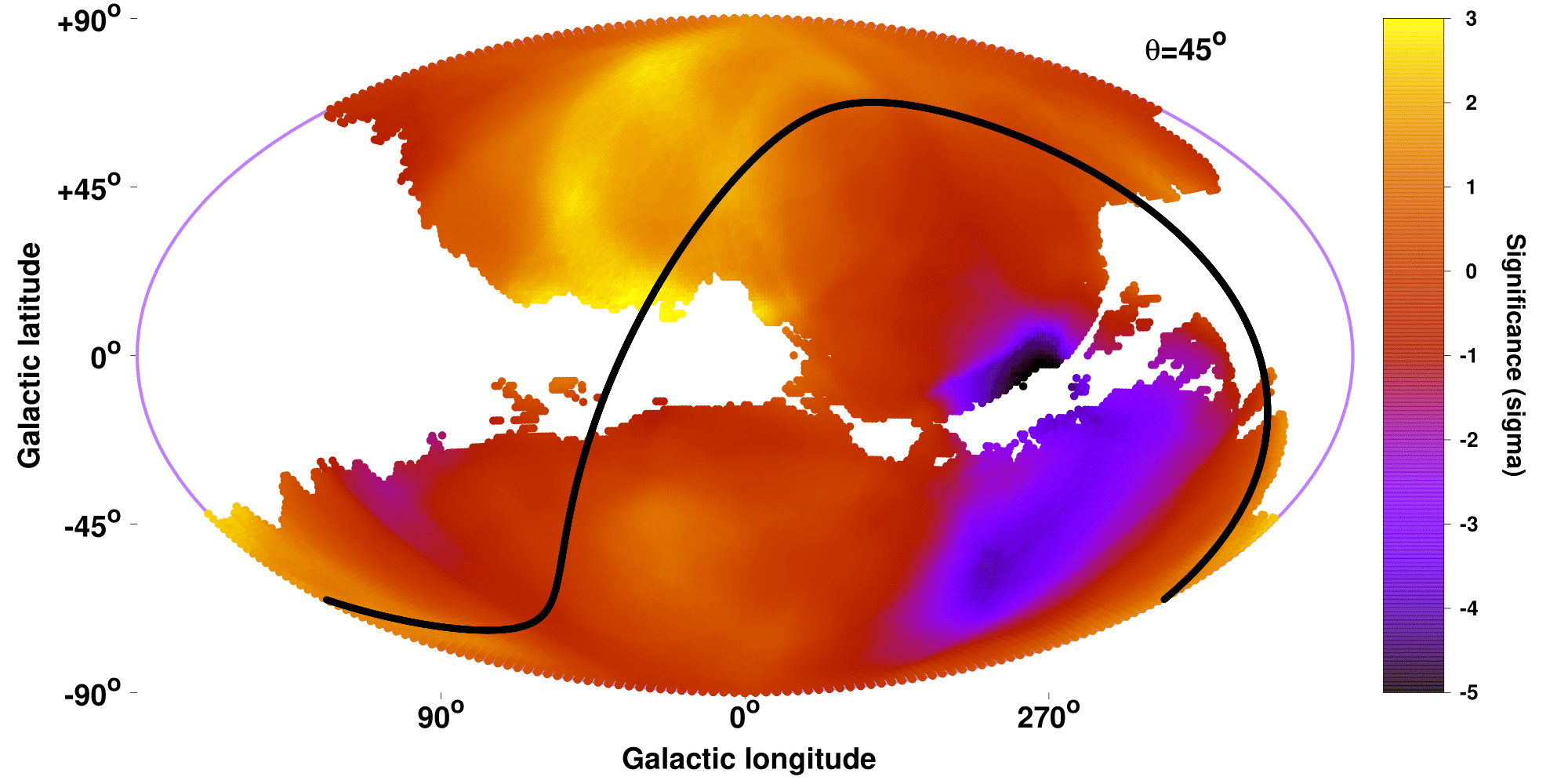}
               \caption{Statistical significance of the deviation of every sky region compared to the rest of the sky as a function of the position in the extragalactic sky. The maps are created with cones of $\theta =90^{\circ}$ (top left), $\theta=75^{\circ}$ (top right), $\theta =60^{\circ}$ (bottom left) and $\theta =45^{\circ}$ (bottom right, only regions with $\geq 35$ clusters are shown). The value of every point is extracted by using all the clusters in the same cone. All maps have the same color scale ($-5\sigma, 3\sigma$). The minus ("-") sign indicates that the corresponding sky region has a lower $A$ than the rest of the sky.}
        \label{sigma_cones}
\end{figure*}

To begin with, we choose a scanning cone with $\theta=90^{\circ}$, meaning we divide the sky in all the possible hemisphere\footnote{Where by "hemisphere" we mean any half of the sky and not "Northern", "Southern" etc.} combinations. The lowest number of clusters in any hemisphere is 109 toward the $(l,b)=(150^{\circ}, -2^{\circ})$ direction, with 204 clusters located in the opposite hemisphere. Constraining $A$ for every hemisphere, one obtains the $A$ and significance color maps displayed in the top left panels of Fig. \ref{fig7} and Fig. \ref{sigma_cones} respectively. 

As shown in the plots, there is mainly one low $A$ region within $\sim 20^{\circ}$ from $(l,b)\sim (270^{\circ}, -5^{\circ})$ with a rather strong behavior. There is also one high $A$ peak. It is noteworthy that the two most extreme regions in the map (deep purple and bright yellow) are located close to the Galactic plane (within $\sim 20^{\circ}$), where there are no observed clusters. The clusters toward the Galactic center in particular seem to be overluminous compared to other sky regions. Of course the color differences are not visually strong here since the color scale was chosen based on the largest deviations, appearing in later maps.

In detail, the most extreme hemispheres are found at $(l,b)=(272^{\circ}, -8^{\circ})$ with $A=1.062\pm 0.048$ and at $(l,b)=(47^{\circ}, +22^{\circ})$ with $A=1.236\pm 0.047$. The angular separation between them is 135$^{\circ}$  while they deviate by 2.59$\sigma$ ($16\pm 6\%$) from each other. Although they are not completely independent, the contribution of the common clusters is not the same to both subsamples due to the applied statistical weighting based on the distance of every cluster from the center of the cone. 
The most extreme dipole (2 independent subsamples separated by $180^{\circ}$ in the sky) appears at $(l,b)=(230^{\circ}, -20^{\circ})$ with a significance of 1.90$\sigma$. This dipole is separated by 75$^{\circ}$ from the CMB dipole, although a dipole interpretation is obviously not reflecting the maximum apparent anisotropies in that case.

\subsubsection{$\theta=75^{\circ}$ cone}

If an anisotropy toward one direction exists, the clusters lying close to that direction would be the most affected ones and as we move further away from that direction, the anisotropic effect on clusters would fade. Therefore, such anisotropic behaviors are better studied if one uses smaller solid angles in the sky. To this end, we decrease the radius of the scanning cone first to $\theta=75^{\circ}$. Indeed, the fluctuations of $A$ as well as the significance of the anisotropies increase, while the general behavior of the directional anisotropies in the $\theta=75^{\circ}$ map however remains relatively unchanged compared to the previous map with a larger cone. The results are displayed in the top right panels of Fig. \ref{fig7} and Fig. \ref{sigma_cones}.  

For the $\theta=75^{\circ}$ cones, $A$ varies from $A=0.999\pm 0.050$ at $(l,b)=(274^{\circ}, -22^{\circ})$ to $A=1.288\pm 0.061$ toward $(l,b)=(17^{\circ}, -9^{\circ})$. These two regions are separated by $99^{\circ}$ and deviate from each other by $3.64\sigma$ ($26\pm 7\%$). Furthermore, the most significant dipole that appears is the one centered at $(l,b)=(263^{\circ}, -21^{\circ})$ at $3.22\sigma$, 68$^{\circ}$ away from the CMB dipole. 

\subsubsection{$\theta=60^{\circ}$ cone}

Further decreasing the size of the solid angles, we use $\theta=60^{\circ}$ cones. We see that the behavior of the sky regions suffers some changes whilst staying generally consistent with the previous results. The most prominent change is the existence of a low $A$ region close to $(l,b)\sim (120^{\circ}, +20^{\circ})$, although its statistical significance (as displayed in Fig. \ref{sigma_cones})  is lower than the other, main low $A$ region since it only contains $\sim 45$ clusters. Another change in the $60^{\circ}$ map is that the brightest part of the sky is shifted toward $(l,b)\sim(170^{\circ},-10^{\circ})$. However, as one can clearly see in Fig. \ref{sigma_cones}, the most statistically significant region with a high normalization remains in the same area as in the previous cases, namely toward $(l,b)=(34^{\circ}, +4^{\circ})$ with 78 clusters and $A=1.346\pm 0.069$.

At the same time, the lowest normalization value $A=0.940\pm 0.051$ is located at $(l,b)=(281^{\circ}, -16^{\circ})$ (84 clusters). The most extreme regions deviate from each other by $4.73\sigma$ ($36\pm 8\%$), which constitutes a considerably strong tension. 
The most extreme dipole in this case is found toward $(l,b)=(260^{\circ}, -36^{\circ})$ with a statistical significance of $3.77\sigma$. 

If we now exclude these two most extreme low and high $A$ regions and their 159 individual clusters from the rest of the sky, we are left with 154 clusters. Performing the fit on these clusters, we obtain $A=1.138\pm 0.048$. We see that the rest of the sky is at a $2.49\sigma$ tension with the bright region toward $(l,b)=(34^{\circ}, +4^{\circ})$, and at a $2.85\sigma$ tension with the faint region toward $(l,b)=(281^{\circ}, -16^{\circ})$. Thus, the anisotropic behavior of the faint region is somewhat more statistically significant than the behavior of the bright region.

\subsubsection{$\theta=45^{\circ}$ cone}

The last cone we use has $\theta=45^{\circ}$. Since there are many regions mostly close to the Galactic plane, with fewer clusters than needed in order to obtain a trustworthy result, we enforce an extra criterion. We only consider regions with $\geq 35$ clusters\footnote{Arbitrary low limit number that provides a satisfactory balance between number of regions available and sufficient insensitivity to outliers}. The $A$ and $\sigma$ maps are shown in the bottom right panels of Fig. \ref{fig7} and Fig. \ref{sigma_cones} respectively. The white regions show the regions without enough clusters for a reliable fit. The most extreme regions are found toward $(l,b)=(280^{\circ}, +1^{\circ})$ (42 clusters) and $(l,b)=(32^{\circ}, +14^{\circ})$ (40 clusters) with $A=0.822\pm 0.067$ and $A=1.413\pm 0.095$. The statistical discrepancy between the rises to 5.08$\sigma$ ($53\pm 10\%$) being the most statistically significant result up to now. 

Additionally, the most extreme dipole is centered at $(l,b)=(255^{\circ}, -53^{\circ})$ with a significance of $4.22\sigma$. However, it should be beared in mind that many regions that appeared to have the maximum dipoles for other cones are excluded now due to low number of clusters. This could lead the maximum dipole to shift toward lower Galactic latitudes on the low normalization side. Moreover, due to the low number of clusters in these regions the results are more sensitive to outliers, especially when these outliers are located close to the center of the regions where they have more statistical weight than other clusters. Nevertheless, the large statistical tension cannot be neglected.


Excluding once again the two extreme regions from the rest of the sample we are left with 234 clusters which have a best-fit of $A=1.107\pm 0.041$. Thus, they are in a $2.96\sigma$ tension with the brightest region and in a $3.63\sigma$ with the faintest region. Once again, the region toward $(l,b)=(280^{\circ}, +1^{\circ})$ seems to be more anisotropic than the one toward $(l,b)=(24^{\circ}, +16^{\circ})$.

\subsubsection{Overview of results}

As a summary of the above, we identify the clear existence of a region with galaxy clusters appearing systematically fainter than expected based on their temperature measurements. This region is roughly located at $(l,b)\sim (277^{\circ}\pm 5^{\circ}, -11^{\circ}\pm 12^{\circ}$). On the contrary, the systematically brightest region is found toward $(l,b)\sim (32^{\circ}\pm 15^{\circ}, +8^{\circ}\pm 17^{\circ})$. Their angular separation in the sky is $\sim 115^{\circ}$. The statistical tension between these two regions rises significantly while narrower cones are considered, reaching $\sim5\sigma$ for the smaller cones. The same is true for the dipole anisotropies, going up to $\sim 4\sigma$.
Interestingly enough, the same behavior for this sky patch is also detected for ACC and XCS-DR1 (see Sect. \ref{joint_an}). 
Another interesting trend is the systematically bright region at $(l,b)\sim (175^{\circ}\pm 15^{\circ}, +5^{\circ}\pm 20^{\circ}$, which appears to have the same behavior in all three maps with the larger scanning cones. Unfortunately, not enough available clusters lie there for the $45^{\circ}$ map to return reliable results. 

The most statistically significant dipole anisotropy is consistently found toward $(l,b)\sim (253^{\circ}\pm 13^{\circ}, -32^{\circ}\pm 15^{\circ})$, lying $\sim 30^{\circ}\pm 25^{\circ}$ away from the systematically fainter sky region. Finally, the correlation of these results with the CMB dipole is not strong since the faintest regions of our analysis are found $\sim 55^{\circ}-75^{\circ}$ away from the CMB's corresponding dipole end, while the strongest anisotropic dipoles of the $L_{\text{X}}-T$ relation are located $\sim 70^{\circ}-85^{\circ}$ away from the CMB one. 

\section{Possible X-ray and cluster-related causes and consistency of anisotropies}\label{possible_causes}

Galaxy clusters are complex systems where many aspects of physics come into play when one wishes to analyze them. Thus, we have to investigate if the apparent anisotropies are caused by any systematic effects. 
With a purpose of trying to identify the reason behind these strong $L_{\text{X}}-T$ anisotropies, we perform an in-depth analysis using different subsamples of the 313 clusters which are chosen based on their physical properties. 
If the best-fit $L_{\text{X}}-T$ relation of galaxy clusters significantly differs for clusters with different physical parameters (e.g., low and high $T$ or $Z$ clusters, different $N_{\text{Htot}}$ values etc.), a nonuniform sky distribution of such clusters could create artificial anisotropies.

\subsection{Excluding galaxy groups and low-$z$ clusters}\label{low_T_z_clusters}

It has been shown that the low-$T$ clusters (mainly galaxy groups) can sometimes exhibit a slightly different $L_{\text{X}}-T$ behavior compared to the most massive and hotter systems \citep[e.g.,][and references therein]{lovisari}. We wish to test if this possibly different behavior has any effects on the apparent anisotropies.

Hence, we first excluded all the systems below $T\leq 2.5$ keV. Moreover, all the clusters within $\sim 130$ Mpc ($z\leq 0.03$) were excluded in order to avoid the peculiar velocity effects on the measured redshift (the vast majority of these clusters are already excluded based on the $T\leq 2.5$ keV limit). This resulted in the exclusion of 67 objects. 

We applied the necessary correction to convert our heliocentric redshifts to "CMB frame" redshifts. This conversion is not expected to cause any significant changes in our results for two reasons. Firstly, the spatial distribution of our sample is rather uniform, therefore only $\sim 25\%$ of this subsample's clusters are located within $30^{\circ}$ from the CMB dipole for which this correction might have a notable impact. Secondly, due to the low-$z$ cut we apply here, the CMB frame redshift correction is much smaller than the cosmological recession velocity ($\lesssim 4\%$). Hence, the final propagated correction to the $L_{\text{X}}$ values is far less than the observed anisotropies. Nevertheless, we transformed the redshifts for the sake of completeness.

When we fit the $L_{\text{X}}-T$ relation to all the 246 clusters with $T> 2.5$ keV and $z>0.03$, we obtain the following best-fit values: 
\begin{equation}
A=1.114^{+0.047}_{-0.041}, \ B=2.096\pm 0.078, \ \text{and}\ \sigma_{\text{int}}=0.218\ \text{dex}.
\end{equation}
It is noteworthy that $A$ and $B$ remain unchanged compared to the case where all the 313 clusters are considered. This indicates that a single power law model can be an efficient option for fitting our sample. The most clear difference of this subsample fitting is the decrease of the intrinsic scatter by $10\%$. The total scatter also goes down by the same factor ($\sigma_{\text{tot}}$=0.236 dex).  The $3\sigma$ solution spaces for the entire sample and for these 246 clusters are displayed in Fig. \ref{bulk3}, being entirely consistent.
 
\begin{figure}[hbtp]
               \includegraphics[width=0.51\textwidth, height=5cm]{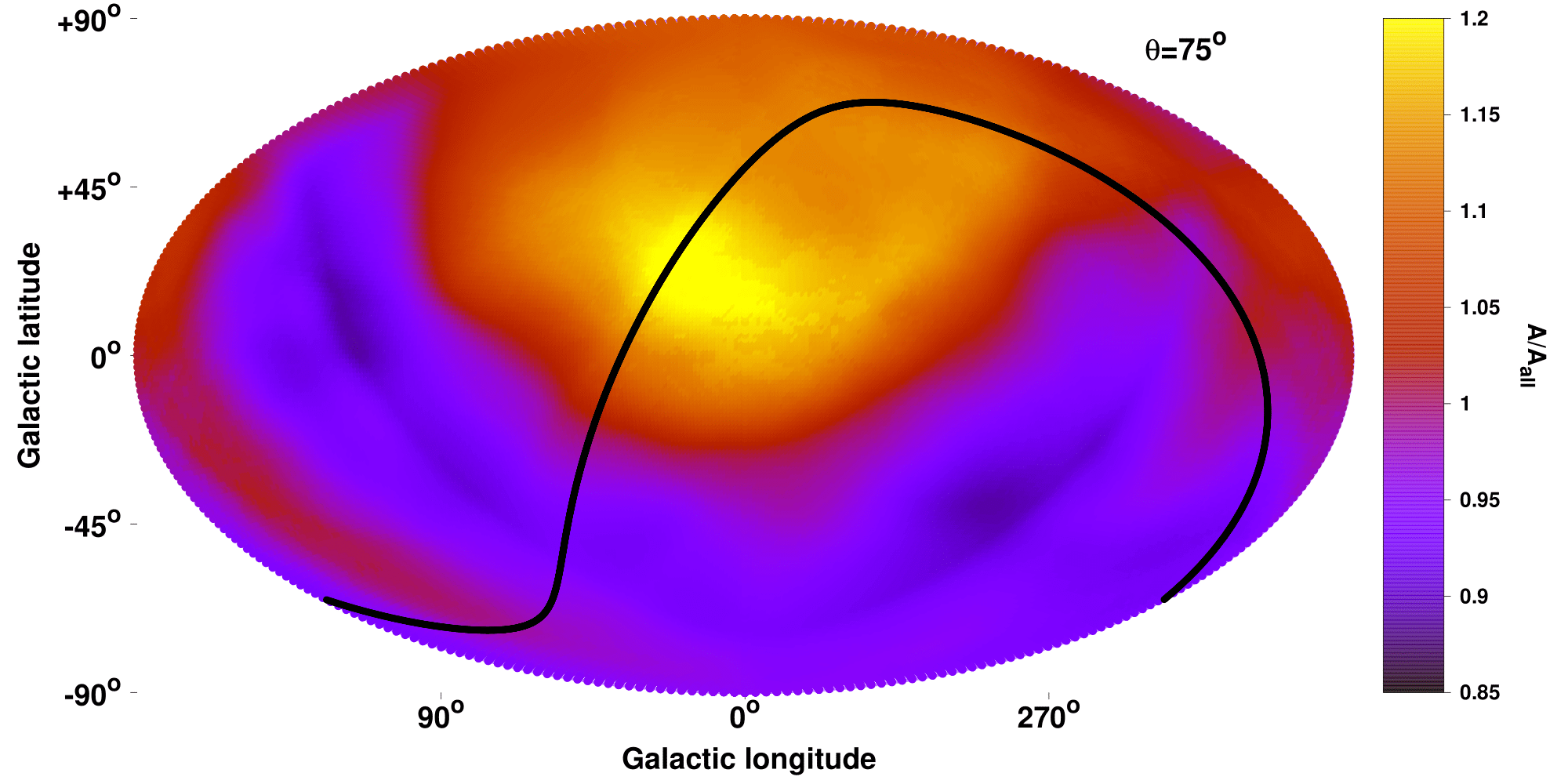}
               \includegraphics[width=0.51\textwidth, height=5cm]{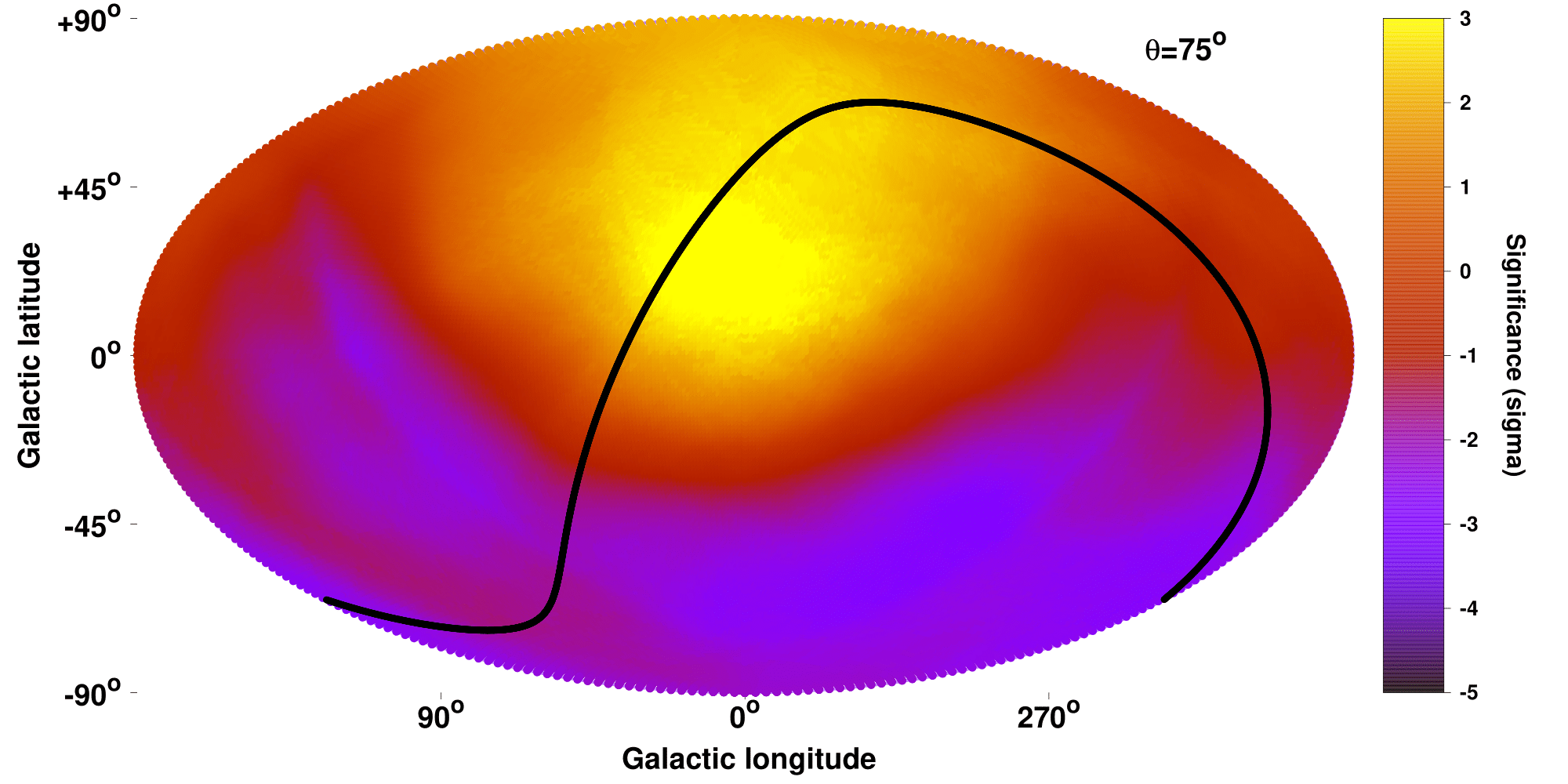}
               \caption{Normalization $A$ of the $L_{\text{X}}-T$ relation (top) and statistical significance of the deviation of every sky region compared to the rest of the sky (bottom) as functions of the position in the extragalactic sky for $\theta=75^{\circ}$ when only the 246 clusters with $T>2.5$ keV and $z>0.03$ are used, as well as CMB frame redshifts.}
        \label{bulk}
\end{figure}

Performing the 2D scanning of the full sky using $\theta=75^{\circ}$, the $A$ map shown in the bottom panel of Fig. \ref{bulk} is produced. A similar pattern with the previous maps persist, although there are some changes. The main differences are that the --statistically insignificant-- bright region toward $(l,b)\sim(170^{\circ}, -10^{\circ})$ vanishes whilst the behavior of the faint region toward $(l,b)\sim (120^{\circ}, +10^{\circ})$ seems to be amplified. Despite of that, the most statistically significant low-$A$ regions approximately remains in the sky patch that was found before, toward $(l,b)=(288^{\circ}, -35^{\circ})$ with $A=1.016\pm 0.045$ (110 clusters). The most extreme high-$A$ sky region is again consistent with our previous findings, lying at $(l,b)=(10^{\circ}, +16^{\circ})$ with $A=1.371\pm 0.061$ (113 clusters). The statistical discrepancy between these two results is $4.68\sigma$ ($32\pm 7 \%$), not being alleviated by the exclusion of these groups and local clusters. Their angular separation in the sky is $93^{\circ}$. The most extreme dipole for this map is found toward $(l,b)=(196^{\circ}, -34^{\circ})$ with $3.27\sigma$, shifted compared to the previously found most extreme dipole regions by $\sim 45^{\circ}\pm 28^{\circ}$.

\begin{figure}[hbtp]
               \includegraphics[width=0.51\textwidth, height=5cm]{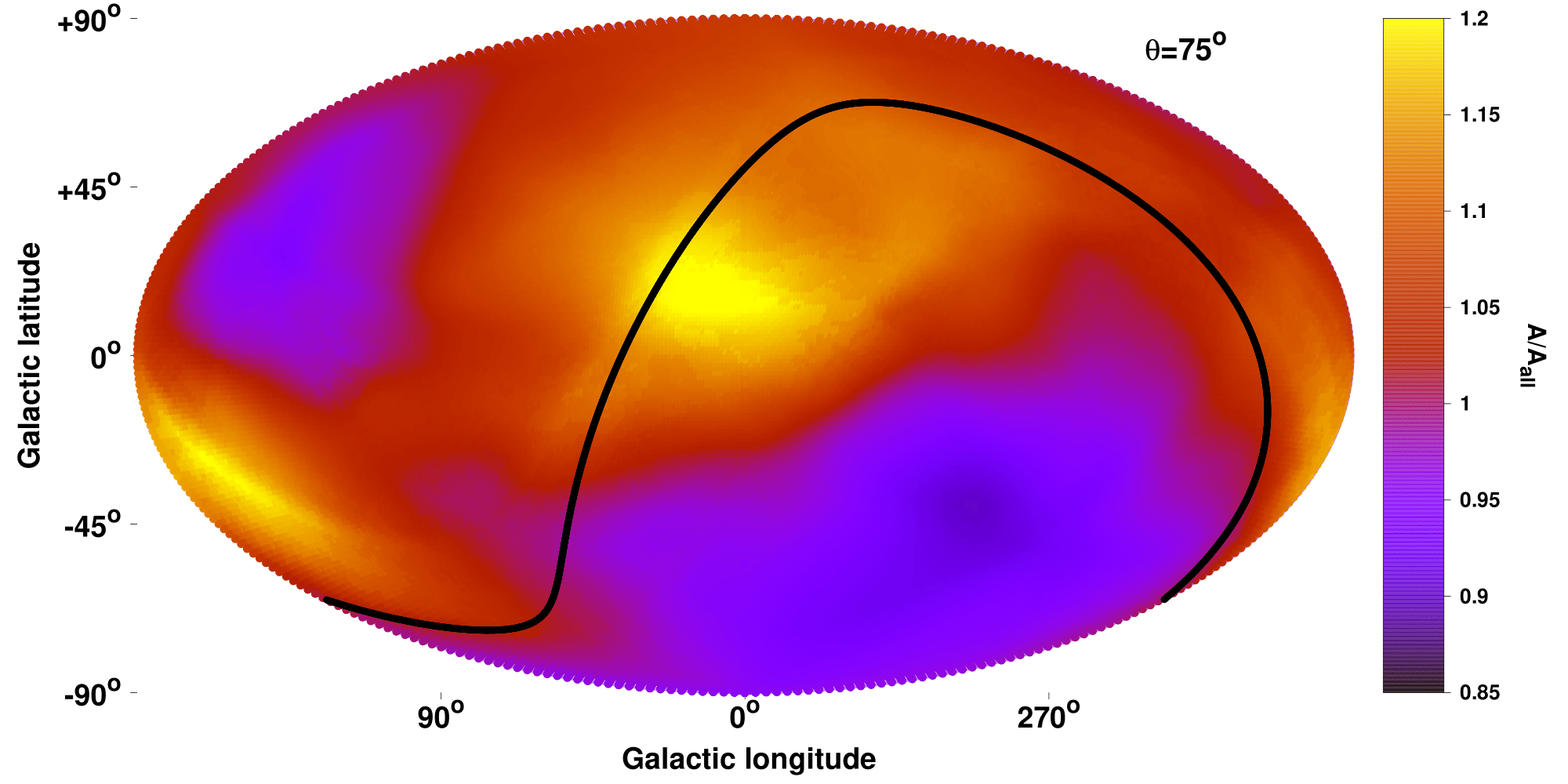}
               \includegraphics[width=0.51\textwidth, height=5cm]{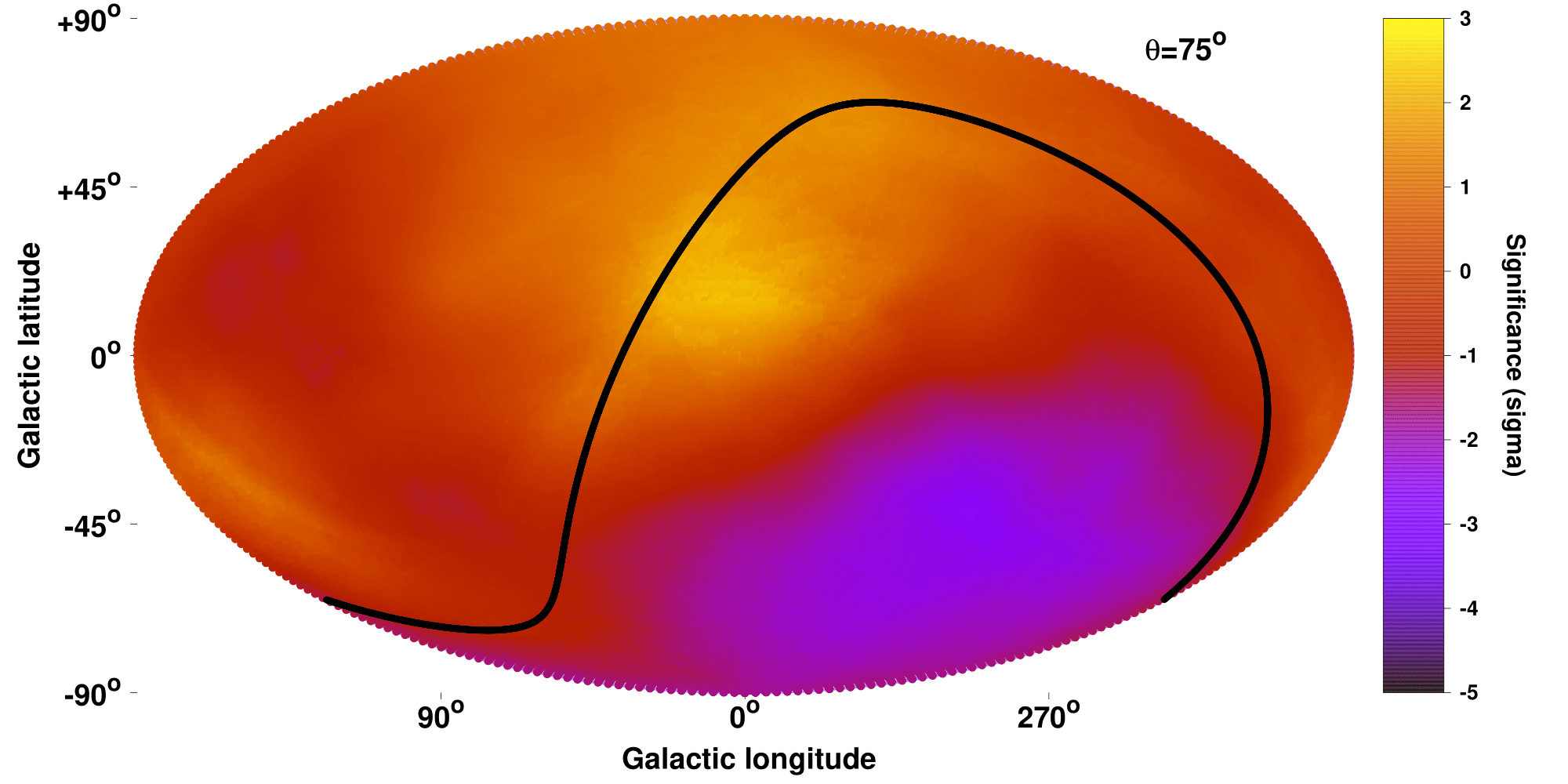}
               \caption{Same as in Fig. \ref{bulk} for the 198 clusters with with $T>3$ keV and $z>0.05$.}
        \label{bulk2}
\end{figure}

To further scrutinize the effects of low-$T$ systems on our results, as well as the effects of local clusters and their peculiar velocities, we wish to restrict our sample even more by expanding the lower limits of $T$ and $z$. To this end, we excluded all the 115 objects with $T<3$ keV or $z<0.05$ ($\sim 210$ Mpc). This left us with 198 clusters. The best-fit results are:

\begin{equation}
A=1.172^{+0.053}_{-0.046}, \ B=2.049\pm 0.077, \ \text{and}\ \sigma_{\text{int}}=0.205\ \text{dex}.
\end{equation}
The best-fit $L_{\text{X}}-T$ relation slightly changes compared to the full sample results, but remains consistent within $1.1\sigma$. At the same time, $\sigma_{\text{int}}$ further decreases, being $15\%$ lower than the full sample's $\sigma_{\text{int}}$. In Fig. \ref{bulk3}, the comparison between the $3\sigma$ solution spaces for the full sample and for these 198 clusters is displayed. In the panel of Fig. \ref{bulk2} the $A$ map is displayed for this subsample of clusters, with a $\theta=75^{\circ}$ cone. The significance map is shown in the bottom panel of the same figure.

The behavior of $A$ throughout the sky remains consistent with the previous results, even after excluding more low-$T$ clusters and using only clusters with $z>0.05$ with CMB-frame $z$ values. The lowest $A=1.081\pm 0.054$ is found toward $(l,b)=(286^{\circ}, -36^{\circ})$ (85 clusters) while the highest $A=1.445\pm 0.070$ is located toward $(l,b)=(9^{\circ}, +15^{\circ})$ (91 clusters). The statistical tension between these two results is $4.12\sigma$ ($31\pm 8\%$). The most extreme dipole on the other hand is centered toward $(l,b)=(223^{\circ}, -47^{\circ})$ with a relatively low significance of $2.27\sigma$. This highlights the fact that the most extreme behavior in the sky is not found in a dipole form, and this becomes more obvious as we go to higher redshifts.
Consequently, it is quite safe to conclude that this anisotropic behavior is caused neither by the galaxy groups or the local clusters nor by the use of heliocentric redshifts.

\subsection{Different cluster metallicities} \label{metallicity}


A slightly nonsimilar behavior of the $L_{\text{X}}-T$ relation for varying metallicities of clusters can be expected mainly due to two factors. Firstly, in the parent catalogs from which our cluster sample has been constructed, the conversion of the count-rate to flux was done by using a fixed metal abundance of $0.3\ Z_{\odot}$. When the true metallicity of a cluster deviates from this fixed value, small biases can propagate in the flux and luminosity determination. In general, the measured luminosity of clusters with $Z>0.3\ Z_{\odot}$ might be eventually slightly underestimated. However, this overestimation is only minimal. For instance, for $\Delta Z\sim 0.4\ Z_{\odot}$ between fixed and true $Z$, the final flux changes by $\sim 0-2\%$, where the exact change depends on the other cluster parameters, such as the temperature.

The second and most important factor is that clusters with higher $Z$ values tend to be intrinsically brighter when the rest of the physical parameters are kept constant. This can be shown through an \textit{apec} model simulation in XSPEC. Even for a small deviation of $\Delta Z\sim 0.1\ Z_{\odot}$ the flux of a cluster can fluctuate by $\gtrsim 17\%$ for a cluster with $T\lesssim 1$ keV, while this fluctuation becomes only $\lesssim 1\%$ for a cluster with $T\gtrsim 8$ keV. Therefore, a randomly different metallicity distribution between different sky regions could in principle cause small anisotropies. However, in order for the observed anisotropies to be purely caused by that, strong inhomogeneities in the metallicity distribution should exist, which, if detected, would be a riddle of its own.

\subsubsection{Core metallicities within $0-0.2\times R_{500}$}

Galaxy clusters do not show a single metallicity component. Since we wish to focus first on the effects that a varying metallicity could have on the luminosity, we consider the metallicity of the core of the cluster ($Z_{\text{core}}$, where by "core" we mean $0-0.2\times R_{500}$) from where the bulk of the X-ray luminosity comes from. It is also expected that the clusters with the higher $Z_{\text{core}}$ values would have a higher fraction of cool-core members, which are generally more luminous than non cool-core clusters for the same $T$ \citep[e.g.,][]{mittal}.

In order to investigate the behavior of the $L_{\text{X}}-T$ relation as a function of the metallicity of the galaxy clusters, we divided our sample into three subsamples based on their $Z_{\text{core}}$ value. Our only criterion for this division was the equal number of clusters in each subsample. These subsamples are 105 clusters with $Z_{\text{core}}\leq 0.452\ Z_{\odot}$, 104 clusters with $0.452\ Z_{\odot}<Z_{\text{core}}\leq 0.590\ Z_{\odot}$ and 104 clusters with $Z_{\text{core}}>0.590\ Z_{\odot}$. For each subsample, we perform the fitting letting $A$ and $B$ to vary. The following results are not particularly sensitive to the exact $Z_{\text{core}}$ limits. 

\begin{figure}[hbtp]
               \includegraphics[width=0.5\textwidth, height=6cm]{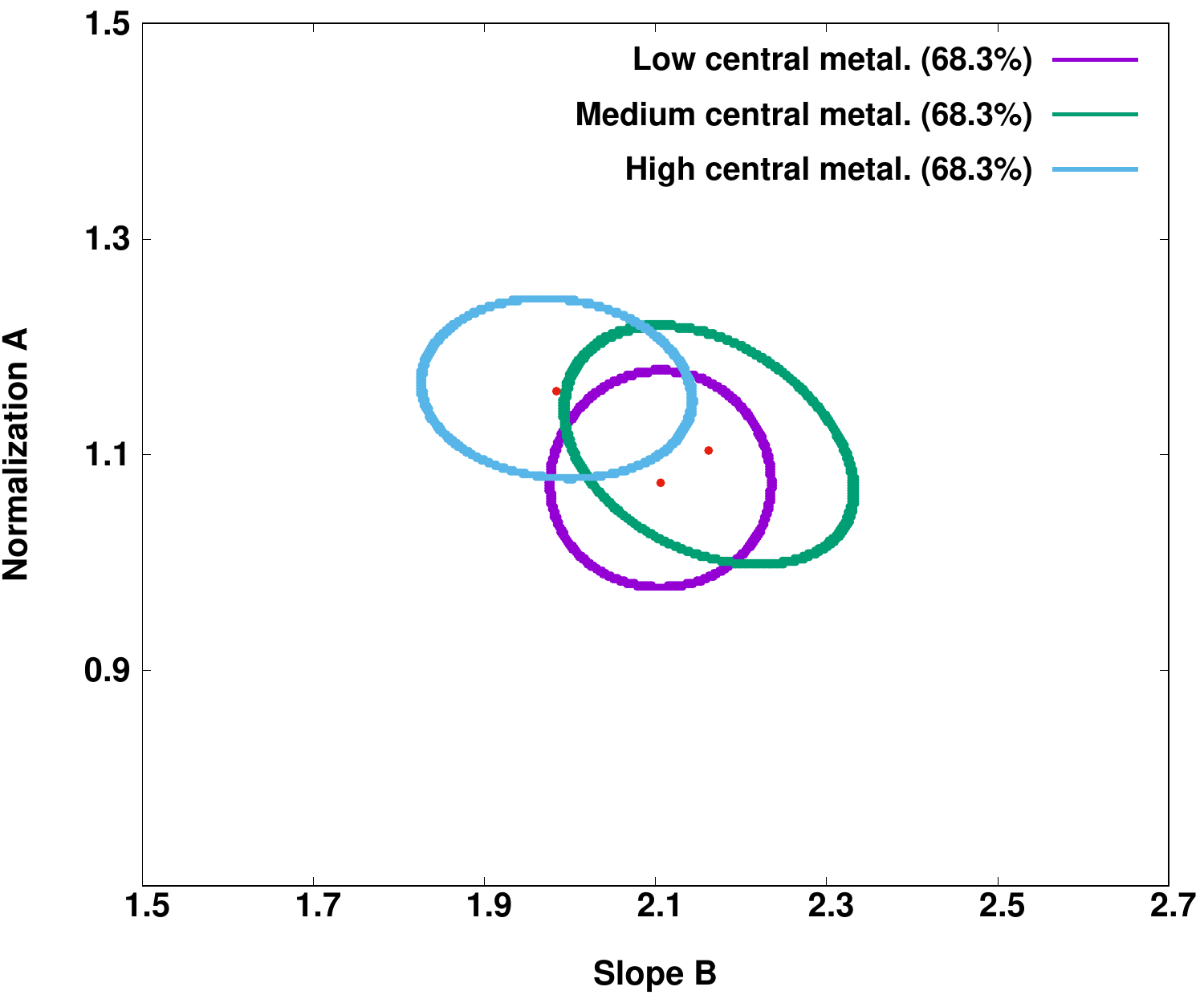}
               \caption{$1\sigma$ (68.3\%) confidence levels of the normalization and slope of the $L_{\text{X}}-T$ relation as derived for the 105 clusters with $Z_{\text{core}}\leq 0.452\ Z_{\odot}$ (purple), the 104 clusters with $0.452\ Z_{\odot}<Z_{\text{core}}\leq 0.590\ Z_{\odot}$ (green) and the 104 clusters with $Z_{\text{core}}>0.590\ Z_{\odot}$ (cyan).}
        \label{ab-metal-core}
\end{figure}

The 1$\sigma$ solution spaces for each subsample are shown in Fig. \ref{ab-metal-core}. One can see that all the three subsamples share a very similar $L_{\text{X}}-T$ solution. The maximum statistical deviation of $\sim 1.12\sigma$ is found between the two subsamples with the lowest and highest $Z_{\text{core}}$, with the latter being slightly more luminous on average. Furthermore, the intrinsic scatter for the two subsamples with the lower $Z_{\text{core}}$ is $\sigma_{\text{int}}\sim 0.260$ dex while for the high-$Z_{\text{core}}$ subsample is $\sigma_{\text{int}}\sim 0.197$ dex.

\begin{figure}[hbtp]
               \includegraphics[width=0.51\textwidth, height=5cm]{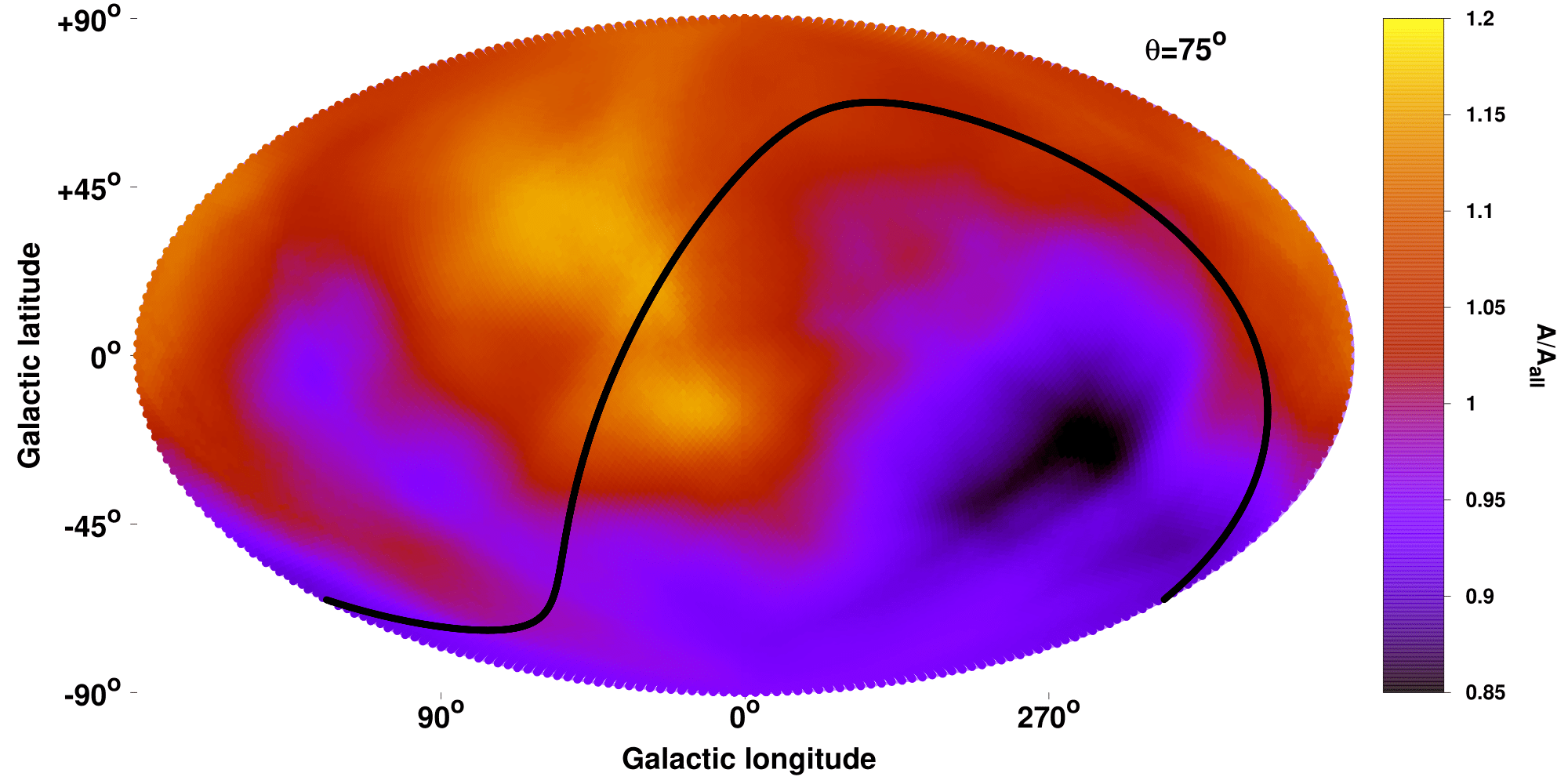}
               \includegraphics[width=0.51\textwidth, height=5cm]{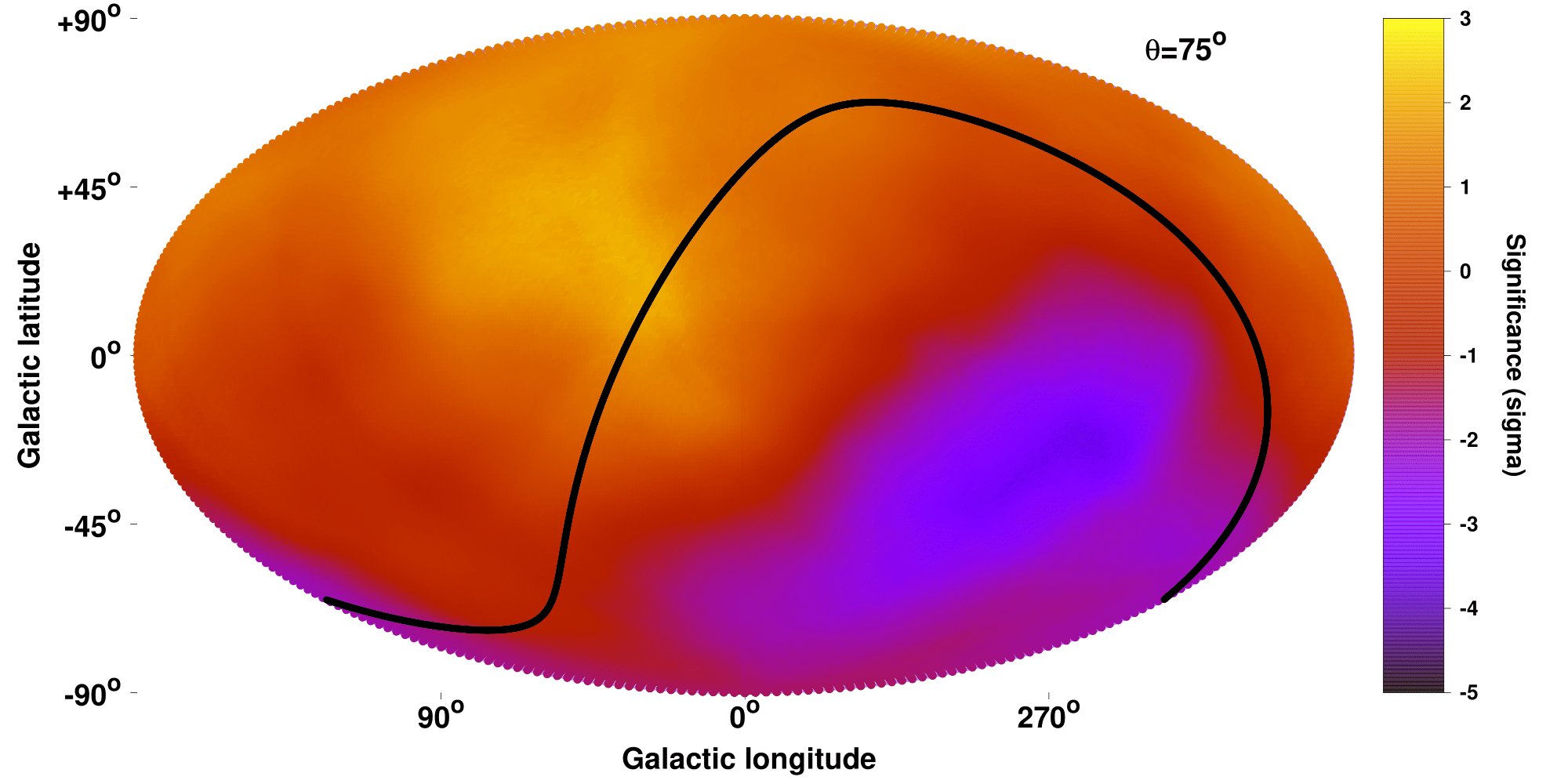}
               \caption{Same as in Fig. \ref{bulk} for the 209 clusters with $Z_{\text{core}}<0.590\ Z_{\odot}$.}
        \label{ab-metal-core-maps}
\end{figure}

Although it is not expected that the high $Z_{\text{core}}$ subsample would cause any apparent $A$ anisotropies with a possibly nonhomogeneous spatial sky coverage, for the sake of completeness we excluded all the 104 clusters with $Z_{\text{core}}>0.590\ Z_{\odot}$ and scanned the sky again with a $75^{\circ}$ radius cone. The produced $A$ and significance maps are illustrated Fig. \ref{ab-metal-core-maps}. The obtained directional behavior of $A$ completely matches the results of the full sample. The lowest $A=0.927\pm0.064$ and highest $A=1.274\pm0.071$ are found toward  $(l,b)=(264^{\circ}, -18^{\circ})$ (83 clusters) and  $(l,b)=(30^{\circ}, +23^{\circ})$ (88 clusters) respectively. Their deviation is $3.63\sigma$ ($32\pm 9\%$), staying unchanged despite the smaller number of available clusters. The most extreme dipole is found toward  $(l,b)=(261^{\circ}, -20^{\circ})$ with $3.41\sigma$ significance.

\subsubsection{Outer metallicities within $0.2-0.5\times R_{500}$}\label{outer_metal}

The metallicity $Z_{\text{out}}$ of the $0.2-0.5\times R_{500}$ annulus might not affect the final $L_{\text{X}}$ as strongly as the core metallicity. However, it could in principle correlate with the measured temperature of a galaxy cluster since these two quantities were fitted simultaneously. To check if there is an inconsistent $L_{\text{X}}-T$ behavior based on $Z_{\text{out}}$ we follow the same procedure as for $Z_{\text{core}}$, dividing the full sample into three subsamples similarly with before. These subsamples are 105 clusters with $Z_{\text{out}}\leq 0.320\ Z_{\odot}$, 104 clusters with $0.320\ Z_{\odot}<Z_{\text{out}}\leq 0.426\ Z_{\odot}$ and 104 clusters with $Z_{\text{out}}>0.426\ Z_{\odot}$. 

\begin{figure}[hbtp]
               \includegraphics[width=0.5\textwidth, height=6cm]{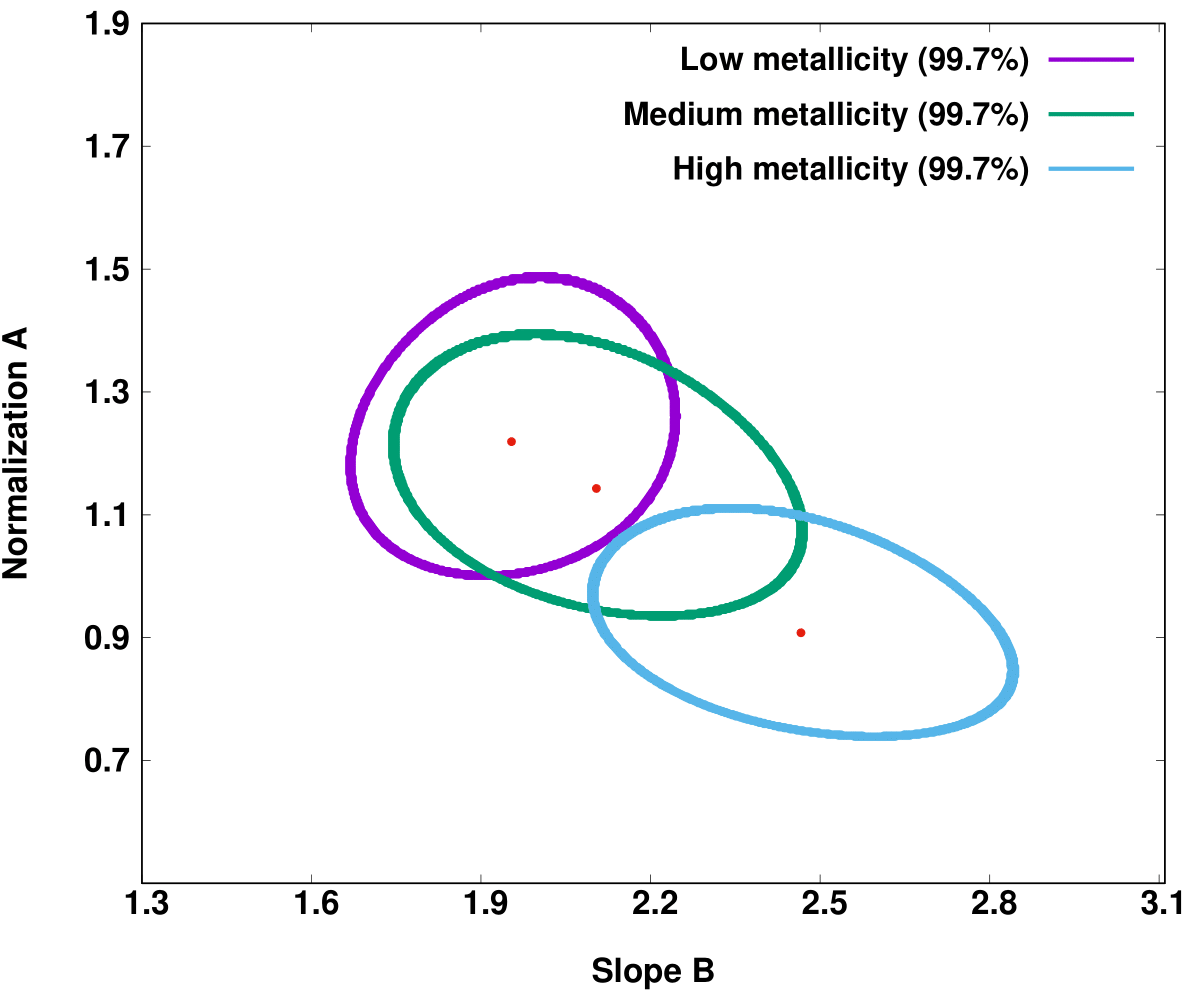}
                \includegraphics[width=0.5\textwidth, height=6cm]{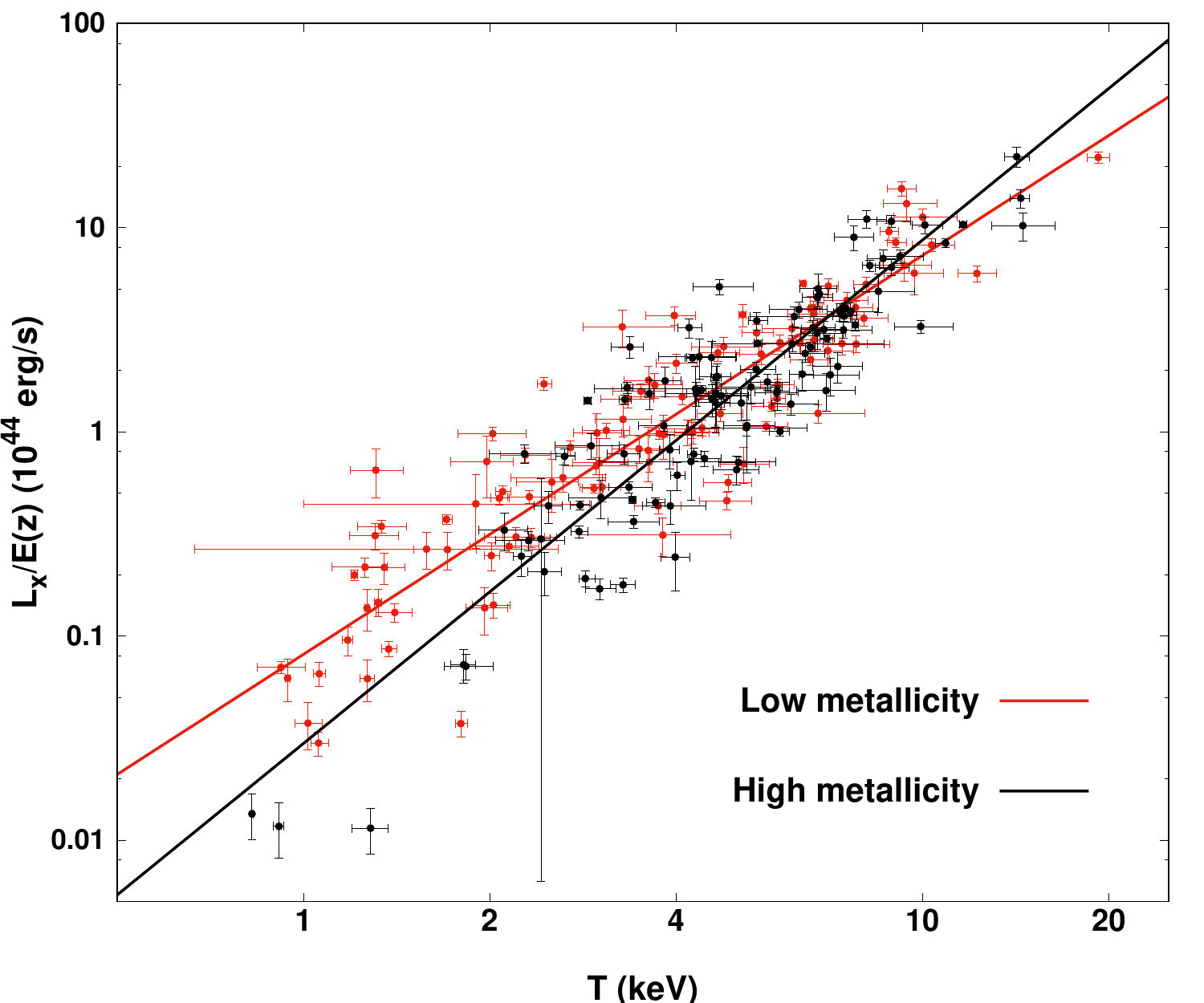}
               \caption{\textit{Top}: $3\sigma$ (99.7\%) confidence levels of the normalization and slope of the $L_{\text{X}}-T$ relation as derived for the 105 clusters with $Z_{\text{out}}\leq 0.320\ Z_{\odot}$ (purple), the 104 clusters with $0.320\ Z_{\odot}<Z_{\text{out}}\leq 0.426\ Z_{\odot}$ (green) and the 104 clusters with $Z_{\text{out}}>0.420\ Z_{\odot}$ (cyan). \textit{Bottom}: $L_{\text{X}}-T$ relation for the 105 clusters with $Z_{\text{out}}\leq 0.320\ Z_{\odot}$ (red) and for the 104 clusters with $Z_{\text{out}}>0.420\ Z_{\odot}$ (black) with their best-fit models.}
        \label{ab-metal}
\end{figure}

In the top panel of Fig. \ref{ab-metal} the $99.7\%$ ($3\sigma$) solution spaces for the three subsamples are shown. It is obvious that the 104 clusters with the highest $Z_{\text{out}}$ share a significantly different $L_{\text{X}}-T$ solution than the 105 clusters with the lowest $Z_{\text{out}}$. The statistical deviation between these two subsamples is $\sim 4.3\sigma$. However, as shown in the bottom panel of Fig. \ref{ab-metal}, the main source of deviation are the local, low temperature groups. Excluding objects with $T<2$ keV and $z<0.02$, the deviation between low and high $Z$ clusters drops to $2.5\sigma$, which is still a nonnegligible tension.
At the same time, the medium $Z_{\text{out}}$ subsample seems to be consistent with the low $Z$ subsamples while also being in tension with the high $Z$ clusters. Furthermore, the intrinsic scatter remains similar for all three $Z_{\text{out}}$ subsamples ($\sim 0.230-0.245$ dex).

\begin{figure}[hbtp]
               \includegraphics[width=0.51\textwidth, height=5cm]{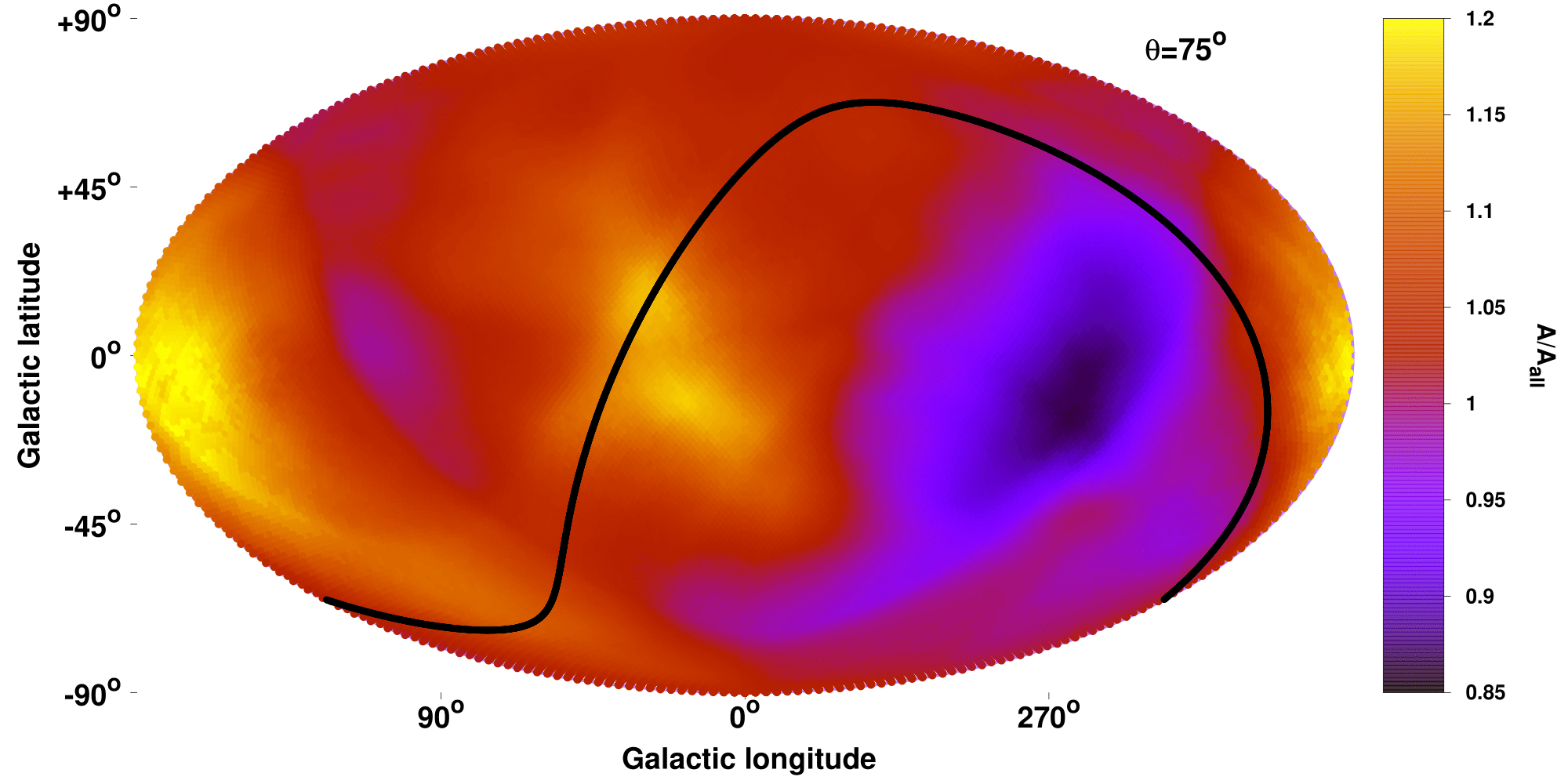}
               \includegraphics[width=0.51\textwidth, height=5cm]{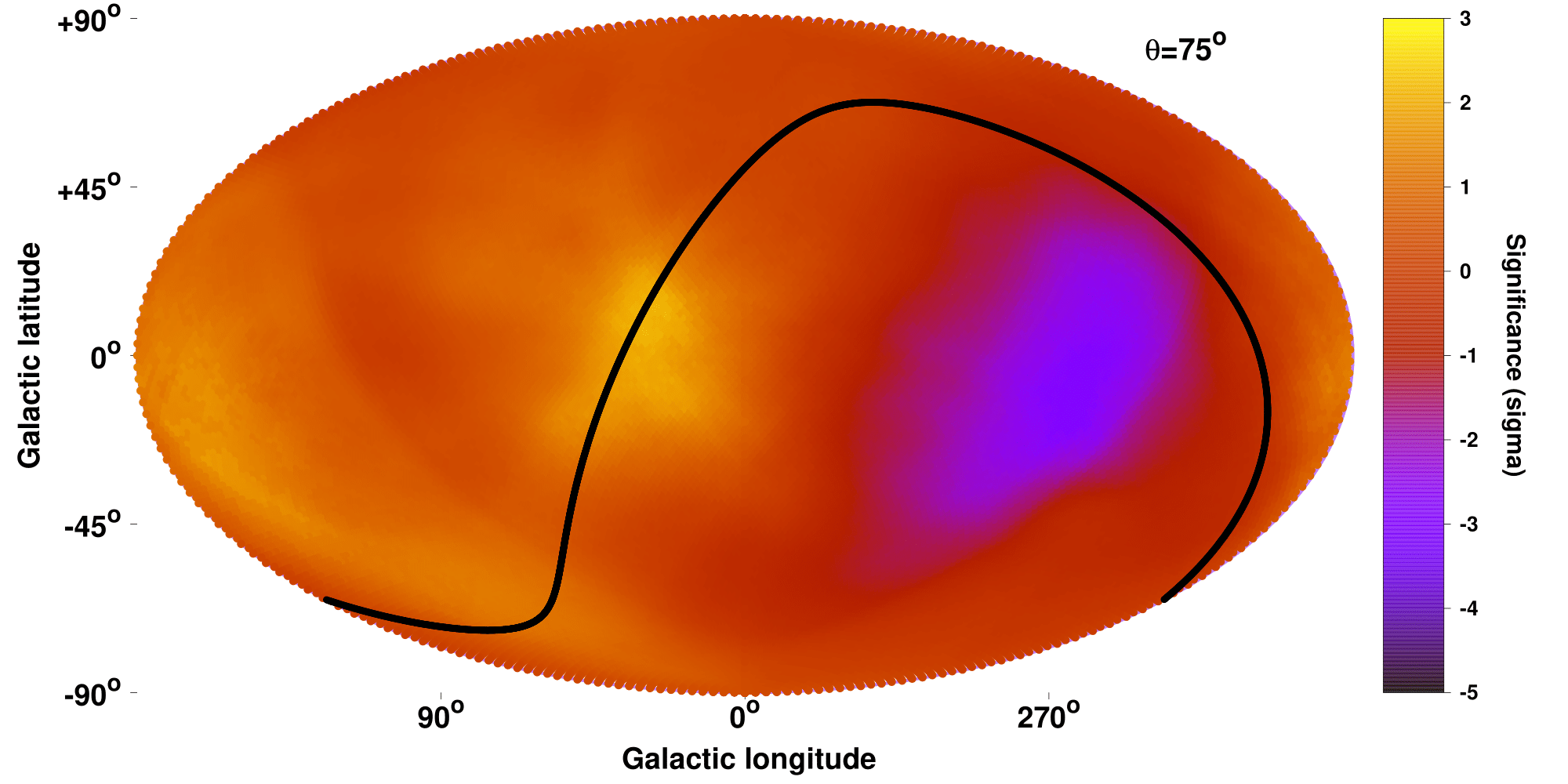}
               \caption{Same as in Fig. \ref{bulk} for the 209 clusters with $Z_{\text{out}}<0.426\ Z_{\odot}$. }
        \label{ab-metal-maps}
\end{figure}

With the purpose of examining whether the strong $A$ anisotropies are affected in any way by these $Z$-dependent different $L_{\text{X}}-T$ behaviors, we once again excluded all the 104 clusters with $Z_{\text{out}}>0.426\ Z_{\odot}$ and performed the usual sky scanning with a $75^{\circ}$ radius cone. In Fig. \ref{ab-metal-maps} the results are displayed.

The similarity with the full sample $\theta=75^{\circ}$ result is striking. The lowest and highest $A$ directions are $(l,b)=(270^{\circ}, -14^{\circ})$ (80 clusters) and $(l,b)=(24^{\circ}, +15^{\circ})$ (93 clusters). The direction $(l,b)=(174^{\circ}, -12^{\circ})$ (58 clusters) is actually brighter by $\sim3\%$ but its statistical significance is lower, which is similar to the results of previous maps. The $A$ values of the most extreme regions are $A=1.035\pm 0.069$ and $A=1.390\pm 0.066$ respectively with a tension of $3.72\sigma$ ($30\pm 8\%$), not relieved despite the exclusion of the $Z_{\text{out}}$ subsample with the significantly different $L_{\text{X}}-T$ behavior. The most extreme dipole is located toward $(l,b)=(265^{\circ}, -16^{\circ})$ but with a lower statistical significance of $2.26\sigma$. Consequently, the derived anisotropies persist when the high $Z$ clusters are excluded, while the significance of the dipolar anisotropy drops by $\sim 1\sigma$ compared to the full sample results.

\subsection{Absorption correction} \label{galaxy_H}

Another possible systematic effect resulting in the observed anisotropies could be the inaccurate treatment of the $N_{\text{Htot}}$ column density correction in our \textit{apec} model. This could lead to systematic differences in the $L_{\text{X}}-T$ values of clusters in directions with different $N_{\text{H}}$. Since also the most extreme regions always lie within $35^{\circ}$ from the Galactic plane, we have to ensure that the apparent anisotropies are not caused by such effects. There are two main cases for which a systematic bias could be introduced through the absorption correction and they are described in the following subsections. 

\subsubsection{Consistency throughout $N_{\text{Htot}}$ range}\label{nh_range_cons}

The first case is that the $N_{\text{Htot}}$ value does not trace the true absorption consistently throughout the full $N_{\text{Htot}}$ range. Thus, clusters in regions with different amounts of hydrogen get a systematically different boost in their $L_{\text{X}}-T$ values after the applied correction. 


This can be easily checked by comparing the $L_{\text{X}}-T$ scaling relation for the clusters with low and high $N_{\text{Htot}}$. To this end, we divided our sample into three subsamples of equal size based on their $N_{\text{Htot}}$ values. These samples are the 105 clusters with $N_{\text{Htot}}\leq 2.53\times 10^{20}/$cm$^2$, the next 104 clusters with $N_{\text{Htot}}\leq 5.16\times 10^{20}/$cm$^2$ and finally the 104 clusters with $N_{\text{Htot}}> 5.16\times 10^{20}/$cm$^2$. 

\begin{figure}[hbtp]
               \includegraphics[width=0.49\textwidth, height=6cm]{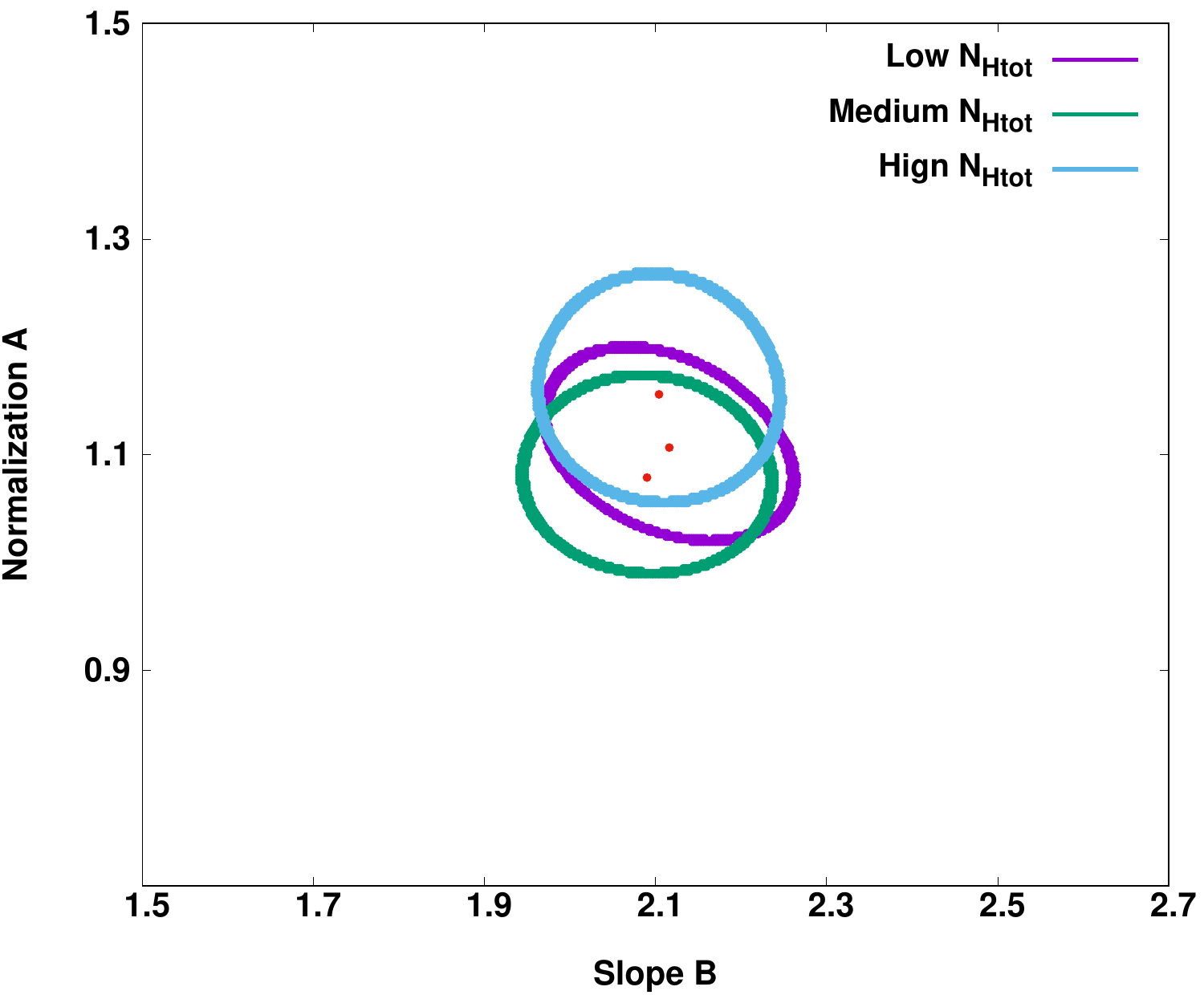}
               \caption{$1\sigma$ confidence levels (68.3\%) of the normalization and slope of the $L_{\text{X}}-T$ relation for  $N_{\text{Htot}}\leq 2.53\times 10^{20}/$cm$^2$ (purple),  $2.53\times 10^{20}/$cm$^2<N_{\text{Htot}}\leq 5.16\times 10^{20}/$cm$^2$ (green) and $N_{\text{Htot}}> 5.16\times 10^{20}/$cm$^2$ (cyan).}
        \label{fig10}
\end{figure}

We fit the full $L_{\text{X}}-T$ relation for these three independent subsamples. As shown in the top panel of Fig. \ref{fig10}, clusters in high and low $N_{\text{H}}$ regions show completely consistent $L_{\text{X}}-T$ behaviors with each other. The only noticeable difference between the three subsamples is their intrinsic scatter. Going from the low to the high $N_{\text{Htot}}$ subsamples, the intrinsic scatter is $\sigma_{\text{intr}}=0.214,0.242$ and $0.258$ dex respectively. This is not surprising since the high $N_{\text{Htot}}$ clusters undergo stronger corrections based on the molecular hydrogen column densities of W13. However, one should not forget that these molecular hydrogen values are approximations and thus some random scatter around the true values is expected, which then propagates to the $L_{\text{X}}$ values. In any case, this does not constitute any source of $N_{\text{Htot}}$-related bias since the overall $A$ and $B$ behavior is similar for different $N_{\text{Htot}}$ values (see also Sects. \ref{correl_sect} and \ref{corr_resid}).

\begin{figure}[hbtp]
               \includegraphics[width=0.51\textwidth, height=5cm]{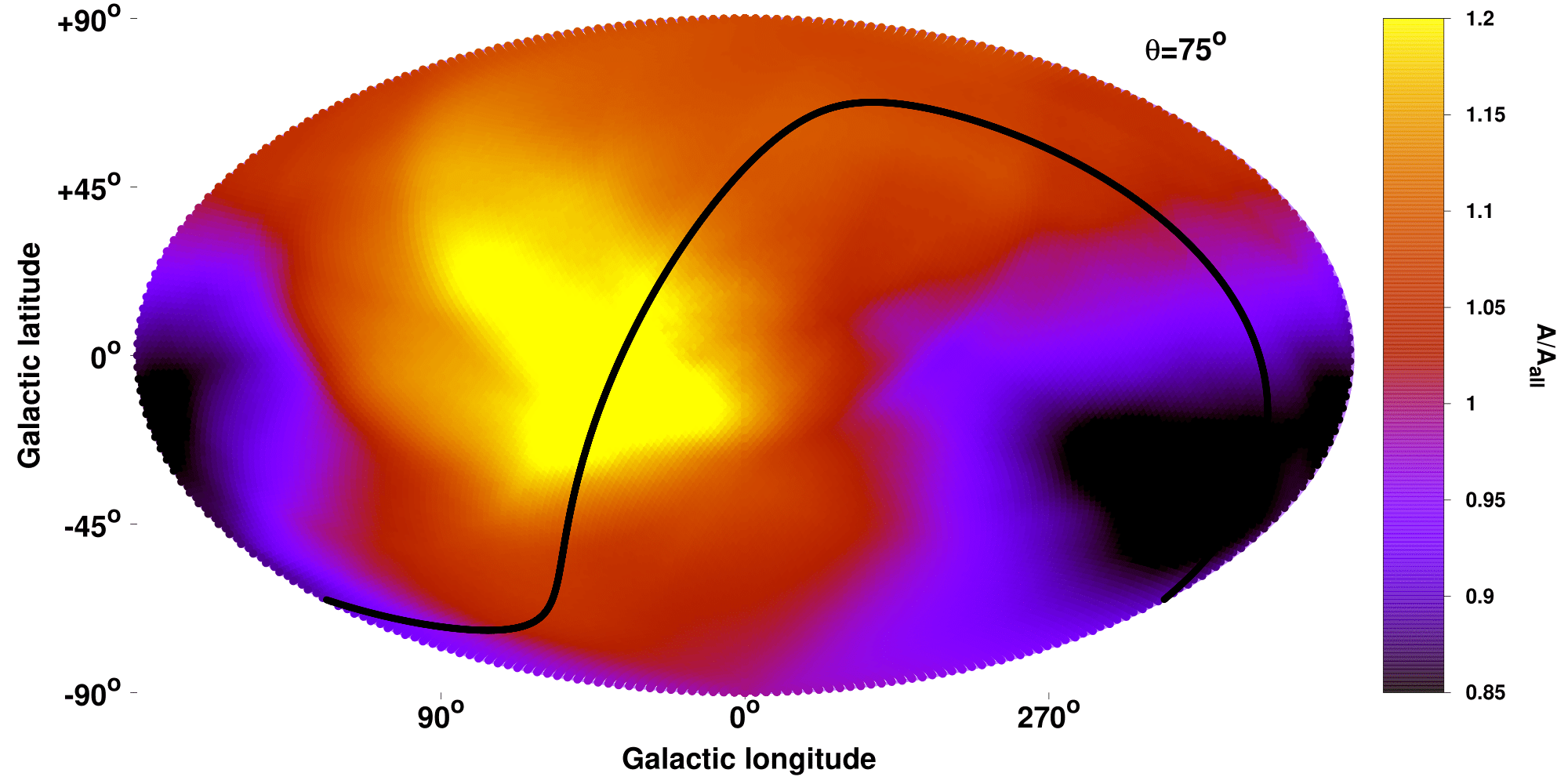}
               \includegraphics[width=0.51\textwidth, height=5cm]{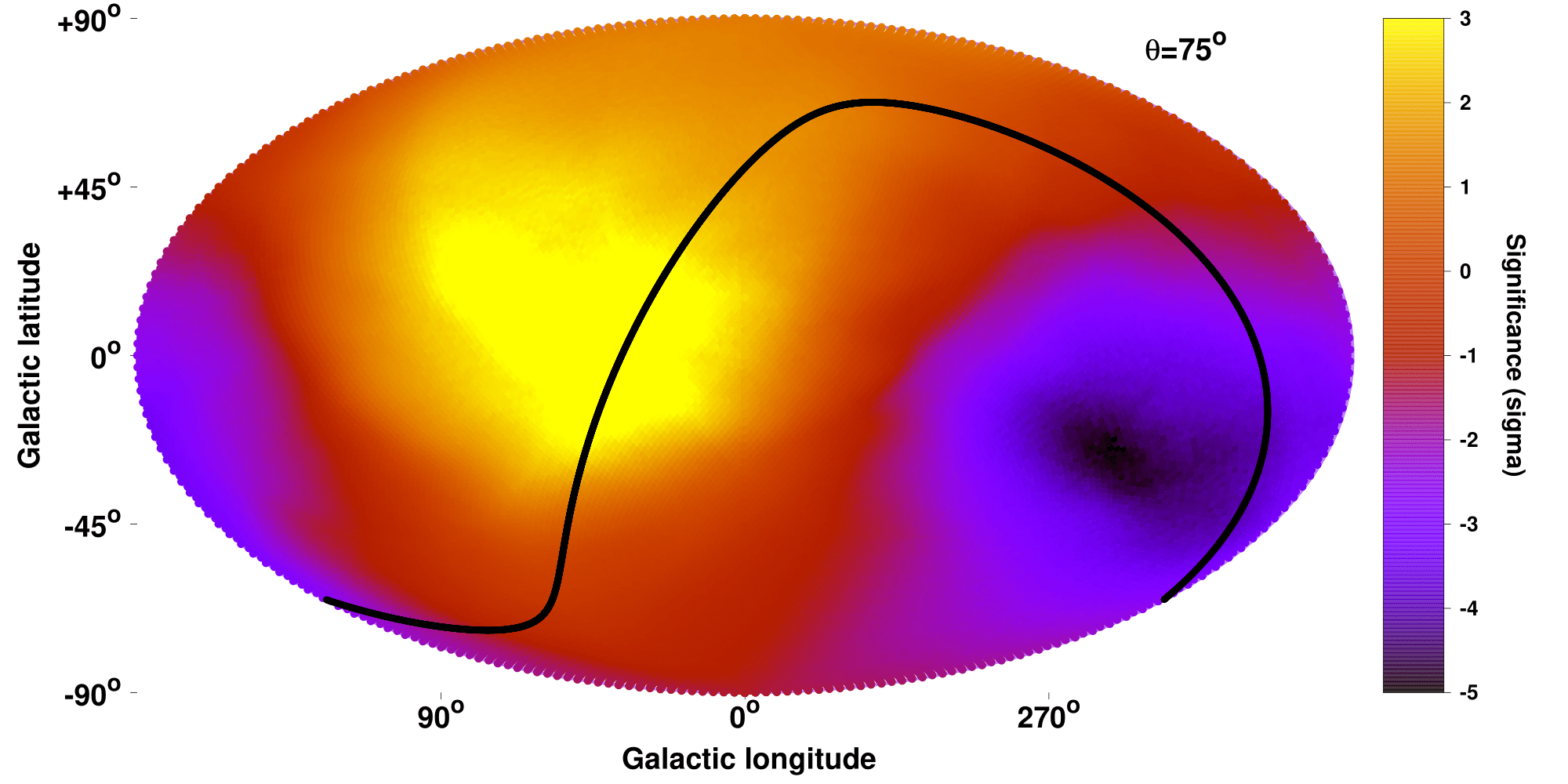}               
               \caption{Same as in Fig. \ref{bulk} for the 209 clusters with $N_{\text{Htot}}<5.16\times 10^{20}/$cm$^2$. }
        \label{nh-maps}
\end{figure}

As an extra test, we excluded the 104 clusters with the highest absorption ($N_{\text{Htot}}>5.16\times 10^{20}/$cm$^2$) and repeated the 2D sky scanning with $\theta=75^{\circ}$ in order to see if we observe the same anisotropies. The results are shown in Fig. \ref{nh-maps}. The previously detected anisotropic behavior persists, with the lower $A=0.879\pm 0.059$ (65 clusters) found toward $(l,b)=(242^{\circ}, -27^{\circ})$, which is consistent within $\sim 30^{\circ}$ from the previous findings. 
The brightest part of the sky remains unchanged compared with the full sample case, namely toward $(l,b)=(35^{\circ}, -15^{\circ})$ (93 clusters) with  $A=1.368\pm 0.069$. These two regions share a statistical tension of $5.39\sigma$ ($45\pm 8 \%$), the most statistically significant result we found up to now. The most extreme dipole anisotropy on the other hand is found toward $(l,b)=(221^{\circ}, -33^{\circ})$ with $3.55\sigma$.

Subsequently, the detected $>3.5-4\sigma$ apparent anisotropies not only do not result due to the different amounts of absorbing material throughout the sky and its effects on X-ray photons, but they significantly increase to a $>5\sigma$ level when the 104 clusters with the highest absorption are excluded. This is mostly due to the decrease of the intrinsic scatter of the clusters left, which leads to a decrease in the final $A$ uncertainties. 

\subsubsection{Extra absorption from undetected material or varying metallicity of the Galactic material} \label{extra_nh_var_metal}

The second case is that the exact amount of X-ray absorbing material is not accurately known and a higher or lower absorption correction is needed than the one applied. Such problems could occur for example if not all the absorbing material in the line of sight of a galaxy cluster has been detected by the radio surveys such as LAB, either because it is outside of the velocity range of the radio survey or for other unknown reasons (e.g., more than expected hydrogen in ionized or molecular form).

Another possible reason could be the varying metal abundance of the ISM throughout the Galaxy. The applied X-ray absorption correction is mostly applied as this: the amount of hydrogen detected is used as a proxy for the total amount of absorbing material that exists toward a given direction. The elements of this material that contribute the most in the absorption of the X-ray photons are helium\footnote{When we refer to metals from now on, helium is also included for convenience.} and metals such as oxygen, neon, silicon etc. Based on the detected  $N_{\text{Htot}}$ value, a Solar metal abundance is assumed for the Galactic interstellar medium (ISM) in every direction in order to quantify the number of metals absorbing X-ray radiation. However, throughout the Galaxy the true metal abundance might diverge from this approximation since there are metal-rich and metal-poor regions. Consequently, the same amount of detected hydrogen could correspond to different amounts of X-ray absorption from metals, which is not taken into account by our current absorption correction models. It needs to be checked if the apparent anisotropies could in principle be caused by such effects.

In order to test this, one can estimate the needed absorption using two ways. Firstly, one can calculate the necessary "true" $N_{\text{Htot}}$ in order to fully explain the observed anisotropies. Secondly, one can fit the extracted X-ray cluster spectra and leave $N_{\text{Htot}}$ to vary. Then, the obtained best-fit $N_{\text{Htot}}$ can be compared with the ones we use, which come from W13.

\paragraph{Necessary $N_{\text{Htot}}$ to fully explain $L_{\text{X}}-T$ anisotropies} \label{extra_nh}

Any existence of X-ray absorbing "dark clouds" in certain parts of the sky could potentially explain the observed anisotropies. Approximately quantifying how much extra (or less) hydrogen column density would cause such an effect, one could see if this value can realistically be missed by radio surveys. Since any such clouds is unlikely to cover significantly large portions of the sky (more than $90^{\circ}$ width) it is more appropriate to first look for them in the smallest radius cones. 

To this end, we consider the region with the lowest normalization $A$ for the $\theta=45^{\circ}$ cone at $(l,b)=(280^{\circ}, +1^{\circ})$ which includes 42 clusters. Its $A$ value is $29\pm 7 \%$ lower than the rest of the sky (the fitting for the rest of the sky is performed without any distance weight) and therefore its clusters would need to be more luminous by the same degree in order to be consistent with an isotropic behavior. To quantify how much extra $N_{\text{Htot}}$ is required to make these clusters more luminous by $\sim29$\% and explain the apparent anisotropy, we performed the following: 

We selected 25 clusters from the region of interest with varying temperatures and metallicities. For each cluster we used an \textit{apec} model in XSPEC, reproducing the current absorbed $L_{\text{X}}$ value. For several $N_{\text{Htot}}$ values we found the new unabsorbed $L_{\text{X}}$. For the same clusters, we refit their X-ray spectra and constrained the new $T$ for the same $N_{\text{Htot}}$ values as above. For every cluster and for every $N_{\text{Htot}}$ change compared to the W13 values, we thus knew the relative change of $L_{\text{X}}$ and $T$ compared to the standard values. Next, we were able to find the average relative change of $L_{\text{X}}$ and $T$ for every tested $N_{\text{Htot}}$. Of course this change is not identical for every cluster since it depends on the exact $T$ and $Z$. However, the actual average value of this change ($\sim 20\%$) is much larger than its variation between clusters with different properties ($\pm 4\%$). Assuming the slope to be $B=2.102$, we could obtain the relative change of $A$ for every sky region based on the tested $N_{\text{Htot}}$ values. Consequently, we found how much extra or less $N_{\text{Htot}}$ one would actually need toward the apparently anisotropic sky regions in order to explain their behavior.


We find that an extra $N_{\text{Htot}}=3.3\pm 0.9\times 10^{20}/$cm$^2$ is required in order to make the above-mentioned low $A$ region consistent with the rest of the sky. For its 37 clusters, the average $N_{\text{Htot}}$ is $\sim 7\times 10^{20}$/cm$^2$. Thus, the final $N_{\text{Htot}}$ which would explain the low $A$ value of this region is $\sim 48\%$ larger than its current value. If we express this difference in terms of metal abundance of the existing Galactic ISM (and not just larger amounts of ISM material), the absorbing elements toward that direction should have a metallicity of $Z\sim 1.5\ Z_{\odot}$ to create such apparent anisotropies due to extra absorption.

The same analysis for the bright region toward $(l,b)=(24^{\circ}, +16^{\circ})$ (which deviates by $31\pm 10 \%$ from the rest of the sky) yields that $3.5\pm 1.1\times 10^{20}/$cm$^2$ less hydrogen would be needed toward that direction. This would mean that we falsely applied a higher absorption correction, systematically increasing the unabsorbed luminosities of the clusters lying in that part of the sky. For these 42 clusters included in that cone, the average total hydrogen column density is $\sim 6\times 10^{20}$/cm$^2$ (the actual individual values vary significantly within $0.9-20.6\times 10^{20}$/cm$^2$). Therefore, $\sim 60\%$ less absorbing material should exist toward this direction, in order for the $A$ value to match the rest of the sky. This seems considerably unlikely. In terms of varying metallicity, the hydrogen cloud that was detected there should be metal-poor ($Z\sim 0.4\ Z_{\odot}$) to explain the obtained discrepancy. 

This seems rather unlikely since it has been shown that toward the central bulges of spiral galaxies, and the Milky Way specifically, the metallicity of the ISM is expected to be higher \citep{boissier,schonrich,spina} than the metallicity in regions further away from the Galactic center. While these studies focus more on the Galactic plane and we do not use any clusters within $20^{\circ}$ from the latter, they indicate that the high-$A$ regions are expected to be more metal-rich than the low-$A$ regions, instead of $\sim 3-4$ times more metal-poor (which could potentially explain the anisotropies). Since X-ray absorption models do not account for these effects that could potentially bias the extracted cluster properties, further testing will be needed in the future. Following the same reasoning for the $\theta=60^{\circ}$ cones results in quantitatively very similar results\footnote{We should note here that if the $N_{\text{Htot}}$ of the extreme e.g., low$-A$ region, was indeed changed, then the needed offset of the opposite bright region would decrease, since the overall $L_{\text{X}}-T$ best-fit would shift closer to the bright region $L_{\text{X}}-T$. However, this effect would not change the overall conclusion since it is rather weak.}. 


Another possible explanation for the behavior of the low-$A$ regions would be the existence of nearby dwarf galaxies \citep[e.g.,][]{dwarf_gal}, that contain sufficient amounts of X-ray absorbing material to cause such dimming to the clusters, for which we do not account for. However, these systems would need to fulfill some conditions such as having a large apparent size in the sky and containing absorbing material not detected by LAB. For the latter to happen, the absorbing material would need to either have a line-of-sight velocity outside of the LAB range or its hydrogen content to be limited compared to the existing metals (as explained before). Even though this "hidden" absorption by nearby galactic systems would still not explain the behavior of the bright, high-$A$ regions, the statistical significance of the latter would drop since clusters from other parts of the sky would see an increase in their $L_{\text{X}}$.

As an overview of this analysis, we see that such large differences between the detected and the true amount of $N_{\text{Htot}}$ (if this is the only reason behind the apparent anisotropies), are relatively difficult to occur, but definitely worth further checking. The necessary metallicities of the ISM to explain the behavior of the anisotropic regions seem quite unlikely as well, since one would expect oversolar metallicities close to the Galactic center, and not undersolar ones.

\paragraph{Free to vary $N_{\text{H}}$ results from literature}

A direct way to check if any of the above cases seems possible to explain our results is to try to estimate the absorption using only the X-ray spectra independently of the $N_{\text{Htot}}$ measurements that were used above. This can be done by leaving the $N_{\text{Htot}}$ parameter free to vary when fitting the cluster spectra. Comparing these estimations to the W13 $N_{\text{Htot}}$ values, one can see if there is a systematic difference for sky regions that show extreme $A$ behavior, indicating lower or higher absorption than the one previously adopted. These potential differences can reflect either differences in the actual amount of the ISM material as calculated before or differences of the true metallicity of that material, compared to the universally-assumed Solar one. Of course if different instruments are used for this estimation (e.g., \textit{Chandra} and \textit{XMM-Newton}), calibration issues must be taken into account.

In our case, the spectral fitting was performed only with a fixed $N_{\text{Htot}}$ as described in previous sections. The results for a varying $N_{\text{Htot}}$ will be presented in future work. For now, we use the $N_{\text{Htot}}$ measurements as obtained by \citet{lovisari19} who fit the X-ray spectra of 207 nearby galaxy groups and clusters, determining their metallicity radial profiles using only \textit{XMM-Newton} observations. We assume that the determination of $N_{\text{Htot}}$ does not strongly depend on the physical properties of the fitted cluster spectra, and thus only the sky coordinates of each object is of interest for our test. There are 142 overlapping clusters between these 207 clusters and our 313 clusters. 

\begin{figure}[hbtp]
               \includegraphics[width=0.4\textwidth, height=6cm]{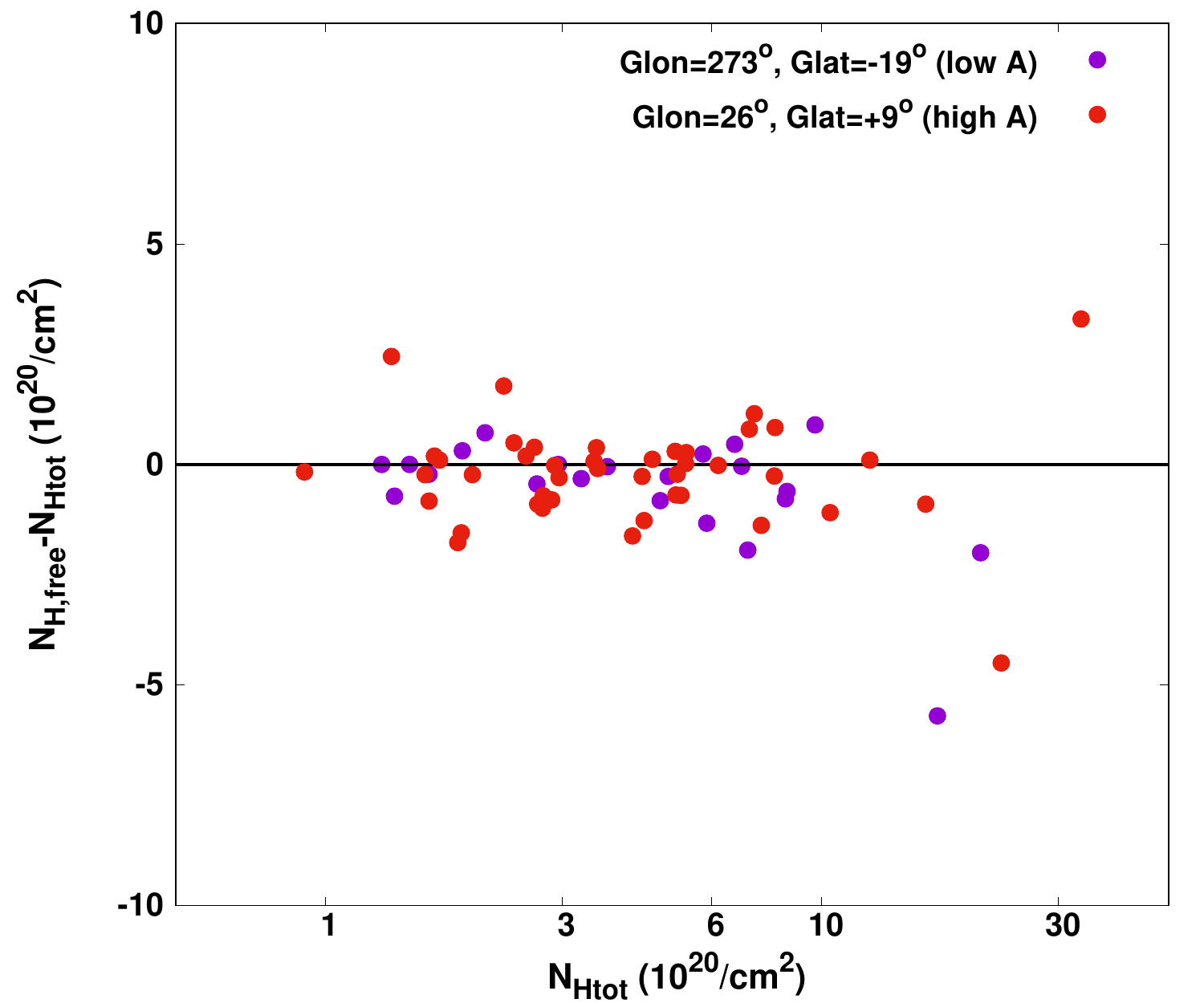}
               \caption{Difference between $N_{\text{Htot}}$ as obtained from the X-ray spectra fit in \citet{lovisari19} and in W13. The difference is displayed as a function of the W13 $N_{\text{Htot}}$ value for two sky regions. Within $45^{\circ}$from $(l,b)\sim (273^{\circ}, -19^{\circ}$ and from $(l,b)\sim (26^{\circ}, +9^{\circ})$.}
        \label{nhfree}
\end{figure}

We plot the difference $D=N_{\text{H,free}}-N_{\text{Htot}}$ as a function of $N_{\text{Htot}}$ for two regions: a cone with $45^{\circ}$ radius centered at $(l,b)\sim (273^{\circ}, -19^{\circ}$) and the same cone centered at $(l,b)\sim (26^{\circ}, +9^{\circ}$. The selected coordinates are the average values of all the results from the analysis up to now. The results are illustrated in Fig. \ref{nhfree}.

As one can see, there are no significant differences of $D$ between the two extreme regions\footnote{Systematics such as the specific abundance table used or calibration issues are expected to affect both regions in the same way (since the physical properties of the clusters are similar). Thus, even if the absolute value of $D$ changes after all possible corrections, it would still be similar for both regions while one would need very different values to explain the apparent anisotropies. }. Moreover, the existing deviations between the free absorption and the previously assumed one are not large enough to explain the apparent anisotropies. For the low-$A$ region, the median value of $D$ is $D_{\text{med}}=-0.28\times 10^{20}/$cm$^2$, while it would need to be $D\gtrsim +3\times 10^{20}/$cm$^2$ to alleviate the existing statistical tension. For the high-$A$ region, the result is $D_{\text{med}}=-0.22\times 10^{20}/$cm$^2$, while the anisotropies could be explained if $D\lesssim -4\times 10^{20}/\ $cm$^2$.

All in all, using the X-ray cluster spectra of \citet{lovisari19} as a first indication we see that the true, total absorption, does not seem to significantly deviate from the adopted absorption from W13. If a significant deviation between true and "currently-measured" absorption is detected in the future, it could potentially explain the observed anisotropies of the $L_{\text{X}}-T$ relation in the sky. For now though, such an explanation seems unlikely.

\subsubsection{Absorption from the Magellanic system}

The Magellanic system is comprised of the Large Magellanic Cloud (\object{LMC}) galaxy, the Small Magellanic Cloud (\object{SMC}) galaxy, the Magellanic Stream (MS), the Magellanic Bridge (MB) and finally the Leading Arm (LA). All these objects are known to contain sufficient amounts of neutral hydrogen which could potentially interfere with our measurements if not taken into account. However, the very vast majority of the hydrogen of the Magellanic system is well within the velocity range for which the $N_{\text{Htot}}$ values are extracted (velocity range of LAB). Thus, it should already be included in the results of the Sect. \ref{galaxy_H} and taken into account during the correction of the $L_{\text{X}}-T$ values for the absorption. 

In addition, it has been shown \citep[e.g.,][and references therein]{magellanic,choudhury} that the Magellanic system is metal-poorer ($\sim 10\%-50\%$) than the Solar metallicity assumed in the absorption correction models. As a result, the latter will overestimate the absorption effects caused by the Magellanic system\footnote{For the detected hydrogen only, since the undetected is not taken into account}, eventually overestimating the unabsorbed $L_{\text{X}}$ values of the clusters in these regions. Combining with the fact that the Magellanic system mostly covers sky regions in which clusters appear to be fainter than expected (low $A$ anisotropies), one sees that is unlikely that the Magellanic system has any effects on our anisotropic results. Nevertheless, we examine all the above-mentioned components to see where they lie in the sky and if they correlate with the anisotropic behavior we observe. 


The LMC is located at  $(l,b)\sim (281^{\circ}, -33^{\circ})$, within the low normalization regions, and moving away from us with +262 km/s \citep{veloc_lmc}. This velocity implies that the LAB survey would have not detected only the neutral hydrogen with a peculiar velocity of $\geq +140$ km/s compared to the LMC center, and toward our line of sight. Moreover, its $N_{\text{HI}}$ distribution is peaked close to the stellar population, covering a "circle" with a $\sim3^{\circ}-4^{\circ}$ radius in the sky, centered at the above coordinates \citet{magellanic} (thereafter D16). From the 313 galaxy clusters we use, only two are within 15$^{\circ}$ from LMC, but only one is dimmer than expected based on its temperature (however within the intrinsic scatter limits). The $N_{\text{Htot}}$ value toward the LMC as given by W13 is $3 \times 10^{21}$/cm$^2$. Therefore, based on all the above, it is safe to conclude that the LMC system does not bias our analysis since we would need multiple systems to be affected by that and appear underluminous.

The SMC is located at $(l,b)\sim (303^{\circ}, -44^{\circ})$ (where $N_{\text{Htot}}=3 \times 10^{21}$/cm$^2$), further away than LMC but still moderately close the low normalization regions. Its line of sight velocity compared to us is 145.6 km/s, therefore the greatest parts of its hydrogen components are expected to have been accounted for from the LAB survey. Its angular size is $\sim 50\%$ of LMC and the $N_{\text{HI}}$ distribution still seems to be mostly concentrated within its optical counterpart (D16). From our sample, only two clusters are within 10$^{\circ}$ of SMC and five are within 15$^{\circ}$. From these five clusters, three have minimal random residuals from the best-fit $L_{\text{X}}-T$ relation for the whole sample, one is up-scattered and the last one is low-scattered. Thus, once again we can safely assume (mainly because of its low relative velocity and the normal $L_{\text{X}}-T$ behavior of the few clusters) that SMC does not cause any significant bias to our results.

The MS extends over 100$^{\circ}$ on the sky, starting from LMC and spreading toward the south Galactic pole. Then it moves up to Galactic latitudes of $b\sim -40^{\circ}$, for $l\sim 100^{\circ}$ (Fig. 1 in D16) covering 2700 deg$^2$ in total. However, its $N_{\text{HI}}$ density linearly decreases more than 20$^{\circ}$away from LMC. Generally, MS is not expected to affect our results for two reasons. Firstly, the fraction of the sky it covers does not seem to correlate with low $L_{\text{X}}-T$ normalization regions, since it lies at fairly low Galactic latitudes and to positions where high normalization regions are also located. Secondly, according to D16, its velocity range varies within $-450$ km/s to $+180$ km/s, so practically all of its hydrogen component is expected to be accounted for in the LAB survey.

The MB connects the LMC and SMC systems and unlike MS, it contains a stellar component as well. Its central coordinates are $(l,b)\sim (294^{\circ}, -37^{\circ})$ and there are only three clusters within 15$^{\circ}$ of these coordinates. However, all these three clusters are already included in the 10$^{\circ}$ circles of LMC and SMC we considered before. The main velocity of MB with respect to us is $\sim 225$ km/s, therefore it should be included in the LAB results.

Finally, the LA extends in the opposite direction compared to the MS, starting from LMC and extending up to the northern Galactic hemisphere (for $l\sim 250^{\circ}-280^{\circ}$). Its angular size is $\sim 60^{\circ}$ (D16). It is the least massive component of the Magellanic System while its velocity range is relatively constant, $\sim 180-270$ km/s. The fact that a large part of the LA lies within 20$^{\circ}$ of the Galactic plane, where no galaxy clusters exist in our sample, combined with its velocity range, indicates that no bias can occur for our results.

\subsection{Systematics, selection effects, and correlation of results with cluster properties}\label{systemat_corr}

Some cluster properties are usually associated with potential systematic effects. For instance, one might expect that clusters with a lower RASS exposure time might be generally up-scattered and vice versa. This is due to the fact that brighter clusters are more likely to be detected than fainter ones for the same exposure time. On the other hand, when the RASS exposure time is large enough, fainter clusters should also meet the detection thresholds set by the parent catalogs. If such a systematic indeed exists it might translate to higher and lower $A$ values for the two cases respectively, creating artificial anisotropies. As shown in the next sections, this has no impact on our results.

Similar systematics might occur near the X-ray flux limit of our sample. For a similar temperature, intrinsically brighter clusters are more likely than other clusters to overpass the flux limit and be included in the final sample (Malmquist bias). Thus, if this applies to our sample and an excess of such clusters exist within a sky region, this will possibly result to higher $A$.  As again shown in the following sections, our sample and analysis do not suffer of such effects. 

A third possible systematic is the detection of clusters in high $N_{\text{Htot}}$ regions since the difference between $N_{\text{Htot}}$ and $N_{\text{HI}}$ can have an impact on the selection of every X-ray flux-limited sample (including the parent catalogs). These selections are based on the unabsorbed flux, corrected only for $N_{\text{HI}}$. This flux will be underestimated for clusters lying in high $N_{\text{Htot}}$ and therefore they might not overcome the flux threshold set by each sample. A $\gtrsim 10\%$ underestimation is expected for regions with $N_{\text{Htot}}-N_{\text{HI}}>4\times10^{20}/$cm$^2$. The Galactic plane is usually excluded from such cluster selection processes. Nevertheless, there are still many sky regions with high $N_{\text{Htot}}$ within which clusters could be missed. This could cause unaccounted selection biases and affect the completeness of the samples (which is not important to our study as explained before).

On the other hand, the high $N_{\text{Htot}}$ clusters that overcome the flux limit of a sample might be intrinsically brighter in average (in order to be detected even though their estimated flux is biased low). This is "revealed" only when we correct their $L_{\text{X}}$ for the molecular absorption. As a result, these clusters might be upscattered in the $L_{\text{X}}-T$ plane and potentially jeopardize our results. In Sect. \ref{galaxy_H} we showed however that there is no such bias in our analysis. In the next sections we provide further evidence for this.

\subsubsection{Correlations between $A$ and subsample average parameters} \label{correl_sect}

As a generalization of the above, the apparent anisotropies in the behavior of the $L_{\text{X}}-T$ relation could be in principle caused by different cluster subpopulations in the different regions. In practice, this would mean that these subpopulations might have different (average) physical properties, leading to the derived directional behavior of $A$. To investigate this, we perform a bootstrap resampling analysis. We drew $10^5$ random subsamples of 65 clusters (typical number for the $\theta=60^{\circ}$ cones) independently of the direction. For every subsample, we find the best-fit $A$ and the weighted mean of the temperature, redshift, core and outer metallicity, flux, luminosity, $N_{\text{Htot}}$, intrinsic scatter and RASS exposure time. Thus, we can study if any correlation between the average values and $A$ exists. Such a correlation, combined with a different parameter distribution in the most extreme regions could (at least partially) explain the observed anisotropies.
Additionally, we created another $10^5$ subsamples with a random number of clusters (between 35 and 170) in order to test if a correlation between the number of data and $A$ exists.

In order to check for any possible correlations, we plot the $A$ value against all these parameters parameters for every one of the $10^5$ subsamples. The most characteristic of these plots are displayed in Fig. \ref{correl}, while the rest can be found in the Appendix (Fig. \ref{all_correl}). We also calculate the Pearson's correlation coefficient $r_{\text{corr}}$ given by Eq. \ref{pearsons}.

\begin{equation}
r_{\text{corr}}=\dfrac{n\sum x_i y_i-\sum x_i\sum y_i}{\sqrt{n\sum x_i^2-\left(\sum x_i\right)^2}\sqrt{n\sum y_i^2-\left(\sum y_i\right)^2}},
\label{pearsons}
\end{equation}
where $x$ and $y$ are the two quantities for which we wish to study their correlation, and $n=10^5$ is the number of subsamples used.
Depending on the obtained value of $r_{\text{corr}}$ one can assess if there is indeed some correlation between the value of $A$ and some average property of the different cluster subsamples.  

\begin{figure*}[hbtp]
               \includegraphics[width=0.47\textwidth, height=6cm]{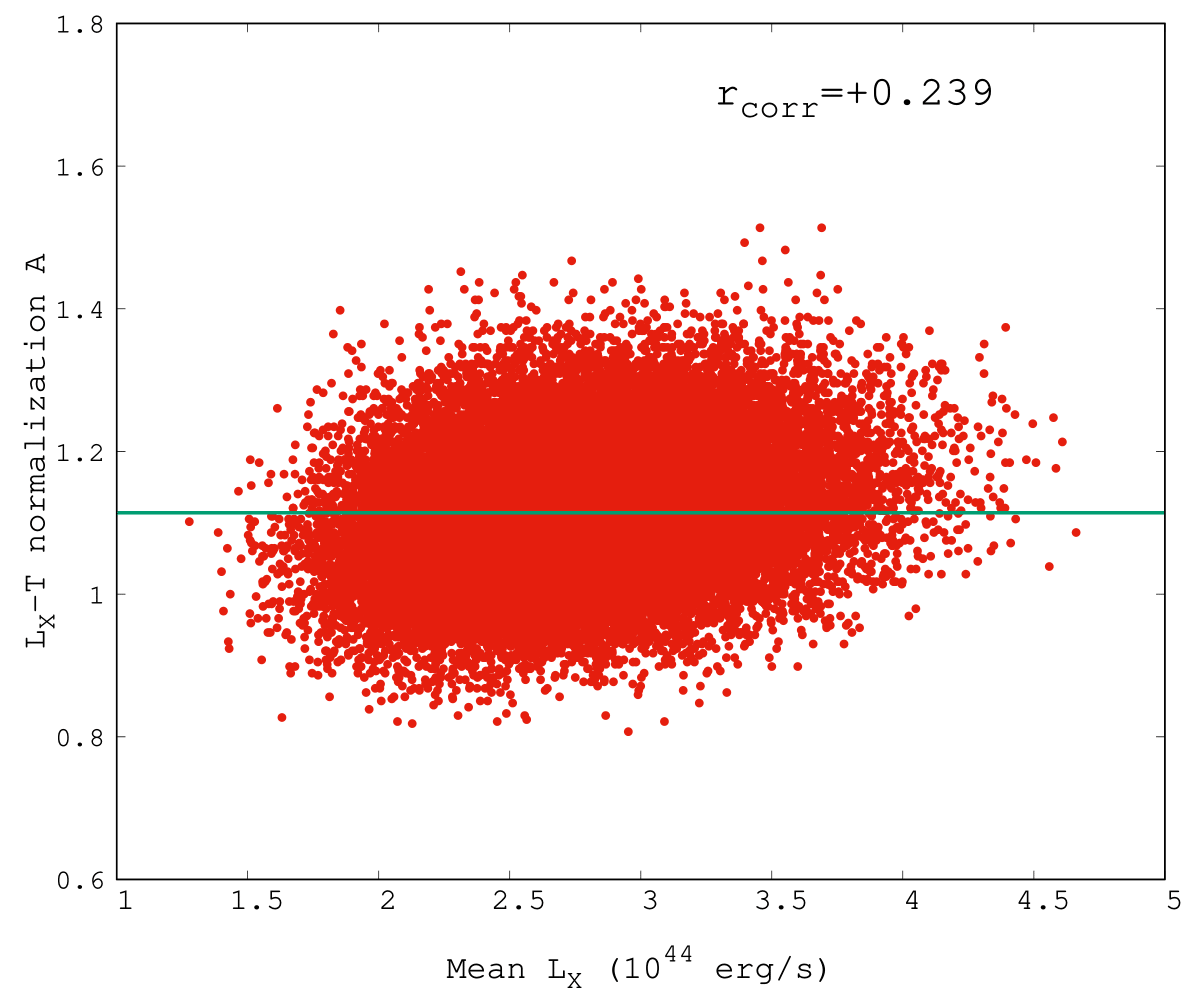}
               \includegraphics[width=0.47\textwidth, height=6cm]{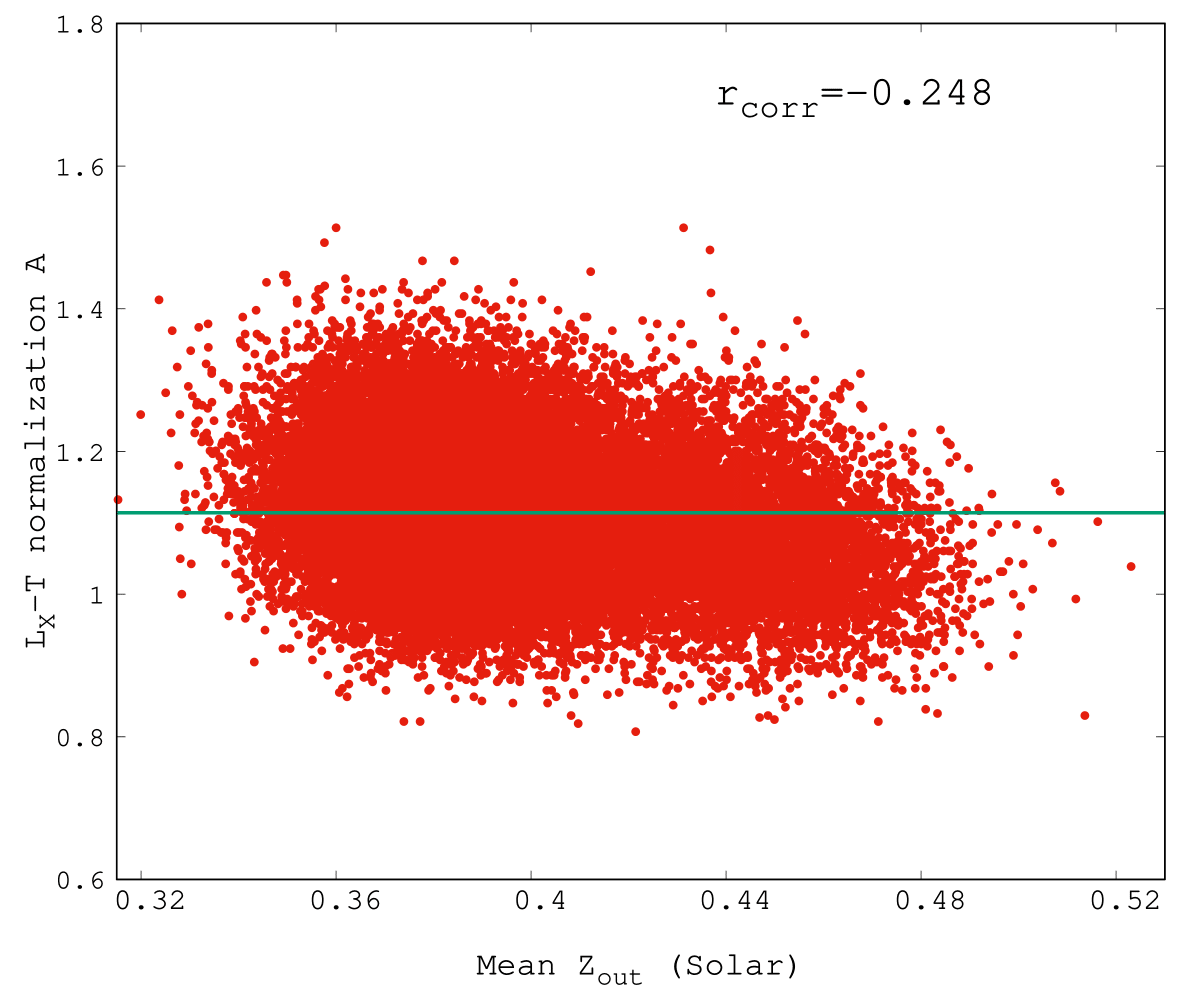}
               \caption{Correlation between the best-fit $A$ value and the average $L_{\text{X}}$ (left) and $Z_{\text{out}}$ (ICM metallicity as measured within $0.2-0.5\ R_{500}$, right) as obtained for every one of the random $10^5$ subsamples of 65 clusters. The Pearson's correlation coefficient is also displayed.}
        \label{correl}
\end{figure*}

The only physical parameters that seem to mildly correlate with the behavior of the $L_{\text{X}}-T$ normalization $A$ are the average subsample luminosity $(r_{xy}=+0.239\pm0.006)$ and the average $0.2-0.5\times R_{500}$ metallicity ($Z_{\text{out}}$, $(r_{\text{corr}}=-0.248\pm0.005)$.  As shown, the best-fit $A$ tends to slightly increase for clusters with a higher average $L_{\text{X}}$, which is expected. Subsamples with randomly more up-scattered clusters in the $L_{\text{X}}-T$ plane will naturally return a higher $A$. Hence, pure randomness can produce similar small anisotropies, which are highly unlikely to explain the observed spatial $L_{\text{X}}-T$ anisotropies (see additionally Sect. \ref{boot_section}).

If randomly up- and down-scattered $L_{\text{X}}$ values were not the reason behind the weak $A-L_{\text{X}}$ correlation, then a similar (positive) correlation would exist between $A$ and average $T$. However, there is no correlation between these parameters as shown in Fig. \ref{all_correl} ($r_{\text{corr}}=+0.005\pm 0.006$). Here we should also note that the weighted $T$ average of the highest and lowest $A$ regions is similar, shifting between $\sim 5.1$ keV and $\sim 5.7$ keV for both of them, depending on the cone radius. Hence, we can safely conclude that the observed anisotropies do not arise due to any different $T$ distribution.

On the other hand, a slight decrease in $A$ can be seen with an increasing average $Z_{\text{out}}$. One can clearly conclude though that the obtained anisotropies of $A$ cannot be attributed to this mild correlation since the highest and lowest $A$ regions have similar average $Z_{\text{out}}$ values ($Z_{\text{out}}\sim 0.40\ Z_{\odot}$).

A rather weak correlation is observed between $A$ and average redshift $(r_{xy}=+0.198\pm0.007)$, although the brightest and faintest regions have again a very similar average $z$ ($\sim 0.09-0.1$ for both). This indicates that the exact choice of the redshift evolution parametrization in the $L_{\text{X}}-T$ relation (e.g., $E(z)^{-1}$) is not particularly important, since any parametrization would approximately have the same effect in the two most anisotropic regions. Indeed, if one tries different $x$ priors for the $E(z)^x$ term, the significance of the final anisotropies fluctuates only by $\pm0.1\sigma$ compared to the used self-similar case (Appendix, Sect. \ref{Ez_prior}).

Another weak correlation of $A$ is observed with $\sigma_{\text{int}}$ ($(r_{\text{corr}}=-0.194\pm 0.006$). However, this trend cannot explain the apparent anisotropies since $\sigma_{\text{int}}$ does not strongly differ for the highest and lowest $A$ regions (0.10 dex and 0.13 dex\footnote{Here we remind the reader that $\sigma_{\text{int}}$ is this case is reduced because of the increase of the statistical uncertainties, due to the random weighting of the clusters as explained earlier in the paper. } respectively for the $\theta=60^{\circ}$ cones). No correlation is observed between the $A$ value and $N_{\text{Htot}}$, no. of clusters, RASS exposure time, $Z_{\text{core}}$ and flux.

As an overview, no correlation of $A$ with an average parameter (including systematics) can explain the apparent anisotropies. In the future, the correlation of $A$ with combinations of these average parameters will be explored as a possible explanation behind the discrepancies, even if this seems unlikely based on the results up to now.

\subsubsection{$L_{\text{X}}-T$ fitting residuals as functions of cluster properties}\label{corr_resid}

As a further, secondary check, we tested the correlation between the cluster properties and their $\log{L_{\text{X}}}$ residuals compared to the overall best-fit $L_{\text{X}}-T$ model\footnote{This test is equivalent to the bootstrap analysis correlation test in the previous section, but we include it for the sake of completeness}. As expected, similar results with the bootstrap analysis were obtained. Thus, to avoid repetition we do not go into a detailed presentation of all the results, but instead focus on the ones usually related to systematic biases. The full discussion is found in Sect. \ref{correl_append}. 



In a nutshell, no strong systematic behavior of the $\log{L_{\text{X}}}$ residuals is observed for varying RASS exp. time, $N_{\text{Htot}}$ and $z$ (Fig. \ref{residuals}). A mild systematic behavior of the residuals exists in terms of the flux and the statistical uncertainties ($\sigma_{\text{stat}}$) of the clusters. However, this has no effect in the derived anisotropies since the strongly anisotropic sky regions have similar flux and $\sigma_{\text{stat}}$ distributions. Finally, the residuals versus the outer cluster metallicity are also displayed in the same figure since there is a mild systematic behavior between these quantities. We already showed that this does not significantly affect our results in previous sections.

\subsection{Fixed slope vs free slope}\label{free_slope}

In our analysis until now we fixed the slope to its best-fit value for every subcategory of clusters, before we study the spatial anisotropies of $A$. This choice is motivated by the fact that $B$ does not significantly fluctuate throughout the sky for the 1D analysis (Fig. \ref{slope-glon}, similar results obtained for a 2D scanning). Moreover, a significant correlation between $A$ and $B$ is not expected, due to the pivot point of the $L_{\text{X}}-T$ relation being close to the median $T$. To investigate the possible biases that a fixed $B$ introduces to our analysis, we perform the following:

Case 1: we scan the sky using $\theta=75^{\circ}$ while we treat $B$ as a nuisance parameter. We allow $B$ to vary simultaneously with $A$, within its $2\sigma$ limits from its overall best-fit value. We then marginalize over $B$ to study the spatial behavior of $A$. The $1\sigma$ uncertainties of $A$ are again extracted based on the $\Delta \chi^2\leq 1$ limits since there is only one parameter of interest.

Case 2: We repeat the procedure but this time we allow $B$ to vary freely. We again study the $A$ anisotropies and quantify the statistical significance using the $\Delta \chi^2\leq 2.3$ limits (2 parameters of interest) for the $1\sigma$ parameter uncertainties.


For Case 1, the maximum anisotropy is found between the regions $(l,b)\sim (272^{\circ}, -21^{\circ})$ ($A=0.977\pm 0.050$) and $(l,b)\sim (26^{\circ}, -13^{\circ})$ ($A=1.274\pm 0.062$). The statistical significance of the tension is $3.74\sigma$ ($27\pm 7\%$), slightly larger than before despite the marginalization over $B$. One sees that the results are entirely equivalent to the case where $B$ is kept fixed, in terms of both statistical significance and direction. This strongly demonstrates the robustness of our method and the independence of the $A$ constraints from $B$.

\begin{figure*}[hbtp]
               \includegraphics[width=0.51\textwidth, height=5cm]{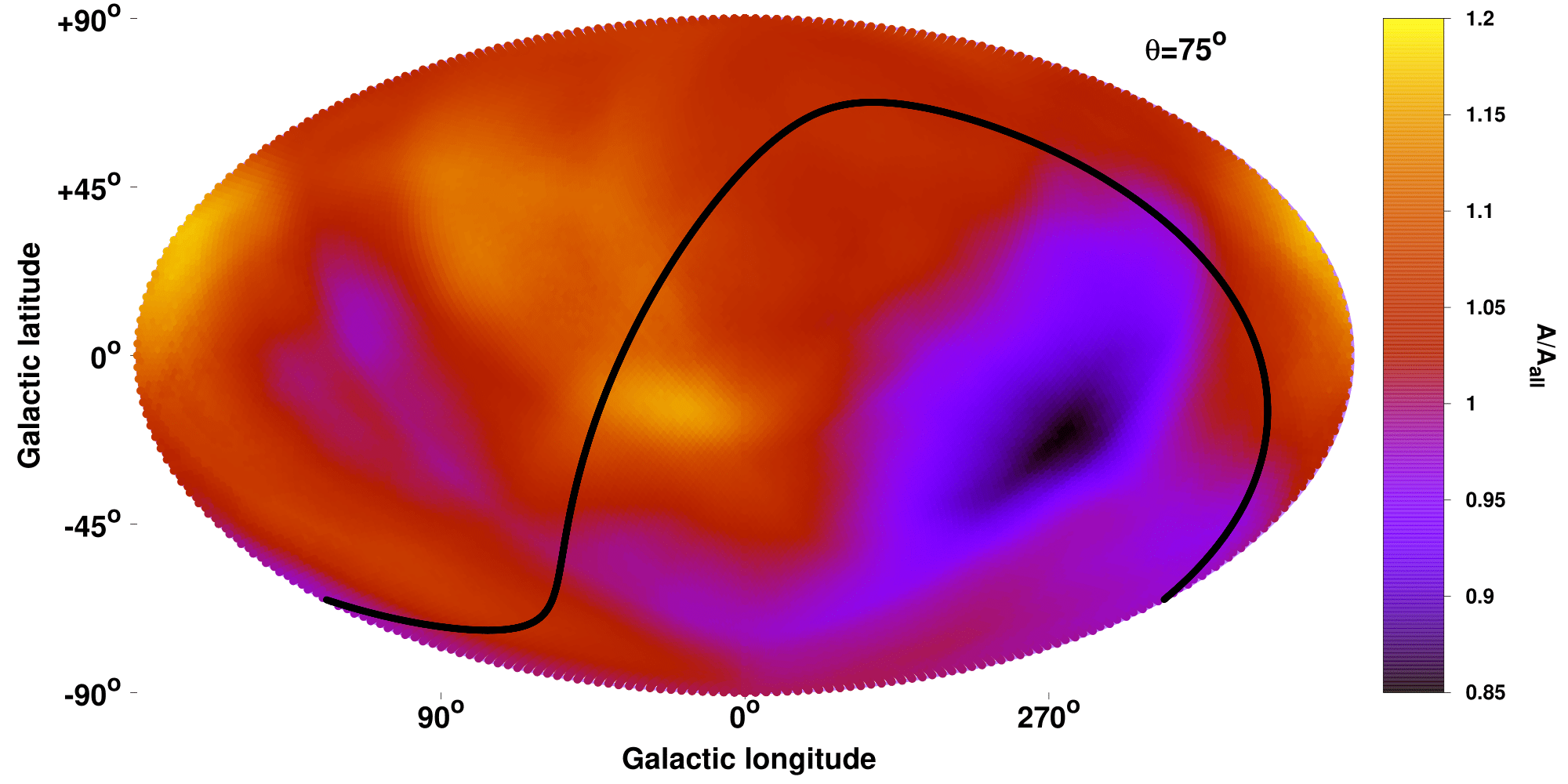}
               \includegraphics[width=0.47\textwidth, height=6cm]{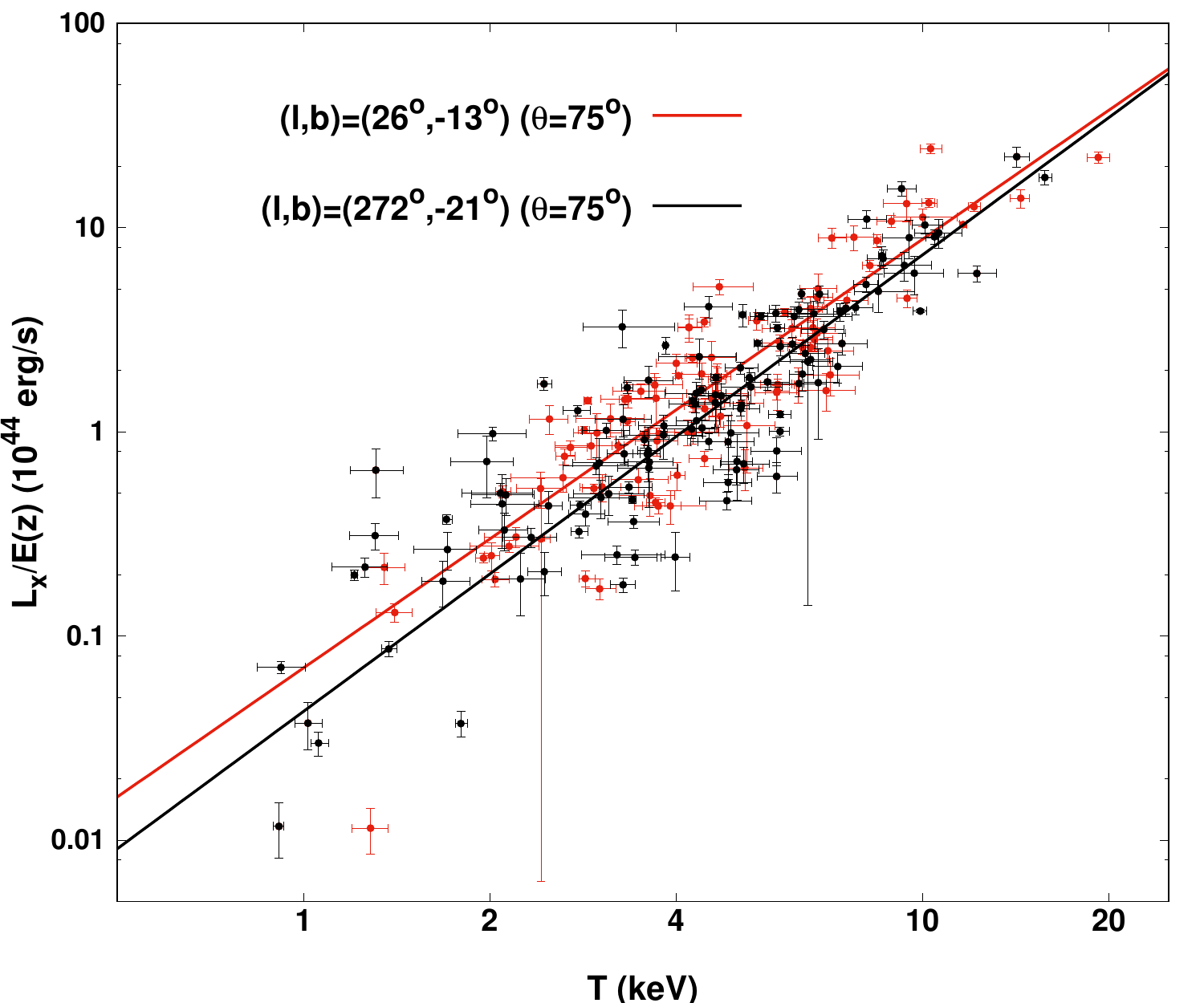}
               \caption{\textit{Left:} Best-fit normalization $A$ of the $L_{\text{X}}-T$ relation for every sky region over $A_{\text{all}}$ as a function of the position in the extragalactic sky when the slope $B$ is left completely free to vary. The cone size used is $\theta=75^{\circ}$. \textit{Right:} $L_{\text{X}}-T$ relation for the 136 clusters within $75^{\circ}$ from $(l,b)\sim (26^{\circ}, -13^{\circ})$ (red) and for the 124 clusters within $75^{\circ}$ from $(l,b)\sim (272^{\circ}, -21^{\circ})$ (black). Their best-fit models are displayed as solid lines.}
        \label{free_slope_fig}
\end{figure*}

For Case 2, the $A$ map is portrayed in Fig. \ref{free_slope_fig} (top panel). The spatial fluctuations of $A$ slightly intensify ($\sim 34\%$ between the most extreme values) and its directional pattern remains the same as when $B$ is kept fixed. This once more illustrates that the derived $A$ sky pattern does not depend on the true $B$ values of the different sky regions.

As already shown for the 1D analysis, the fluctuations of $B$ are smaller ($\sim 19\%$) than the ones of $A$. Every sky region is consistent within $<2\sigma$ with the rest of the sky, making the behavior of $B$ fairly consistent throughout the sky. The largest $A$ anisotropy is found toward the same regions as for Case 1 (drifting by $<6^{\circ}$) and now slightly drops to $2.78\sigma$ ($30\pm 11\%$). This is due to the enlarged uncertainties obtained from the 2-parameter $\Delta \chi^2\leq 2.3$ limits. These two regions return very similar slope values ($B= 2.256\pm 0.110$ and $B=2.109\pm 0.119$ for lowest and highest $A$ respectively). Their $L_{\text{X}}-T$ plot is displayed in the bottom panel of Fig. \ref{free_slope_fig}.



These results confirm that the choice of keeping the slope fixed for the bulk of our analysis does not introduce any biases in the directional behavior of $A$ or the statistical significance of the observed anisotropies.

\section{Cosmological constraints} \label{cosmology}

Many reasons that could potentially lead to a biased anisotropic behavior of the $L_{\text{X}}-T$ relation were tested until now. These tests explored the possibility that the apparent anisotropies could appear due to systematic differences of the subsamples in different patches of the sky or that unknown effects could influence the observed X-ray photons coming from specific extragalactic regions. Furthermore, we tested if possible systematics, such as RASS exposure time, Malmquist bias close to the flux limit etc., could bias our results. The observed anisotropies seem to be consistent and are not significantly alleviated by such tests. 

During our analysis up to now, we assumed fixed cosmological parameters toward all the directions in the sky when deriving the normalization and slope of the $L_{\text{X}}-T$ relation. On the contrary, one can reasonably assume that the physics within the ICM of galaxy clusters that determine the correlation between $L_{\text{X}}$ and $T$ should be the same regardless of the direction. As a result, the true normalization and slope of the $L_{\text{X}}-T$ relation should not depend on the coordinates and should be fixed to their best-fit values. 

Consequently, the last thing to be checked is if any apparent anisotropies could occur because of an anisotropic Hubble expansion. In practice, this would mean that the luminosity distance would differ toward varying directions for a fixed $z$. These differences can be expressed in terms of the cosmological parameter $H_0$ which enters in the luminosity distance through the conversion of the X-ray flux to luminosity. 

To explain the behavior of faint $L_{\text{X}}-T$ regions we would need a higher $D_L$ for the same $z$ (thus higher $L_{\text{X}}$). For $z\lesssim 0.3$ it is known that $D_{\text{L}}\approx \dfrac{1}{H_0}[z+\dfrac{z^2}{2}(1-q_0)+O(z^3)]$, where $q_0$ is the deceleration parameter. Hence, a lower $H_0$, implying a lower current expansion rate, would return a higher $D_L$ for a fixed $z$. The same would be true for a more negative $q_0$, implying a higher acceleration rate\footnote{With this paragraph we aim to make clear to the reader that a lower expansion rate is equivalent with a higher acceleration rate, which might seem counter-intuitive.}.

One could also study the directional behavior of $\Omega_{\text{m}}$, but for low redshift objects (like the clusters we use), $D_{\text{L}}$ is not very sensitive to this parameter. Therefore, large deviations from region to region would be needed in order to explain the anisotropies (see results of M18). Moreover, $\Omega_{\text{m}}$ variations would have a different effect to higher and lower redshift (and thus temperature) clusters, changing both the normalization and the slope of the $L_{\text{X}}-T$ relation. Other effects that have to be taken into consideration in such a case is the higher (lower) matter density of the Universe toward different directions, leading to more (less) structures. Structured environment can alter the behavior of the $L_{\text{X}}-T$ relation as we have shown in M18. Based on that, a robust directional study of  $\Omega_{\text{m}}$ is not ideal for our sample and method.
On the other hand, the effect of $H_0$ on $D_{\text{L}}$ does not depend on $z$ and hence variations of smaller amplitude than $\Omega_{\text{m}}$ could result in the observed anisotropies. Also, since $H_0$ variations will have the same effect on the $L_{\text{X}}$ of every cluster independently of its $z$ (and thus $T$), the slope of the $L_{\text{X}}-T$ will remain unchanged.

Therefore, we fix $A=1.114$ and $B=2.102$ and fit $H_0$ as the only free parameter (together with $\sigma_{\text{intr}}$) as described in Sect. \ref{fitting}.
It should be noted that here we investigate the relative change of $H_0$ due to spatial anisotropies of the cosmic expansion, and absolute values of $H_0$ are arbitrary.

\begin{figure}[hbtp]
               \includegraphics[width=0.51\textwidth, height=5cm]{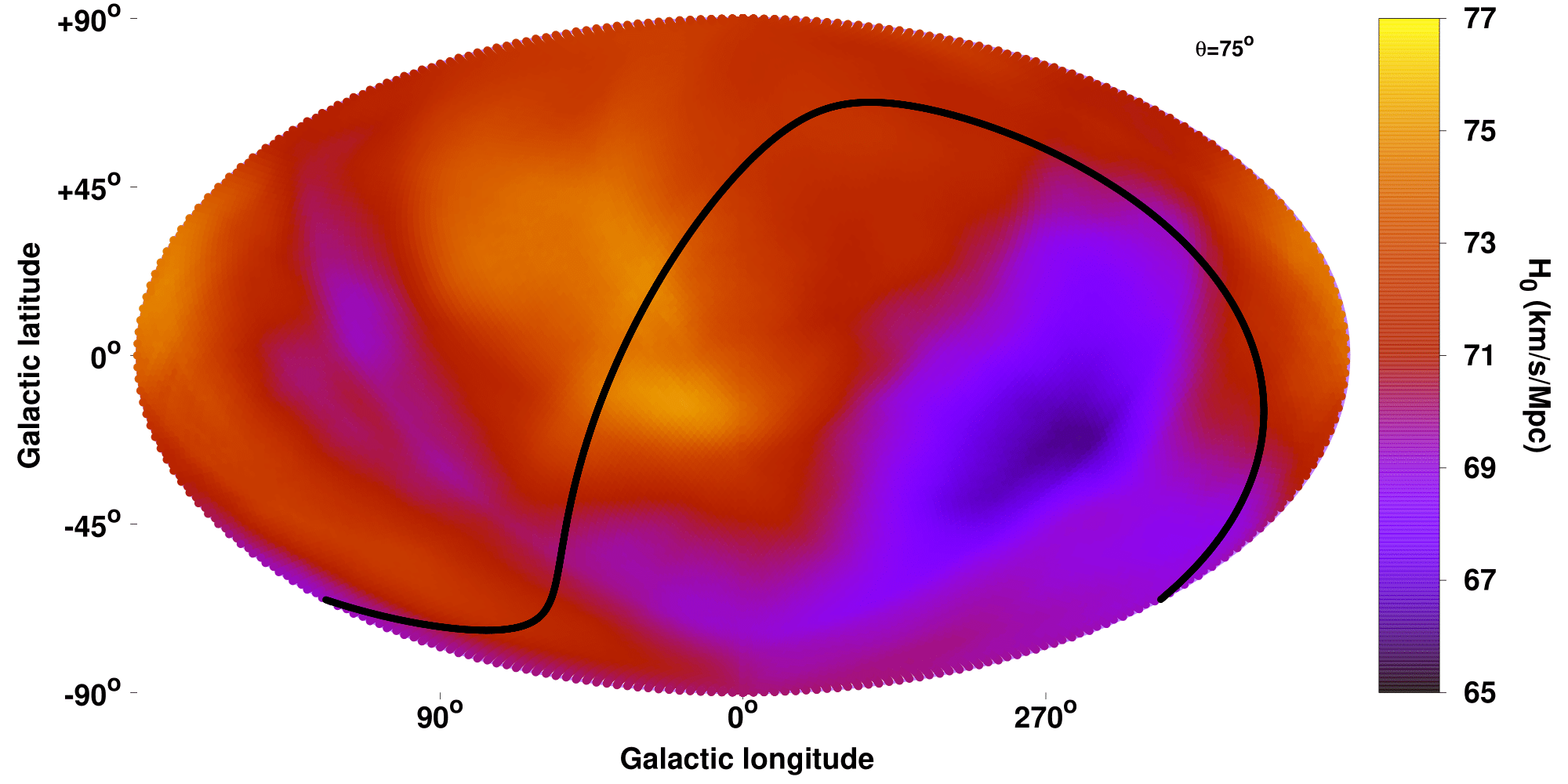}
               \caption{Best-fit $H_0$ value as obtained through the $L_{\text{X}}-T$ relation as a function of the position in the extragalactic sky for $\theta =75^{\circ}$ cones using all the 313 clusters in our sample.}
        \label{hubble-eehif}
\end{figure}

The $H_0$ map as produced using $\theta=75^{\circ}$ is portrayed in Fig. \ref{hubble-eehif}. As one can see the $H_0$ and $A$ maps show exactly the same behavior for the reasons explained above. We do not plot the significance map in this case since it is identical to the significance map of $A$ using $\theta=75^{\circ}$, as shown in Fig. \ref{sigma_cones}.

The apparently maximum acceleration direction is found toward $(l,b)=(274^{\circ}, -22^{\circ})$ with $H_0=66.20\pm 1.72$ km/s/Mpc while the most extreme opposite behavior is found at $(l,b)=(17^{\circ}, -9^{\circ})$ with $H_0=75.17\pm 1.81$ km/s/Mpc. Their deviation from each other is at 3.59$\sigma$ ($13\pm 4\%$). The most extreme dipole is centered at $(l,b)=(263^{\circ}, -21^{\circ})$ with a $3.15\sigma$ significance. One sees that these three directions completely match the directions for the normalization analysis using the 75$^{\circ}$ radius cones, highlighting the ability of the normalization of the $L_{\text{X}}-T$ relation to trace possible cosmological anisotropies. Moreover, the sigma values also match the ones from the normalization map as expected.

\section{Combination with ACC and XCS-DR1} \label{joint_an}

\subsection{$H_0$ results for each sample}\label{acc_xcs}

\begin{figure*}[hbtp]
               \includegraphics[width=0.51\textwidth, height=5cm]{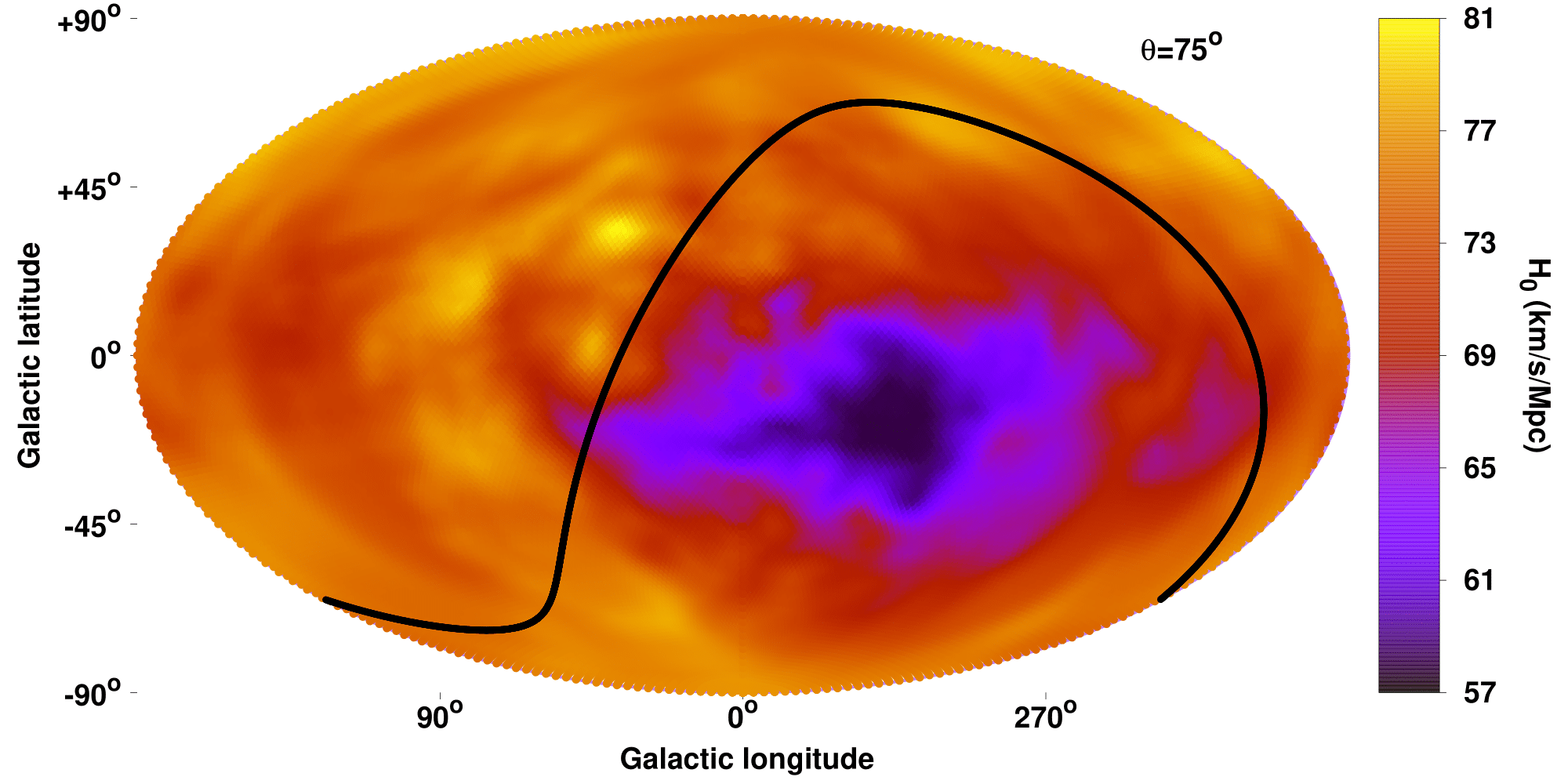}
                \includegraphics[width=0.51\textwidth, height=5cm]{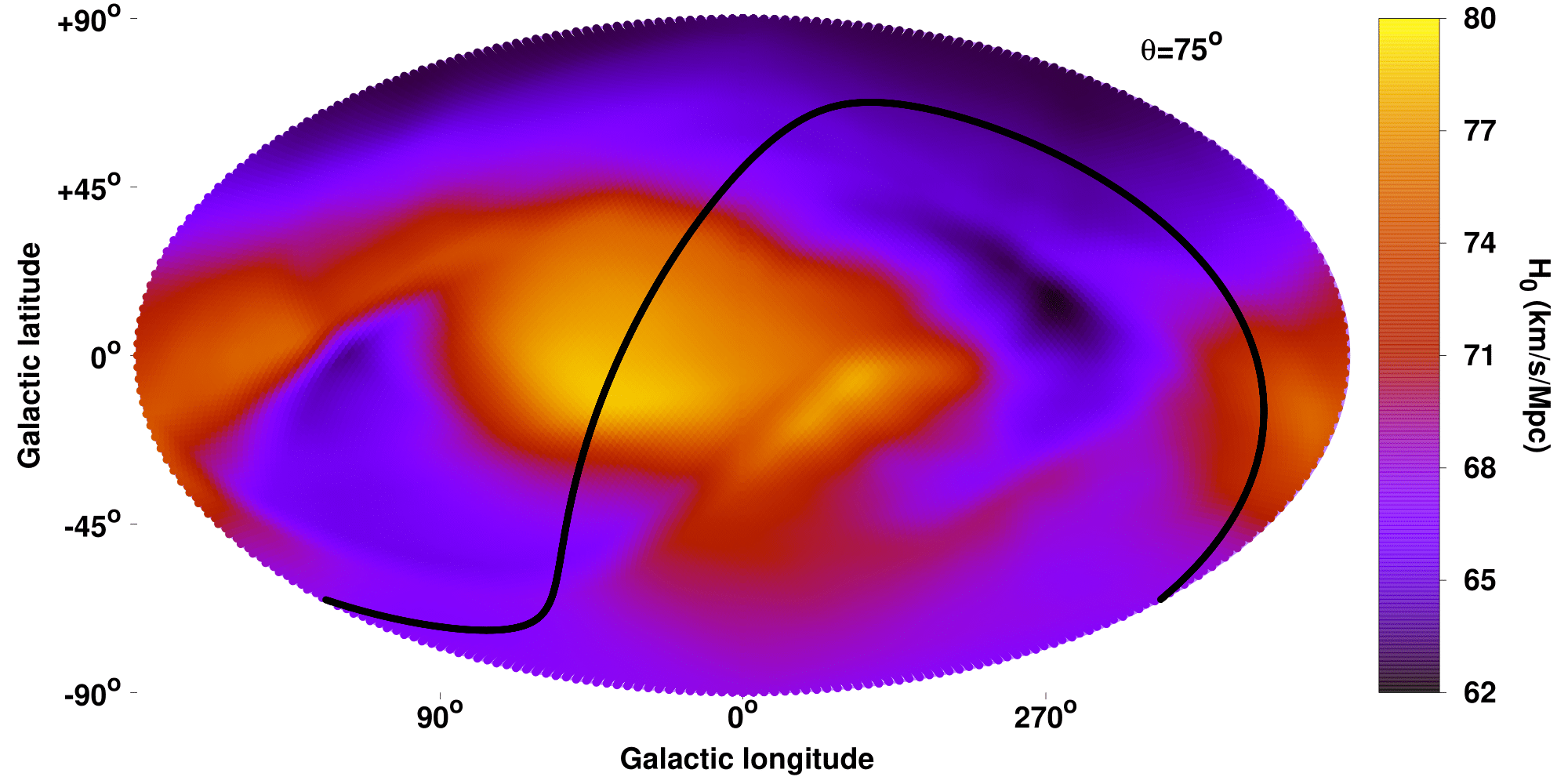}
                \includegraphics[width=0.51\textwidth, height=5cm]{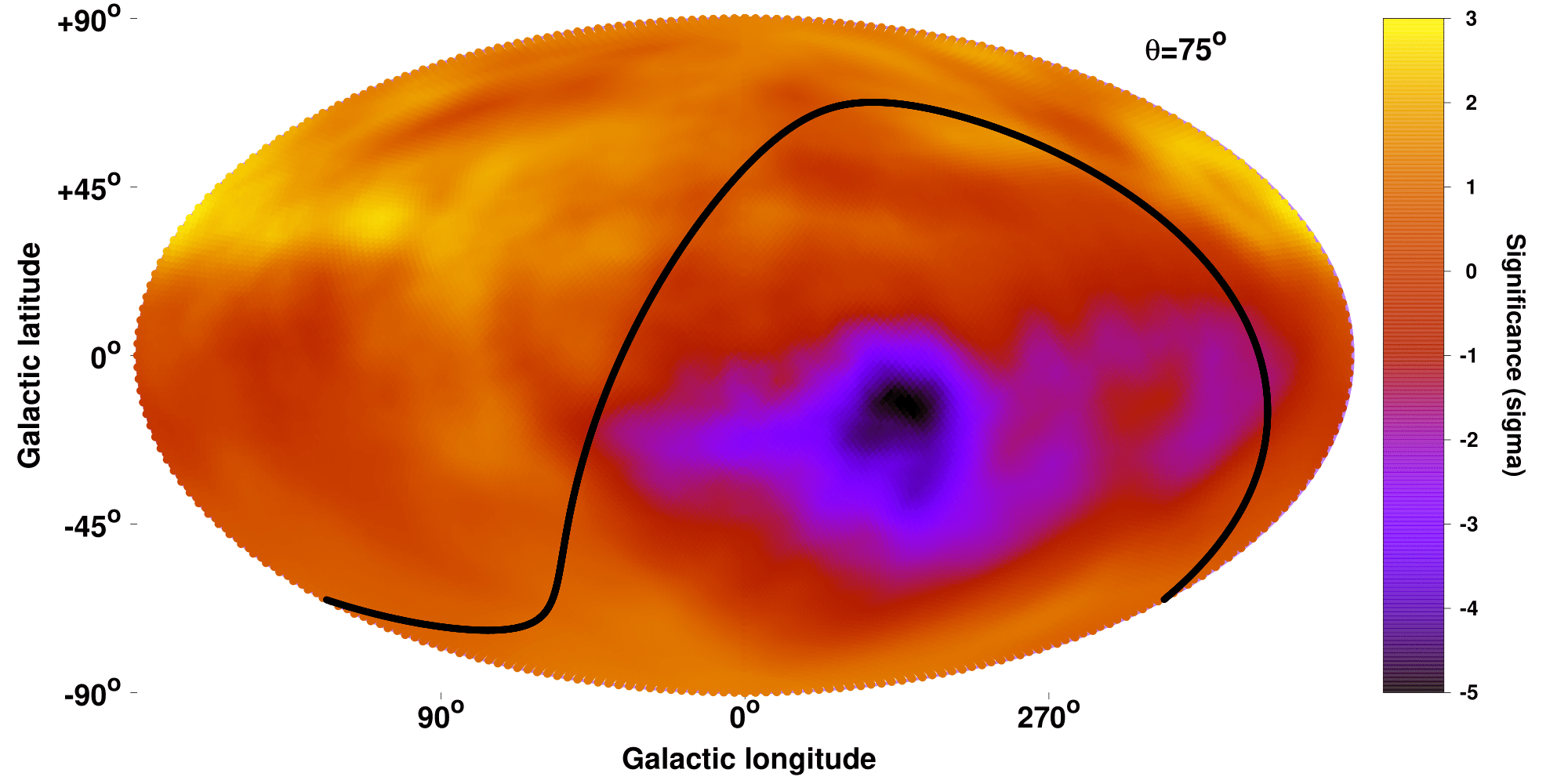}
                \includegraphics[width=0.51\textwidth, height=5cm]{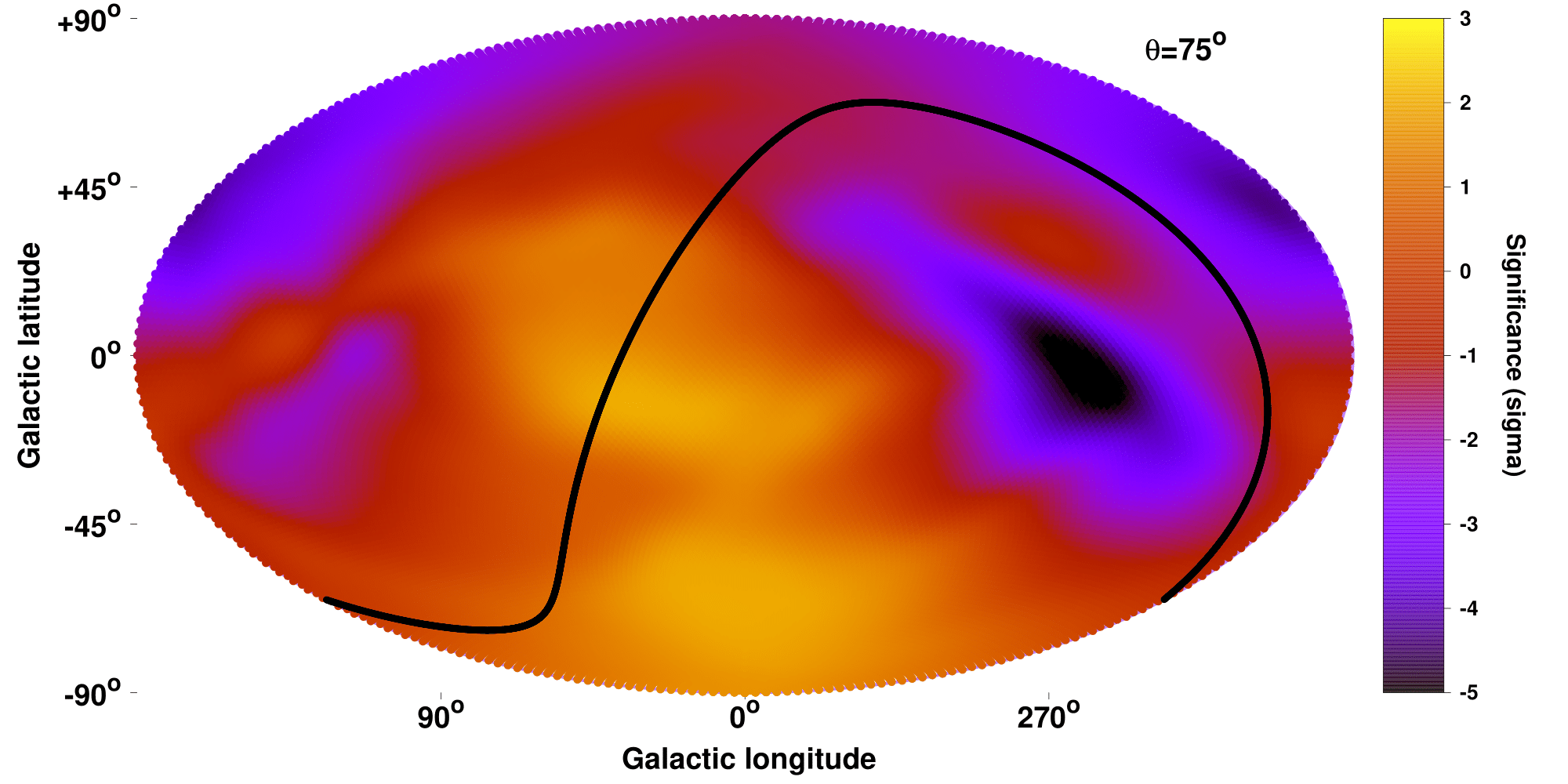}
               \caption{\textit{Top:} Best-fit $H_0$ value as a function of the position in the extragalactic sky for $\theta =75^{\circ}$ cones for ACC (left) and XCS-DR1 (right). \textit{Bottom:} Significance map of the anisotropy between every sky region and the rest of the sky for ACC (left) and XCS-DR1 (right).}
        \label{acc-xcs}
\end{figure*}

For the ACC and XCS-DR1 samples only the 1D analysis is presented in M18. In order to see if the behavior of the $L_{\text{X}}-T$ relation for these two samples is comparable to the one of our sample in the 2D space, we repeat the analysis described in this paper using these two samples. Prior to the analysis, the 104 common clusters between our sample and ACC are excluded from the latter. Since the focus of the paper is on the cluster sample we build and use, here we only present the results for $\theta=75^{\circ}$. Nevertheless, we cannot use more narrow cones to either sample. This is due to the fact that ACC does not have enough clusters (168 in total, after the exclusion of the 104 common clusters) for such small cones, while the spatial distribution of the XCS-DR1 clusters is not entirely uniform. This results in the number of clusters falling below 30 for many regions when 60$^\circ$ cones are used. The results for both samples can be seen in Fig. \ref{acc-xcs}.

For ACC the highest and lowest $H_0$ (brightest and faintest respectively in terms of $A$) regions are at $(l,b)=(77^{\circ}, +15^{\circ})$ and $(l,b)=(317^{\circ}, -14^{\circ})$ respectively. These two directions are relatively consistent with the general behavior of our sample, with a $\sim 40^{\circ}$ separation compared to the results of our sample. We remind the reader that the two samples are completely independent. The $H_0$ values of these extreme regions are $H_0=78.76\pm4.15$ km/s/Mpc and $H_0=58.12\pm2.68$ km/s/Mpc deviating by $4.18\sigma$ ($30\pm7\%$). Their angular separation is $122^{\circ}$, which is similar to the ones for the extreme regions of our sample. The most extreme dipole is found toward $(l,b)=(327^{\circ}, -21^{\circ})$ with a $3.68\sigma$. It is noteworthy that the low $H_0$ is much more statistically significant in this sample than the high $H_0$ region, indicating a monopole anisotropy. One obtains similar results for ACC when $B$ is left free to vary within its 2$\sigma$ limits as a nuisance parameter, but with a decreased statistical significance. In that case, the statistical significance of the anisotropy slightly drops to $3.12\sigma$ (from $4.18\sigma$ for a fixed $B$) toward similar sky directions.

For XCS-DR1 the most extreme regions are located at $(l,b)=(31^{\circ}, +25^{\circ})$ (brightest) and $(l,b)=(281^{\circ}, +24^{\circ})$ (faintest) separated by $117^{\circ}$. Their respective $H_0$ values are $H_0=77.91\pm2.20$ km/s/Mpc and $H_0=63.56\pm 2.32$ km/s/Mpc deviating by $4.52\sigma$ ($21\pm5\%$). One can see that this discrepancy is larger than the one in our sample or the one obtained from ACC. However, XCS-DR1 has some properties that might lead to overestimating the anisotropies between different sky region. For instance, overluminous clusters tend to have smaller statistical uncertainties (M18), and when these clusters are in the center of the cones (higher statistical weight), this can lead to artificially high $H_0$ (or $A$). Thus, one has to be conservative when interprenting the statistical significance of the anisotropies found in the XCS-DR1 sample.

Interestingly, the direction for the lowest $H_0$ (which corresponds to the maximum cosmic expansion rate) is separated only by 28$^{\circ}$ from the CMB dipole. Since XCS-DR1 is a high redshift sample (median $z\sim0.35$), naively one would not expect any effects on the XCS-DR1 results due to the peculiar velocity of the Solar System compared to the CMB frame and therefore there is no obvious reason why these two directions should be close. The most extreme dipole for XCS-DR1 is located toward $(l,b)=(211^{\circ}, +14^{\circ})$ with a $2.75\sigma$ significance.

Now we allow $B$ to vary within its best-fit $2\sigma$ limits. Some changes are observed, although the general directional behavior of $H_0$ remains relatively consistent. The statistical significance of the maximum anisotropies significantly decreases from $4.52\sigma$ to $2.82\sigma$. This is due to the fact that the median $T=2.7$ keV of the XCS-DR1 sample is smaller than the pivot point (4 keV) of the $L_{\text{X}}-T$ relation, and thus $A$ and $B$ values are more correlated than for our sample (or ACC). These small differences between the results of the two cases can be avoided if one chooses the pivot point to be $\sim 2.7$ keV for the XCS-DR1 modeling. Despite of these small alternations, the most extreme region is still found toward $(l,b)\sim (292^{\circ}, +23^{\circ})$, only $\sim 10^{\circ}$ away from the previously found direction, and still with a $\sim 3\sigma$ significance.

\subsection{Combining the $H_0$ results for the three samples}

Remarkably, ACC and XCS-DR1 roughly agree with our sample on their $L_{\text{X}}-T$ anisotropic behavior despite the fact that they do not share any common clusters. While at first sight it might seem that the $H_0$ maps of ACC and XCS-DR1 look different, the location of their most extreme regions is still consistent within $\sim 40^{\circ}-55^{\circ}$. 

In total, they contain 842 different galaxy clusters. Consequently, any constraints on the fitted parameters would be much stronger if we combined them. While, the normalization values of the three samples are quite different (cluster populations, used energy range for $L_{\text{X}}$, $T$ constrain method etc. vary significantly), $H_0$ is a global parameter that should not depend on specific samples or even cosmological probes. The normalization and slope values of the three different samples can be set in such way so the best-fit $H_0$ value considering the entire sample is $H_0=70$ km/s/Mpc. Nevertheless, we see that the three samples return a different $H_0$ range. As shown before ACC and XCS-DR1 show a larger variation of $H_0$ ($\pm \sim 20\%$) than our sample ($\pm \sim 9\%$). This correlates with the larger scatter of the other two samples and it can be attributed to randomness (since the $H_0$ uncertainties of ACC and XCS-DR1 are $\sim 2-3$ times larger than the ones of our sample), reasons that we have not yet identified or a combination of the above (the significance however remains similar for the three samples).

\begin{figure*}[hbtp]
               \includegraphics[width=1.05\textwidth, height=9cm]{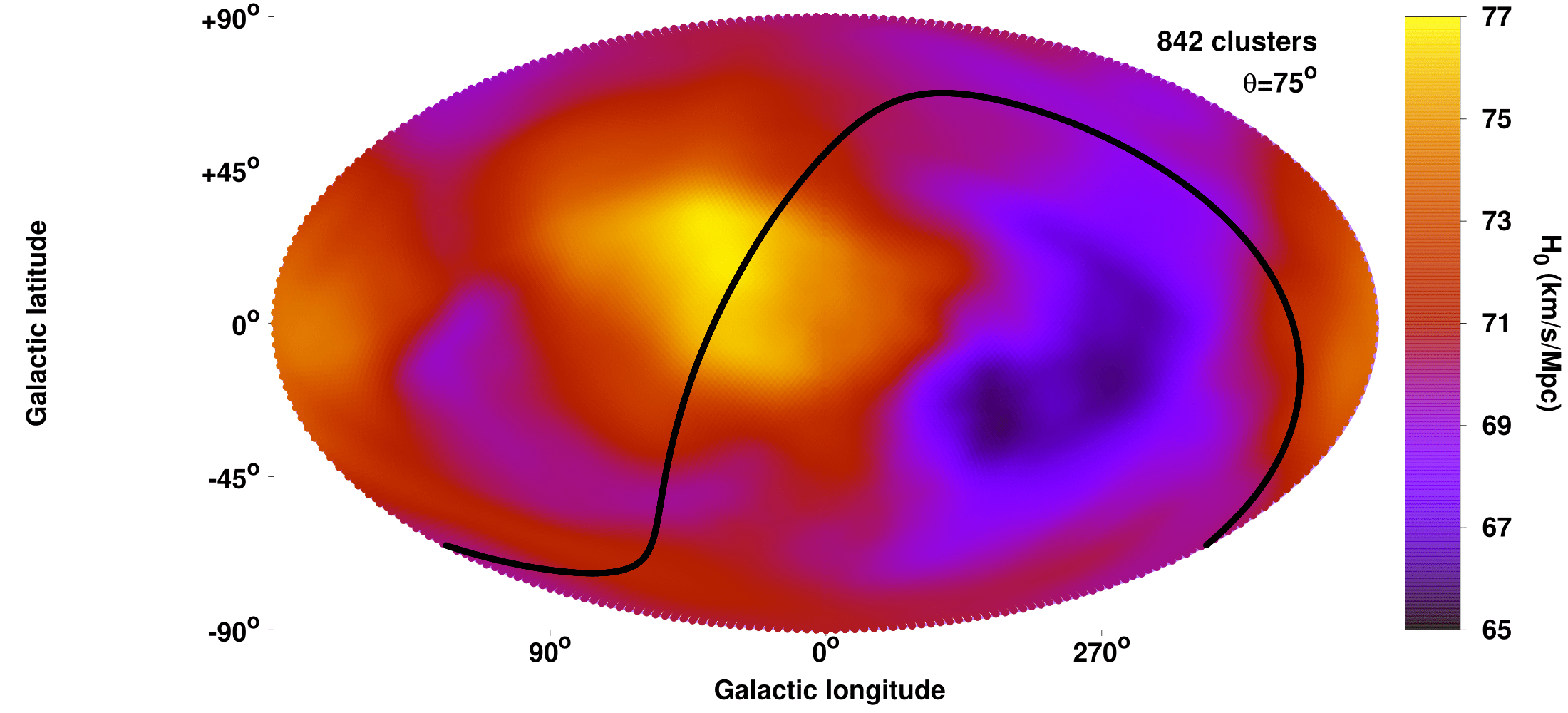}
               \includegraphics[width=1.05\textwidth, height=9cm]{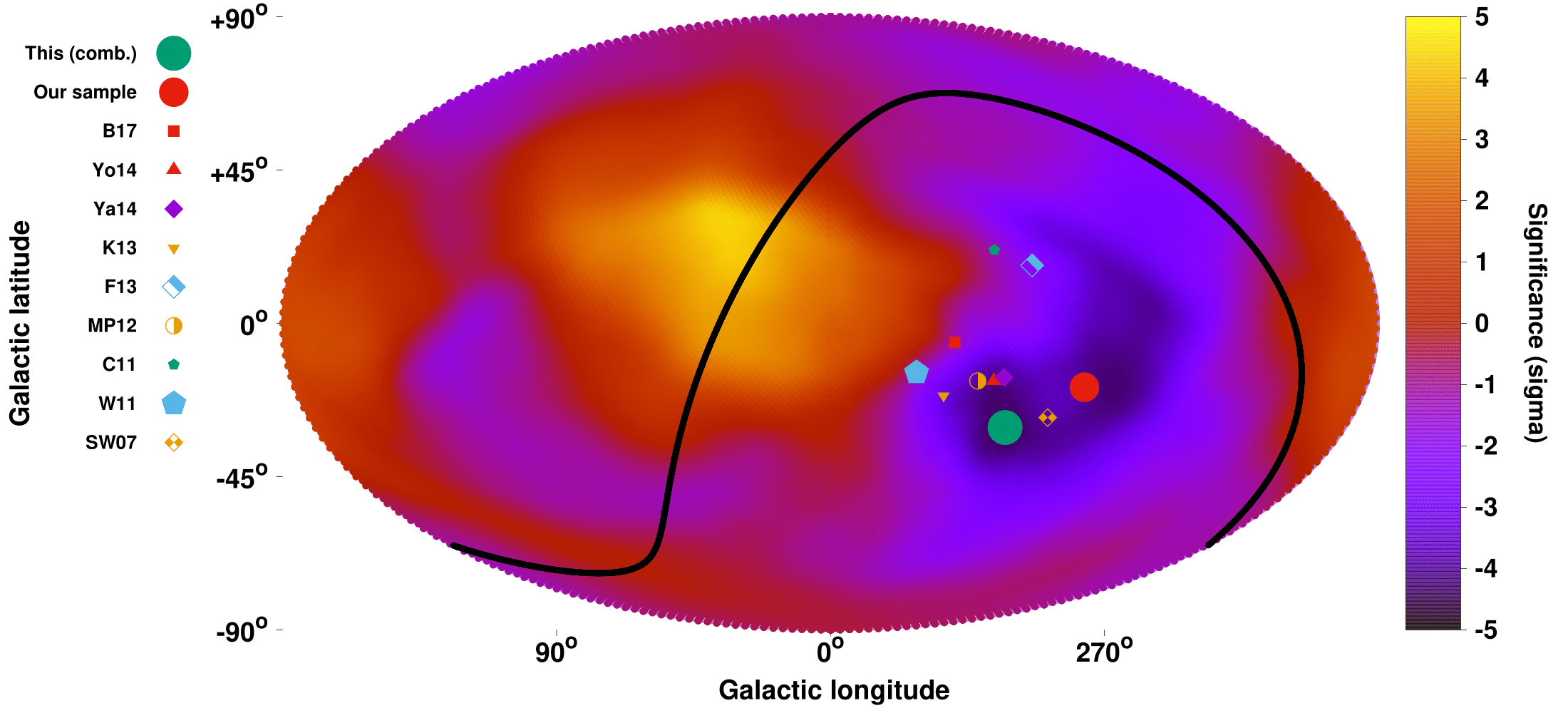}
               \caption{Combined results of $H_0$ as obtained through the $L_{\text{X}}-T$ relation using all three independent samples (this work's sample, ACC and XCS-DR1), as a function of the position in the extragalactic sky for $\theta =75^{\circ}$ cones. \textit{Top}: Most likely $H_0$ value for every sky region. \textit{Bottom:} Combined significance map of the anisotropy between every sky region and the rest of the sky. We note that the color scale ($-5\sigma, +5\sigma$) is wider than the other significance maps since the amplitude of anisotropies is larger in this case. The most anisotropic directions as found in our analysis and other studies are overplotted. Larger symbols correspond to higher statistical significance. The order of the symbols (studies) follow the same order as in Table \ref{studies}.}
        \label{fig9}
\end{figure*}

By performing the $H_0$ scanning analysis, one obtains three different and independent estimations of the likelihood of the $H_0$ parameter for every region. Multiplying these three likelihoods gives us the combined most likely $H_0$ value for every region in the sky. In order to consistently use the three samples, we use the smallest possible cone radius (75$^{\circ}$) for which we have enough data for all three catalogs in any cone, and we use the same parameter fitting range ($H_0\in [50,90]$ km/s/Mpc) as well. Therefore, the $H_0$ map displayed in the left panel of Fig. \ref{fig9} is obtained, while the significance map is shown in the bottom panel of the same figure (we also overplot the results of other studies, as discussed in Sect. \ref{other_studies} and Table \ref{studies}).

From the combined $H_0$ results, the lowest value $H_0=65.20\pm 1.48$ km/s/Mpc occurs toward $(l,b)=(303^{\circ}, -27^{\circ})$ (237 clusters) while the highest value $H_0=76.64\pm 1.41$ km/s/Mpc is found at $(l,b)=(34^{\circ}, +26^{\circ})$ (302 clusters). Therefore, the null isotropy hypothesis between these two regions is rejected with a remarkable significance of $5.59\sigma$ ($16\pm3\%$). The angular separation of these two regions is $103^{\circ}$. On the other hand, the strongest dipole occurs toward $(l,b)=(265^{\circ}, -20^{\circ})$ (57$^{\circ}$ away from the CMB dipole) with a significance of $4.06\sigma$.

We repeat the joint analysis considering the obtained $H_0$ results from every sample when $B$ was left free to vary as a nuisance parameter. As expected, the overall behavior of $H_0$ persists with some limited changes. The statistical significance of the maximum anisotropy drops to $4.55\sigma$ (from $5.59\sigma$), and is found between $(l,b)\sim (312^{\circ}, -21^{\circ})$ and $(l,b)\sim (45^{\circ}, +21^{\circ})$. Consequently, the choice of keeping $B$ fixed slightly overestimates the exact statistical significance of our findings but does not affect the general conclusion.

All these results demonstrate clearly that the similar anisotropies in all three independent samples are extremely unlikely to be random and that there is an underlying reason causing the $L_{\text{X}}-T$ relation to show a strong directionally depended behavior. 

\section{Discussion}\label{discuss}

The significance of cosmic isotropy for the standard cosmological paradigm is undisputed. Designing scrutinizing methods to test this hypothesis is vital since much new information about the Universe can be revealed through such tests. 

One can assume that the isotropic expansion of the Universe holds, but a cosmological probe could still consistently show a significantly anisotropic behavior. This could result in the identification of yet unknown factors with a surprisingly strong impact on the data collection, analysis, or both. Since these factors are not accounted for in previous studies using similar wavelengths (e.g., X-rays) or the same astrophysical objects, these biases could in principle extrapolate to many aspects of relative research fields. 

For instance, the anisotropy of the $L_{\text{X}}-T$ scaling relation found in this paper could have multiple implications for other studies using X-ray galaxy clusters or other X-ray objects. Since the strong anisotropies do not strongly depend on the specific sample, X-ray satellite etc. the vastly more probable scenario is that the underlying reason is not a sample-specific systematic. 

Moreover, the amplitude and direction of the anisotropies are preserved even after excluding several cluster subcategories, such as low-$T$ systems, local clusters, clusters with high absorption, metal-rich clusters, high flux ones etc. Thus, different subpopulations of clusters toward different directions is not a likely explanation as well. Additionally, possible biases due to selection effects do not explain the findings. Therefore, if this is eventually proven to be caused by an unknown (extra)Galactic effect acting on X-ray photons, previously published results would need modifications correcting for this effect. 

Such an example would be the galaxy cluster masses obtained through the X-ray luminosity-mass scaling relation $L_{\text{X}}-M$. If we assume a typical scaling relation slope of $L_{\text{X}}\sim M^{1.5}$, then the masses of the clusters toward the faint regions of our analysis would be underestimated by $\sim 10-20\%$ while the clusters in the bright regions would end up with masses overestimated by the same amount. As a result, the cosmological parameters obtained via the halo mass function could be biased if the sky coverage of the clusters is not uniform. Even in the latter case, the scatter of the final results would increase. It is characteristic that without applying any statistical weighting in the clusters and performing the sky scanning using $\theta =60^{\circ}$, $\sim 72\%$ of the subsamples in the different directions show a lower $\sigma_{\text{int}}$ than the full sample results. This indicates the potential increase of the scatter in X-ray scaling relations when the full sky is used as if galaxy clusters were showing the same behavior everywhere. 
Possibly biased results when the used samples do not cover the full sky homogeneously can clearly occur to any other studies as well, if these use measured X-ray luminosities (or temperatures) of galaxy clusters. 

Another useful test would be to study the dependance of these anisotropies on the exact energy range. This will be particularly helpful in order to check if the observed anisotropies could be the result of absorption effects, such as strong variations in the galactic ISM metallicity, metal-rich nearby dwarf galaxies etc. However, in Sect. \ref{extra_nh_var_metal} we showed that this is unlikely, but further testing is needed. Nevertheless, checking if these $L_{\text{X}}-T$ anisotropies also appear in the hard X-ray band alone, where the absorbing effects are minimal, would provide us with valuable information about their exact nature. This will be feasible with the upcoming eROSITA all-sky survey. Here we should remind the reader that while we only use the $0.1-2.4$ keV energy range for $L_{\text{X}}$, the ACC and XCS-DR1 samples (which also show similar anisotropies), use the bolometric energy range. 

In the $L_{\text{X}}$ measurements used for this study the cluster cores are not excluded, since this is very difficult to do with \textit{ROSAT} data due to its large PSF. It has been shown however that core-excised luminosities scatter less in their scaling with temperature \citep[e.g.,][etc.]{markev,pratt,maughan}. Such values would be optimal for our analysis since a lower scatter in the $L_{\text{X}}-T$ relation would decrease the uncertainties of the derived $A$ values. This could eventually allow the detection of spatial anisotropies with an even higher statistical significance and strengthen our results. This will be possible with \textit{eROSITA} data and with possible future \textit{XMM-Newton} and \textit{Chandra}-based samples that provide core-excised luminosity values.

The summary of the best-fit $A$, $B$ and $\sigma_{\text{int}}$ values is shown in Table \ref{results}. The directions of the most extreme regions for every subsample, together with the statistical significance of the anisotropic signal between these two regions and the direction and significance of the most anisotropic dipole are shown in Table \ref{anisot_results}.

\begin{table*}[hbtp]
\caption{\small{Best-fit normalization $A$ and slope $B$ values of the $L_{\text{X}}-T$ relation with their 1$\sigma$  ($68.3\%$) uncertainties. The results for different examined subsamples are displayed as well as for the full sample. Also, the XCS-DR1 and ACC results are displayed, where for the latter the $T-L_{\text{X}}$ fitting is performed (denoted by *) as described in M18. The intrinsic and total scatter are also shown for comparison.}}
\label{results}
\begin{center}
\renewcommand{\arraystretch}{1.3}
\small
\begin{tabular}{ c  c  c  c  c }
\hline \hline

Clusters (No.) & $A$ & $B$ & $\sigma_{\text{int}}$ (dex) & $\sigma_{\text{tot}}$ (dex) \\
\hline \hline

 & &\ \  Our sample & & \\ \hline

All (313) &\ $1.114^{+0.044}_{-0.040}$\ & \ $2.102\pm 0.064$ \ & $0.239$ & $0.262$ \\ \hline
$T>2.5$ keV, $z>0.03$ (246) &\  $1.114^{+0.047}_{-0.041}$\ & $2.096\pm 0.078$\ & $0.218$ & $0.236$ \\ 
$T>3$ keV, $z>0.05$ (198) &\ $1.172^{+0.053}_{-0.046}$\ & \ $2.049\pm 0.077$\ & $0.205$ & $0.228$ \\ 
$Z_{\text{core}}\leq 0.590\ Z_{\odot}$ (209) &$1.135^{+0.060}_{-0.051}$\ & \ $2.034\pm 0.082$\ & $0.252$ & $0.270$ \\ 
$Z_{\text{out}}\leq 0.426\ Z_{\odot}$ (209) &$1.197^{+0.058}_{-0.051}$\ & \ $2.006\pm 0.073$\ & $0.231$ & $0.254$ \\ 
$N_{\text{Htot}}\leq 5.16\times 10^{20}/$cm$^2$ (209) & $1.084^{+0.052}_{-0.046}$\ & \ $2.082\pm 0.072$\ & $0.226$ & $0.249$ \\ 

 \hline
 
   & &\ Other samples& &\\ \hline
XCS-DR1 (364) &$1.315^{+0.088}_{-0.079}$\ & \ $2.462\pm 0.086$\ & $0.206$ & $0.379$ \\
ACC* (168) &\ $2.660^{+0.243}_{-0.190}$\ & \ $3.635\pm 0.135$ & $0.101$ ($\sigma_{T|L_{\text{X}}}$) &  $0.119$ ($\sigma_{T|L_{\text{X}}}$) \\

\end{tabular}
\end{center}
\end{table*}

The consistent value of the slope throughout the different subsamples is noteworthy. The largest difference ($\sim 1\sigma$) is found between the $Z_{\text{out}}\leq 0.426\ Z_{\odot}$ and the $N_{\text{Htot}}\leq 7.37\times 10^{20}/$cm$^2$ subsamples. On the other hand, $A$ deviates by $\sim 2\sigma$ between the $T>3$ keV, $z>0.05$ and the $N_{\text{Htot}}\leq 7.37\times 10^{20}/$cm$^2$ subsamples, while it is quite consistent between the rest. As expected, the lowest scatter is found for the subsamples with the highest $T$ and $z$. On the contrary, the subsample where the high $Z_{\text{core}}$ clusters were excluded returns the largest scatter, still consistent though with the other subsamples. We should also note here the significantly lower total scatter of our sample against XCS-DR1 and ACC (after converted to $L_{\text{X}}-T$ scatter).

\begin{table*}[hbtp]
\caption{\small{Directions of the most statistically significant lowest and highest $A$ and $H_0$ sky regions are displayed together with their statistical deviation from one another. Additionally, the direction and statistical significance of the most anisotropic dipole is displayed. The results are shown for the same subsamples as in Table \ref{results}. For the results labeled as "Case 1" and "Case 2"  see Sect. \ref{free_slope}.}}
\label{anisot_results}
\begin{center}
\renewcommand{\arraystretch}{1.1}
\small
\begin{tabular}{ c  c  c  c  c  c}
\hline \hline

Clusters ($\theta$)  & Lowest $A/H_0$ ($l,b$) & Highest $A/H_0$ ($l,b$) & Anisotropy significance & Max. dipole ($l,b$) & Dipole significance \\
\hline \hline
 & &\ \  Our sample & & & \\ \hline

All ($90^{\circ}$) &\ $(272^{\circ}, -8^{\circ})$\ & \ $(47^{\circ}, +22^{\circ})$ \ & $2.59\sigma$ & $(230^{\circ}, -20^{\circ})$ & $1.90\sigma$\\ 
All ($75^{\circ}$) & $(274^{\circ}, -22^{\circ})$ & \ $(17^{\circ}, -9^{\circ})$\ & \ $3.64\sigma$ \ & $(263^{\circ}, -21^{\circ})$ & $3.21\sigma$ \\
All ($60^{\circ}$) &\ $(281^{\circ}, -16^{\circ})$\ & \ $(34^{\circ}, +4^{\circ})$\ & $4.73\sigma$ & $(260^{\circ}, -36^{\circ})$\ & \ $3.77\sigma$\\
All ($45^{\circ}$) &\ $(280^{\circ}, +1^{\circ})$\ & \ $(32^{\circ}, +14^{\circ})$\ & $5.09\sigma$ & $(255^{\circ}, -53^{\circ})$ & $4.22\sigma$\\  \hline
All ($75^{\circ}$, varying $B$ - Case 1) &\ $(272^{\circ}, -21^{\circ})$\ & \ $(26^{\circ}, -13^{\circ})$\ & $3.74\sigma$ & $(262^{\circ}, -21^{\circ})$ & $3.29\sigma$\\ 
All ($75^{\circ}$, varying $B$ - Case 2) &\ $(269^{\circ}, -17^{\circ})$\ & \ $(23^{\circ}, -10^{\circ})$\ & $2.78\sigma$ & $(262^{\circ}, -22^{\circ})$ & $2.36\sigma$\\  \hline
$T>2.5$ keV, $z>0.03$ ($75^o$) &\  $(288^{\circ}, -35^{\circ})$\ & \ $(10^{\circ}, +16^{\circ})$\ & $4.68\sigma$ & $(194^{\circ}, -34^{\circ})$ & $3.27\sigma$\\ 
$T>3$ keV, $z>0.05$ ($75^o$) &\ $(286^{\circ}, -36^{\circ})$\ & \ $(9^{\circ}, +15^{\circ})$\ & $4.12\sigma$ & $(223^{\circ}, -47^{\circ})$ & $2.27\sigma$\\ 
$Z_{\text{core}}\leq 0.59\ Z_{\odot}$ ($75^o$) &\ $(264^{\circ}, -18^{\circ})$\ & \ $(30^{\circ}, +23^{\circ})$\ & $3.63\sigma$ & $(261^{\circ}, -20^{\circ})$ & $3.41\sigma$\\ 
$Z_{\text{out}}\leq 0.426\ Z_{\odot}$ ($75^o$) &\ $(270^{\circ}, -14^{\circ})$\ & \ $(24^{\circ}, +15^{\circ})$\ & $3.72\sigma$ & $(265^{\circ}, -16^{\circ})$ & $2.26\sigma$\\ 
$N_{\text{Htot}}\leq 5.16\times 10^{20}/$cm$^2$ ($75^o$) &\ $(242^{\circ}, -27^{\circ})$\ & \ $(35^{\circ}, -15^{\circ})$\ & $5.39\sigma$ & $(221^{\circ}, -33^{\circ})$ & $3.55\sigma$\\

 \hline
 
  & &\ Other samples& & &\\ \hline
ACC ($75^{\circ}$) &\ $(314^{\circ}, -17^{\circ})$\ & \ $(77^{\circ}, +15^{\circ})$\ & $4.18\sigma$ & $(327^{\circ}, -21^{\circ})$ & $3.68\sigma$\\
XCS-DR1 ($75^{\circ}$) &\ $(281^{\circ}, +24^{\circ})$\ & \ $(31^{\circ}, +25^{\circ})$\ & $4.52\sigma$ & $(211^{\circ}, +14^{\circ})$ & $2.75\sigma$\\ \hline
Our sample+ACC+XCS ($75^{\circ}$) &\ $(303^{\circ}, -27^{\circ})$\ & \ $(34^{\circ}, +26^{\circ})$\ & $5.59\sigma$ & $(265^{\circ}, -20^{\circ})$ & $4.06\sigma$\\ 
Same (varying $B$ - Case 1) &\ $(312^{\circ}, -21^{\circ})$\ & \ $(45^{\circ}, +21^{\circ})$\ & $4.55\sigma$ & $(271^{\circ}, -15^{\circ})$ & $3.32\sigma$\\ \hline

\end{tabular}
\end{center}
\end{table*}

Generally, as $\theta$ decreases, the statistical significance of the results increases, as we are able to pinpoint the anisotropies more effectively. However, the amount of available data is not yet enough to use even narrower angles. This will change with future surveys such as the upcoming all-sky \textit{eROSITA} survey which will provide us with a larger number of observed clusters with temperature measurements \citep{borm}. Nevertheless, the existence and consistency of these apparent $L_{\text{X}}-T$ anisotropies are already on solid ground granting these results, especially when one combines all three independent samples with a $>5\sigma$ anisotropy emerging from this. This holds even when the slope is left free to vary (within a limited range) from region to region, and then marginalized over. It is also quite interesting that the maximum anisotropic directions in almost every tested case seem to prefer an angular separation of $\sim 80^{\circ}-120^{\circ}$ instead of a dipole form. The most extreme observed dipole anisotropies have a statistical significance of $\sim 4\sigma$, with an angular distance of $\sim 50^{\circ}-100^{\circ}$ from the CMB dipole direction. At the same time, the faintest parts of the maps are slightly closer ($\sim35^{\circ}-90^{\circ}$) to the corresponding end of the CMB dipole. 

Here we should discuss some possible reasons for caution when one interprets the large statistical significance of the observed anisotropies. Firstly, while we have tested a large number of potential X-ray and cluster-related reasons and systematics that might cause such a spatially inconsistent behavior, we only tested them one by one. If one takes into account two or more such reasons simultaneously the statistical tension might decrease. Although it seems improbable that the observed anisotropies can be attributed purely to such effects (since three independent samples show similar behavior), one cannot discard the possibility of an overestimation of the anisotropies due to the (unchecked) combination of systematics. Secondly, the derived statistical significance of the results is based on the $\Delta \chi^2$ limits of the fit. While the applied bootstrap method returns similar results (see Sect. \ref{boot_section}), one still has to consider the so-called cosmic variance. To do so, one can use Monte Carlo simulations to draw similar samples from an inputted isotropic universe and, following the same method as in this paper, check how often such large anisotropies appear. This will be done in future work.

\subsection{Comparison with other studies}\label{other_studies}

Except for identifying previously unknown factors that can significantly affect the determination of physical parameters of astrophysical objects as discussed above, testing the isotropy of the Universe has of course another aspect as well. If many independent cosmological probes agree on a similar anisotropic direction and amplitude, while all known biases have been accounted and corrected for, then the hypothesis of cosmic isotropy should be reconsidered. This could eventually lead to a major shift in the standard cosmological model. 

The direction we identify as the one with the maximum acceleration (or minimum expansion rate as explained before) if the anisotropies were indeed only of cosmological origin, agrees well with many other studies that used SNIa and other probes to look for possible anisotropies in the Hubble expansion. Several examples of such studies are shown in Table \ref{studies}, together with their the most anisotropic directions and their significance.

\begin{table*}[hbtp]
\caption{\small{Several examples of different probes and methods indicating similar anisotropic results to ours.}}
\label{studies}
\begin{center}
\tiny
\begin{tabular}{ c  c  c  c  c | c}
\hline \hline

Reference  & Used method & Maximum & Significance & Angular distance from & Comments\\
 & & anisotropy ($l,b$) &  & our combined results & \\
\hline \hline

This work   & Clusters $L_{\text{X}}-T$ ($\theta=75^{\circ}$) & $(303^{\circ}, -27^{\circ})$&  $5.59\sigma$ & $-$ & Combination of all samples \\ 
This work   & Clusters $L_{\text{X}}-T$ ($\theta=60^{\circ}$)& $(281^{\circ}, -16^{\circ})$ & $4.73\sigma$ & $23^{\circ}$ & Our sample \\ \hline
  \citet{Bengaly2017}  & Infrared galaxies  & $(323^{\circ}, -5^{\circ})$ & $p=0.064$ &  $22^{\circ}$ & \\
 \citet{yoon}  & Infrared galaxies  & $(310^{\circ}, -15^{\circ})$ & $\sim 2.5\sigma$ &  $13^{\circ}$ & \\
  \citet{yang}  & SNIa  & $(307^{\circ}, -14^{\circ})$ & $p=0.046$ &  $13^{\circ}$ & \\
  \citet{kalus}  & SNIa  & $(325^{\circ}, -19^{\circ})$ & $95\% $ &  $22^{\circ}$ & $z<0.2$ SNIa\\
  \citet{feindt}  & SNIa  & $(298^{\circ}, +15^{\circ})$ & $p=$0.010 &  $41^{\circ}$ & $z<0.035$ SNIa, probably bulk flow\\
  \citet{mariano}  & SNIa+Quasars  & $(315^{\circ}, -15^{\circ})$ & $\sim 99\%$ &  $16^{\circ}$ & \\
  \citet{colin2011}   & SNIa & $(309^{\circ}, +19^{\circ})$& $p=$0.054 & $45^{\circ}$ & $z<0.06$ SNIa, probably bulk flow\\ 
   \citet{webb}   & Quasars & $(334^{\circ}, -13^{\circ})$& $4.2\sigma$ & $33^{\circ}$ & $0.22<z<4.18$, \\ 
  \citet{schwarz}  & SNIa  & $(290^{\circ}, -24^{\circ})$ & $>95\%$ &  $11^{\circ}$ & $z<0.2$ SNIa\\ \hline

\end{tabular}
\end{center}
\end{table*}

Generally, it is usual that the anisotropies found in SNIa come mostly from $z\lesssim 0.1$ and they are attributed to local bulk flows, arising due to the Shapley supercluster at $(l,b)\sim(306^{\circ}, +30^{\circ})$ with $z\sim0.04-0.05$. We should note however that the anisotropic results of SNIa strongly depend on the used sample since studies that have been performed with the latest SNIa compilations tend to find consistency with isotropy as discussed in Sect. \ref{intro}. Moreover, the rather inhomogeneous SNIa coverage of the sky can create problems in the search of a preferred cosmological axis.

Within an isotropic FLRW background the directions of peculiar velocities are expected to be randomly distributed. However, a coherent bulk flow toward a massive structure due to gravitational attraction, it would affect the redshifts of local objects in a systematic way. If not taken into account, the luminosity distance (calculated through $z$) of clusters would be over or underestimated depending on their position in the sky. This could inevitably lead to apparent anisotropies arising from local probes.

Although these local flow motions are not expected to extent beyond $\sim 200h^{-1}$ Mpc, the studies shown in Table \ref{studies} (among others) detect bulk flows (or anisotropies) further away than this scale and with amplitudes which are hard to explain within $\Lambda$CDM. This detection is performed by different independent probes. The statistical significance however decreases compared to local probes due to the limited number of data in certain sky patches. An example of studying the scale of bulk flows is given in \citet{carrick} who find a $5\sigma$ bulk flow of $\sim 160$ km$/s$ extending over $200h^{-1}$ Mpc toward $(l,b)\sim (304^{\circ}, +6^{\circ})$.  The structures that could fully explain such a bulk flow motion have not been identified yet. Moreover, the direction of the anisotropies of more distant probes tends to converge with the one from the CMB dipole, but often with a slightly larger amplitude.

The consistency of the apparent anisotropies beyond $\sim 210h^{-1}$ Mpc ($z>0.05$) can be also seen in our results, where the tension with the null hypothesis of isotropy does not  decrease. Another effective test could be to perform our $L_{\text{X}}-T$ anisotropy analysis with clusters at $z>0.2$, beyond the effects of the recent large-scale bulk flow detections. Currently there are not enough data for such a test though, but this is expected to change with the upcoming all-sky eROSITA survey. Finally, if the only reason behind the anisotropies we observe in the $L_{\text{X}}-T$ behavior was local or cosmic coherent flow motions, one would expect to retrieve mostly dipole anisotropies, whether we have shown that anisotropies separated by $\sim 90^{\circ}-120^{\circ}$ are more significant in our analysis. However, a more in-depth testing is needed to draw safe conclusions about this scenario.

\subsection{Statistical significance validation by bootstrapping} \label{boot_section}

In order to further investigate the statistical significance of our results and if they could be attributed to pure chance we perform a bootstrap resampling analysis. We consider two cases: 

In the first case, we used all the 313 clusters covering the whole sky. We drew $10^5$ random subsamples of the same size as the region we want to test its significance. We assigned random statistical weights in the drawn clusters\footnote{These weights follow the average $1/\cos{(...)}$ distribution of weights applied throughout the sky scanning method.} to simulate the method we use during the sky scanning analysis. There, the weights were assigned based on the distance of every cluster from the center of the scanning cone. This test demonstrates how often our cluster sample can reproduce such low or high $A$ values randomly and independently of the direction, when having the same number of clusters as in the extreme regions.

In the second case, we excluded the subsample of interest and performed the $10^5$ resamplings based on the rest of the clusters. This way, we can estimate how many times the extreme result of the excluded subsample can occur randomly from data in other directions. The random statistical weighting is used here as well. 

Both cases also offer a direct comparison with the deviations occurring from the $\Delta \chi^2$ limits, from which the reported statistical significance for every result comes from. 
In order to have minimal overlapping between the $10^5$ realizations, we choose to perform this analysis for the results occurring for $\theta=60^{\circ}$. The number of clusters in the extreme regions is small enough so there is no significant overlapping, while it is large enough to be relatively insensitive to strong outliers.

Drawing and analyzing $10^5$ subsamples of 84 clusters from the full sample, we find that only $0.68\%$ of the results have a lower $A$ than the one found for the $(l,b)=(281^{\circ}, -16^{\circ})$ direction ($A=0.940\pm 0.051$). This corresponds to a $p-$value of $p=0.007$ for the null hypothesis, or in a Gaussian significance of $2.71\sigma$. Now we repeat the analysis with a subsample size of 78 clusters, same as the brightest region for a $\theta=60^{\circ}$ cone toward $(l,b)=(34^{\circ}, +4^{\circ})$. We find that $10\%$ $(p=0.010, 1.65\sigma)$ of the results have a higher $A\leq1.346$ compared to the aforementioned bright region. Therefore, the statistical significance of the fainter region toward $(l,b)=(281^{\circ}, -16^{\circ})$ is much higher in that case. The statistical deviation of these two extreme regions based on the based on the $\Delta \chi^2$ limits as shown in Eq. \ref{sigma_sig} is $4.73\sigma$ (Table \ref{anisot_results}). As found from the bootstrap resampling method however is  $\sim 3.9\sigma$, slightly decreased but still significant. Finally, the most probable value for these $10^5$ realizations is $A\sim 1.118\pm 0.115$, which is consistent with the results of the full sample fitting.

For the second case, we first excluded the 84 clusters within $60^{\circ}$ from $(l,b)=(281^{\circ}, -16^{\circ})$ and we only considered the rest 229 clusters. Following the same procedure as before, we find that only $0.32\%$ of the subsamples have $A\leq 0.940$ ($p=0.003,\ 2.95\sigma$), the same result as the one we obtained from $\Delta \chi^2$ limits in Sect. \ref{extra_nh}. Doing the same for the 78 clusters (full sample except these 78) toward $(l,b)=(34^{\circ}, +4^{\circ})$ we see that an $A\geq 1.346$ value is reproduced only for $0.45\%$ of the 100000 subsamples ($p=0.005,\ 2.85\sigma$), again consistent with our previous findings. In this case where only clusters from the rest of the sky are considered, the probability of the high $A$ result to occur randomly drops significantly compared to the case where the full sample is used. This indicates that these 78 clusters strongly affect the bootstrap results when all 313 clusters are used.

\begin{figure*}[hbtp]
               \includegraphics[width=0.49\textwidth, height=6cm]{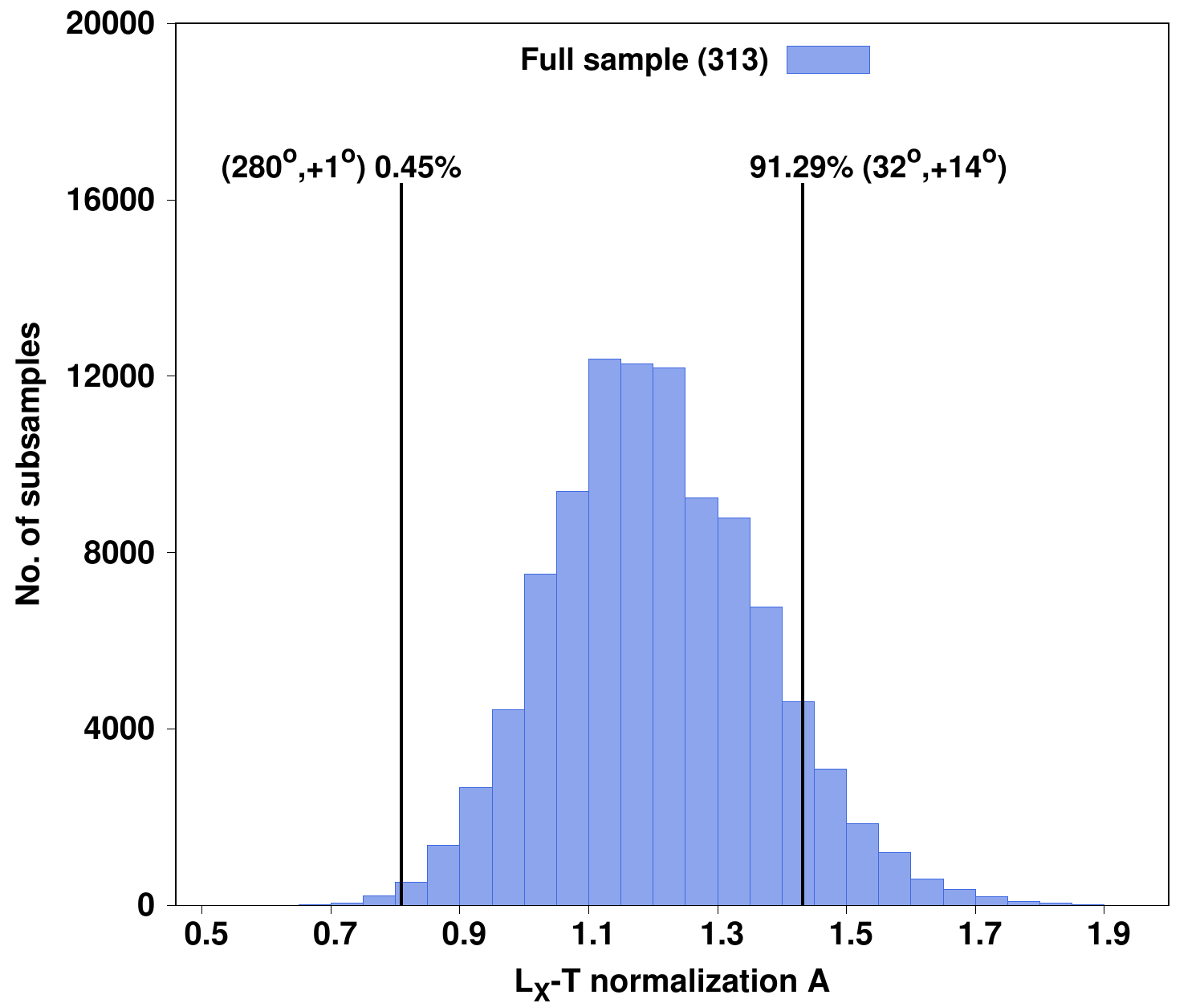}
               \includegraphics[width=0.49\textwidth, height=6cm]{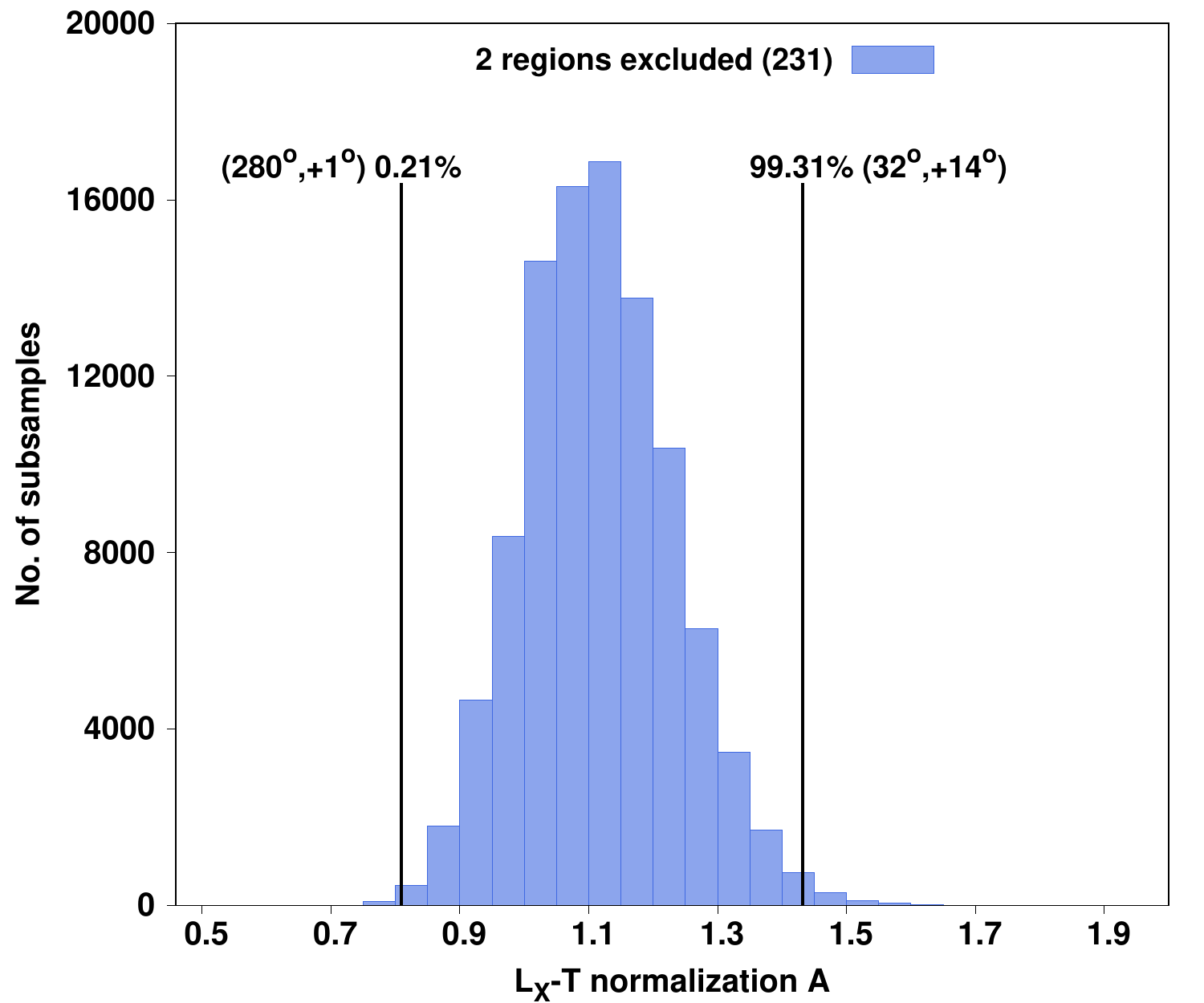}
               \caption{$L_{\text{X}}-T$ normalization results for $10^5$ realizations of 40 clusters randomly drawn from the full sample (left panel) and from the 234 clusters left after the exclusion of the two extreme $\theta=45^{\circ}$ regions at $(l,b)=(280^{\circ}, +1^{\circ})$ and $(l,b)=(32^{\circ}, +14^{\circ})$ (right panel). The statistical significance of these two regions is also displayed with the black vertical lines.}
        \label{bootstrap}
\end{figure*}

Repeating the analysis for the results when $\theta=45^{\circ}$, first we used the full sample with all the 313 clusters. For the results described in Sect. \ref{2d_analysis}, we obtain a probability of $0.45\%$ for the lowest $A$ value to occur randomly from the whole sample, and a probability of $8.71\%$ for the highest $A$ value. The deviation between these two results is $\sim 3.9 \sigma$. This is displayed in the left panel of Fig. \ref{bootstrap}.

Excluding these two extreme subsamples one at a time as we did in the second case before, the probability decreases to $0.09\%$ ($p=9\times 10^{-4},\ 3.33\sigma$) for the faintest region and to $0.67\%$ for the brightest region. The results are once again consistent with the deviations obtained from the $\Delta \chi^2$ in Sect. \ref{extra_nh}. 

If one excludes both subsamples simultaneously and only considers the 234 left, it results in a probability of $0.21\%$ and $0.69\%$ of the null hypothesis to reproduce the $A$ results of the faintest and brightest regions respectively. The result can be seen in the right panel of Fig. \ref{bootstrap}. The deviation of the two extreme results is at $\sim 5.4\sigma$, slightly higher from the $5.08\sigma$ value predicted from the $\Delta \chi^2$ limits.

Finally, as discussed in Sect. \ref{scan}, in order to ensure that the artificially low values of $\sigma_{\text{int}}$ do not affect our results, we repeated the bootstrap analysis without including the $\sigma_{\text{int}}$ term in our model (Eq. \ref{eq3}). For that we also found the $A$ value of the extreme regions without accounting for $\sigma_{\text{int}}$. The significance of the bootstrap results remains high, decreasing only by $\sim 5-12\%$ for every case (e.g., from $\sim 3.9\sigma$ to $\sim 3.7\sigma$ for the $\theta=60^{\theta}$ case and from $\sim 5.4\sigma$ to $\sim 4.8\sigma$ for the  $\theta=45^{\circ}$) since outliers are now more likely to cause extreme $A$ behaviors.

This analysis strongly demonstrates the high statistical significance of the results. Also, it is shown that the apparent anisotropies are very unlikely to be attributed to randomness, as well as that the statistical deviations obtained through $\Delta \chi^2$ limits match the ones from bootstrapping.

\section{Conclusions}\label{conclusions}

In this work, we constructed and analyzed a new, large homogeneously selected X-ray galaxy cluster sample of 313 objects, with the purpose of probing the anisotropic behavior of the $L_{\text{X}}-T$ scaling relation as first found in \citet{Migkas18} (M18). Through the strong correlation between the X-ray luminosity and temperature and the null hypothesis that the $L_{\text{X}}-T$ behavior must be similar throughout the sky, one can probe the existence of up-to-now unknown factors affecting the behavior of X-ray photons, galaxy clusters or both, for different sky directions. Furthermore, one can estimate how isotropic the Hubble expansion seems to be by constraining the cosmological parameters for different sky patches. This can be done due to the inclusion of the cosmological parameters in the X-ray flux-luminosity conversion, where we take advantage of the fact that the determination of the temperature is cosmology-independent. A necessary requirement however is to verify that no underlying unknown systematics exist, affecting the X-ray observations and the galaxy cluster scaling relations in particular. 

We tested the consistency of the $L_{\text{X}}-T$ relation for different directions by scanning the full sky using cones of different sizes, and quantify deviations in terms of the normalization parameter $A$, or the Hubble constant $H_0$. A consistent and strong directional behavior of these parameters emerged. Dividing the sky into hemispheres, we first found that the hemisphere with its pole located at $(l,b)=(272^{\circ}, -8^{\circ})$ seems to be fainter (lower $A$ or lower $H_0$) compared to the opposite hemisphere at a $2.58\sigma$ level. With our cluster sample having a quite uniform spatial distribution we could pinpoint apparent anisotropies more effectively with narrower cones. Using cones with $75^{\circ}$ down to $45^{\circ}$ radius we found that the sky region toward $(l,b)\sim (277^{\circ}, -11^{\circ})$ systematically returns a lower $A/H_0$ compared to the sky region toward $(l,b)\sim (32^{\circ}, +15^{\circ})$ with a significance of $\sim 3.6-5\sigma$ ($99.97-99.9999\%$). The main bulk of the deviations though come from the faint region rather than being balanced between the two extreme regions.

Surprisingly, the maximum dipole form anisotropies are systematically weaker by $\sim 0.4-0.9\sigma$ compared to these $\sim 110^{\circ}$ anisotropies, although still significant. Moreover, the region close to $(l,b)\sim (170^{\circ}, +15^{\circ})$ is also systematically brighter with values comparable to the $(l,b)\sim (25^{\circ}, +4^{\circ})$  region but with lower significance due to fewer clusters. 

We examined multiple reasons, mostly related to galaxy cluster physics, X-ray analysis and systematic biases, that could provide us with an explanation about the derived anisotropies. For instance, the $L_{\text{X}}-T$ behavior of different cluster population was studied. We found that clusters in low absorption regions show the same behavior with clusters in high absorption regions after the proper corrections have been applied. Excluding the latter subcategory, the anisotropies remain. Moreover, excluding galaxy groups and clusters with $T>3$ keV and redshifts of $z>0.05$ do not significantly affect our results, as can be seen in Table \ref{anisot_results}. Dividing our cluster sample according to the metallicity values of our clusters (both core and outer regions) and performing the sky scanning also does not seem to explain our findings. The same is true if one allows the slope to vary within limits during the sky scanning, and then marginalizes over the slope values. We also checked if our analysis is biased by selection effects related to the RASS exposure times of the clusters, the applied flux limit and high molecular hydrogen regions, not finding any indication for such effects. However, all these tests were done one at a time. One can argue that a combination of such effects may partially decrease the high statistical significance of the anisotropies. Of course this is still a presumption since the full magnitude of the anisotropies seems unaffected by the different tests, but it is worth checking in future work.

Furthermore, we discussed the possibility of extragalactic, metal-rich systems causing X-ray absorption that is not accounted for in the LAB survey. The case where the true metal abundance in the Galaxy's ISM shows strong spatial variations, possibly biasing the applied absorption correction and causing these anisotropies, was also discussed. Even though we showed that the last two cases are unlikely to be the reason behind the apparent $L_{\text{X}}-T$ anisotropies, it is worth checking if this behavior persists also in the case where $L_{\text{X}}$ is only measured in the hard X-ray band. In these photon energies the X-ray absorption is not significant and one would not expect any anisotropies caused by such effects. The eROSITA all-sky survey would be a great tool that will allow us to test that.

As a final test, we created $10^5$ random bootstrap subsamples and investigated the correlation of the average properties of their clusters with the best-fit $A$ value. No strong correlation was found, while the most extreme regions tended to have similar average properties. This bootstrapping method we used further verified the statistical significance of our results, while it hints to the faint sky region as the most statistically unique one. In future work, simulated isotropic samples similar to the one in this work will be used to test the frequency with which such strong anisotropies appear.

Some useful by-products of our analysis have to do with the general $L_{\text{X}}-T$ scaling relation behavior, such as the decrease in the scatter for higher $T$ and $z$ clusters, the slightly larger scatter of low core metallicity clusters compared to the rest, the $\sim 3\sigma$ discrepancy in the $L_{\text{X}}-T$ slope when \textit{Chandra} or \textit{XMM-Newton} were used, the excellent agreement between X-ray and optical redshifts, as well as the strong $L_{\text{X}}-T$ inconsistency between clusters with low and high metallicities within the $0.2-0.5\times R_{500}$ annulus.

When our sample is combined with the ACC and XCS-DR1 samples as used in M18, we see that their sky behavior agree well with each other even without having even one common cluster among them. Moreover, the fact that the observations of the three samples come mostly from three different telescopes and the sample have been compiled by different authors and sharing different properties (such as the $z$ distribution) should be kept in mind. Creating a full-sky $H_0$ map using the 842 individual clusters included in these three catalogs, a $\sim 5.5\sigma$ anisotropy was obtained between the sky regions toward $(l,b)\sim (303^{\circ}, -27^{\circ})$ ($H_0\sim 65$ km/s/Mpc) and $(l,b)\sim (34^{\circ}, +26^{\circ})$ ($H_0\sim 77$ km/s/Mpc). These values were obtained keeping the slope fixed. When the slope is free to vary one obtains similar results at a $\sim 4.5\sigma$ level. This could either mean that indeed the explanation of the anisotropies might be of cosmological origin (including strong bulk flows) or that there is a hidden (extra)Galactic factor that affects X-ray cluster measurements independently of the used sample. The direction of the anisotropies strongly correlates with results from other independent probes as shown in Table \ref{studies}.

The assumption of the isotropic nature of X-ray galaxy cluster scaling relation is common, even though this had not been observationally tested and confirmed before. The possible discovery of systematics which X-ray cluster studies do not account for until now, could considerably alter the way X-ray scaling relations are used and interpreted. If this anisotropic behavior persists in other X-ray wavelengths as well, it could indicate that also other X-ray astronomy studies might need readjustments.

On the other hand, the cosmic isotropy still remains an ambiguous topic since several independent cosmological probes have been found to have an anisotropic behavior recently. While there are results not reporting any significant anisotropies, others claim to detect $\sim 2-3\sigma$ anisotropies either in the local Universe ($z\lesssim 0.1$) or to larger distances. To assess this question, independent methods such as the $L_{\text{X}}-T$ test are needed to be applied and their results to be compared. If no biases are identified as the reason behind the anisotropies we observe and other probes seem to consistently agree, then the explanation might indeed be of cosmological origin. Such examples would be an anisotropic dark energy nature leading to different expansion rates for different directions in the late Universe, coherent bulk flow motions up to certain cosmic scales affecting the cosmological redshift measurements etc. Irrelevantly of the actual reason, studies dealing with X-ray cluster measurements are potentially affected from our findings.

\begin{acknowledgements}  
We would like to thank the anonymous referee for their valuable feedback and their constructive comments that helped solidify the robustness of our results. KM is a member of the Max-Planck International School for Astronomy and Astrophysics (IMPRS) and of the Bonn-Cologne Graduate School for Physics and Astronomy (BCGS), and thanks for their support. GS acknowledges support through NASA Chandra grants GO4-15129X, GO5-16137X, and AR9-20013X. THR, FP and MERC acknowledge support from the German Aerospace Agency (DLR) with funds from the Ministry of Economy and Technology (BMWi) through grant 50 OR 1514. LL acknowledges support from NASA through contracts 80NSSCK0582 and 80NSSC19K0116.
This research has made use of the NASA/IPAC Extragalactic Database (NED)
which is operated by the Jet Propulsion Laboratory, California Institute of Technology, under contract with the National Aeronautics and Space Administration.
\end{acknowledgements}

\newpage

\bibliographystyle{aa} 
\bibliography{XXX}          


\newpage

\appendix

\section{Extra tests}\label{extra_tests}

\subsection{Effect of redshift evolution parametrization}\label{Ez_prior}

During this analysis, we choose a fixed prior of $E(z)^{-1}$ for the redshift scaling of the $L_{\text{X}}-T$ relation as shown in Eq. \ref{eq2}. If the true scaling is not self-similar and if two separate subsamples have a different redshift distribution, artificial normalization anisotropies might be induced. As already discussed in Sect. \ref{correl_sect} though, the two most anisotropic sky regions (brightest and faintest) share a similar redshift distributions and are not expected to be affected by such possible biases. Also, the cluster redshifts are relatively low and thus $E(z)$ does not rise to high values in order to significantly affect our results.

Nevertheless, we wish to test the dependance of our results on the exact $E(z)$ prior selection. To this end, we repeat the $\theta=75^{\circ}$ analysis for the full sample, for four different cases, $E(z)^{-2}$, $E(z)^{-1.5}$, $E(z)^{-0.5}$ and for no redshift evolution (keeping $B$ fixed to the best-fit value obtained for every case separately). The location of the most extreme regions, both faintest and brightest, fluctuate only by $<9^{\circ}$. The statistical significance for the anisotropy between these regions maximizes for the $E(z)^{-2}$ case ($3.75\sigma$) and decreases gradually to $3.57\sigma$ for the case without any redshift evolution. The $A$ and the significance maps for all four $E(z)$ scenarios do not practically differ from the maps shown in the upper left panels of Fig. \ref{fig7} and Fig. \ref{sigma_cones} respectively and thus we do not display them.

\subsection{Correlations and systematics}\label{correl_append}

\begin{figure*}[hbtp]
               \includegraphics[width=0.33\textwidth, height=4cm]{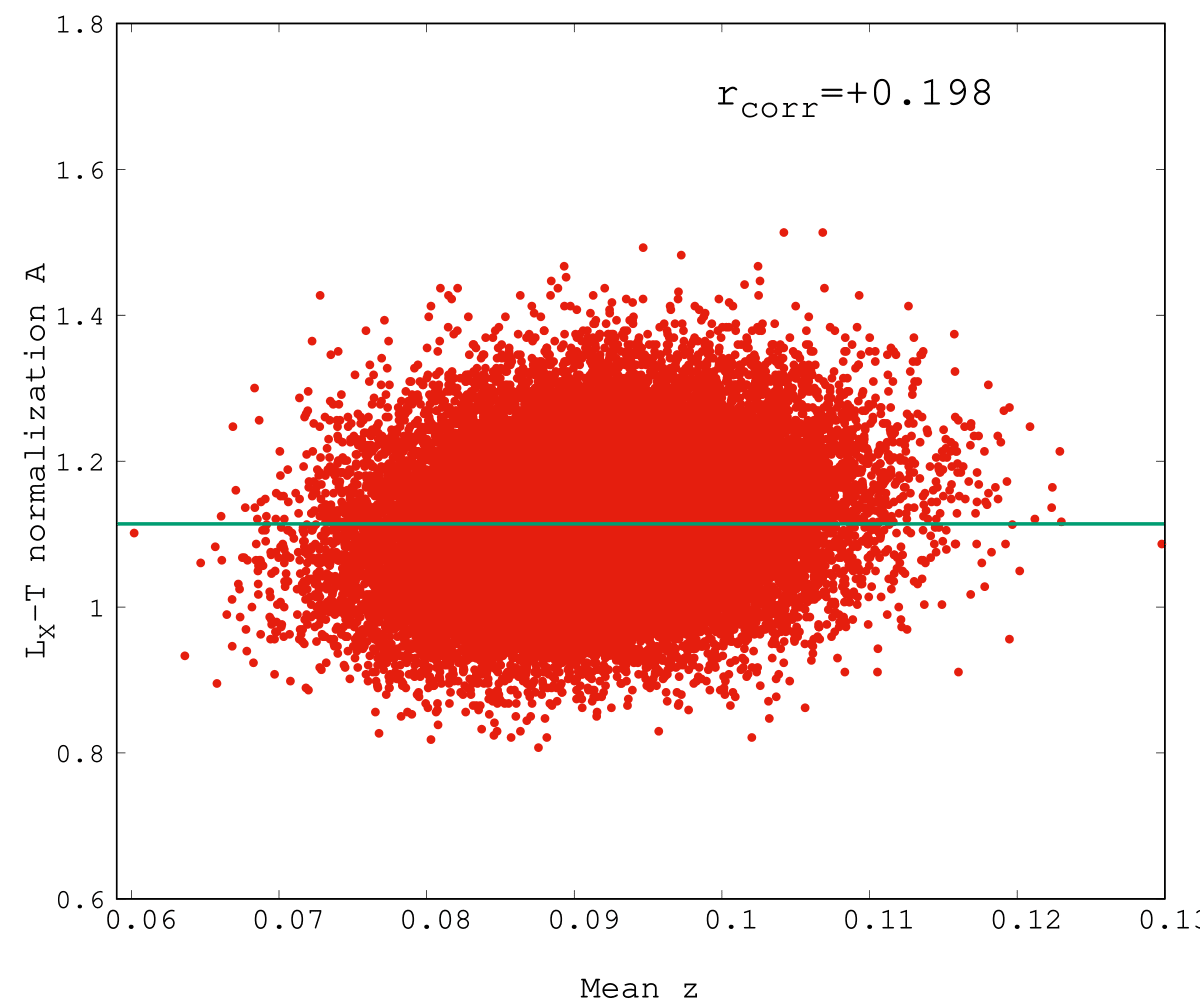}
               \includegraphics[width=0.33\textwidth, height=4cm]{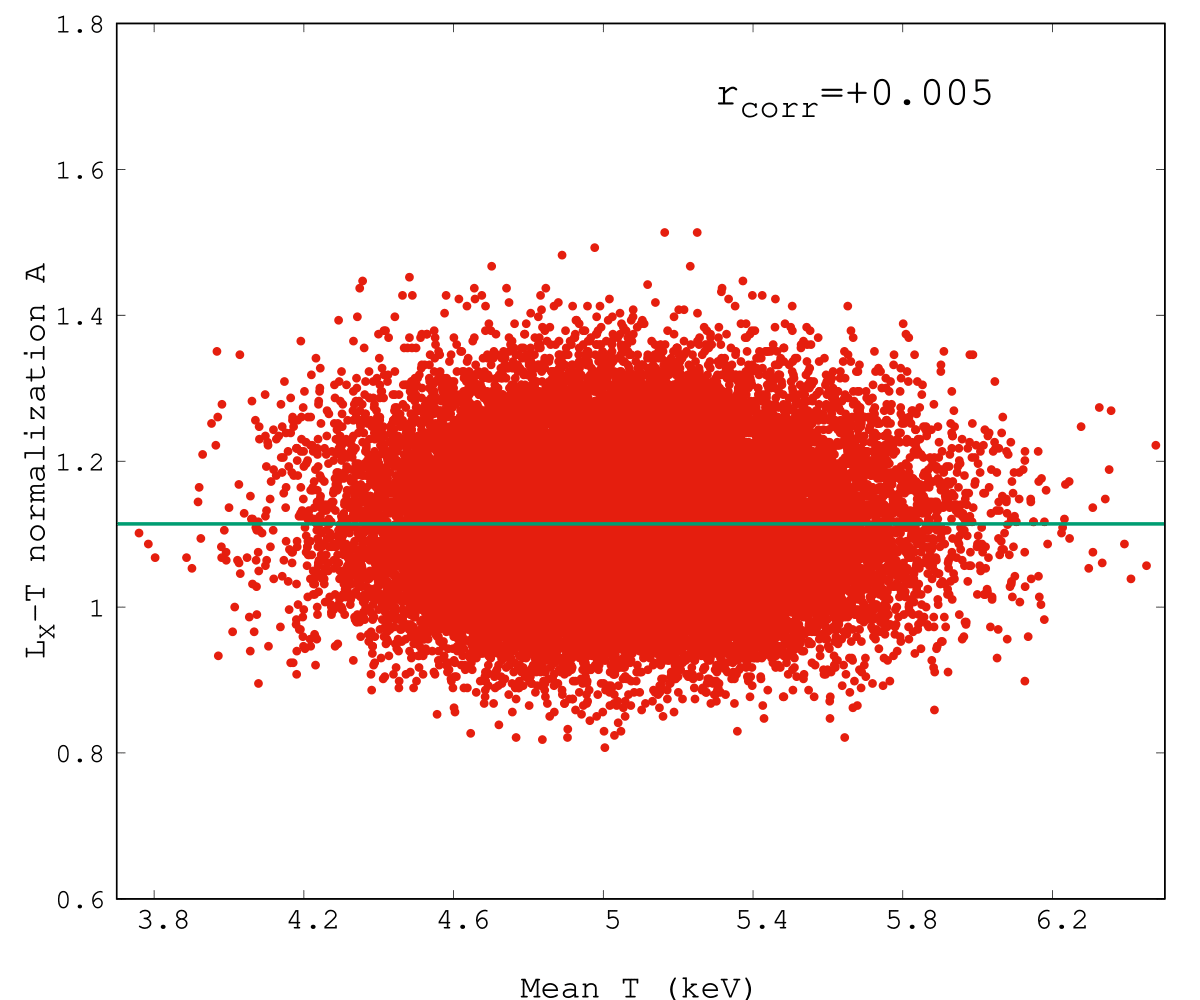}
               \includegraphics[width=0.33\textwidth, height=4cm]{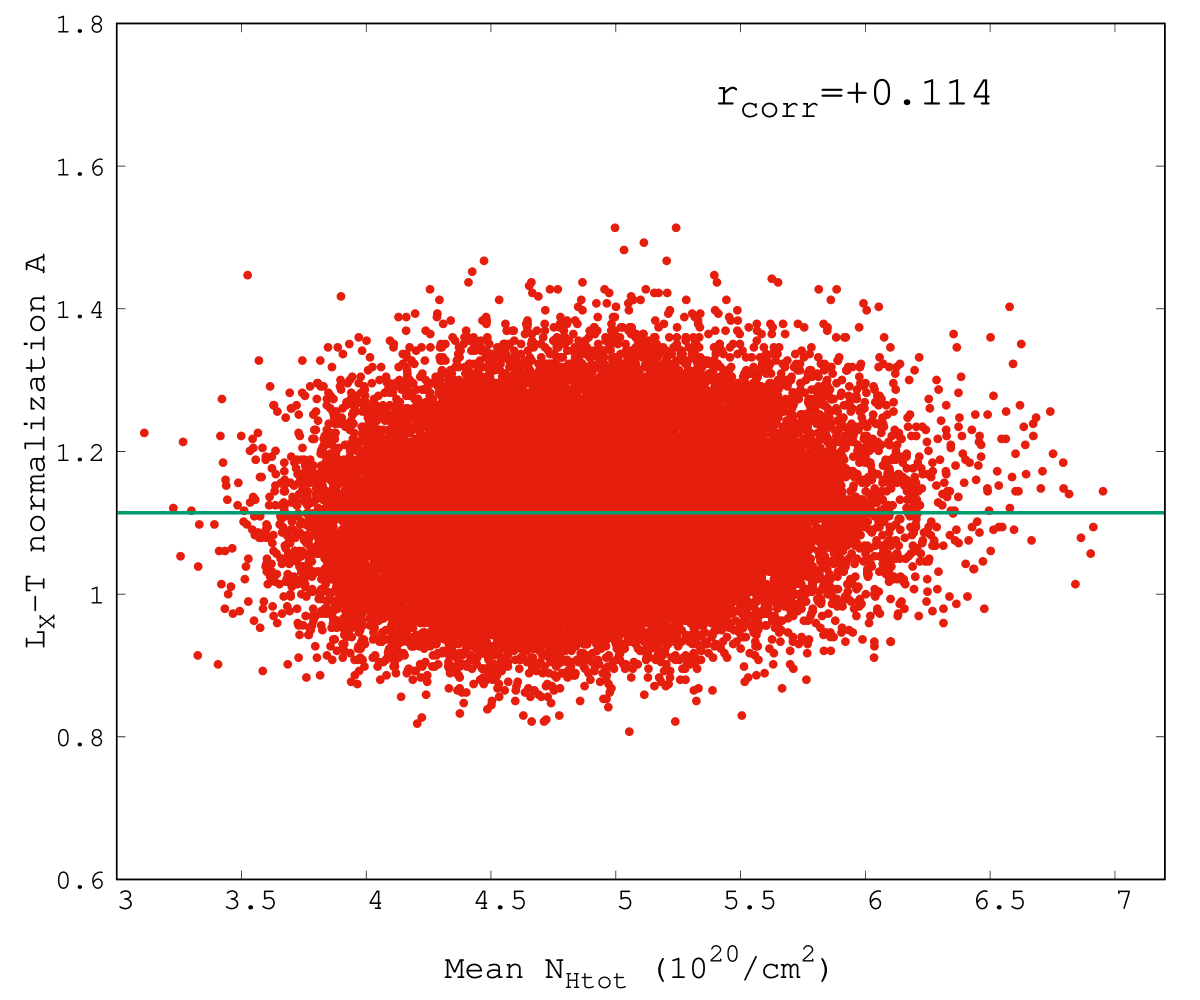}
               \includegraphics[width=0.33\textwidth, height=4cm]{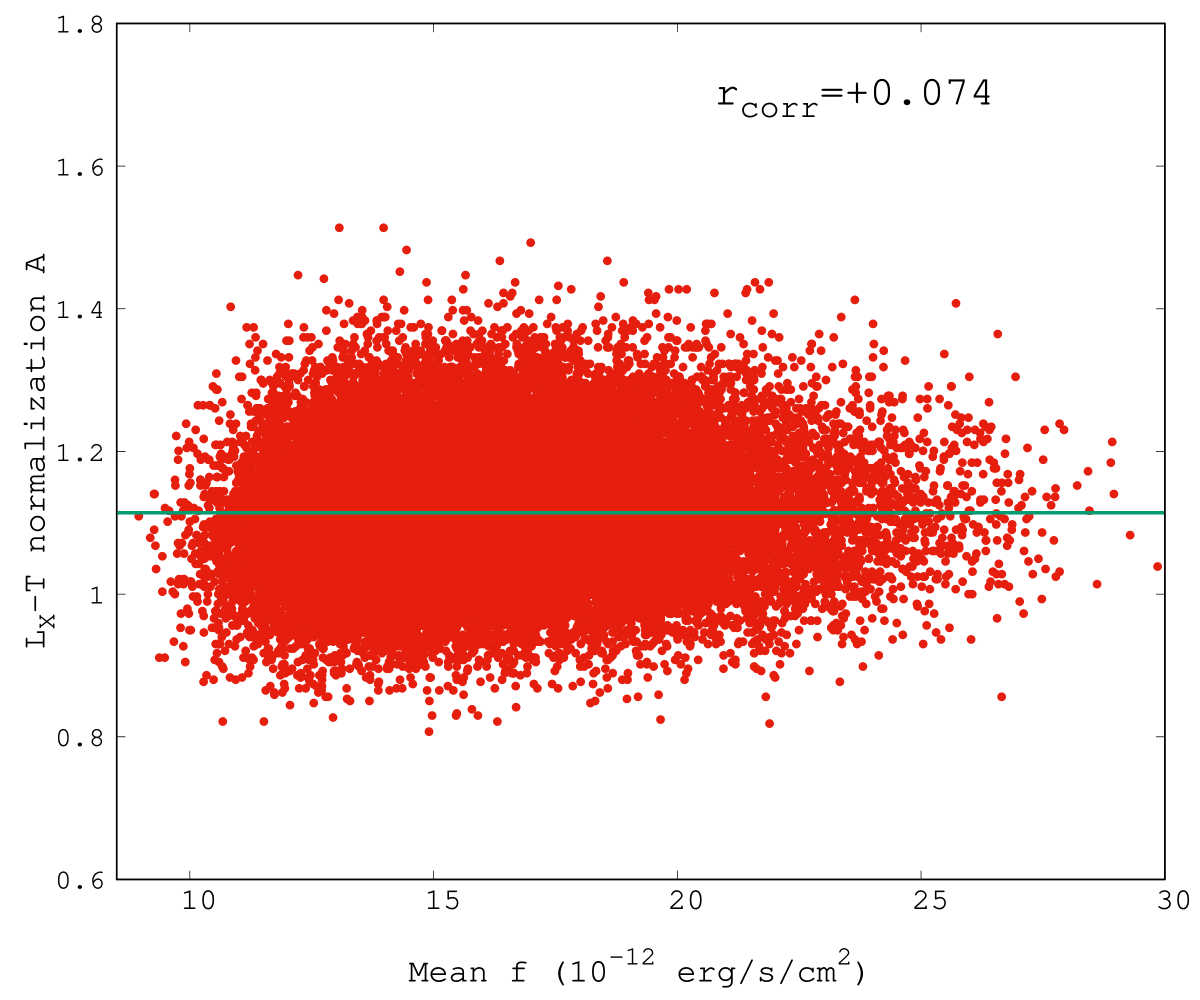}
               \includegraphics[width=0.33\textwidth, height=4cm]{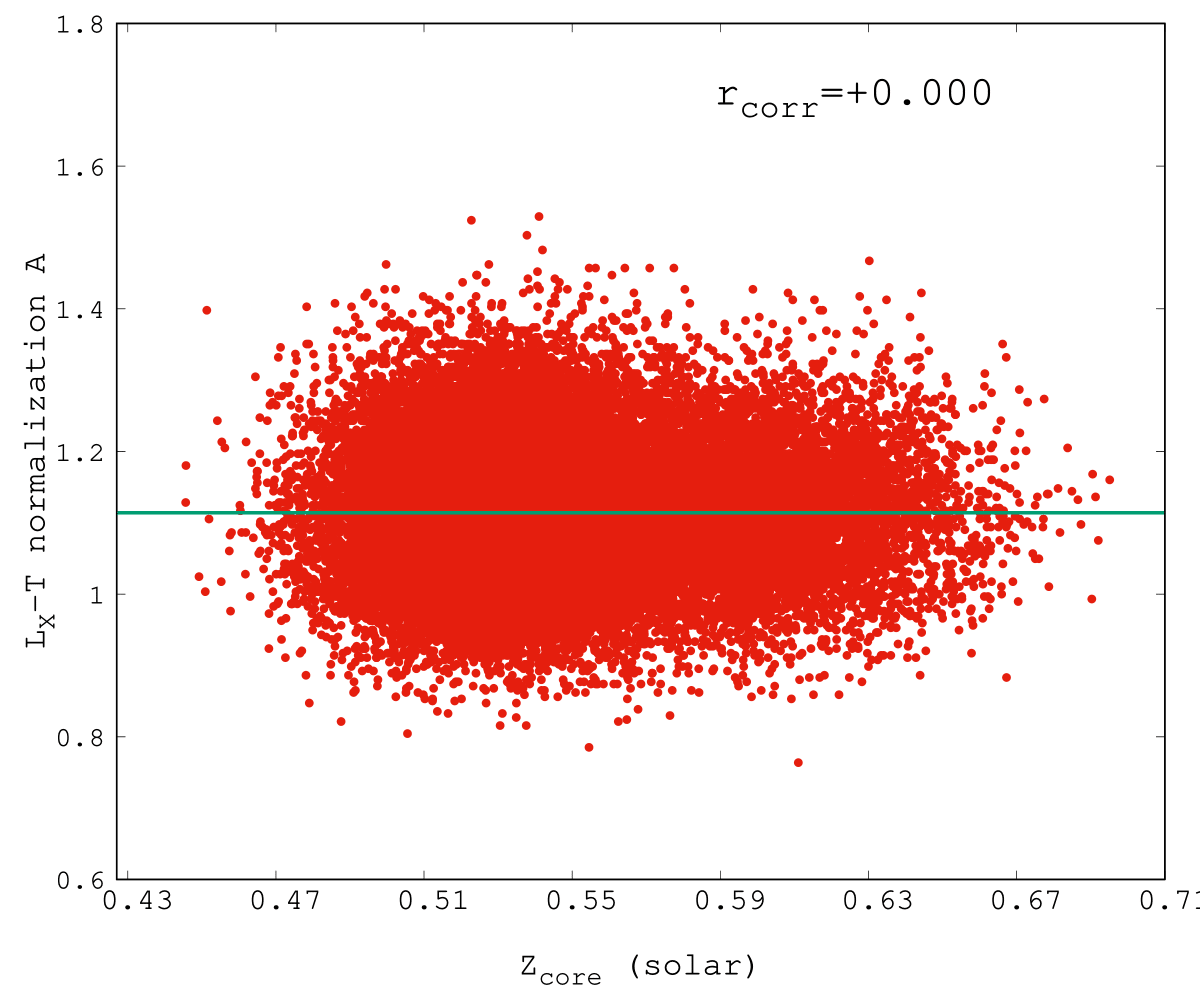}
               \includegraphics[width=0.33\textwidth, height=4cm]{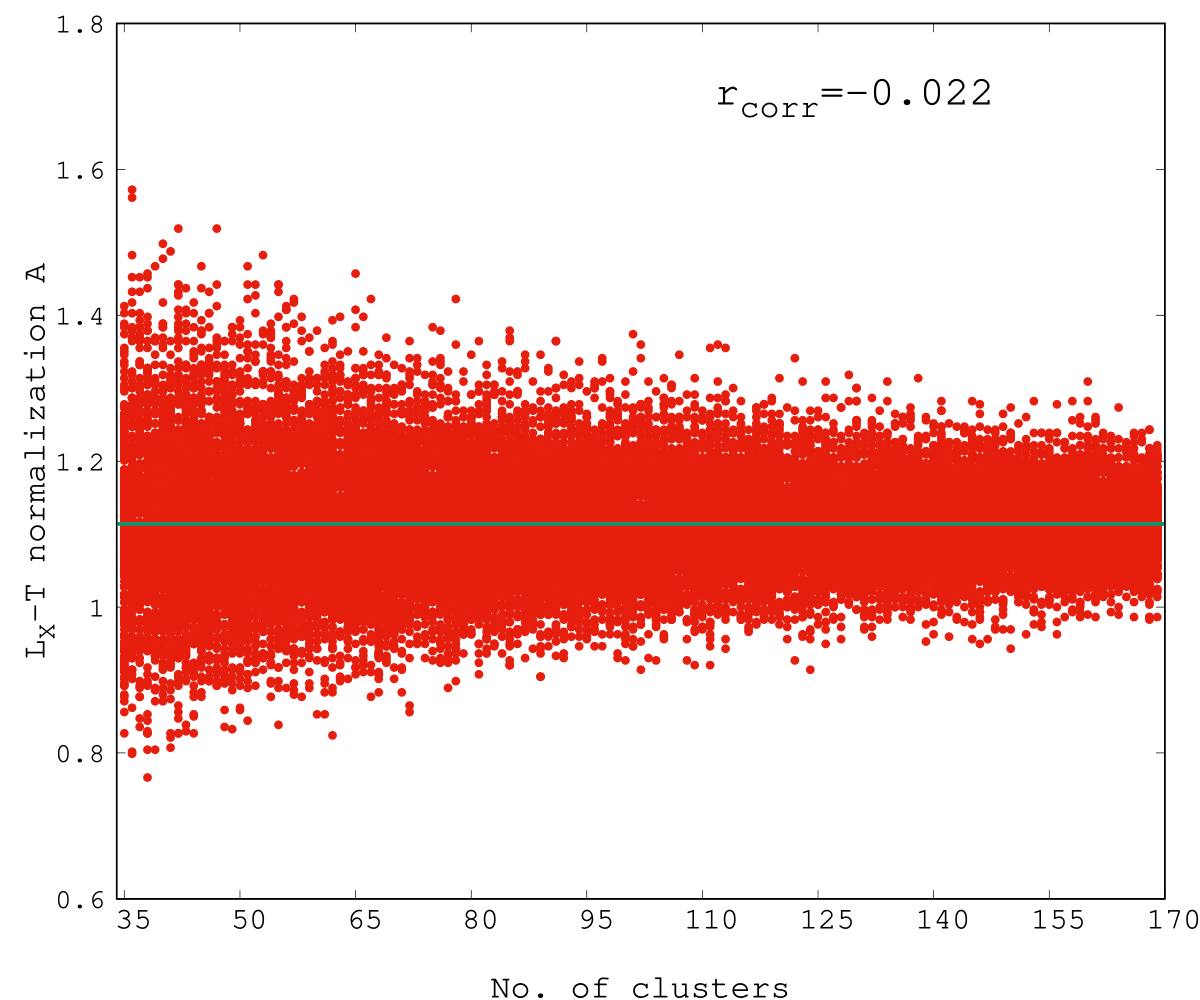}
               \includegraphics[width=0.33\textwidth, height=4cm]{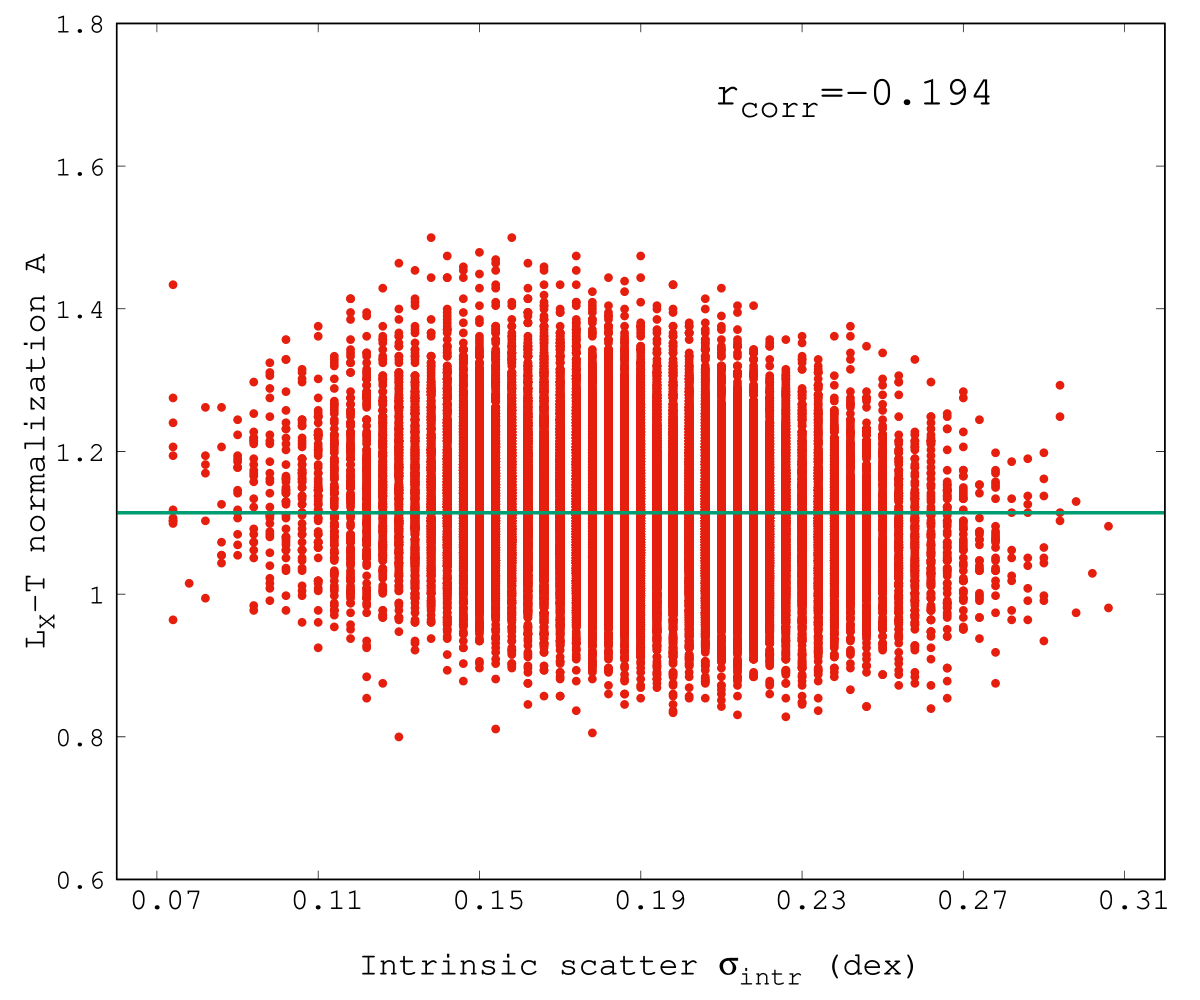}
               \includegraphics[width=0.33\textwidth, height=4cm]{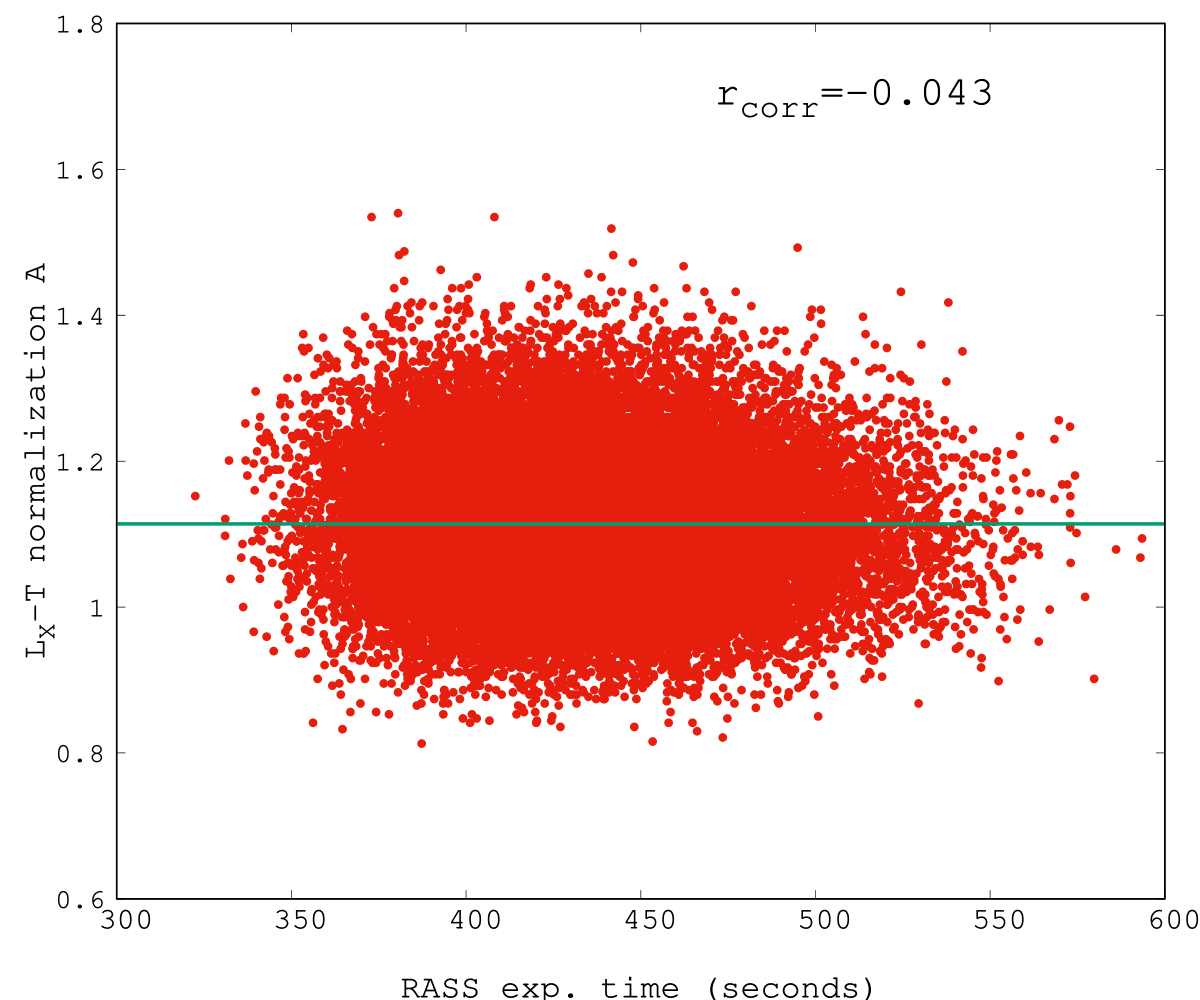}
               \caption{Correlation between the best-fit $A$ value and the average parameters of the subsamples as obtained for every one of the $10^5$ random subsamples. The correlation coefficient is also displayed in every plot. The parameters, moving from left to right and from top to bottom are: redshift, temperature, total hydrogen column density, flux, core metallicity, number of clusters, intrinsic scatter and RASS exposure time.}
        \label{all_correl}
\end{figure*}

As discussed in Sect. \ref{correl_sect} we examine the possible correlations that might exist between $A$ and the average physical parameters of every region subsample. In Fig. \ref{all_correl} the rest of the correlations are shown. None of the physical parameters seems to have a significantly enough correlation with $A$ to explain the observed anisotropies.

As also explained in Sect. \ref{correl_sect}, we look for any possible systematic behavior between the properties of our 313 clusters and their logarithmic luminosity ($\log{L_{\text{X}}}$) residuals from the overall best-fit model. To this end, we fit the behavior of the residuals against every cluster property. Thus, we can quantify the significance of the possible deviation from the case of no systematic behavior. This method cannot be applied to the bootstrap realizations since the significance depends on the number of data points, and one creates as many realizations (data points) as one wishes.

\begin{figure*}[hbtp]
               \includegraphics[width=0.33\textwidth, height=4cm]{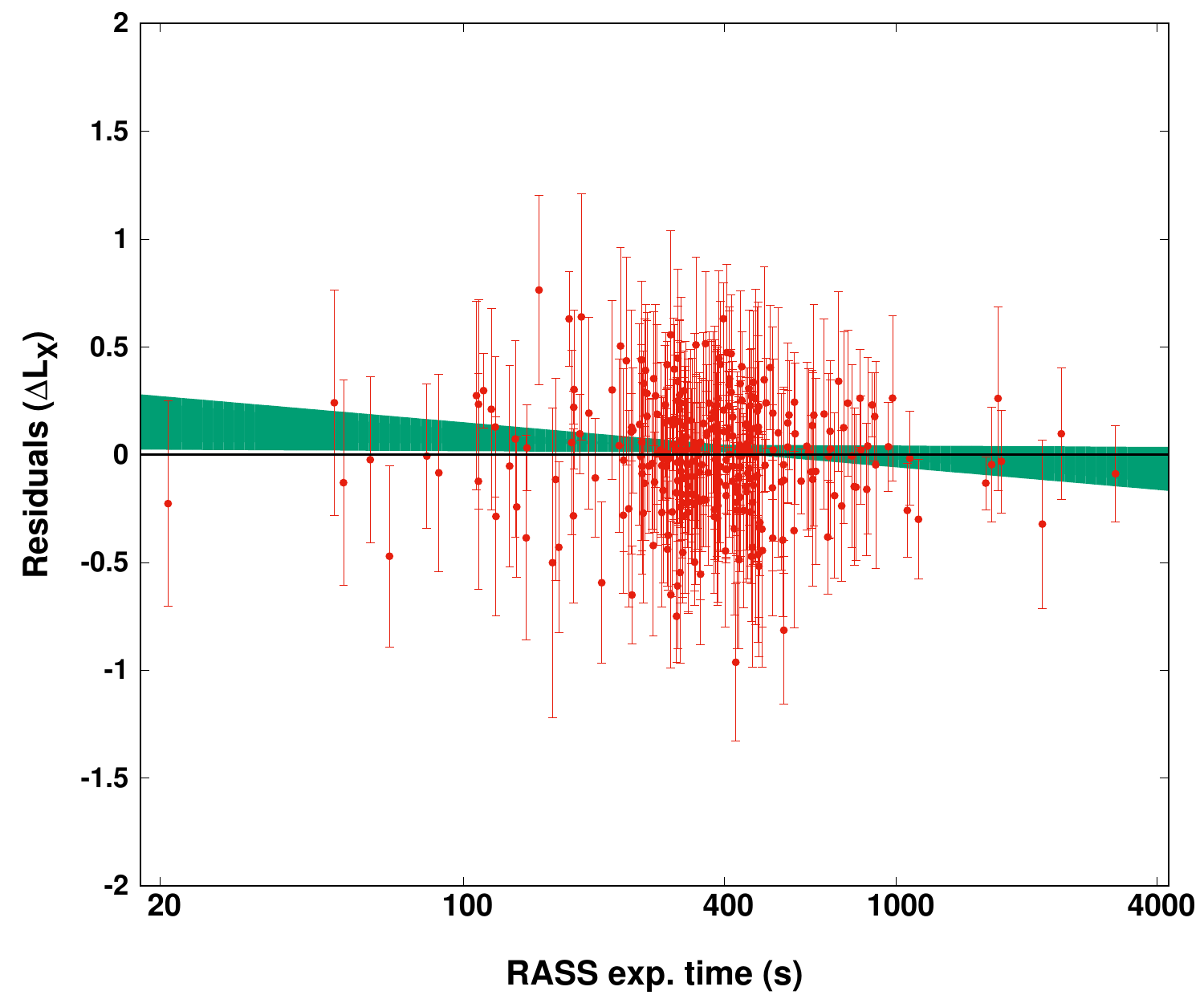}
               \includegraphics[width=0.33\textwidth, height=4cm]{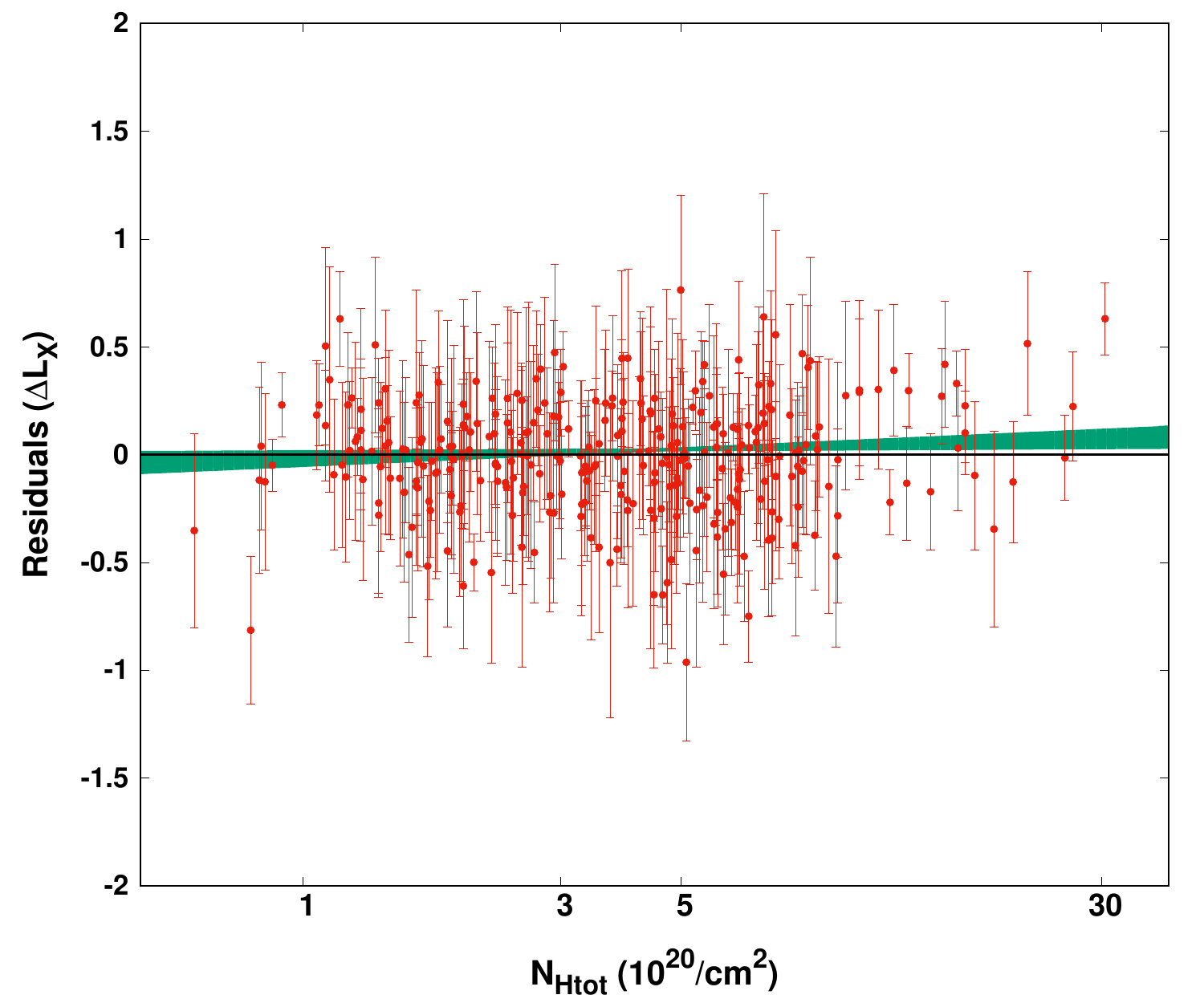}
               \includegraphics[width=0.33\textwidth, height=4cm]{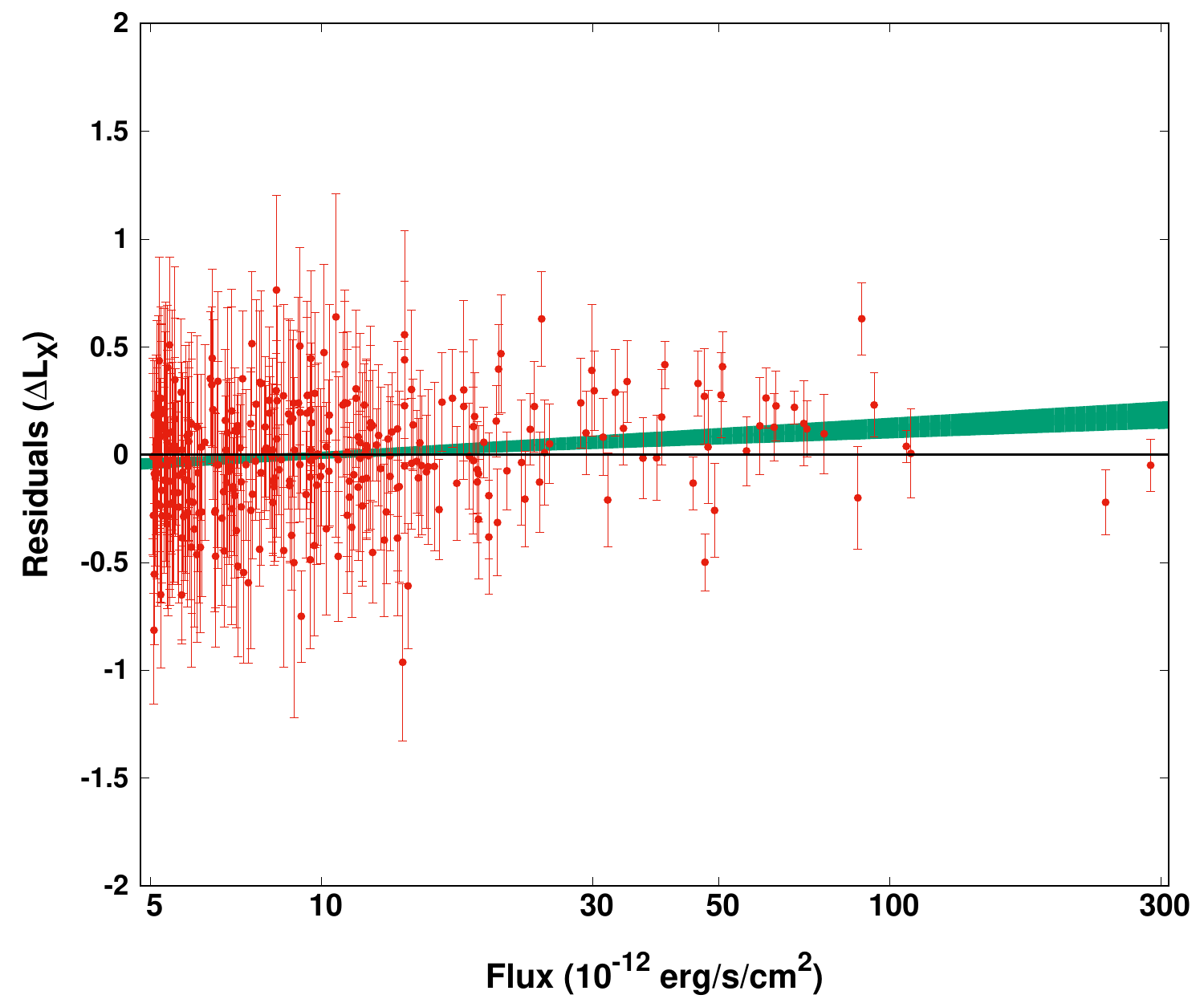}
               \includegraphics[width=0.33\textwidth, height=4cm]{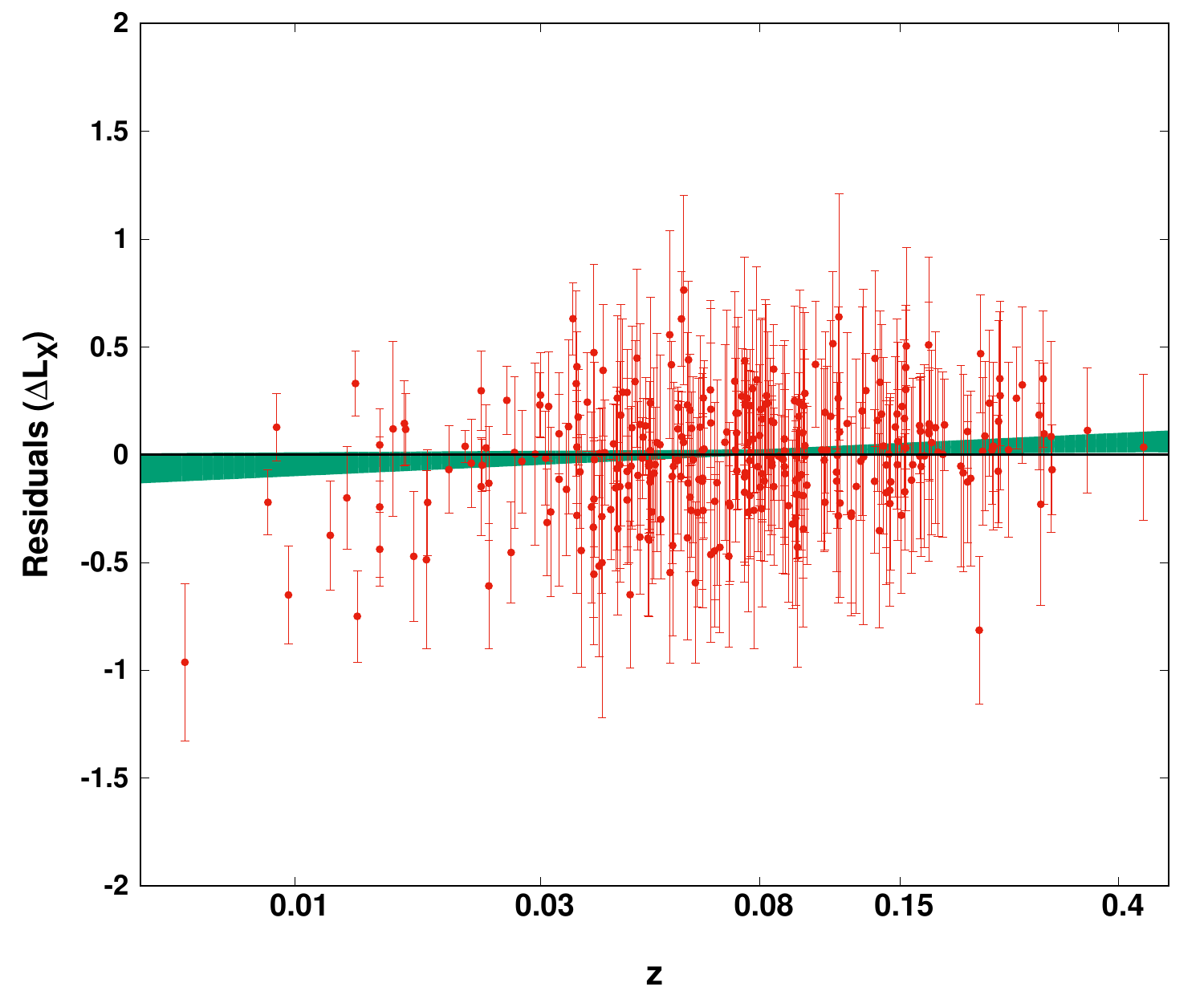}
               \includegraphics[width=0.33\textwidth, height=4cm]{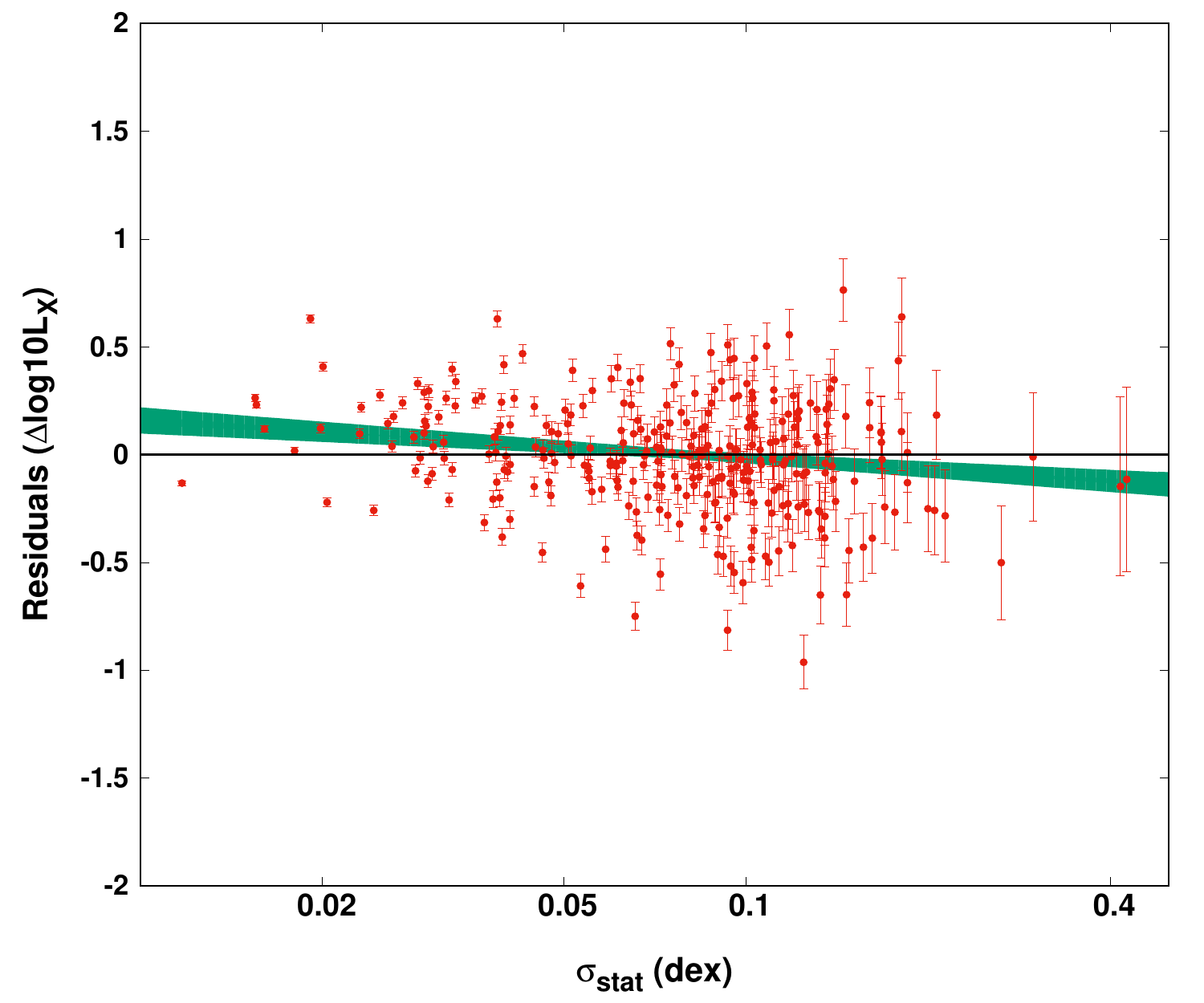}
               \includegraphics[width=0.33\textwidth, height=4cm]{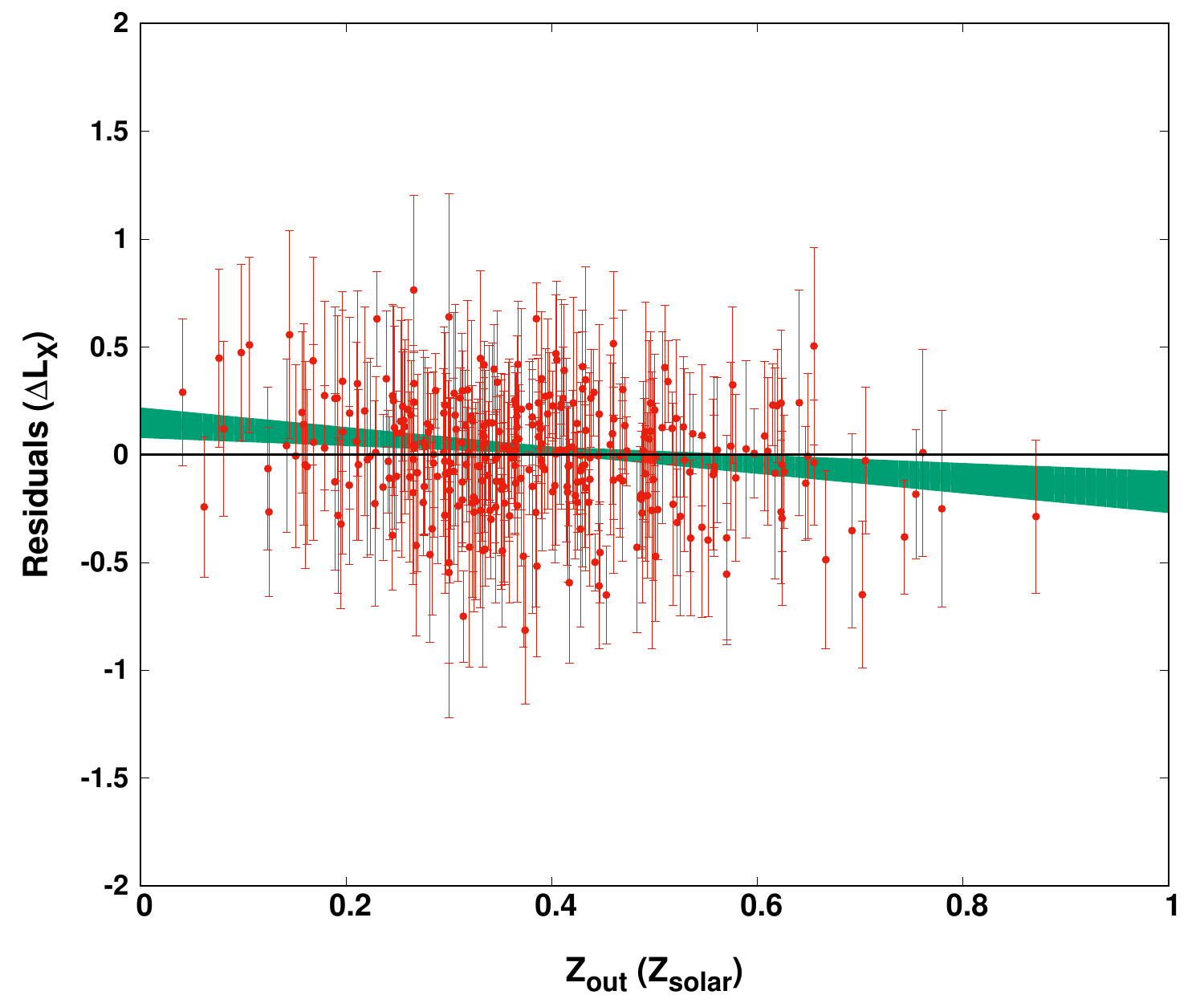}
               \caption{$\log{L_{\text{X}}}$ residuals of the 313 clusters as a function of their RASS exposure time (upper left), total hydrogen column density (upper center), flux (upper right), redshift (bottom left), statistical uncertainty (bottom center) and outer metallicity (bottom right). The best-fit relation between the residuals and these quantities is also plotted with its 1$\sigma$ uncertainty (green area). The black line represents the best-fit model for the full sample against which the residuals are calculated.}
        \label{residuals}
\end{figure*}

The $\log{L_{\text{X}}}$ residuals as a function of various cluster properties are displayed in Fig. \ref{residuals}.

No systematic behavior of the cluster $L_{\text{X}}-T$ residuals arises for different RASS exposure times. This also holds true for the absorption correction measure $N_{\text{Htot}}$. The only noticeable feature there is toward large $N_{\text{Htot}}$ values where the clusters seem to be weakly upscattered compared to the overall $L_{\text{X}}-T$ best fit model. The effect of this in our observed anisotropies has been quantified in Sect. \ref{nh_range_cons}. There we found that if one excludes these clusters, the significance of the anisotropies actually increases, since these clusters lie in relatively low-$A$ sky regions. The residuals seem to be consistent throughout the $z$ range as well, with the exception of $z\lesssim 0.025$. In Sect. \ref{low_T_z_clusters} we extensively show that excluding these clusters neither alleviate the statistical tension between the oppositely anisotropic regions nor changes their sky direction.

A mild systematic behavior can be seen for high flux clusters, being upscattered in average. This is the opposite behavior than the one expected due to selection biases. In low fluxes, one can see that the residuals are randomly distributed. This limited number of upscatered high flux clusters would only affect our anisotropy results if they were not randomly distributed in the sky (which they are). Despite of that, we excluded the 37 clusters with $f>2.6\times 10^{-11}$erg/s/cm$^2$ (after which this systematic behavior becomes clear) and repeated the sky scanning process with $\theta=75^{\circ}$.  The maximum anisotropy actually increased from $3.64\sigma$ to $3.91\sigma$ (for this cone size), and is found between the regions  $(l,b)=(272^{\circ},-26^{\circ})$ and $(l,b)=(39^{\circ},-7^{\circ})$.


Another mild systematic behavior is observed for the clusters with low statistical uncertainties ($\sigma_{\text{stat}}=\sqrt{\sigma _{\log{L}}^2+{B^2\times \sigma _{\log{T}}}^2}$, with $B=2.102$), as they tend to be intrinsically brighter than average. Based on the distance-weighing method we follow during the sky scanning process, when such a low $\sigma_{\text{stat}}$ bright cluster is close to the center of a cone, the best-fit $A$ value of that cone can be biased high to roughly match the behavior of this particular cluster. Consequently larger anisotropies might be obtained. However, this effect is limited in this work due to the inclusion of the intrinsic scatter term $\sigma_{\text{intr}}$ in our model.

To test this, we excluded the 39 clusters with $\sigma_{\text{stat}}<0.035$ dex and repeated the analysis for $\theta=75^{\circ}$. While the most anisotropic low-$A$ region was found again toward $(l,b)=(272^{\circ},-18^{\circ})$, the brightest region shifted toward $(l,b)=(75^{\circ},+22^{\circ})$. The statistical significance of their in-between anisotropy slightly decreased to $2.85\sigma$ (from $3.64\sigma$). This small change is expected since for this test we discard from our sample the clusters with the best-quality measurements, marginally increasing the uncertainties of the derived $A$. Despite of that one sees that the significance of the anisotropies remains high. If we repeat the test for the $\theta=60^{\circ}$ cones, the maximum anisotropy found is $3.76\sigma$ (from $4.73\sigma$ initially).  

Finally, the clusters with high metallicities in the $0.2-0.5$ $R_{500}$ annulus appear to be systematically fainter. This result and its effects on the apparent anisotropies have been extensively discussed in Sect. \ref{outer_metal}.

\subsection{\textit{Chandra}-only clusters}

In order to make sure that the anisotropic behavior of the $L_{\text{X}}-T$ relation is not the result of a systematic bias coming between \textit{Chandra} and \textit{XMM-Newton} clusters (even if we calibrate the temperatures properly as described in the paper), we reproduce some of the $A$ color maps using only the 237 \textit{Chandra} clusters.

\begin{figure}[hbtp]
               \includegraphics[width=0.49\textwidth, height=6cm]{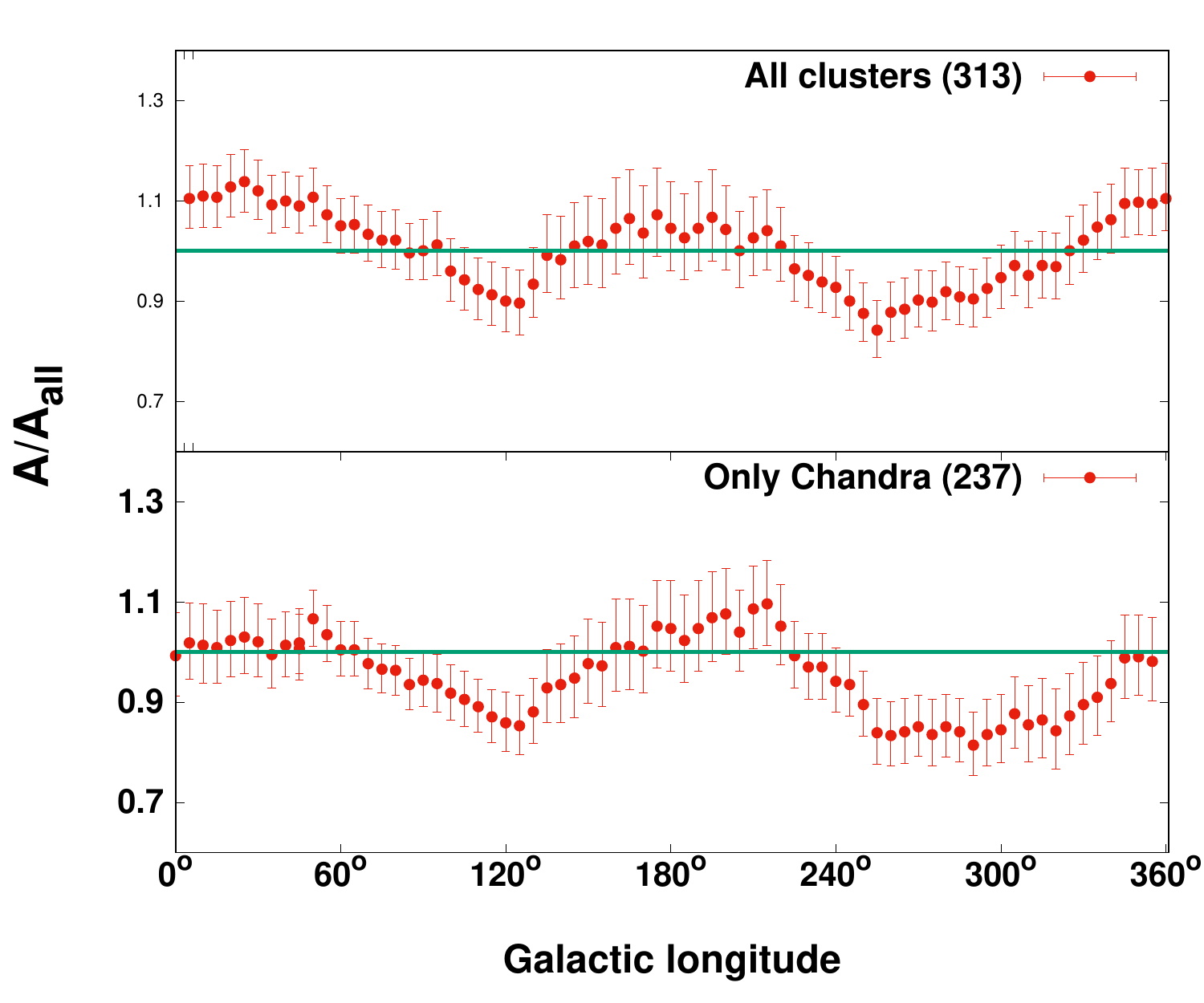}
               \caption{Normalization of the $L_{\text{X}}-T$ relation as a function of the Galactic longitude for all the 313 clusters (top) and for the 237 clusters with \textit{Chandra} temperatures (bottom). The green lines represents the best-fit values for the full samples}
        \label{chandra}
\end{figure}

\begin{figure}[hbtp]
               \includegraphics[width=0.51\textwidth, height=5cm]{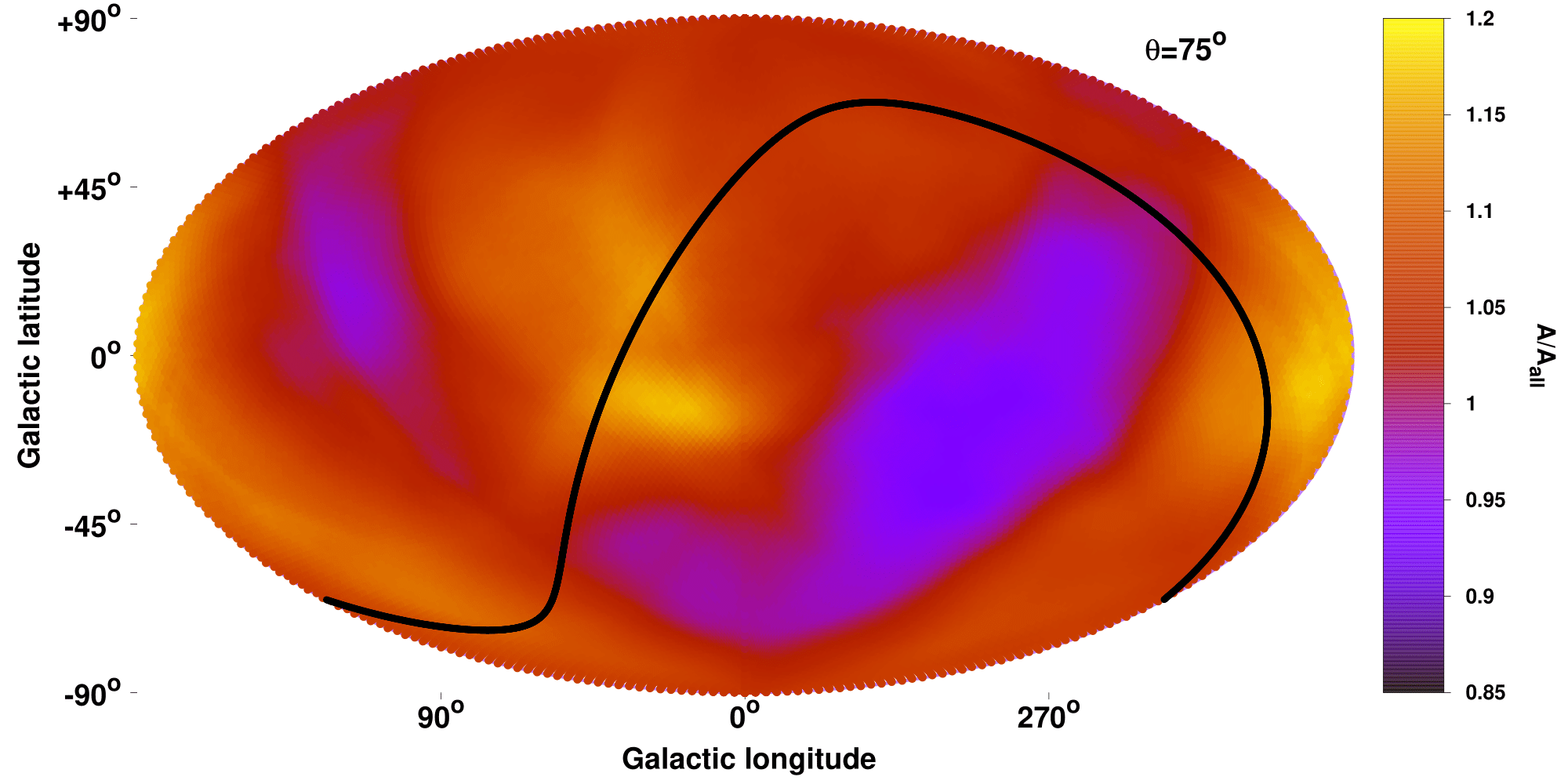}
               \includegraphics[width=0.51\textwidth, height=5cm]{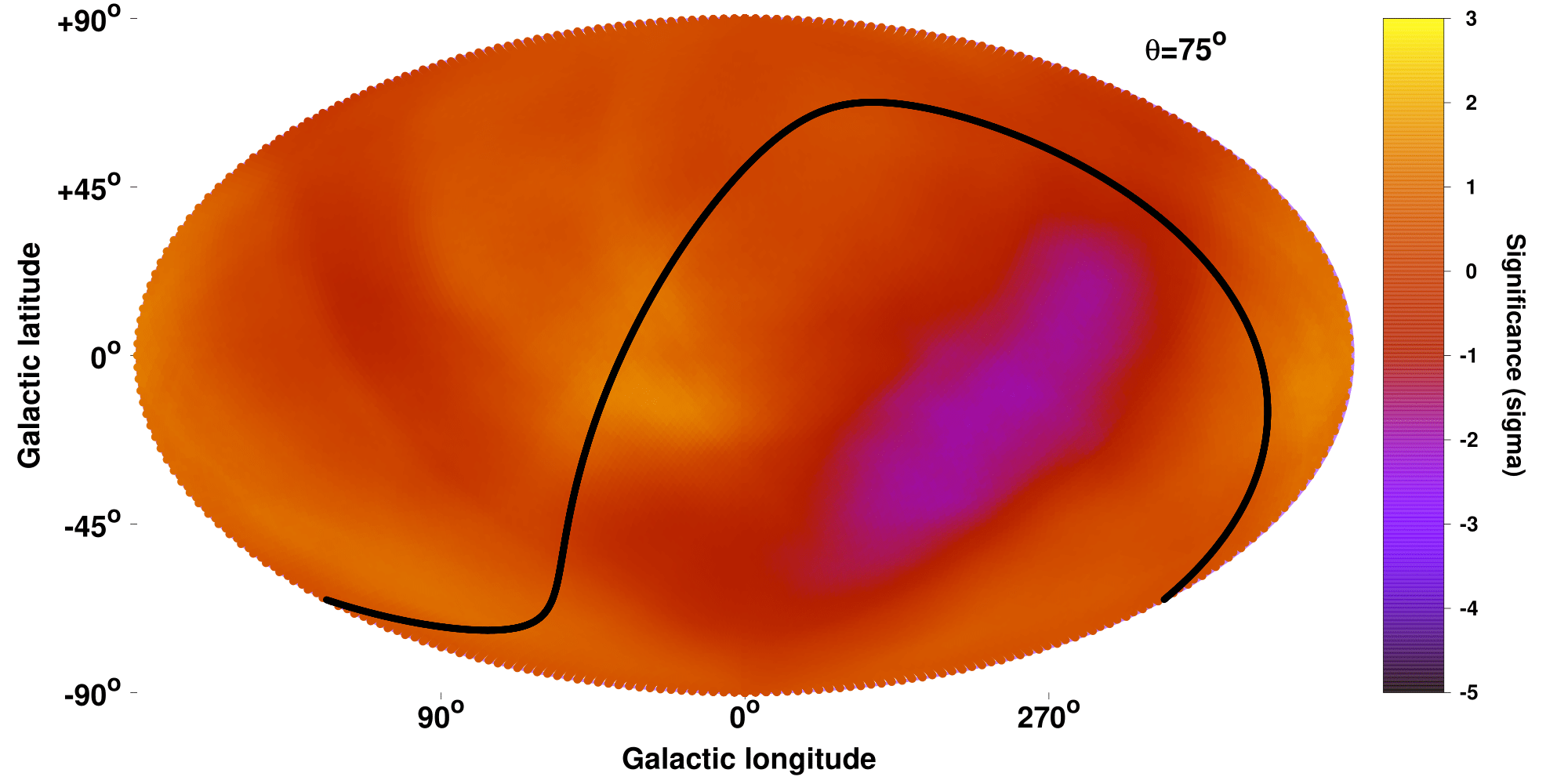}

               \caption{Same as in Fig. \ref{bulk} for the 237 clusters with \textit{Chandra} temperatures.}
        \label{chandra-maps}
\end{figure}
In Fig. \ref{chandra} and Fig. \ref{chandra-maps} is shown that both the 1D and the 2D analysis yield similar results to the full sample. In the 2D map the faint regions tend to shift to lower Galactic latitude. However, the lowest $A$ is found toward $(l,b)=(294^{\circ}, -34^{\circ})$ which is only $10^{\circ}$ away from the combined lowest result found when all three independent samples were used. The location of the brightest regions is at $(l,b)=(25^{\circ}, -11^{\circ})$. The statistically deviation between the two most extreme regions is $2.82\sigma$, somewhat decreased compared to the full sample ($3.64\sigma$) but not relieved.

\subsection{Optical redshifts only}

\begin{figure}[hbtp]
               \includegraphics[width=0.49\textwidth, height=5cm]{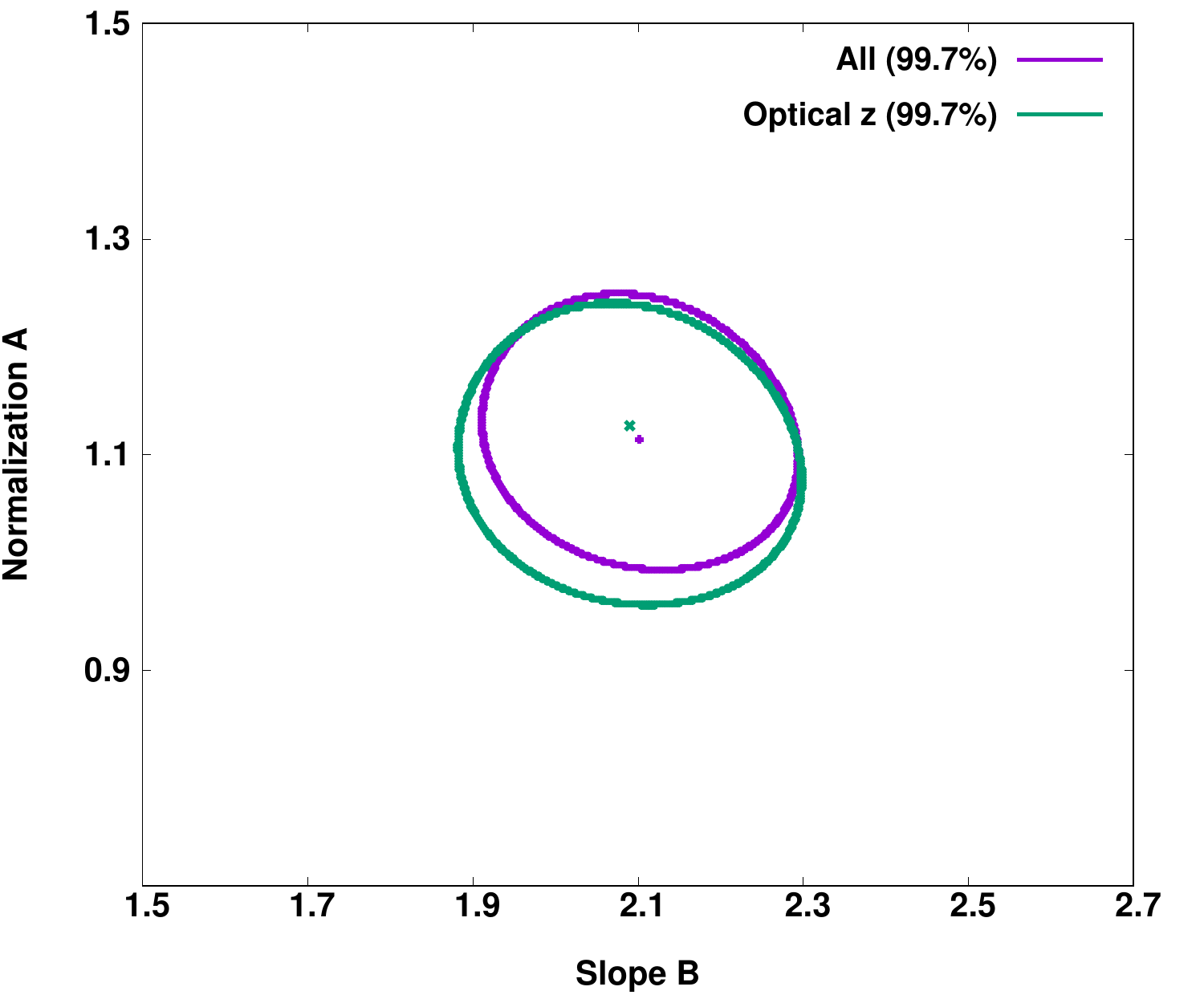}
               \caption{$3\sigma$ confidence levels (99.7\%) of the normalization and slope of the $L_{\text{X}}-T$ relation for the full sample (purple) and for the 271 clusters with optical redshifts only (green).}
        \label{optical-z}
\end{figure}

Finally, we check if the use of X-ray redshifts affect our results somehow. This is not expected to happen since there is an excellent agreement between the two types of redshifts as discussed in Sect. \ref{redshifts}. 

As shown in Fig. \ref{optical-z}, the $A$ and $B$ solution space remains identical when we considered all the 313 clusters or only the 271 clusters with optical $z$. Consequently, clusters with X-ray $z$ do not bias the results and agree well with the rest.

\subsection{Solution space excluding low $T$ systems}

\begin{figure*}[hbtp]
               \includegraphics[width=0.45\textwidth, height=6cm]{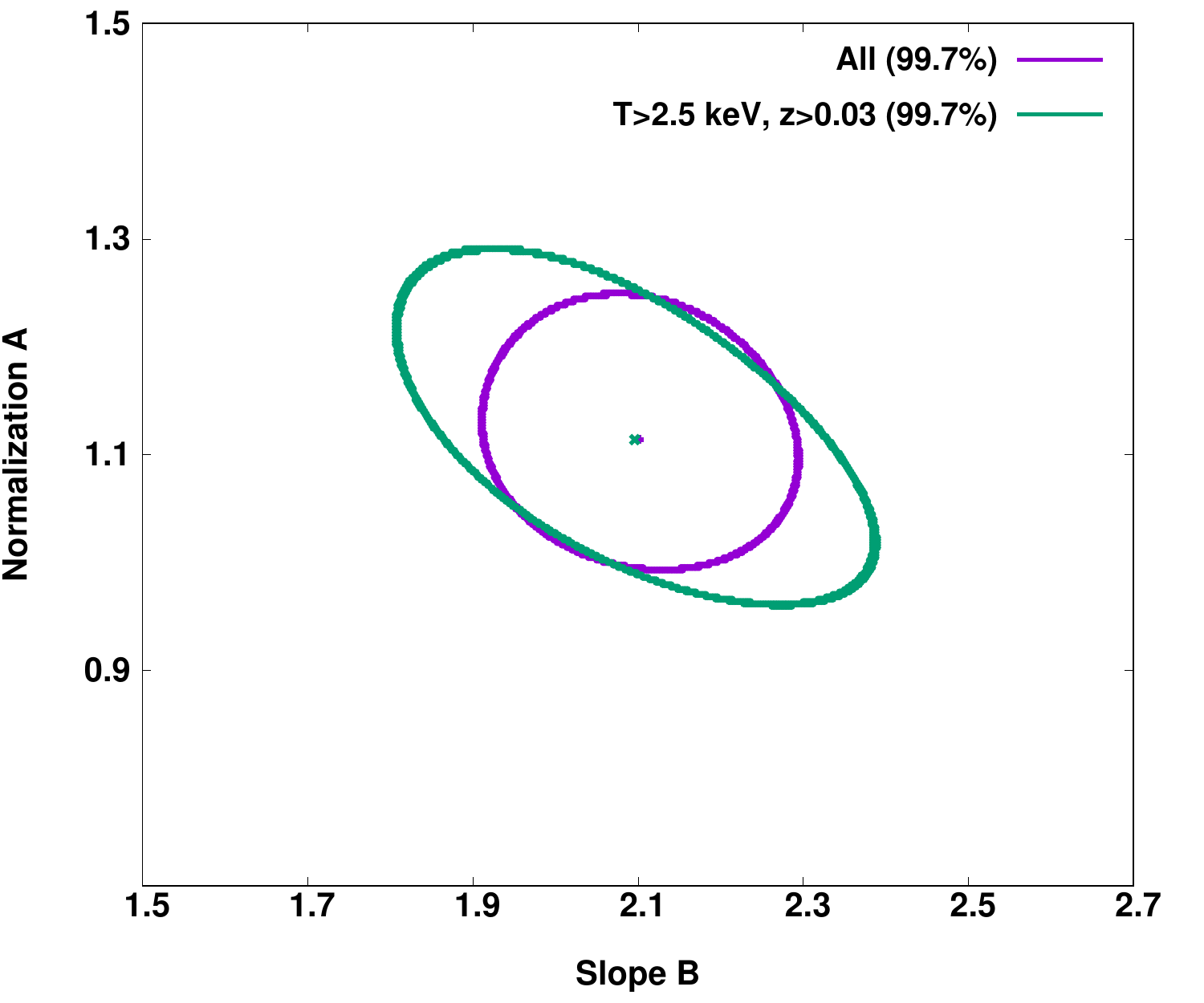}
               \includegraphics[width=0.45\textwidth, height=6cm]{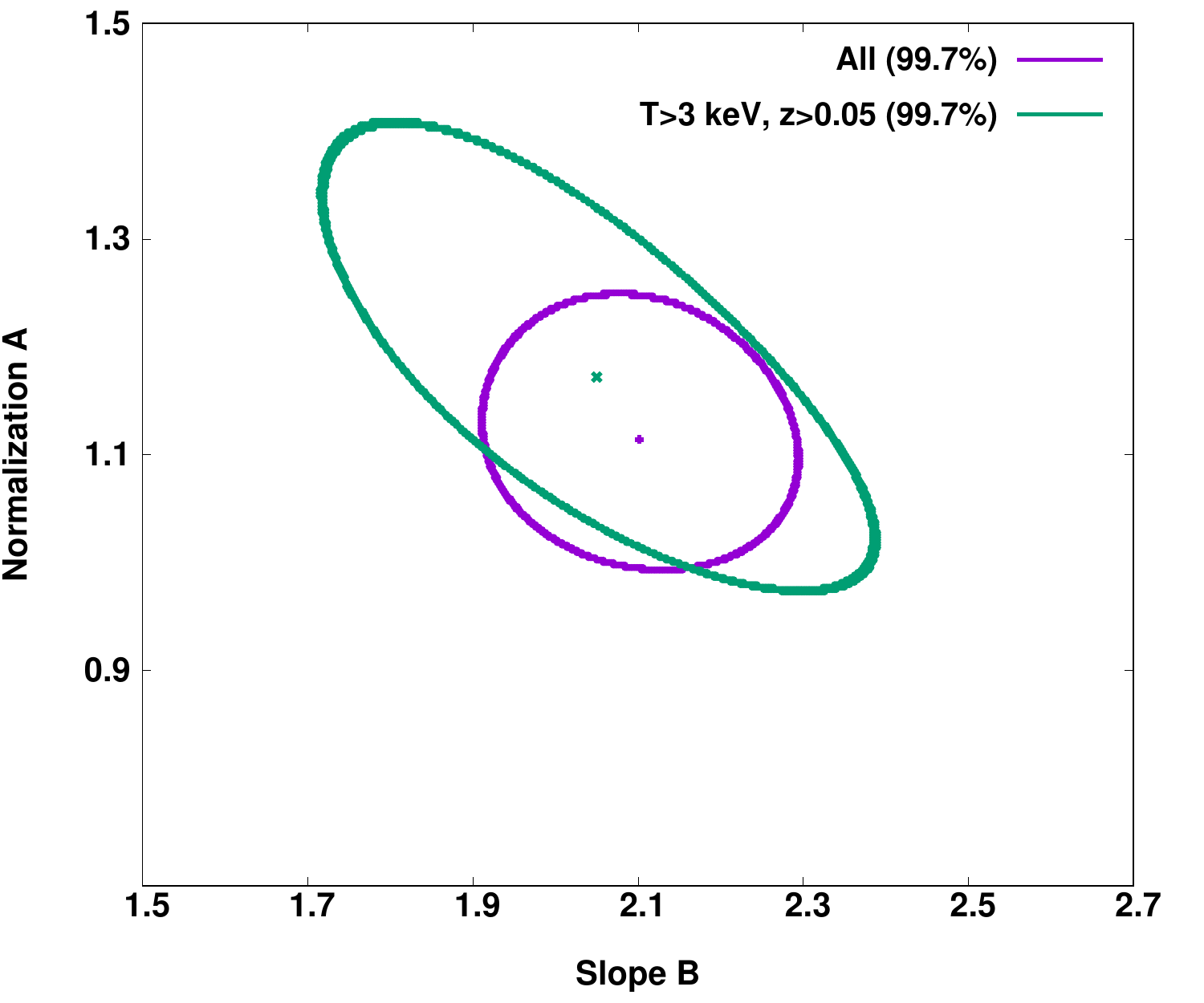}
               \caption{$3\sigma$ (99.7\%) confidence levels of the normalization and slope of the $L_{\text{X}}-T$ relation as derived using the full sample (purple) and only clusters with $T>2.5$ keV and $z>0.03$ (green, left) and $T>3$ keV and $z>0.05$ (green,right). }
        \label{bulk3}
\end{figure*}

If one excludes the low $T$ and low $z$ clusters, one can see that they do not strongly affect the overall $L_{\text{X}}-T$ solution of the sample. This is shown in Fig. \ref{bulk3} where the $3\sigma$ solution spaces are shown for the cases where we exclude all the clusters with $T<2.5$ keV and $z<0.03$ (left) and $T<3$ keV and $z<0.05$ (right), compared to the solution of the full sample.
 
While in the first case the two solutions are entirely consistent, in the second case they are consistent within $\sim 1\sigma$. Therefore, adding to all the tests done in the main sections of the paper, we can safely conclude that these systems do not affect our anisotropic findings.

\subsection{REFLEX vs NORAS $L_{\text{X}}-T$ behavior}

Our sample consists mainly of clusters from the REFLEX (185 clusters, 59\%) and NORAS (105, 34\%) clusters. A possible systematic difference between the two catalogs in their cluster population or in the flux measurements could artificially create apparent anisotropies and bias our findings. Here we should note that the two catalogs were constructed by the same team and analyzed in a similar way, so naively significant discrepancies should not be expected.

\begin{figure}[hbtp]
               \includegraphics[width=0.45\textwidth, height=6cm]{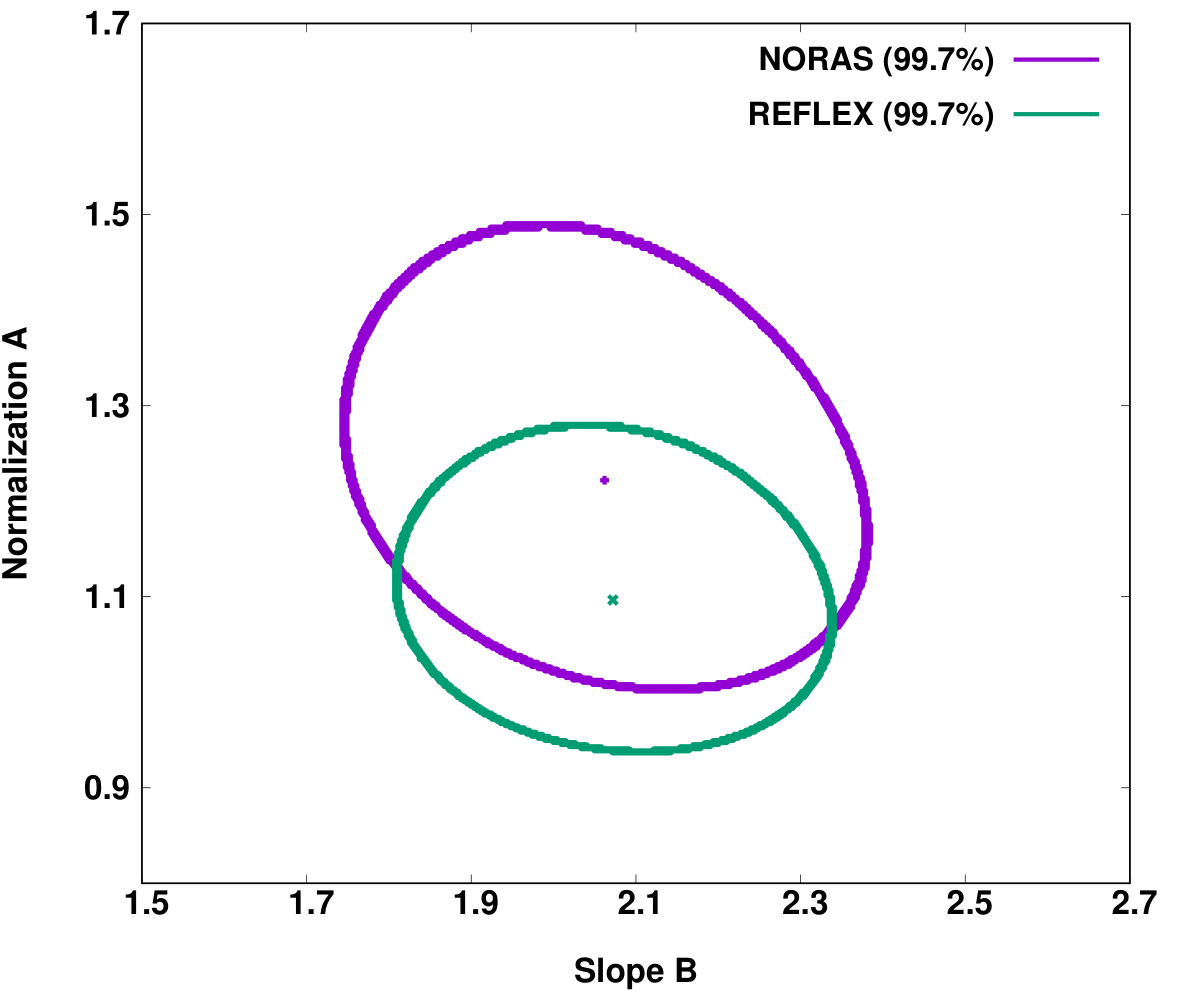}
               \caption{$3\sigma$ (99.7\%) confidence levels of the normalization and slope of the $L_{\text{X}}-T$ relation as derived using the 105 clusters coming from NORAS (purple) and the 185 clusters coming from REFLEX (green). }
        \label{ref-nor}
\end{figure}

In order to test the consistency of the $L_{\text{X}}-T$ behavior of the clusters coming from these two catalogs, we fit $A$ and $B$ for both subsamples, and compare the results. The 3$\sigma$ contour plots for the two subsamples are shown in Fig. \ref{ref-nor}.

The best-fit results are consistent for the two subsamples at a $1.4\sigma$ level. It is clear that this discrepancy is not the reason behind the observed anisotropies when the full sample is used. The NORAS clusters seem to be slightly more luminous than the REFLEX clusters, but this seems to be due to the existence of the strongest low-$A$ anisotropic region in the REFLEX part of the sky (south ecliptic hemisphere). If one excludes the clusters within $25^{\circ}$ from the lowest $A$ sky direction as found for the $\theta=75^{\circ}$ cones (Table \ref{anisot_results}), then the discrepancy between the two subsamples drops to $0.8\sigma$, which is negligible. 

\subsection{Systematic temperature differences between \textit{Chandra} and \textit{XMM-Newton}}

\begin{figure}[hbtp]
               \includegraphics[width=0.4\textwidth, height=6cm]{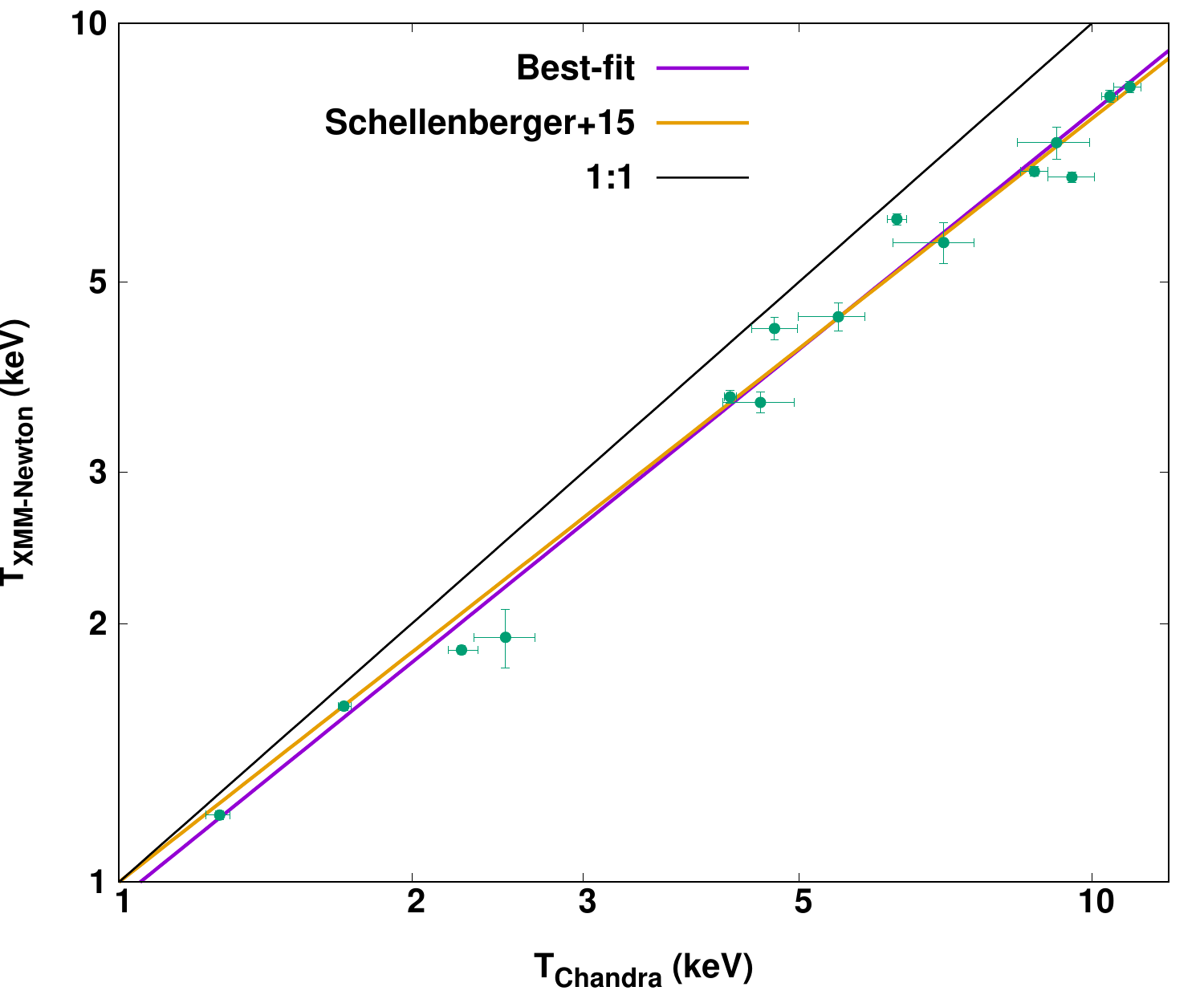}
               \caption{Comparison between the measured temperatures of 15 clusters using both \textit{Chandra} and \textit{XMM-Newton} data. The best-fit line for the relation between the two temperatures is shown (purple) together with the derived relation of S15 (orange) where more clusters were used. Also, the equality line is displayed (black).} 
        \label{temp_corr}
\end{figure}

As explained in Sect. \ref{temperat}, the \textit{Chandra} and \textit{XMM-Newton} telescopes show systematic differences in the temperature determination. In order to consistently use the measurements from both telescopes, one has to take this into account. To this end, we converted all the temperatures measured with \textit{XMM-Newton} into "Chandra" temperatures, using the relation found in S15. To verify that this relation sufficiently describes the needed conversion for our sample as well, we measured the temperature of 15 clusters with both instruments and compare the results, which are shown in Fig. \ref{temp_corr}.

As shown, the conversion relation found by S15 using 64 clusters is consistent with our results and thus used for the necessary temperature conversions. The statistical uncertainties of the S15 best-fit relation as well as the given scatter are taken into account in the final converted temperature values we use.

\subsection{Isotropic $L_{\text{X}}$ processing throughout catalogs}\label{Lx_fraction}

The $L_{\text{X}}$ values have gone through several steps of processing (RASS to REFLEX/NORAS/eBCS to MCXC to our values). If the values suffered an anisotropically biased analysis during this multiprocessing, this would propagate to our results.

\begin{figure}[hbtp]
               \includegraphics[width=0.51\textwidth, height=5cm]{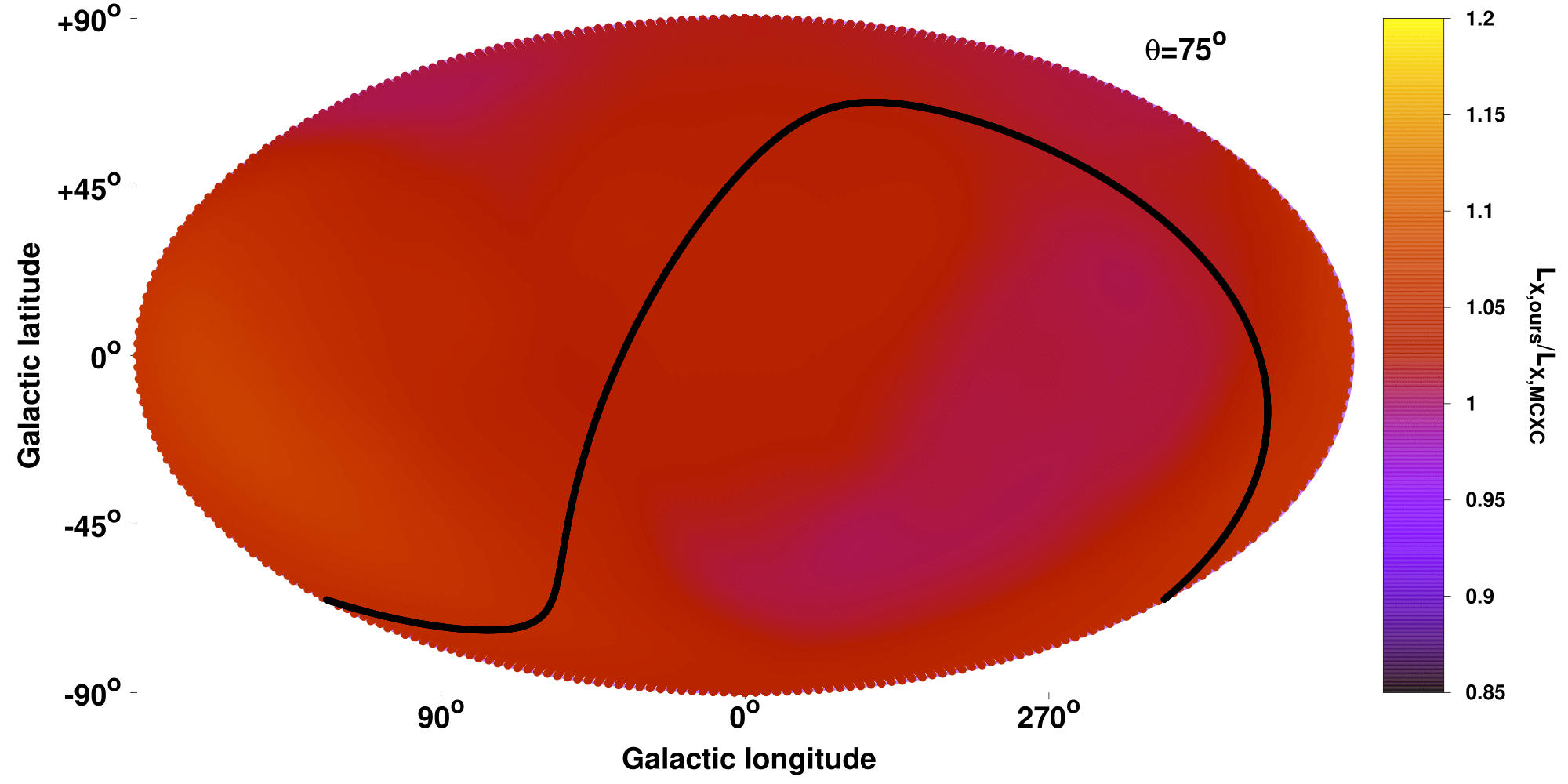}
               \includegraphics[width=0.51\textwidth, height=5cm]{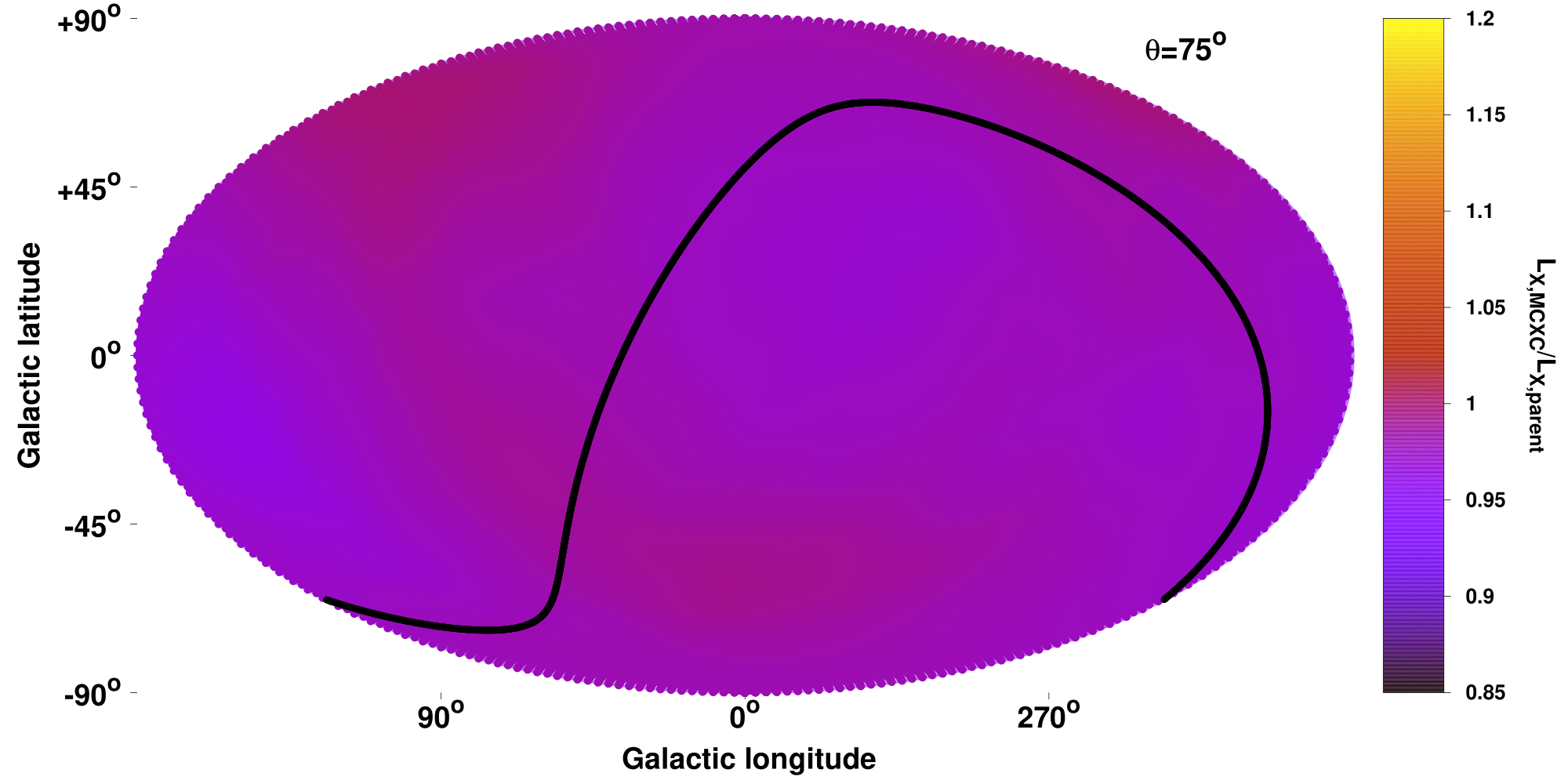}
               \caption{\textit{Top panel:} Fraction of the luminosity values $L_{\text{X,ours}}$ used in this analysis over the values coming from MCXC ($L_{\text{X,MCXC}}$) over the full extragalactic sky. All the 313 clusters of our sample were used. The same distance-weighting was used as for the main $A$ analysis. The color scale is the same as for the $A/A_{\text{all}}$ maps throughout the paper. \textit{Bottom panel:} Same as in top panel, for the fraction of $L_{\text{X,MCXC}}$ over the luminosity values coming from the parent catalogs $L_{\text{X,parent}}$.} 
        \label{Lx_frac_fig}
\end{figure}

Firstly, we need to ensure that the $L_{\text{X}}$ corrections we applied to the respective MCXC values did not introduce any artificial anisotropy. For this purpose, we check the directional behavior of the fraction between our luminosity estimated $L_{\text{X, ours}}$ and the MCXC $L_{\text{X, MCXC}}$. This is done with the same methodology as the $A$ scanning of the sky, for $\theta=75^{\circ}$ cones. Each cluster is assigned a statistical weight based on its distance from the center of each cone and the average $L_{\text{X, ours}}/L_{\text{X, MCXC}}$ is obtained. The produced map is displayed in the upper panel of Fig. \ref{Lx_frac_fig}. In order to directly compare with the observed anisotropies of the $L_{\text{X}}-T$, the same color scale is used.

One can see that the corrections we applied to the $L_{\text{X, MCXC}}$ values did not introduce any spatial anisotropies. The lowest fraction $L_{\text{X, ours}}/L_{\text{X, MCXC}}=1.005$ is found toward $(l,b)\sim (320^{\circ}, -46^{\circ})$ while the highest fraction $L_{\text{X, ours}}/L_{\text{X, MCXC}}=1.071$ is found toward $(l,b)\sim (147^{\circ}, -15^{\circ})$.

Next, we test the isotropy of the processing step from the parent catalogs to MCXC. We follow the same procedure as before, using the 313 clusters of our sample. As shown in the bottom panel of Fig. \ref{Lx_frac_fig}, the MCXC homogenization of the original $L_{\text{X, parent}}$ is greatly isotropic. The lowest fraction $L_{\text{X, MCXC}}/L_{\text{X, parent}}=0.964$ is found toward $(l,b)\sim (148^{\circ}, -17^{\circ})$ while the largest $L_{\text{X, MCXC}}/L_{\text{X, parent}}=0.999$ is found toward $(l,b)\sim (156^{\circ}, +54^{\circ})$.

Thus, no anisotropic bias was introduced going from the original catalogs to our sample. The last step to be tested is the original $L_{\text{X}}$ measurement from REFLEX, NORAS and eBCS using the RASS data. Such a procedure clearly cannot be checked unless we remeasure the cluster fluxes from the RASS data ourselves. However, there is no obvious reason why such a directional behavior would exist in the original analysis, especially since for the vast majority of clusters ($\sim 88\%$ of the sample) the analysis was conducted in a self-consistent way by the same authors (REFLEX/NORAS). 

\subsection{\textit{ROSAT} vs \textit{XMM-Newton} $L_{\text{X}}$ measurements}\label{rosat_vs_xmm}

\begin{figure}[hbtp]
               \includegraphics[width=0.4\textwidth, height=6cm]{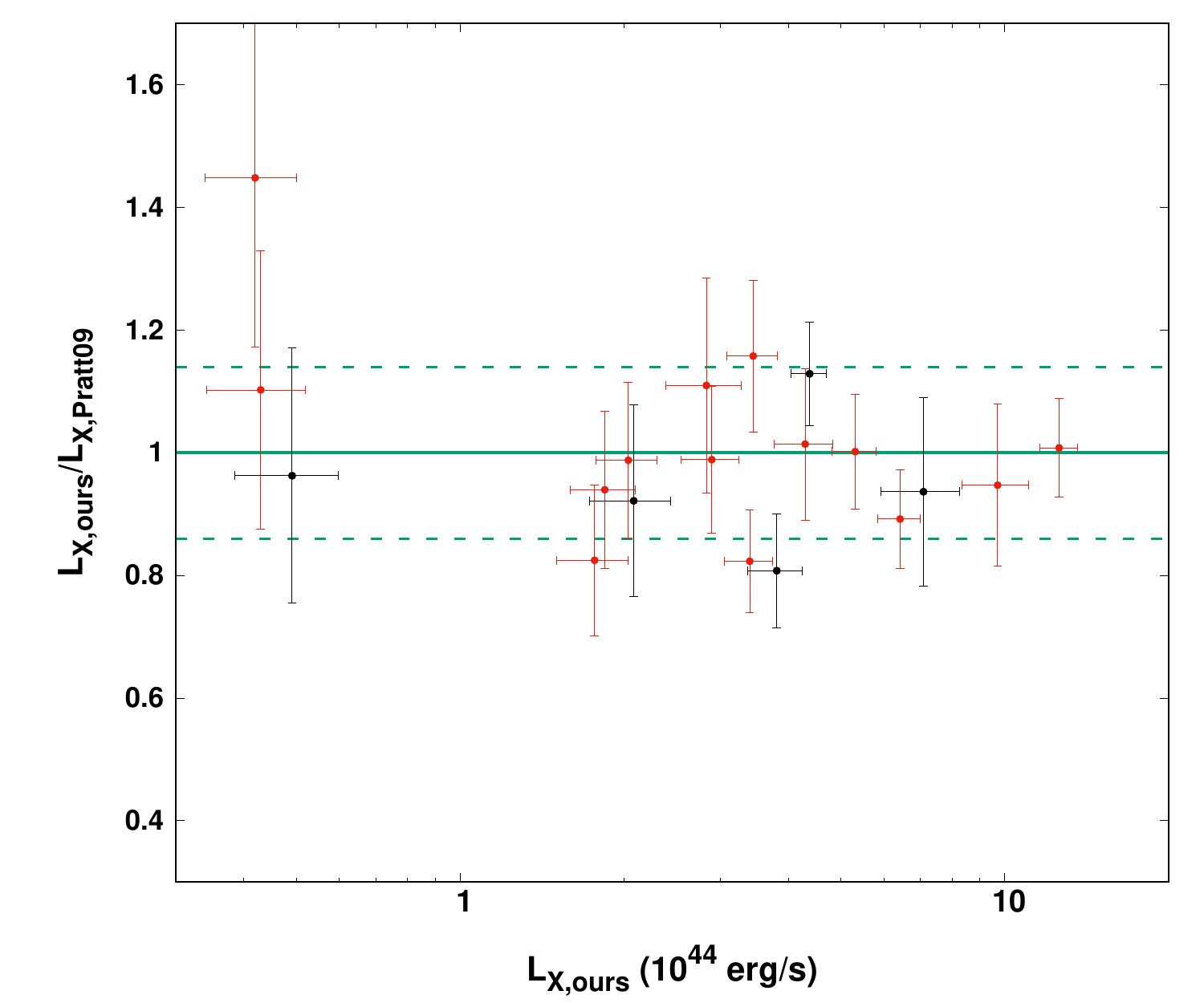}
               \caption{Fraction of the $L_{\text{X,ours}}$ used in this analysis over the ones derived in \citet{pratt} ($L_{\text{X,Pratt09}}$) for the 19 common clusters between the two samples. The clusters lying in statistically significant low $A$ regions are displayed with black. Also, the equality line is displayed (solid green) with its 1$\sigma$ scatter (dashed green).} 
        \label{pratt_comp}
\end{figure}

It has already been discussed that \textit{ROSAT} and \textit{XMM-Newton} return consistent $L_{\text{X}}$ values for the same clusters. As an additional test, we compare our $L_{\text{X,ours}}$ values with the one derived by \citet{pratt} ($L_{\text{X, Pratt09}}$, \textit{XMM-Newton} values) for the 19 common clusters between the two samples. For that, we calibrated our values using the same $z$ as for $L_{\text{X, Pratt09}}$. The comparison is portrayed in Fig. \ref{pratt_comp}.

The weighted mean for the $L_{\text{X,ours}}/L_{\text{X, Pratt09}}$ fraction is $1.001\pm0.150$ and highlights that the luminosity measurements values based on the two different telescopes and studies agree with each other. There is no cluster more than $2\sigma$ away from the 1:1 line. Finally, the five clusters that are located in low-$A$ regions in our analysis do not show a different behavior (weighted mean $L_{\text{X,ours}}/L_{\text{X, Pratt09}}=0.954\pm 0.116$ ) compared to the rest of the clusters.

\section{$R_{500}$ and temperature dependence on cosmology}

The most useful feature of the X-ray galaxy cluster $L_{\text{X}}-T$ scaling relation for cosmological isotropy studies is that the determination of the temperature is insensitive to cosmology. The only way that $T$ can be affected by cosmological parameters is through the angular diameter distance $D_A$ and the apparent size of $R_{500}$. The latter is used to select the area from which the spectrum is extracted and the X-ray cluster parameters are constrained. Below we show that the way $R_{500}$ is determined and used in our work is almost independent of changes in cosmological parameters (in particular $H_0$ which we fit), which propagates to the $T$ determination.

The apparent size of $R_{500}$ which we use, is in arcmin. Thus, it is equal to the physical size of $R_{500}$ in Mpc over $D_A$. Moreover, the $R_{500}^{\text{Mpc}}$ is derived based on the $L_{\text{X}}\sim E(z)^{7/3}M_{500}^{1.64}$ relation of \citet{arnaud2} and the fact that $M_{500}\sim R_{500}^3 \ H_0^2E(z)^2$.
With the measured redshift $z$ of the clusters remaining unchanged and the cosmological parameters $\Omega_{\text{m}}$ and $\Omega_{\Lambda}$ fixed to global values (as done in Sect. \ref{cosmology}), this can be written as a function of $H_0$ as shown in Eq. \ref{r500}.

\begin{equation}
\begin{aligned}
R_{500}^{\text{arcmin}}=\dfrac{R_{500}^{\text{Mpc}}}{D_A}\sim \left(\dfrac{M_{500}}{H_0^2}\right)^{1/3}\frac{1}{D_A}\sim \dfrac{L_{\text{X}}^{0.203}}{H_0^{0.667}}\frac{1}{D_A}.
\label{r500}
\end{aligned}
\end{equation}

The luminosity $L_{\text{X}}$ depends on $H_0$ only through the luminosity distance $D_L\sim 1/H_0$. This dependance writes as $L_{\text{X}}\sim D_L^2\sim 1/H_0^2$. Moreover, it also holds that $D_A\sim 1/H_0$. Plugging these two relation in Eq. \ref{r500} results in:

\begin{equation}
\begin{aligned}
R_{500}^{\text{arcmin}}\sim \dfrac{H_0^{-0.406}}{H_0^{0.667}} H_0 \implies R_{500}^{\text{arcmin}}\sim H_0^{-0.073}.
\label{r500-2}
\end{aligned}
\end{equation}
Consequently, a $20\%$ in $H_0$, which is similar to the $H_0$ deviations we obtain in Sect. \ref{cosmology}, it would only cause a $\sim 1\%$ change in $R_{500}^{\text{arcmin}}$ with a similar change in the measured $T$. At the same time, it would cause a $\sim45\%$ change in $L_{\text{X}}$. Additionally, due to the above, the angular radius within which $L_{\text{X},500}$ is measured does not significantly change as well. Thus, one can safely neglect the impact of $H_0$ anisotropies on the measured flux through the selection of the apparent radius. 

All these strongly demonstrate the usefulness of the $L_{\text{X}}-T$ relation for cosmic anisotropies studies.


\clearpage

\section{Table of galaxy cluster data.}

\begin{table*}[ht!]
\caption{\small{Properties of the 313 clusters used in this work. Columns: (1) Cluster name. (2) Redshift (X-ray redshifts noted with "*", redetermined redshifts based on optical spectroscopic data noted with "**"}. (3) Galactic longitude ($^{\circ}$). (4) Galactic latitude ($^{\circ}$). (5) Temperature within $0.2-0.5\ R_{500}$ (keV).  (6) X-ray luminosity within $R_{500}$ for the 0.1-2.4 keV energy range ($10^{44}$ erg/s). (7) Uncertainty of X-ray luminosity (\%). (8) X-ray flux ($10^{-12}$ erg/s/cm$^2$). (9) Neutral + molecular hydrogen column density ($10^{20}$/cm$^2$). (10) Metal abundance within $0.2-0.5\ R_{500}$ ($Z_{\odot}$). (11) Instrument used for analysis.  }
\label{tab2}
\begin{center}
\renewcommand{\arraystretch}{1.3}
\small
\begin{tabular}{ c  c  c  c c c c c c c c }
\hline \hline

Cluster  & $z$ & $l$ & $b$ & $T$  &  $L_{\text{X}}$ & $\sigma _{L_{\text{X}}}$  & $f$ & $N_{\text{Htot}}$ & $Z$ & Instrument  \\
(1) & (2) & (3) & (4) & (5) & (6) & (7) & (8) & (9) & (10) & (11) \\
\hline \hline

Coma                		&		0.023	&	57.227	& 	87.993	& $	7.410^{+0.049}_{-0.046}$  &      3.457	&  12.4	&	288.00	&	0.88	&  	$ 0.417^{+0.021}_{-0.018}  $ &	  Chandra     \\
A3526***            		&		0.009	&	302.399	&	21.561	& $	3.401^{+0.041}_{-0.043}$  &      0.466	&  3.9	&	240.00	&	12.20	&	$ 0.436^{+0.022}_{-0.013}  $ &	 Chandra      \\
A3571               		&		0.039	&	316.320	&	28.561	& $	7.362^{+0.158}_{-0.171}$  &      3.932	&  10.0	&	109.00	&	5.08	&	$ 0.597^{+0.040}_{-0.042}  $ &	 Chandra      \\
A1367               		&		0.021	&	235.084	&	73.015	& $	3.812^{+0.011}_{-0.010}$  &      1.083	&  6.0	&	107.00	&	1.89	&	$ 0.333^{+0.022}_{-0.018}  $ &	  Chandra     \\
A2199               		&		0.030	&	62.931	&	43.694	& $	4.036^{+0.052}_{-0.045}$  &      1.908	&  2.6	&	93.90	&	0.91	&	$ 0.365^{+0.015}_{-0.011}  $ &	  Chandra     \\
2A0335              		&		0.035	&	176.261	&	-35.054	& $	3.219^{+0.052}_{-0.046}$  &      3.031	&  3.1	&	89.30	&	30.50	&	$ 0.385^{+0.027}_{-0.024}  $ &	 Chandra      \\
\hline

\end{tabular}
\end{center}
\end{table*}

\clearpage

\begin{table}[h!]
\ContinuedFloat
\caption{\small{Properties of the 313 clusters used in this work (continued). }}

\begin{center}
\renewcommand{\arraystretch}{1.3}
\small
\begin{tabular}{ c  c  c  c c c c c c c c }
\hline \hline

Cluster  & $z$ & $l$ & $b$ & $T$  &  $L_{\text{X}}$ & $\sigma _{L_{\text{X}}}$  & $f$ & $N_{\text{Htot}}$ & $Z$ & Used instrument \\
(1) & (2) & (3) & (4) & (5) & (6) & (7) & (8) & (9) & (10) & (11) \\
\hline \hline

A2597               		&		0.085	&	65.341	&	-64.854	& $	4.443^{+0.096}_{-0.113}$  &      3.593	&  5.8	&	20.50	&	2.75	&	$ 0.344^{+0.028}_{-0.027}  $ &	  Chandra     \\
A2634               		&		0.031	&	103.479	&	-33.086	& $	3.706^{+0.130}_{-0.117}$  &      0.457	&  5.1	&	20.40	&	6.20	&	$ 0.522^{+0.052}_{-0.051}  $ &	 Chandra      \\
A1650               		&		0.084	&	306.676	&	61.062	& $	5.724^{+0.085}_{-0.082}$  &      3.461	&  6.1	&	20.30	&	1.43	&	$ 0.323^{+0.023}_{-0.016}  $ &	  Chandra     \\
A3376               		&		0.047	&	246.519	&	-26.084	& $	5.891^{+0.212}_{-0.234}$  &      1.027	&  4.8	&	19.70	&	5.84	&	$ 0.743^{+0.118}_{-0.112}  $ &	 Chandra      \\
Zw1215.1+0400       		&		0.077	&	282.503	&	65.186	& $	7.600^{+0.383}_{-0.326}$  &      2.768	&  5.4	&	19.70	&	1.88	&	$ 0.423^{+0.113}_{-0.109}  $ &	  Chandra     \\
A0133               		&		0.057	&	149.524	&	-84.154	& $	4.251^{+0.061}_{-0.059}$  &      1.463	&  6.7	&	19.30	&	1.65	&	$ 0.364^{+0.032}_{-0.040}  $ &	  Chandra     \\
\hline

\end{tabular}
\end{center}
\end{table}

\clearpage

\begin{table}[h!]
\ContinuedFloat
\caption{\small{Properties of the 313 clusters used in this work (continued). }}

\begin{center}
\renewcommand{\arraystretch}{1.3}
\small
\begin{tabular}{ c  c  c  c c c c c c c c }
\hline \hline

Cluster  & $z$ & $l$ & $b$ & $T$  &  $L_{\text{X}}$ & $\sigma _{L_{\text{X}}}$  & $f$ & $N_{\text{Htot}}$ & $Z$ & Used instrument \\
(1) & (2) & (3) & (4) & (5) & (6) & (7) & (8) & (9) & (10) & (11) \\
\hline \hline

MKW8                		&		0.026	&	355.494	&	54.791	& $	2.853^{+0.104}_{-0.066}$  &      0.194	&  8.7	&	12.30	&	2.68	&	$ 0.447^{+0.047}_{-0.043}  $ &	  Chandra     \\
A1750               		&		0.085	&	322.606	&	59.485	& $	4.565^{+0.362}_{-0.274}$  &      2.143	&  11.6	&	12.20	&	2.67	&	$ 0.339^{+0.229}_{-0.140}  $ &	  Chandra     \\
CAN010*             		&		0.045	&	151.856	&	-75.045	& $	2.520^{+0.260}_{-0.310}$  &      0.579	&  28.7	&	12.20	&	1.99	&	$ 0.247^{+0.175}_{-0.124}  $ &	  Chandra     \\
A1307               		&		0.083	&	243.664	&	67.753	& $	5.386^{+0.405}_{-0.297}$  &      2.109	&  8.2	&	12.10	&	3.26	&	$ 0.445^{+0.150}_{-0.135}  $ &	  Chandra     \\
A2426               		&		0.098	&	49.685	&	-49.489	& $	5.888^{+0.534}_{-0.498}$  &      2.862	&  8.7	&	12.00	&	4.80	&	$ 0.142^{+0.181}_{-0.100}  $ &	  Chandra     \\
A2734               		&		0.062	&	19.563	&	-80.985	& $	4.405^{+0.208}_{-0.177}$  &      1.076	&  9.0	&	12.00	&	1.45	&	$ 0.242^{+0.088}_{-0.086}  $ &	  Chandra     \\
\hline

\end{tabular}
\end{center}
\end{table}

\clearpage

\begin{table}[h!]
\ContinuedFloat
\caption{\small{Properties of the 313 clusters used in this work (continued). }}

\begin{center}
\renewcommand{\arraystretch}{1.3}
\small
\begin{tabular}{ c  c  c  c c c c c c c c }
\hline \hline

Cluster  & $z$ & $l$ & $b$ & $T$  &  $L_{\text{X}}$ & $\sigma _{L_{\text{X}}}$  & $f$ & $N_{\text{Htot}}$ & $Z$ & Used instrument \\
(1) & (2) & (3) & (4) & (5) & (6) & (7) & (8) & (9) & (10) & (11) \\
\hline \hline

A1800               		&		0.075	&	40.626	&	77.139	& $	4.716^{+0.412}_{-0.393}$  &      1.270	&  8.2	&	9.54	&	1.20	&	$ 0.262^{+0.241}_{-0.147}  $ &	  Chandra     \\
S0780*              		&		0.235	&	340.910	&	35.074	& $	9.430^{+1.130}_{-0.780}$  &     14.755	&  18.2	&	9.44	&	10.10	&	$ 0.179^{+0.103}_{-0.179}  $ &	 Chandra      \\
S0810               		&		0.073	&	319.494	&	-27.513	& $	3.283^{+0.360}_{-0.327}$  &      1.192	&  18.1	&	9.42	&	7.10	&	$ 0.203^{+0.041}_{-0.041}  $ &	 XMM-Newton   \\
A1285*              		&		0.095	&	275.200	&	43.877	& $	6.400^{+0.260}_{-0.310}$  &      2.007	&  17.7	&	9.39	&	3.88	&	$ 0.487^{+0.126}_{-0.106}  $ &	  Chandra     \\
S0753               		&		0.013	&	319.611	&	26.536	& $	1.796^{+0.044}_{-0.037}$  &      0.038	&  14.4	&	9.21	&	6.68	&	$ 0.314^{+0.023}_{-0.025}  $ &	 Chandra      \\
A0007               		&		0.107	&	113.289	&	-29.710	& $	4.674^{+0.389}_{-0.321}$  &      2.565	&  9.1	&	9.17	&	5.45	&	$ 0.157^{+0.136}_{-0.157}  $ &	 Chandra      \\
\hline

\end{tabular}
\end{center}
\end{table}

\clearpage

\begin{table}[h!]
\ContinuedFloat
\caption{\small{Properties of the 313 clusters used in this work (continued). }}

\begin{center}
\renewcommand{\arraystretch}{1.3}
\small
\begin{tabular}{ c  c  c  c c c c c c c c }
\hline \hline

Cluster  & $z$ & $l$ & $b$ & $T$  &  $L_{\text{X}}$ & $\sigma _{L_{\text{X}}}$  & $f$ & $N_{\text{Htot}}$ & $Z$ & Used instrument \\
(1) & (2) & (3) & (4) & (5) & (6) & (7) & (8) & (9) & (10) & (11) \\
\hline \hline

A3301               		&		0.054	&	242.414	&	-37.413	& $	4.832^{+0.462}_{-0.465}$  &      0.471	&  10.0	&	7.29	&	2.23	&	$ 0.301^{+0.078}_{-0.047}  $ &	  XMM-Newton  \\
Zw1021.0+0426       		&		0.285	&	239.399	&	47.957	& $	9.254^{+0.524}_{-0.480}$  &     17.966	&  8.2	&	7.26	&	2.70	&	$ 0.239^{+0.074}_{-0.104}  $ &	  Chandra     \\
A0990               		&		0.144	&	165.071	&	54.123	& $	8.044^{+0.761}_{-0.700}$  &      3.858	&  8.2	&	7.22	&	0.85	&	$ 0.189^{+0.152}_{-0.123}  $ &	  Chandra     \\
S0805               		&		0.015	&	332.248	&	-23.600	& $	1.017^{+0.057}_{-0.047}$  &      0.038	&  26.1	&	7.22	&	8.24	&	$ 0.062^{+0.025}_{-0.019}  $ &	Chandra       \\
NGC1132             		&		0.023	&	176.442	&	-51.078	& $	1.179^{+0.023}_{-0.022}$  &      0.097	&  16.2	&	7.19	&	6.70	&	$ 0.266^{+0.029}_{-0.033}  $ &	 Chandra      \\
A0907*              		&		0.164	&	249.368	&	33.268	& $	6.820^{+0.210}_{-0.150}$  &      5.121	&  9.3	&	7.12	&	6.67	&	$ 0.471^{+0.070}_{-0.073}  $ &	 Chandra      \\
\hline

\end{tabular}
\end{center}
\end{table}

\clearpage

\begin{table}[h!]
\ContinuedFloat
\caption{\small{Properties of the 313 clusters used in this work (continued). }}

\begin{center}
\renewcommand{\arraystretch}{1.3}
\small
\begin{tabular}{ c  c  c  c c c c c c c c }
\hline \hline

Cluster  & $z$ & $l$ & $b$ & $T$  &  $L_{\text{X}}$ & $\sigma _{L_{\text{X}}}$  & $f$ & $N_{\text{Htot}}$ & $Z$ & Used instrument \\
(1) & (2) & (3) & (4) & (5) & (6) & (7) & (8) & (9) & (10) & (11) \\
\hline \hline

A3854               		&		0.149	&	8.457	&	-56.330	& $	6.156^{+0.758}_{-0.656}$  &      3.449	&  10.7	&	5.84	&	1.25	&	$ 0.210^{+0.150}_{-0.113}  $ &	  Chandra     \\
A1553               		&		0.165	&	285.589	&	72.733	& $	7.650^{+0.817}_{-0.631}$  &      4.189	&  10.9	&	5.83	&	2.29	&	$ 0.558^{+0.237}_{-0.236}  $ &	  Chandra     \\
A0013               		&		0.094	&	72.276	&	-78.455	& $	4.719^{+0.196}_{-0.194}$  &      1.246	&  11.3	&	5.82	&	2.13	&	$ 0.332^{+0.061}_{-0.071}  $ &	  Chandra     \\
MKW11               		&		0.022	&	335.015	&	72.244	& $	1.061^{+0.023}_{-0.022}$  &      0.066	&  13.4	&	5.82	&	2.27	&	$ 0.285^{+0.076}_{-0.053}  $ &	  Chandra     \\
S0868*              		&		0.061	&	22.879	&	-29.105	& $	3.630^{+0.347}_{-0.390}$  &      0.502	&  20.6	&	5.81	&	5.85	&	$ 0.384^{+0.091}_{-0.075}  $ &	 XMM-Newton   \\
RXCJ0510.7-0801     		&		0.220	&	208.598	&	-26.006	& $	8.638^{+0.626}_{-0.492}$  &      7.876	&  10.7	&	5.80	&	8.88	&	$ 0.607^{+0.135}_{-0.158}  $ &	 Chandra      \\
\hline

\end{tabular}
\end{center}
\end{table}

\clearpage

\begin{table}[h!]
\ContinuedFloat
\caption{\small{Properties of the 313 clusters used in this work (continued). }}

\begin{center}
\renewcommand{\arraystretch}{1.3}
\small
\begin{tabular}{ c  c  c  c c c c c c c c }
\hline \hline

Cluster  & $z$ & $l$ & $b$ & $T$  &  $L_{\text{X}}$ & $\sigma _{L_{\text{X}}}$  & $f$ & $N_{\text{Htot}}$ & $Z$ & Used instrument \\
(1) & (2) & (3) & (4) & (5) & (6) & (7) & (8) & (9) & (10) & (11) \\
\hline \hline

A0761*              		&		0.091	&	240.373	&	24.548	& $	4.905^{+0.544}_{-0.509}$  &      1.030	&  14.9	&	5.20	&	5.49	&	$ 0.366^{+0.082}_{-0.092}  $ &	  XMM-Newton  \\
A1348*              		&		0.113	&	277.351	&	47.091	& $	4.631^{+0.446}_{-0.119}$  &      1.601	&  15.0	&	5.19	&	2.98	&	$ 0.428^{+0.050}_{-0.044}  $ &	  XMM-Newton  \\
A1648*              		&		0.075	&	304.978	&	36.175	& $	1.978^{+0.208}_{-0.251}$  &      0.741	&  33.6	&	5.18	&	8.68	&	$ 0.168^{+0.056}_{-0.023}  $ &	 XMM-Newton   \\
RXCJ1742.8+3900*    		&		0.042	&	64.471	&	29.408	& $	1.352^{+0.106}_{-0.102}$  &      0.221	&  17.6	&	5.17	&	3.74	&	$ 0.191^{+0.025}_{-0.012}  $ &	  XMM-Newton  \\
ZwCl3179            		&		0.143	&	228.640	&	53.054	& $	7.817^{+1.035}_{-0.902}$  &      2.881	&  9.4	&	5.14	&	4.08	&	$ 0.228^{+0.268}_{-0.157}  $ &	  Chandra     \\
Zw0959.6+3257       		&		0.050	&	193.759	&	53.241	& $	2.311^{+0.331}_{-0.269}$  &      0.300	&  11.1	&	5.13	&	1.76	&	$ 0.617^{+0.291}_{-0.270}  $ &	  Chandra     \\
\hline

\end{tabular}
\end{center}
\end{table}

\end{document}